\documentclass[a4paper,12pt,twoside,openright]{report}

\usepackage{amsmath}
\usepackage{amssymb}
\usepackage{amsfonts}
\usepackage{wrapfig}
\usepackage{bbold}
\usepackage[dvips]{graphicx}
\usepackage{indentfirst}
\usepackage{fancyhdr}
\usepackage{latexsym}
\usepackage{mathrsfs}
\usepackage{amsfonts}
\usepackage{pstricks}
\usepackage{color}
\usepackage{accents}
\usepackage[center,footnotesize,hang]{subfigure}
\usepackage{cite}
\usepackage{inputenc}
\usepackage{appendix}
\usepackage{url}    
\newcommand{\al}{\alpha}
\newcommand{\be}{\beta}
\newcommand{\ga}{\gamma}

\newcommand{\de}{\delta}
\newcommand{\De}{\Delta}
\newcommand{\ep}{\epsilon}

\newcommand{\te}{\theta}

\newcommand{\la}{\lambda}
\newcommand{\La}{\Lambda}
\newcommand{\om}{\omega}
\newcommand{\sig}{\sigma}

 \def\cK{{\cal K}}
 \def\cM{{\cal M}}
 \def\cO{{\cal O}}

\def\cW{{\cal W}}

\newcommand{\meV}{\;\mathrm{meV}}

\newcommand{\MeV}{\;\mathrm{MeV}}
\newcommand{\GeV}{\;\mathrm{GeV}}
\newcommand{\TeV}{\;\mathrm{TeV}}
\newcommand{\cm}{\;\mathrm{cm}}

\newcommand{\mbb}[1]{\mathbb{#1}}

\newcommand{\mcal}[1]{\mathcal{#1}}

\newcommand{\mrm}[1]{\mathrm{#1}}

\newcommand{\mscr}[1]{\mathscr{#1}}

\newcommand{\emphcap}[1]{\emph{\footnotesize{#1}}}

\newcommand{\hc}{\text{h.c.}}
\newcommand{\mean}[1]{\langle#1\rangle}
\newcommand{\derp}{\partial}
\newcommand{\unity}{\mathbb{1}}
\newcommand\diag{\mathrm{diag}}
\newcommand\tr{\mathrm{tr}}

\newcommand{\im}{\mbb{I}\mrm{m}}
\renewcommand{\Re}{\mbb{R}\mrm{e}}
\renewcommand{\Im}{\mbb{I}\mrm{m}}

\newcommand\dd{\displaystyle}
\newcommand\nn{\nonumber}
\newcommand{\eq}[1]{eq. (\ref{#1})}
\DeclareMathOperator{\Tr}{Tr}
\def\D{\mathrm{d}}

\newcommand{\LL}{\mscr{L}}
\newcommand{\ov}{\overline}

\def\smu{\sigma^{\mu}}

\def\smub{{\ov\sigma}^{\mu}}

\def\snu{\sigma^{\nu}}

\def\snub{{\ov\sigma}^{\nu}}
\def\smn{\sigma^{\mu\nu}}
\def\smnb{{\ov\sigma}^{\mu\nu}}

\newcommand{\msusy}{m_{SUSY}}
\newcommand{\mgut}{M_{GUT}}

\newcommand{\phit}{\varphi_T}
\newcommand{\phis}{\varphi_S}
\newcommand{\xit}{\tilde{\xi}}

\newcommand{\xipp}{\xi^{\prime\prime}}

\def\beq{\begin{equation}}
\def\eeq{\end{equation}}
\def\bea{\begin{eqnarray}}
\def\eea{\end{eqnarray}}
\def\ba{\begin{array}}
\def\ea{\end{array}}
\def\baq{\beq\ba{rcl}}
\def\eaq{\ea\eeq}
\newcommand{\bac}{\beq\begin{array}}
\newcommand{\eac}{\end{array}\eeq}
\def\bet{\begin{tabular}}
\def\eet{\end{tabular}}
\def\bes{\begin{subequations}\bea}
\def\ees{\eea\end{subequations}}

\begin{document}
\clearpage{\pagestyle{empty}\cleardoublepage}
\pagestyle{empty}
\hspace{-2.5cm}
\begin{minipage}[l]{17cm}
\begin{wrapfigure}{l}{4cm}
\vspace{-0.8cm}
   \includegraphics[width=3.5cm]{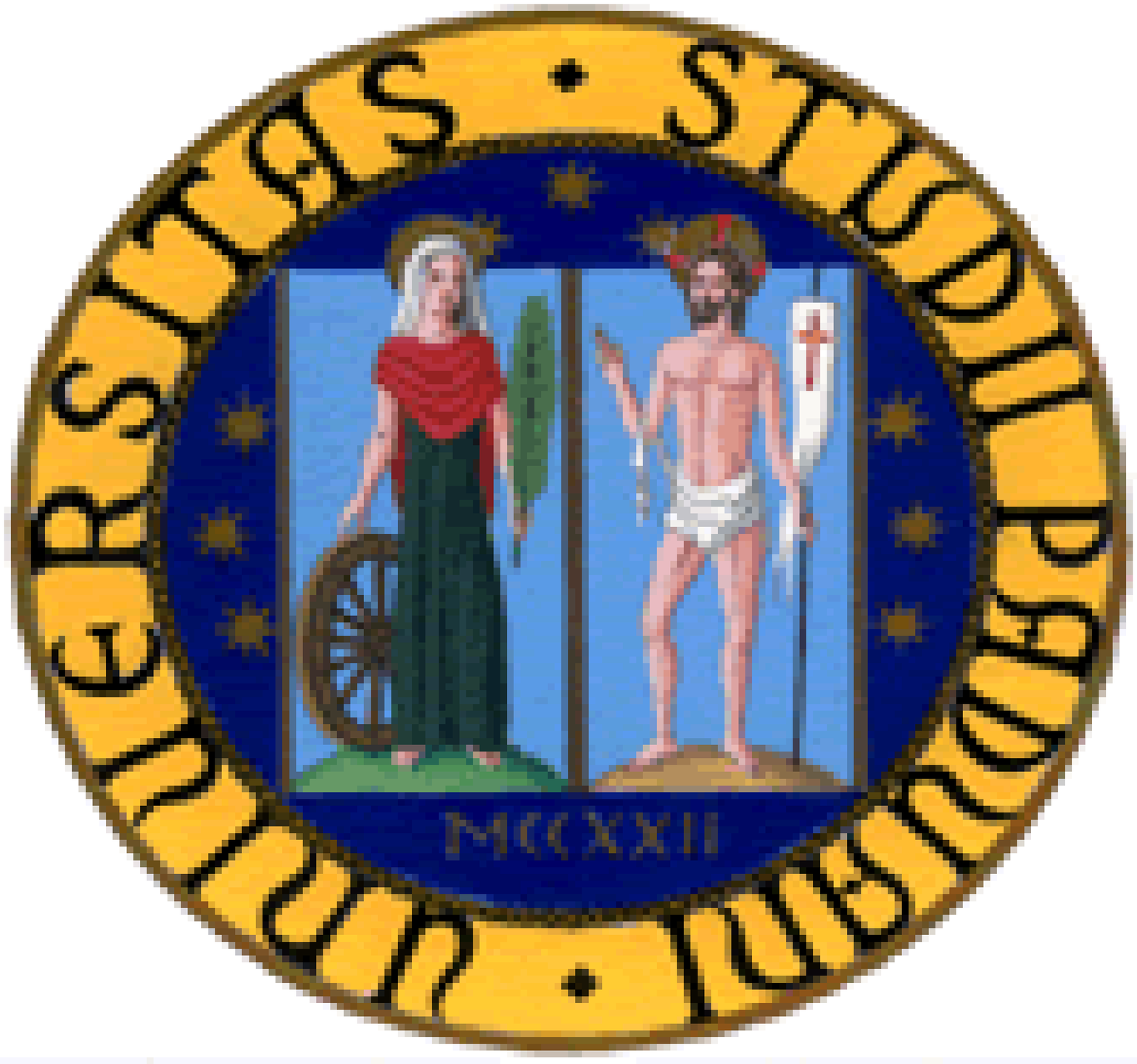}
\end{wrapfigure}
\centering
\Large\textsc{Universit\`{a} degli studi di Padova}\\
\Large\textsc{Facolt\`{a} di Scienze MM.FF.NN.}\\
\Large\textsc{Dipartimento di Fisica ``G. Galilei''}
\end{minipage}
\vspace{1.5cm}
\begin{center}
\large{SCUOLA DI DOTTORATO DI RICERCA IN FISICA}\\
\vspace{0.3cm}
\large{CICLO XXII}\\
\vspace{3cm}
\Huge\textbf{Phenomenology of}\\
\vspace{0.5cm}
\Huge\textbf{Discrete Flavour Symmetries}\\
\vspace{5cm}
\begin{flushleft}
\large{\textsf{Coordinatore:}}\;\;\large{\textsf{Ch.mo Prof. ATTILIO STELLA}}\\
\vspace{1cm}
\large{\textsf{Supervisore:}}\;\;\large{\textsf{Ch.mo Prof. FERRUCCIO FERUGLIO}}
\end{flushleft}
\vspace{0.5cm}
\begin{flushright}
   \large{\textsf{Dottorando:}}\;\;\large{\textsf{Dott. LUCA MERLO}}
\end{flushright}
\vfil
\normalsize{February 1, 2010}
\end{center}

\clearpage{\pagestyle{empty}\cleardoublepage}
\addtolength{\textheight}{1.9cm}
\addtolength{\topmargin}{-2.5cm}
\newpage
\chapter*{Abstract}
\thispagestyle{empty}

The flavour puzzle is an open problem both in the Standard Model and in its possible supersymmetric or grand unified extensions. In this thesis, we discuss possible explanations of the origin of fermion mass hierarchies and mixings by the use of non-Abelian discrete flavour symmetries. We present two realisations in which the flavour symmetry contains either the double-valued group $T'$ or the permutation group $S_4$: the spontaneous breaking of the flavour symmetry produces realistic fermion mass hierarchies, the lepton mixing matrix close to the so-called tribimaximal pattern ($\sin^2\theta_{12}=1/3$, $\sin^2\theta_{23}=1/2$ and $\theta_{13}=0$) and the quark mixing matrix comparable to the Wolfenstein parametrisation.

The exact tribimaximal scheme deviates from the experimental best-fit angles for values at most of the $1\sigma$ level. In the $T'$- and $S_4$-based  models, the symmetry breaking accounts for such discrepancies, by introducing corrections to the tribimaximal pattern of the order of $\lambda^2$, being $\lambda$ the Cabibbo angle. On the experimental side, the present measurements do not exclude $\theta_{13} \sim \lambda$ and therefore, if it is found that $\theta_{13}$ is close to its present upper bound, this could be interpreted as an indication that the agreement with the tribimaximal mixing is accidental. Then a scheme where instead the bimaximal mixing ($\sin^2\theta_{12}=1/2$, $\sin^2\theta_{23}=1/2$ and $\theta_{13}=0$) is the correct first approximation modified by terms of $\mathcal{O}(\lambda)$ could be relevant. This recalls the well-known empirical quark-lepton complementarity, for which $\theta_{12}+\lambda\sim \pi/4$. We present a flavour model based on the spontaneous breaking of the $S_4$ discrete group which naturally leads to the bimaximal mixing at the leading order and, after the introduction of the breaking terms, to $\theta_{13} \sim \lambda$  and $\theta_{12}+\mathcal{O}(\lambda)\sim \pi/4$, which we call ``weak'' complementarity relation.

Masses and mixings are evaluated at a very high energy scale and for a comparison with experimental measurements we illustrate a general analysis on the stability under the renormalisation group running to evolve these observables to low energies.

We consider also the constraints on flavour violating processes arising from introducing a flavour symmetry: in particular we concentrate on the lepton sector, analysing some lepton flavour violating decays and the discrepancy between the theoretical prediction and the experimental measurement of the anomalous magnetic moment of the muon. We develop the study both in the Standard Model scenario and in its minimal supersymmetric extension, using at first an effective operator approach and then a complete loop computation. Interesting hints for the scale of New Physics and for the forthcoming experimental results from LHC are found.

Finally we discuss the impact of an underlining flavour symmetry on leptogenesis in order to explain the baryon asymmetry of the universe.

\clearpage{\pagestyle{empty}\cleardoublepage}
\newpage
\addtolength{\topmargin}{-1cm}
\chapter*{Riassunto della Tesi}
\thispagestyle{empty}

Lo studio del sapore nella fisica particellare \`e tutt'oggi un problema aperto sia nel Modello Standard sia nelle sue estensioni supersimmetriche o grande unificate. In questa tesi, affrontiamo la questione dell'origine della gerarchia di massa nei fermioni e dei loro angoli di mescolamento, utilizzando simmetrie discrete di sapore non Abeliane. In particolare, illustriamo due modelli in cui la simmetria di sapore contiene il gruppo $T'$ o il gruppo $S_4$: la rottura spontanea della simmetria di sapore produce come effetto delle gerarchie di massa realistiche per i fermioni, la matrice di mescolamento leptonica con la cosiddetta struttura tribimassimale ($\sin^2\theta_{12}=1/3$, $\sin^2\theta_{23}=1/2$ e $\theta_{13}=0$) con piccole correzioni e la matrice di mescolamento per i quark che ben si confronta con la parametrizzazione di Wolfenstein.

La struttura tribimassimale presenta delle deviazioni al massimo ad $1\sigma$ dai valori centrali trovati sperimentalmente. Nei modelli basati sui gruppi $T'$ e $S_4$, la rottura della simmetria compensa a queste piccole deviazioni, introducendo delle correzioni alla struttura tribimassimale dell'ordine di $\lambda^2$, dove $\lambda$ rappresenta l'angolo di Cabibbo. Sperimentalmente, le misure attuali non escludono $\theta_{13}\sim\lambda$ e quindi se il valore dell'angolo di reattore risulter\`a vicino al suo attuale limite superiore, questo potrebbe essere  interpretato come un'indicazone che l'accordo con la struttura tribimassimale \`e solo accidentale. In questo caso, la struttura bimassimale ($\sin^2\theta_{12}=1/2$, $\sin^2\theta_{23}=1/2$ and $\theta_{13}=0$) potrebbe essere in prima approssimazione una migliore scelta se poi intervengono delle correzioni dell'ordine di $\lambda$. Questo meccanismo ricorda l'osservazione del tutto empirica per cui $\theta_{12}+\lambda\sim \pi/4$, che va sotto il nome di relazione di complementariet\`a. Studiamo questa alternativa in un modello basato sulla rottura spontanea del gruppo $S_4$ che presenta la struttura bimassimale in prima approssimazione e, dopo l'introduzione dei termini di rottura, $\theta_{13}\sim\lambda$ e  $\theta_{12}+\mathcal{O}(\lambda)\sim \pi/4$, che chiamiamo complementariet\`a ``debole''.

In questi modelli, le masse e gli angoli di mescolamento sono tipicamente studiati a energie molto alte e per il confronto con le misure sperimentali sviluppiamo uno studio sulla stabilit\`a di questi osservabili durante l'evoluzione a bassa scala dovuta al gruppo di rinormalizzazione.

Inotre consideriamo i limiti su processi con violazione di sapore che sorgono dall'uso di una simmetria di sapore: in particolare analizziamo alcuni decadimenti con violazione di sapore leptonico e la discrepanza tra la predizione teorica e la misura sperimentale del momento magnetico anomalo del muone. Sviluppiamo l'analisi sia nel Modello Standard sia nella sua estensione supersimmetrica minimale, usando prima un approccio di Lagrangiana efficace e poi uno studio quantistico a un loop. Troviamo  interessanti indicazioni sulla scala di energia della nuova fisica, specialmente in previsione dei prossimi risultati a LHC.

In fine discutiamo l'impatto dell'introduzione di una simmetria di sapore sulla leptogenesi, utilizzata per spiegare l'asimmetria barionica nell'universo.

\addtolength{\textheight}{-5cm}
\clearpage{\pagestyle{empty}\cleardoublepage}
\addtolength{\topmargin}{4cm}
\newpage\pagestyle{plain}
\pagenumbering{Roman}

\tableofcontents

\clearpage{\pagestyle{empty}\cleardoublepage}
\newpage
\chapter*{Introduction and Outline}
\addcontentsline{toc}{chapter}{Introduction}
\setcounter{equation}{0}
\setcounter{footnote}{3}

The Standard Model of the particle physics is not completely successful in describing nature and its behaviour and neutrinos are the most outstanding proof of this defeat: indeed the solar and atmospheric anomalies find a simple and attractive solution in the oscillations of three massive neutrinos. It is then interesting and fundamental to understand what is the theory which embeds the Standard Model and describes neutrino masses and mixings at the same time.

While global fits on neutrino oscillation experimental data have pointed out a scenario with two large angles and an approximately  vanishing one, by looking at the theoretical developments in the neutrino sector of the last few years, we cannot feel satisfied: there is a so large number of existing models, that we can interpret it as the lack of a unique and compelling theoretical picture. Furthermore, it has not been given yet an answer to several basic questions: why are neutrinos much lighter than charged fermions? which is the absolute neutrino mass scale? which is the correct neutrino spectrum? why are lepton mixing angles so different from those of the quark sector? which is the most probable range for $|U_{e3}|$? is the lepton atmospheric angle maximal? which is the nature of the active neutrinos, Dirac or Majorana? Other similar queries, such those on the number of the fermion generations, on the origin of the lepton and quark mass hierarchies and on the nature of the CP violation, naturally arise regarding the full flavour sector. The lack of a fundamental understanding of all these problems is addressed as the ``flavour puzzle''.

An interesting approach to search for a solution to the flavour problem consists in extending the gauge group of the Standard Model with an additional symmetry acting only on the fermion generations. In literature there are many attempts in this direction with a variegated choice of the symmetry: either continuous or discrete, either Abelian or non-Abelian, either global or local. Since the mixing patterns of leptons and quarks manifest large differences, it seems reasonable to introduce two different flavour symmetries, one for each sector. A common belief among many physicists, however, is that these apparent differences should be explained with a unified description and therefore a valuable task would be to use a unique symmetry, able to describe at the same time the small quark mixings and the (two) large leptonic ones. The closeness of the leptonic atmospheric angle $\theta_{23}$ to the maximal value \cite{NeutrinoData,Fogli:Indication,Maltoni:Indication} gives relevant indications on the symmetry: it is well known \cite{LV_Theorem,FeruglioSymBreaking} that a maximal $\theta_{23}$ is not achievable with an exact realistic symmetry. This forces to study models based on the breaking of the flavour symmetry and a promising choice is based on the non-Abelian discrete group $A_4$, the group of even permutations of four objects. The basic idea of the model is to get, in first approximation and in the basis of diagonal charged lepton mass matrix, the neutrino mixing matrix of the so-called tribimaximal (TB) pattern \cite{HPS} ($\sin^2\theta^{TB}_{12}=1/3$, $\sin^2\theta^{TB}_{23}=1/2$ and $\sin\theta^{TB}_{13}=0$); as a result also the observable lepton mixing matrix develops the tribimaximal structure and it represents a very good approximation of the experimental measurements; then the corrections from the next-to-leading order terms provide perturbations to the angles and in particular a deviation from zero for the reactor angle, in agreement with the recent indication of a positive value for $\theta_{13}$ \cite{Fogli:Indication}. In a series of papers \cite{AF_Extra,AF_Modular,AFL_Orbifold} on the $A_4$ group, it is shown how to get a spontaneous breaking scheme responsible for the tribimaximal mixing, by the use of a convenient assignment of the quantum numbers to the Standard Model particles and the introduction of a suitable set of scalar fields, the ``flavons'', which, getting non-zero vacuum expectation values (VEVs), are responsible for the symmetry breaking. A central aspect of the model building is the symmetry breaking chain: $A_4$ is broken down to two distinct subgroups, which correspond to the low-energy flavour symmetries of the charged leptons and of neutrinos.

When extending such $A_4$-based model to quarks to get a unified description for both the sectors, we find that $A_4$ is not suitable for quarks as it is for leptons: adopting for quarks the same representations as for leptons, the CKM mixing is the unity matrix, but the sub-leading contributions do not provide the right corrections in order to get a realistic quark mixing matrix. Furthermore it is necessary to keep separated leptons and quarks at least at the leading order, this to prevent mutual (possibly dangerous) corrections between the two sectors. A possibility to overcome this problem is to enlarge the symmetry group. We find a promising candidate in $T^\prime$ \cite{FHLM_Tp}, the double covering of $A_4$: this group has three two-dimensional representations more than $A_4$ and the idea is to adopt for leptons the same representations as in \cite{AF_Extra,AF_Modular,AFL_Orbifold} and to use the doublet ones to describe quarks. As a result we manage in keeping under control the interferences between the two sectors, preserving the results of the $A_4$-based model, and, in addition, we get interesting features in the quark sector: the top Yukawa coupling arises from a renormalisable operator, while the other Yukawas come from sub-leading order terms; the vacuum misalignment of the flavons, which justify the symmetry breaking chain, is a natural solution of the minimisation of the scalar potential; two predictions hold between quark masses and the entries of the CKM matrix,
\begin{equation}
\sqrt{\dfrac{m_d}{m_s}}=|V_{us}|\;,\qquad\qquad\sqrt{\dfrac{m_d}{m_s}}=\left|\dfrac{V_{td}}{V_{ts}}\right|\;,
\end{equation}
where the first expression is the well-known Gatto-Sartori-Tonin relation \cite{GST_Relation}.

There is an alternative successful realisation to describe simultaneously leptons and quarks: in \cite{BMM_S4,BMM_SS}, we study a model based on the permutation discrete group $S_4$ which contains $A_4$ as a subgroup and has the same number of elements as $T'$, but different representations. This enables the possibility to describe neutrinos with a different mass matrix, still diagonalised by the tribimaximal mixing, with respect to the $A_4$ model. This leads to a completely new neutrino phenomenology: considering only the leading order contributions, it is in principle possible, even if difficult, to distinguish among the different realisations; unfortunately, the introduction of the higher-order corrections makes the predictions overlap in all the parameter space, apart from very small areas, which will be hard to test in the near future.

All these models indicate a value for the reactor angle very close to zero. However, if the next future neutrino-appearance experiments will find a value for $\theta_{13}$ close to its present upper bound, about the Cabibbo angle $\lambda$, the tribimaximal mixing should be considered as an accidental symmetry. In this case a new leading principle would be necessary. In \cite{AFM_BimaxS4} we use the old idea of the quark-lepton complementary relation \cite{Complementarity}, $\theta_{12}+\lambda\sim \pi/4$, in order to recover a neutrino mixing in agreement with the data, but with a reactor angle close to its present upper bound. We develop a model based on the $S_4$ discrete group in which the PMNS matrix coincides with the bimaximal (BM) mixing \cite{BMmixing} ($\sin^2\theta^{BM}_{12}=1/2$, $\sin^2\theta^{BM}_{23}=1/2$ and $\sin\theta^{BM}_{13}=0$) in first approximation, in the basis of diagonal charged lepton mass matrix; since the BM value of the solar angle exceeds the $3\sigma$ error, large corrections are needed to make the model agree with the data; we naturally constrain the perturbations to get the ``weak'' complementarity relation, $\theta_{12}+\mathcal{O}(\lambda)\sim \pi/4$, and $\sin\theta_{13}\sim\lambda$ in most of the parameter space. In this model we only deal with the lepton sector and in order to include a realistic description of quarks we investigate on a Pati-Salam grand unified model \cite{ABM_PSS4} in which we recover the weak complementary relations and a value for the reactor angle close to $\lambda$. Furthermore we analyse the Higgs scalar potential, providing a natural description for the gauge symmetry breaking steps.\\

In all of the flavour models listed above, mass matrices and mixings are evaluated at a very high energy scale. On the other hand, for a comparison with the experimental results, it is important to evolve the observables to low energies through the renormalisation group running. In general, deviations from high energy values due to this running consist in minor corrections, but the future improvements of neutrino experiments could hopefully bring the precision down to these small quantities. For this reason we discuss \cite{LMP_RGE} the effects of the renormalisation group running on the lepton sector when masses and mixings are the result of an underlying flavour symmetry.\\

Once we consider the predictions of the models based on $A_4$, $T'$ and $S_4$, they all can fit the experimental data. However, comparing the phenomenological results of the various models, we cannot see a clear distinction. In order to find new ways to characterise each model, it would be highly desirable to investigate on other types of observables, not directly related to neutrino properties. In \cite{FHLM_Efficace} we use an effective operator approach to discuss the Lagrangian of the model: it is a very useful tool because it is not necessary to know the particle spectrum above the electroweak energy scale. The simplest scenario consists in the presence of two thresholds: a first very large, $M_{GUT}$, where can live right-handed neutrinos, superheavy gauge bosons, superheavy scalar fields, and where grand unified theories (GUTs) and flavour symmetries find their natural settlement; a second very low, the electroweak scale, at which particle masses and mixing angles have the measured values. While neutrino masses and mixing angles can be interpreted as a result at low-energy of a larger and more complete theory at $M_{GUT}$, it is very difficult from them to get information about the fundamental theory. We need to find some new observables which are not directly related to neutrino properties. A possibility is to introduce an intermediate energy scale, $M$, at about $1-10$ TeV: this corresponds to the presence of some kind of new physics, which we do not specify, at this scale. Other indications, which enforce this choice, come for example from the discrepancy in the anomalous magnetic moment of the muon, the presence of Dark Matter, the convergence to a unique value of the gauge coupling constants and the solution to the hierarchy problem, which all would benefit by the presence of new physics at $1-10$ TeV.

Studying the effective Lagrangian of the model, we can point out the presence of a unique five-dimensional operator, which is responsible for the neutrino masses, and many six-dimensional operators, that represent those new observables we are interesting in: electric dipole moments $d_i$, magnetic dipole moments $a_i$ and lepton flavour violating transitions such as $\mu\to e\gamma$, $\tau\to\mu\gamma$ and $\tau\to e\gamma$. A first distinctive feature of the model is to predict the branching ratios of all the previous decays equal and, as a first result considering the present MEGA bound, the $\tau$ decays are below the future expected sensitivities. Afterwards, constraining the operators with the experimental values or bounds of the corresponding observables, we get interesting bounds on the value of the scale $M$: while the discrepancy in $a_\mu$ indicates a value of about $3$ TeV, very interesting for LHC, $d_e$ and the $BR(\mu\to e\gamma)$ push it up to $10$ TeV in the best case. Other very stringent bounds come from the $4$-fermion operators which fix the lower value for $M$ at about $15$ TeV. For this reason we conclude that these values are probably above the region of interest to explain the discrepancy in $a_{\mu}$ and for LHC.

In a subsequent moment, we specify the kind of new physics that could be at the scale $M$ and we study a supersymmetric version of the effective model. The results seem to be very attractive due to a cancellation in the right-left block of the charged slepton mass matrix: the indication from the discrepancy in $a_\mu$ remains the same (in a low $\tan\be$ regime), but the bound from $BR(\mu\to e\gamma)$ is softened and the final results indicate values for $M$ at a few TeV, which let us explain the discrepancy in $a_\mu$ and a possible positive signal for $\mu\to e\gamma$ at MEG. Finally the model indicates an upper bound for $\theta_{13}$ of few degrees, which is close to the future expected sensitivity.

In \cite{FHLM_LFV,FHM_VEV}, we move from the effective approach to a full supersymmetric scenario. In this way we have a stricter control on the contributions of the observables discussed above and we can investigate in the supersymmetric particle spectrum. Through a detailed calculation of the slepton mass matrices in the physical basis and evaluating the branching ratios for the mentioned lepton flavour violating decays in the mass insertion approximation, we find that their behaviour, expected from the supersymmetric variant of the effective Lagrangian approach, is violated by a single, flavour independent contribution to the right-left block of the slepton mass matrix, associated to the sector necessary to maintain the correct breaking of the flavour symmetry. We also enumerate the conditions under which such a contribution is absent and the original behavior is recovered, though we could not find a dynamical explanation to justify the realisation of these conditions in our model.

Concerning the agreement of our results with the experimental measurements and bounds, assuming a supergravity framework with a common mass scale $m_{SUSY}$ for soft sfermion and Higgs masses and a common mass $m_{1/2}$ for gauginos at high energies, we numerically study the normalised branching ratios of $\ell_i\to \ell_j\gamma$ using full one-loop expressions and explore the parameter space of the model. We find that the branching ratios for $\mu\to e \gamma$, $\tau\to \mu\gamma$ and $\tau \to\ e \gamma$ are all of the same order of magnitude. Therefore, applying the present MEGA bound on $BR(\mu\to e \gamma)$, this implies that $\tau\to \mu\gamma$ and $\tau \to\ e \gamma$ have rates much smaller than the present (and near future) sensitivity. Moreover, still considering the MEGA limit, we find that small values of the symmetry breaking terms and $\tan\beta$ are favoured for $m_{SUSY}$ and $m_{1/2}$ below $1000$ GeV, i.e. in the range of a possible detection of sparticles at LHC. Furthermore, it turns out to be rather unnatural to reconcile the values of superparticle masses necessary to account for the measured deviation in the muon anomalous magnetic moment from the Standard Model value with the present bound on the branching ration of $\mu\to e\gamma$. In our model values of such deviation smaller than $100 \times 10^{-11}$ are favoured.\\

In the last few years there have been a great interest in studying the link between leptogenesis, as an explanation of the baryon asymmetry of the universe, and flavour symmetries: the See-Saw mechanisms explain the smallness of the light neutrino masses, but we need a flavour symmetry in order to predict the mixing angles; the type I is the best known mechanism and in this case the symmetry fixes the spectrum of the heavy right-handed neutrinos and the flavour structure of the Dirac and the Majorana mass matrices. While in a general context there is no relationship between low-energy parameters and $\epsilon$, the CP asymmetry from the heavy right-handed neutrino decays entering the definition of the baryon asymmetry, adding a flavour symmetry there could be a possibility to recover such a connection.

In \cite{ABMMM_Lepto} we provide a strict link among the nature of the PMNS mixing matrix and $\epsilon$: we find a model independent argument for which when the neutrino mixing matrix is mass-independent (it is independent from any mass parameter) then $\epsilon$ vanishes. This fact can be used in all the flavour models which present, in the limit of the exact symmetry, the tribimaximal pattern as well as the bimaximal and the golden-ratio schemes and some cases of the trimaximal one.

When the symmetry is broken, some corrections are introduced and in some special cases, when the number of the new parameters is sufficiently small, it is possible to express $\epsilon$ as a function of some low-energy observables.\\

The thesis is structured as follows. In chapter \ref{Sec:Overview} we first fix the notation and briefly review the main explanations for the light neutrino masses, and after we discuss about fermion masses and mixings in the physical basis, reporting their experimental determinations. Chapter \ref{Sec:FlavourPuzzle} is devoted to the flavour problem and we summarise some well-known approaches to explain the observed data, such as M(L)FV, texture zeros, mass-independent textures and flavour symmetries, focussing on discrete non-Abelian symmetry groups. In chapter \ref{Sec:FlavourModelsTBM} we deal with three different flavour models in which the lepton mixing matrix presents the tribimaximal structure in first approximaion: the first one, which accounts only with the lepton sector, is the well-known Altarelli-Feruglio model based on the $A_4$ group, whose will be recalled the main features; the other two, based on the groups $T'$ and $S_4$, represent possible alternatives to the Altarelli-Feruglio model in which also the quark sector is studied. In chapter \ref{Sec:FlavourModelsBM} we illustrate a flavour model based on the group $S_4$, in which the PMNS matrix corresponds to the bimaximal pattern in first approximations; considering the symmetry breaking corrections, we find the weak complementarity relation and a reactor angle close to the present upper bound. Chapter \ref{Sec:Running} deals with the stability of masses and mixings for leptons under the evolution from high to low energies through the renormalisation group running. In the last two chapters, \ref{Sec:FlavourViolation} and \ref{Sec:Leptogenesis}, we study the impact of an underlying flavour symmetry on flavour violating processes and on leptogenesis, respectively. In particular in chapter \ref{Sec:FlavourViolation} we focus on flavour models based on the group $A_4$, analysing their predictions for some rare decays, such as $\mu\to e\gamma$, $\tau\to e\gamma$ and $\tau\to\mu\gamma$, and the possibility to explain the discrepancy between the Standard Model prediction and the experimental measurement of the anomalous magnetic moment of the muon, through the presence of new physics at $1\div10$ TeV. Furthermore we investigate on the particle spectrum in the case of supersymmetric new physics. In chapter \ref{Sec:Leptogenesis} we present an argument for which $\epsilon$, the CP-violating parameter relevant for leptogenesis, is vanishing when the leptonic mixing matrix corresponds to a mass-independent texture in the exact symmetry phase. Finally in chapter \ref{Sec:Conclusions} we conclude and in the appendices we report details and useful tools.

\clearpage{\pagestyle{empty}\cleardoublepage}

\newpage \pagestyle{fancy}
\setlength{\headheight}{15pt}
\addtolength{\headwidth}{1cm}
\renewcommand{\headrulewidth}{0.4pt}
\textwidth 16.2 cm
\textheight 24 cm
\topmargin -0.3 cm
\renewcommand{\chaptermark}[1]{\markboth{{\sc\chaptername\ \thechapter.\ #1}}{}}
\renewcommand{\sectionmark}[1]{\markright{{\sc\thesection\ #1}}{}}
\cfoot{}
\rhead[\fancyplain{}{\bfseries\leftmark}]{\fancyplain{}{\bfseries\thepage}}
\lhead[\fancyplain{}{\bfseries\thepage}]{\fancyplain{}{\bfseries\rightmark}}
\renewcommand{\theequation}{\arabic{chapter}.\arabic{equation}}
\def\thefootnote{\fnsymbol{footnote}}
\pagenumbering{arabic}\setcounter{page}{1}

\chapter{The Standard Model and Beyond}
\label{Sec:Overview}
\setcounter{equation}{0}
\setcounter{footnote}{3}

\section{The Standard Model and the Neutrino Masses}
\label{Sec:SM}
\setcounter{footnote}{3}

The fundamental particles and their interactions are described by the Standard Model (SM), a quantum field theory in a 4-dimensional relativistic framework based on the gauge group $SU(3)_c\times SU(2)_L\times U(1)_Y$ \cite{SM}. The first term, $SU(3)_c$, refers to the quantum chromodynamics, the theory of strong interactions of coloured particles, such as gluons and quarks. The $SU(2)_L\times U(1)_Y$ term is the group of the electroweak force, which describes the behaviour of the weak gauge bosons, $W^+$, $W^-$ and $Z^0$, as well as the electromagnetic one, the photon $\gamma$, in their mutual interactions and in the presence of fermions.

The constituents of matter are leptons and quarks, which transform as spinors under the Lorenz group. For each spinor, it is possible to define a left-handed (LH) and a right-handed (RH) part, which behave differently under the Standard Model gauge group. For this reason it is usually more convenient to adopt a two-component Weyl spinor representation instead of a four-component Dirac one: they are equivalent indeed a Dirac spinor $\Psi$ is composed of a left-handed Weyl spinor $\Psi_L$ and a right-handed Weyl spinor $\Psi_R$,
\beq
\Psi=\left(
       \begin{array}{c}
         \Psi_L \\
         \Psi_R \\
       \end{array}
     \right)\;.
\eeq
If a Dirac spinor have the left- and right-handed Weyl spinors which satisfy to $\Psi_L=i\sigma^{2*}\Psi_R^*$, it is called Majorana spinor and in this case the two parts are equivalent.

It is also convenient to adopt an other notation: we consider a basis in which all the fields are left-handed and therefore $\Psi$ refers to the true left-handed component and $\Psi^c$ to the charge conjugate of the right-handed part. Using this convention, we clarify the equivalence between the two- and the four-component notations. For example $e$ ($e^c$) denotes the left-handed (right-handed) component of the electron field. In terms of the four-component spinor $\psi^T_e = (e,\,\ov{e}^c)$, the bilinears $\ov{e}\, \snub e $ and $e^c \snu \ov{e}^c$ correspond  to $\ov{\psi}_{e} \ga^\nu P_L \psi_{e}$ and $\ov{\psi}_e \ga^\nu P_R \psi_e $ (where $P_{L,R} = \frac12 (1\mp \ga^5)$) respectively. We  take $\smu \equiv (1,\vec{\sigma})$, $\smub\equiv (1,-\vec{\sigma})$, $\smn  \equiv \frac14 (\smu \snub -\snu\smub)$, $\smnb  \equiv \frac14 (\smub \snu -\snub\smu)$ and $g_{\mu\nu}= \diag(+1, -1, -1, -1)$, where $\vec{\sigma} = (\sigma^1, \sigma^2, \sigma^3)$ are the $2\times 2$ Pauli matrices:
\beq
\sigma^1=\left(
             \begin{array}{cc}
               0 & 1 \\
               1 & 0 \\
             \end{array}
           \right)\qquad
\sigma^2=\left(
             \begin{array}{cc}
               0 & -i \\
               i & 0 \\
             \end{array}
           \right)\qquad
\sigma^3=\left(
             \begin{array}{cc}
               1 & 0 \\
               0 & -1 \\
             \end{array}
           \right)\;.
\eeq
Here the four-component matrix $\gamma^\mu$ is in the chiral basis, where the 2$\times$2 blocks along the diagonal vanish, the upper-right block is given by $\smu$ and the lower-left block is equal to $\smub$.

Leptons and quarks are present in three generations or families and each of them accounts for four left-handed $SU(2)_L$-doublets, one in the lepton sector, $\ell=(\nu,\,e)$, and three in the quark sector, $q=(u,\,d)$, and seven right-handed $SU(2)_L$-singlets, the charged lepton $e^c$ and the up- and down-quarks $u^c$ and $d^c$. The symmetry charge assignments of one such family under the Standard Model gauge group are displayed in table \ref{SM:SMFermions}, along with the symmetry assignments of the Higgs boson $H=(H^+,\,H^0)$. The electric charge  is given by $Q_\text{em}=T_{3L}+Y$, where $T_{3L}$ is the weak isospin and $Y$ the hypercharge.

\begin{table}[h]
\centering
\begin{math}
\begin{array}{|c||cc|ccc|c|}
\hline
&&&&&&\\[-3mm]
 & \ell & e^c & q & u^c & d^c & H \\[3mm]
\hline
&&&&&&\\[-3mm]
SU(3)_c & \bf1 & \bf1 & \bf3 & \bf\overline3 & \bf\overline3 & \bf1 \\[3mm]
SU(2)_L & \bf2 & \bf1 & \bf2 & \bf1 & \bf1 & \bf2 \\[3mm]
U(1)_Y & -1/2 & +1 & +1/6 & -2/3 & +1/3 & +1/2 \\[3mm]
\hline
\end{array}
\end{math}
\caption{\emphcap{Charge assignments of leptons, quarks and the Higgs field under the Standard Model gauge group.}}
\label{SM:SMFermions}
\end{table}

The most general $SU(3)_c\times SU(2)_L\times U(1)_Y$ gauge invariant renormalisable Lagrangian density can be written as follows:
\beq
\LL_\text{SM}=\LL_K+\LL_Y+V\;,
\eeq
where $\LL_K$ contains the kinetic terms and the gauge interactions for fermions and bosons, while $\LL_Y$ referees to the fermion Yukawa terms and $V$ is the scalar potential. The Yukawa Lagrangian can be written as
\beq
\LL_Y = (Y_e)_{ij}\,e^c_iH^\dag\ell_j + (Y_d)_{ij}\,d^c_iH^\dag q_j + (Y_u)_{ij}\,u^c_i\widetilde{H}^\dag q_j + \hc
\label{SM:LagrangianY}
\eeq
where $\widetilde{H}\equiv i\sigma^2H^*$. The Standard Model gauge group prevents direct fermion mass terms in the Lagrangian. However, when the neutral component of the Higgs field acquires a non-vanishing vacuum expectation value (VEV), $\mean{H^0}=v/\sqrt2$ with $v\simeq246$ GeV, the electroweak symmetry is spontaneously broken,
\beq
SU(2)_L\times U(1)_Y\longrightarrow U(1)_{em}\;,
\eeq
and as a result all the fermions, apart from neutrinos, acquire non-vanishing masses:
\beq
\LL_Y = (M_u)\, u^c_i u_j  +  (M_d)_{ij}\, d^c_i d_j + (M_e)_{ij}\, e^c_i e_j \qquad\text{where}\qquad M_i=Y_i \dfrac{v}{\sqrt2}\;.
\label{SM:SMFermionsMasses}
\eeq

Observations of massive neutrinos require an extension of the Standard Model. The minimal variation consists in the introduction of a set of right-handed neutrinos, $\nu^c$, which transform under the gauge group of the Standard Model as $({\bf1},\,{\bf1},\,0)$, i.e. they do not have any interactions with the vector bosons. In this way, it is possible to construct a Yukawa term for neutrinos similar to the up-quark Yukawa:
\beq
(Y_\nu)_{ij}\,\nu^c_i\widetilde{H}^\dag\ell_j+\hc
\eeq
which in the electroweak symmetry broken phase becomes a neutrino Dirac mass term
\beq
(m_\nu)_{ij}\,\nu^c_i\nu_j\qquad\mrm{where}\qquad m_\nu=Y_{\nu}\dfrac{v}{\sqrt2}\;.
\label{SM:mD}
\eeq
According to the observations, $m_\nu\sim0.1$ eV and as a consequence it requires that $Y_\nu\sim 10^{-12}$, which does not find any natural explanation.

An alternative minimal extension of the Standard Model consists in assuming the explicit violation of the accidental global symmetry $L$, the lepton number, at a very high energy scale, $\La_L$. It is then possible to write the Weinberg operator \cite{Weinberg}, a five-dimensional non-renormalisable term suppressed by $\La_L$:
\beq
\cO_5 = \dfrac{1}{2} (Y_\nu)_{ij}\dfrac{(\ell_i\widetilde{H}^*)\,(\widetilde{H}^\dag\ell_j)}{\Lambda_L}\;.
\label{SM:O5operator}
\eeq
When the Higgs field develops the VEV, this produces a neutrino Majorana mass term
\beq
(m_\nu)_{ij} \nu_i\nu_j\qquad\mrm{where}\qquad m_\nu=Y_\nu\dfrac{v^2}{4\La_L}\;.
\eeq
Considering again an average value for the neutrino masses of the order of $0.1$ eV, it implies that $\Lambda_L$ can reach $10^{14\div15}$ GeV, for $(Y_\nu)_{ij}=\cO(1)$. Once we accept explicit lepton number violation, we gain an elegant explanation for the lightness of neutrinos as they turn out to be inversely proportional to the large scale where lepton number is violated.

\subsection{The See-Saw Mechanisms}
\label{Sec:SM:SeeSaw}
\setcounter{footnote}{3}

It is then interesting to investigate on the kind of new physics which accounts for the lepton number violation and generates the Weinberg operator in a renormalisable extension of the Standard Model. Tree-level exchange of three different types of new particles makes the job: right-handed neutrinos, fermion $SU(2)_L$-triplets and scalar $SU(2)_L$-triplets. We summarise in the following these proposals which are known with the name of ``See-Saw'' mechanisms.

\begin{figure}[h!]
 \centering
\includegraphics[width=4.5cm]{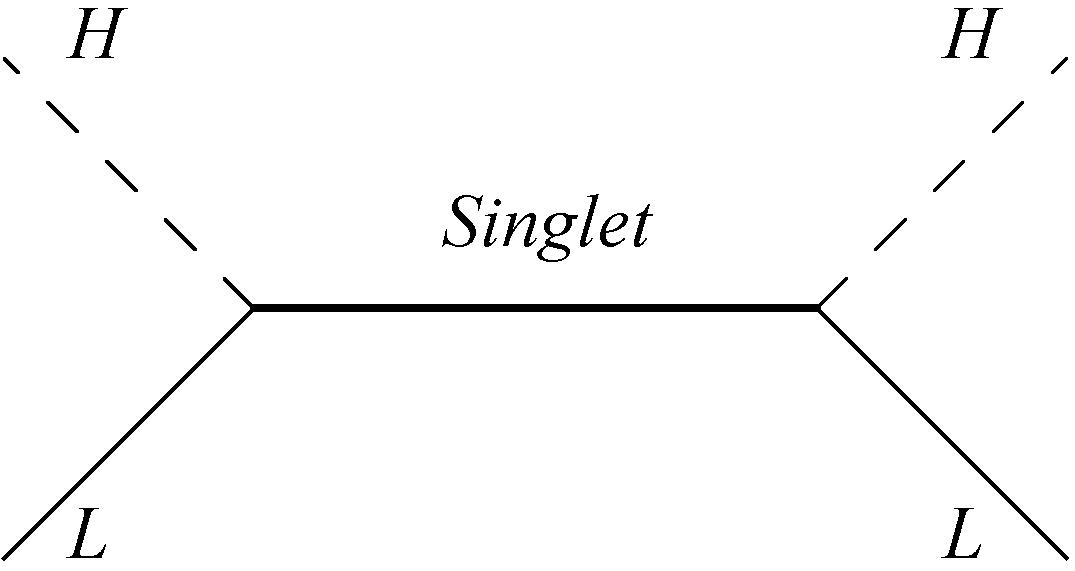}\qquad\qquad
\includegraphics[width=2.4cm]{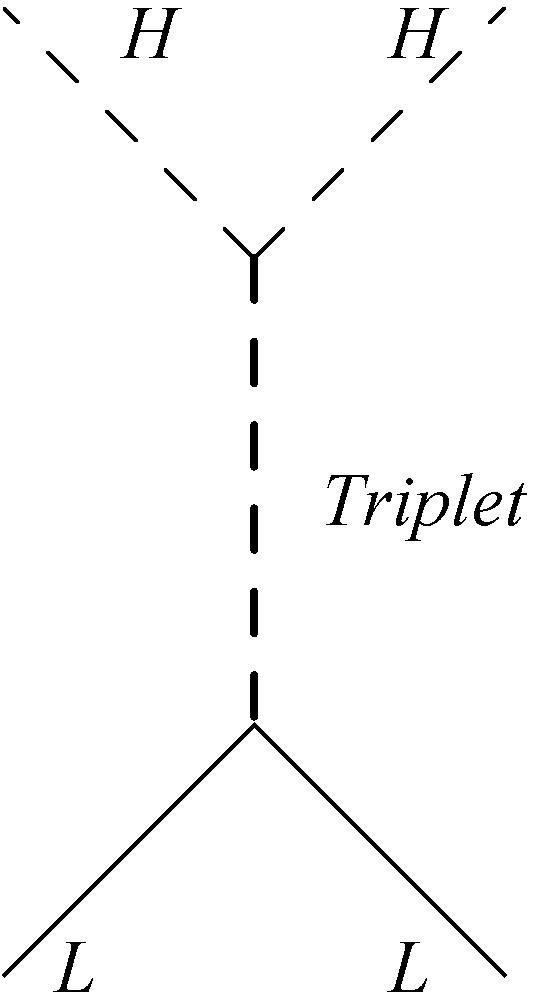}\qquad\qquad
\includegraphics[width=4.5cm]{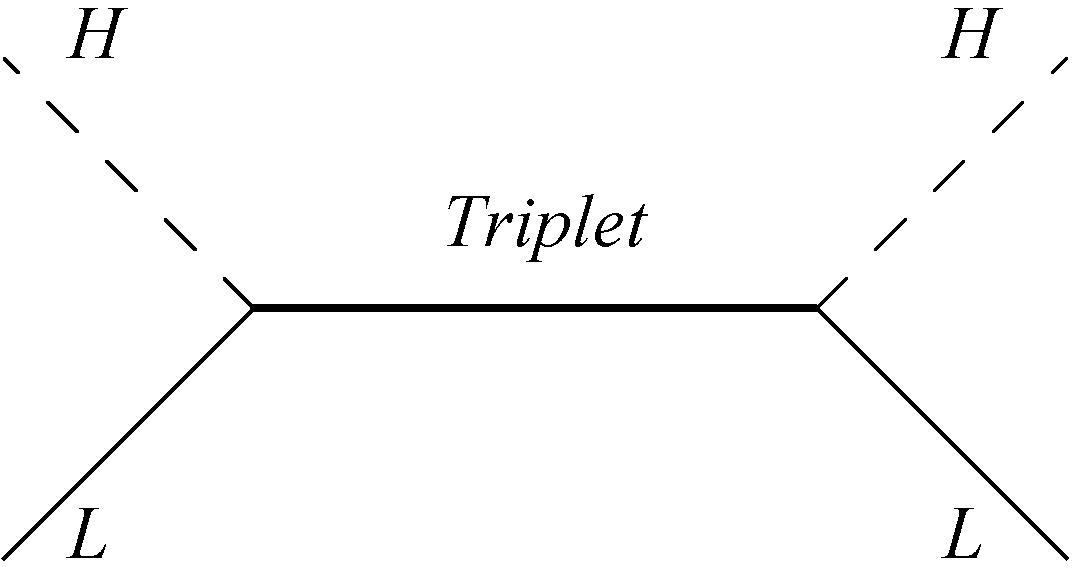}
\caption{\it The Weinberg operators can be produced by tree-level exchange of fermion singlets (type I), scalar triplets (type II) and fermion triplets (type III).}
 \label{fig:See-Saw}
\end{figure}

\subsubsection{Type I See-saw Mechanism}
\setcounter{footnote}{3}

The presence of new fermions with no gauge interactions, which play the role of right-handed neutrinos, is quite plausible because any grand unified theory (GUT) group larger than $SU(5)$ requires them: for example, $\nu^c$ complete the representation $\bf{16}$ of $SO(10)$. As already anticipated they have a Dirac Yukawa interaction $Y_\nu$ with the left-handed neutrinos. Assuming explicit lepton number violation a second term appears, a Majorana mass $M_R$: the Lagrangian can then be written as
\beq
\LL_\text{type I} = (Y_\nu\,)_{ij}\nu^c_i\widetilde{H}^\dag \ell_j+\dfrac{1}{2}(M_R\,)_{ij}\nu^c_i\nu^c_j+\hc\;.
\label{SM:LagrangianTypeI}
\eeq
$Y_\nu$ and $M_R$ are $3\times3$ matrices in the flavour space: $M_R$ is symmetric, while $Y_\nu$ is in general non hermitian and non symmetric. The Dirac mass term originates through the Higgs mechanism as in eq. (\ref{SM:mD}). On the other hand, the Majorana mass term is $SU(3)\times SU(2)_L\times U(1)_Y$ invariant and therefore the Majorana masses are unprotected and naturally of the order of the cutoff of the low-energy theory. The full neutrino mass matrix is a $6\times6$ mass matrix in the flavour space and can be written as
\beq
M_\nu=\bordermatrix{    & \nu & \nu^c \cr
              \nu & 0 & m_D^T \cr
              \nu^c & m_D & M_R}\;,
\eeq
where $m_D=Y_\nu\,v/\sqrt2$. By block-diagonalising $M_\nu$, the light neutrino mass matrix is obtained as
\beq
m_\nu=-m_D^TM_R^{-1}m_D\;.
\eeq
This is the well known type I See-Saw Mechanism \cite{SeeSawI}: the light neutrino masses are quadratic in the Dirac masses and inversely proportional to the large Majorana ones, justifying the lightness of the left-handed neutrinos.

The same result can be derived integrating out the heavy neutrinos which gives a non-renormalisable effective Lagrangian that only contains the observable low-energy fields. Figure \ref{fig:See-Saw} shows the tree-level exchange of $\nu^c$ and the $\cO_5$ operator of eq. (\ref{SM:O5operator}) is easily derived, where $\Lambda_L$ corresponds to the Majorana masses of the right-handed neutrinos.

This construction holds true for any number of heavy $\nu^c$ coupled to the three known light neutrinos. In the case of only two $\nu^c$, one light neutrino remains massless, which is a possibility not excluded by the present data.

\subsubsection{Type II See-saw Mechanism}
\setcounter{footnote}{3}

The type II See-Saw mechanism \cite{SeeSawII} refers to the general scenario in which neutrinos get a mass thanks to the coupling of the lepton $SU(2)_L$-doublet $\ell$ with a scalar $SU(2)_L$-triplet $\Delta\equiv(\delta_1,\,\delta_2,\,\delta_3)$ transforming as $({\bf1},\,{\bf3},\,+1)$ under the Standard Model. It is useful to express the triplet scalar by three complex (electric charge neutral $\Delta^0$, singly charged $\Delta^+$ and doubly charged $\Delta^{++}$) scalars:
\beq
\Delta=\dfrac{1}{\sqrt2}\sum_{i}\sigma^i\Delta_i=\left(
                                             \begin{array}{cc}
                                               \Delta^+/\sqrt2 & \Delta^{++} \\
                                               \Delta^0 & -\Delta^+/\sqrt2 \\
                                             \end{array}
                                           \right)\;,
\eeq
where $\sigma^i$ are the Pauli matrices and $\Delta^0=(\delta_1+i\delta_2)/\sqrt2$, $\Delta^+=\delta_3$ and $\Delta^{++}=(\delta_1-i\delta_2)/\sqrt2$.

The relevant Lagrangian for the type II See-Saw can then be written as
\beq
\begin{split}
\LL_\text{type II} & =\, -\dfrac{1}{2}(Y_\Delta)_{ij}\overline{\ell^c}_i i\sigma^2\Delta\ell_j+\hc\\[3mm]
& =\, -\dfrac{1}{2}(Y_\Delta)_{ij}\nu_i\Delta^0\nu_j +\dfrac{1}{\sqrt2}(Y_\Delta)_{ij}\nu_i\Delta^+e_j
+\dfrac{1}{2}(Y_\Delta)_{ij}e_i\Delta^{++}e_j+\hc\;.
\end{split}
\eeq
The neutrino masses are generated when the neutral component of the scalar triplet develops a VEV, $\mean{\Delta^0}=v_\Delta/\sqrt2$:
\beq
m_\nu = Y_\Delta\dfrac{v_\Delta}{\sqrt2}\;.
\eeq

The same result is achieved integrating out the Higgs triplet: in figure \ref{fig:See-Saw} the relevant tree-level diagram is shown. In this case we get an effective five-dimensional operator similar to $\cO_5$ of eq. (\ref{SM:O5operator}), where $\Lambda_L$ corresponds to $M_\Delta$, the mass of the Higgs triplet. The two pictures are phenomenologically identical since the minimisation of the potential leads  to $v_\Delta \propto v^2/M_\Delta$, as shown below.

The VEV of the Higgs triplet is usually induced by the VEV of the Higgs doublet $H$: indeed in the Lagrangian it is possible to write these two terms
\beq
M_\Delta^2 \tr[\Delta^\dag \Delta] +\dfrac{1}{2}\left(\lambda_\Delta M_\Delta \widetilde{H}^\dag\Delta^\dag H+\hc\right)
\eeq
and when the Higgs doublet develops a non-zero VEV, it induces a tadpole term for $\Delta$ through the last term in the previous equation and as a consequence a VEV for the Higgs triplet is generated,
\beq
\mean{\Delta}\sim\lambda_\Delta \dfrac{v^2}{M_\Delta}\,.
\label{SM:SSII:vevDelta}
\eeq

Note that $v_\Delta$ contributes to the weak boson masses and introduces a deviation of the $\rho$-parameter from the Standard Model prediction, $\rho\simeq1$ at tree-level. The current precision measurements constrain this deviation and consequently the ratio $v_\Delta/v$ \cite{BoundsTypeII}: $\Delta\rho=\rho-1\simeq v_\Delta/v\lesssim0.01$. From eq. (\ref{SM:SSII:vevDelta}) we see that this implies $M_{\Delta}\gtrsim(25 \lambda_{\Delta})$ TeV.

The type II See-Saw involves lepton number violation because the co-existence of $Y_\Delta$ and $\lambda_\Delta$ does not allow a
consistent way of assigning a lepton charge to $\Delta$.

\subsubsection{Type III See-saw Mechanism}
\setcounter{footnote}{3}

In the type III See-Saw mechanism \cite{SeeSawIII}, three fermion $SU(2)_L$-triplets $\Sigma^a$ are added to the Standard Model content.\footnote{It is also possible to introduce only two such fermion triplets: indeed it is the minimal number of $\Sigma$ in order to fit the data with a massless neutrino. With three $\Sigma$, however, it is possible to describe three distinct non-vanishing neutrino masses.} These new particles should be color-singlets and carry hypercharge $0$. Each $\Sigma$ has three components defined as $\Sigma=(\varsigma_1,\,\varsigma_2,\,\varsigma_3)$ and can be written in the fundamental representation of $SU(2)_L$ as
\beq
\Sigma=\sum_i\sigma^i\varsigma_i=\left(\begin{array}{cc}
                                               \Sigma^0/\sqrt2 & \Sigma^{+} \\
                                               \Sigma^- & -\Sigma^0/\sqrt2 \\
                                             \end{array}
                                           \right)\;,
\eeq
where $\Sigma^0=\varsigma_3$ and $\Sigma^{\pm}=(\varsigma_1\mp i\varsigma_2)/\sqrt2$ are electric charge neutral and positive or negative singly charged. The relevant Lagrangian terms have a form that is similar to the singlet-fermion case, but the contractions of the $SU(2)_L$ indices are different:
\beq
\LL_\text{type III}=(Y_\Sigma)_{ai}\widetilde{H}^\dag\Sigma^a\ell_i-\dfrac{1}{2}(M_\Sigma)_{ab} \tr[\Sigma^a\Sigma^b]+\hc\,.
\eeq
Here $Y_\Sigma$ is a $3\times3$ matrix of dimensionless, complex Yukawa couplings. When the electroweak symmetry is broken, both charged leptons and neutrinos mix with the $\Sigma$ components. In the basis $(\nu,\,\Sigma)$, the following mass term can be written
\beq
M_\nu=\bordermatrix{    & \nu & \Sigma \cr
              \nu & 0 & m_D^T \cr
              \Sigma & m_D & M_\Sigma}\;.
\label{SM:SSIII:FullMassTypeIII}
\eeq
where $m_D=Y_\Sigma\,v/\sqrt2$ and $M_\Sigma$ is a $3\times3$ real mass matrix. As in the case of $M_R$, the mass of the right-handed neutrinos in the type I See-Saw mechanism, $M_\Sigma$ is unprotected by any symmetry and it could reach values close to the cutoff of the low-energy theory. The exchange of fermion triplets, as shown in figure \ref{fig:See-Saw}, generates an effective five-dimensional operator similar to $\cO_5$ of eq. (\ref{SM:O5operator}), where $\Lambda_L$ corresponds to $M_\Sigma$, which leads to the neutrino masses:
\beq
m_\nu=-m_D^TM_\Sigma^{-1}m_D\;.
\label{SM:SSIII:nuMassesTypeIII}
\eeq
The same result can be achieved simply diagonalising the matrix in eq. (\ref{SM:SSIII:FullMassTypeIII}).

It is clear that for what concerns the phenomenology of the light neutrinos, the type I and the type III See-Saw mechanisms cannot be distinguished. The two descriptions however may be discriminated by taking into account phenomenological processes involving for example the charged lepton sector that presents a small mixing with the charged components of $\Sigma$. In the basis $(e,\,\Sigma)$, the following mass term can be written
\beq
\bordermatrix{    & e & \Sigma \cr
              e^c & M_e & 0 \cr
              \Sigma & m_D & M_\Sigma}\;.
\label{SM:SSIII:FullMassTypeIIICharged}
\eeq
where $M_e$ is the usual charged lepton mass matrix. The perturbations induced by $\Sigma$ to $M_e$ have the same form and order of magnitude of the neutrino masses in eq. (\ref{SM:SSIII:nuMassesTypeIII}) and therefore they are negligible. However, the presence of the triplets induces some lepton flavour violating decays which could have interesting experimental hints \cite{AbadaTypeIII}.

The type III See-Saw involves lepton number violation because the co-existence of $Y_\Sigma$ and $M_\Sigma$ does not allow a consistent way of assigning a lepton charge to $\Sigma^a$.

\subsection{The Physical Basis and the Mixing Matrices}
\label{Sec:SM:PhysicalBasis}
\setcounter{footnote}{3}

If we simply assume that the lepton number is explicitly violated by the introduction of the Weinberg operator in the Lagrangian of the Standard Model, when the electroweak symmetry is broken, all the fermions develop a mass term:
\beq
\LL_\text{mass}=(M_u)\, u^c_i u_j  +  (M_d)_{ij}\, d^c_i d_j + (M_e)_{ij}\, e^c_i e_j + \dfrac{1}{2} (m_\nu)_{ij} \nu_i \nu_j\;,
\eeq
where $M_i$ and $m_\nu$ are $3\times3$ mass matrices in the flavour space. Counting the number of free parameters, there are 9 complex entries
for each mass matrix, apart for $m_\nu$ which is symmetric and owns only 6. To reduce this amount, we can move to the physical basis in which the kinetic terms are canonical and all the mass matrices are diagonal. In this basis also the Yukawa coupling matrices are diagonal and therefore there is no tree-level flavour changing currents mediated by the neutral Higgs boson. This feature is in general lost extending the Standard Model by introducing multiple Higgs doublets or extra fermions.

We make unitary transformations on the fermions in the family space in order to move to the physical basis. Unitarity of these matrices ensures that the kinetic terms remain canonical. Specifically, we define $V_{u,\,u^c,\,d,\,d^c}$ and $U_{e,\,e^c,\,\nu}$ such that the transformations produce the following diagonal matrices:
\beq
\ba{rclrcl}
V_{u^c}^\dag\,M_u\,V_u&=&\diag(m_u,\,m_c,\,m_t)\;,&\qquad
V_{d^c}^\dag\,M_d\,V_d&=&\diag(m_d,\,m_s,\,m_b)\;,\\[3mm]
U_{e^c}^\dag\,M_e\,U_e&=&\diag(m_e,\,m_\mu,\,m_\tau)\;,&\qquad
U_{\nu}^T\,m_\nu\,U_\nu&=&\diag(m_1,\,m_2,\,m_3)\;.
\ea
\eeq
Experiments showed that quarks and charged leptons have strongly hierarchical masses: the mass of the first families are smaller than those of the second families and the third families are the heaviest. The quark masses are given by \cite{PDG08}\footnote{The u-, d-, and s-quark masses are estimates of so-called ``current-quark masses'', in a mass-independent subtraction scheme such as $\overline{\mathrm{MS}}$ at a scale $\mu\approx 2$ GeV. The c- and b-quark masses are the ``running'' masses, $m(\mu=m)$, in the $\overline{\mathrm{MS}}$ scheme. Only the mass of the t-quark is a result of direct observation of top events.}
\beq
\ba{lll}
m_u=1.5\div3.3\MeV\;, &\qquad m_c=1.27^{+0.07}_{-0.11}\GeV\;, &\qquad m_t=171.2\pm2.1\GeV\;,\\[3mm]
m_d=3.5\div6\MeV\;, &\qquad m_s=104^{+26}_{-34}\MeV\;, &\qquad m_b=4.20^{+0.17}_{-0.07}\GeV\;.
\ea
\eeq
The charged lepton masses are very precisely known and they read \cite{PDG08}
\bea
m_e &=& 0.510998910\pm0.000000013\MeV\;,\nn\\[3mm]
m_\mu &=& 105.658367\pm0.000004\MeV\;,\\[3mm]
m_\tau &=& 1776.84\pm0.17\MeV\nn\;.
\eea
In the neutrino sector the mass hierarchy is much milder and only two mass squared differences have been measured in oscillation experiments.\footnote{There is an indication for the existence of a third independent mass squared difference from the LSND experiment \cite{LNSD}, which could be explained only if an additional (sterile) neutrino is considered. However, the MiniBooNE collaboration \cite{MiniBoone} have not recently confirmed the LSND result.} The mass squared differences are defined as
\beq
\De m^2_{21} = m_2^2-m_1^2 \equiv \De m^2_{sol}\;,\qquad\quad\De m^2_{31} = m_3^2-m_1^2 \equiv \pm\De m^2_{atm}
\label{SM:PhysicalBasisDeltamSol+Atm}
\eeq
and in table \ref{table:OscillationData} we can read the results of two independent global fits on neutrino oscillation data from \cite{Fogli:Indication} and \cite{Maltoni:Indication}. It is interesting to report also the ratio between the two mass squared differences\cite{Maltoni:Indication}:
\beq
r=\dfrac{\De m^2_{sol}}{\De m^2_{atm}}=0.032^{+0.006}_{-0.005}\;.
\eeq

\begin{table}[h!]
\begin{center}
\begin{tabular}{|l||cc|cc|}
\hline
&&&&\\[-2mm]
& \multicolumn{2}{c}{Ref.~\cite{Fogli:Indication}} & \multicolumn{2}{|c|}{Ref.~\cite{Maltoni:Indication}}\\[2mm]
parameter & best fit (\@$1\sig$) & 3$\sig$-interval & best fit (\@$1\sig)$ & 3$\sig$-interval\\[2mm]
\hline
&&&&\\[-2mm]
$\De m^2_{sol}\:[\times10^{-5}\mathrm{eV}^2]$
        & $7.67^{+0.16}_{-0.19}$ & $7.14-8.19$
        & $7.65^{+0.23}_{-0.20}$ & $7.05-8.34$\\[2mm]
$\De m^2_{atm}\: [\times10^{-3}\mathrm{eV}^2]$
        & $2.39^{+0.11}_{-0.08}$ & $2.06-2.81$
        & $2.40^{+0.12}_{-0.11}$ & $2.07-2.75$\\[2mm]
$\sin^2\theta_{12}$
        & $0.312^{+0.019}_{-0.018}$ & $0.26-0.37$
        & $0.304^{+0.022}_{-0.016}$ & $0.25-0.37$\\[2mm]
$\sin^2\theta_{23}$
        & $0.466^{+0.073}_{-0.058}$ & $0.331-0.644$
        & $0.50^{+0.07}_{-0.06}$ & $0.36-0.67$\\[2mm]
$\sin^2\theta_{13}$
        & $0.016^{+0.010}_{-0.010}$ & $\leq$ $0.046$
        & $0.010^{+0.016}_{-0.011}$ & $\leq$ $0.056$\\[2mm]
\hline
\end{tabular}
\end{center}
\caption{\it Neutrino oscillation parameters from two independent global fits \cite{Fogli:Indication, Maltoni:Indication}.}
\label{table:OscillationData}
\end{table}

\begin{figure}[h!]
\centering
\includegraphics[width=5.5cm]{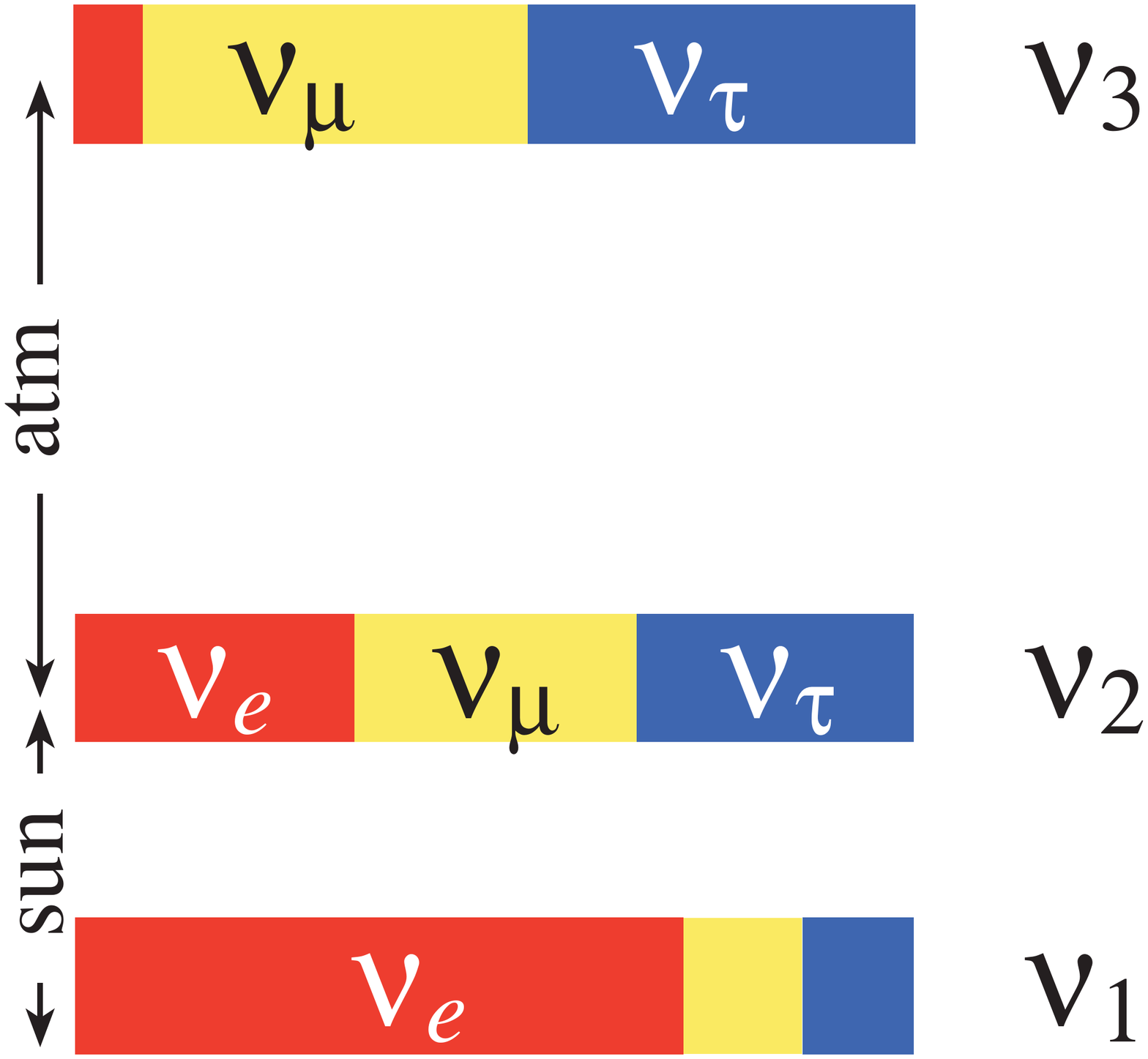}\qquad
\includegraphics[width=5.5cm]{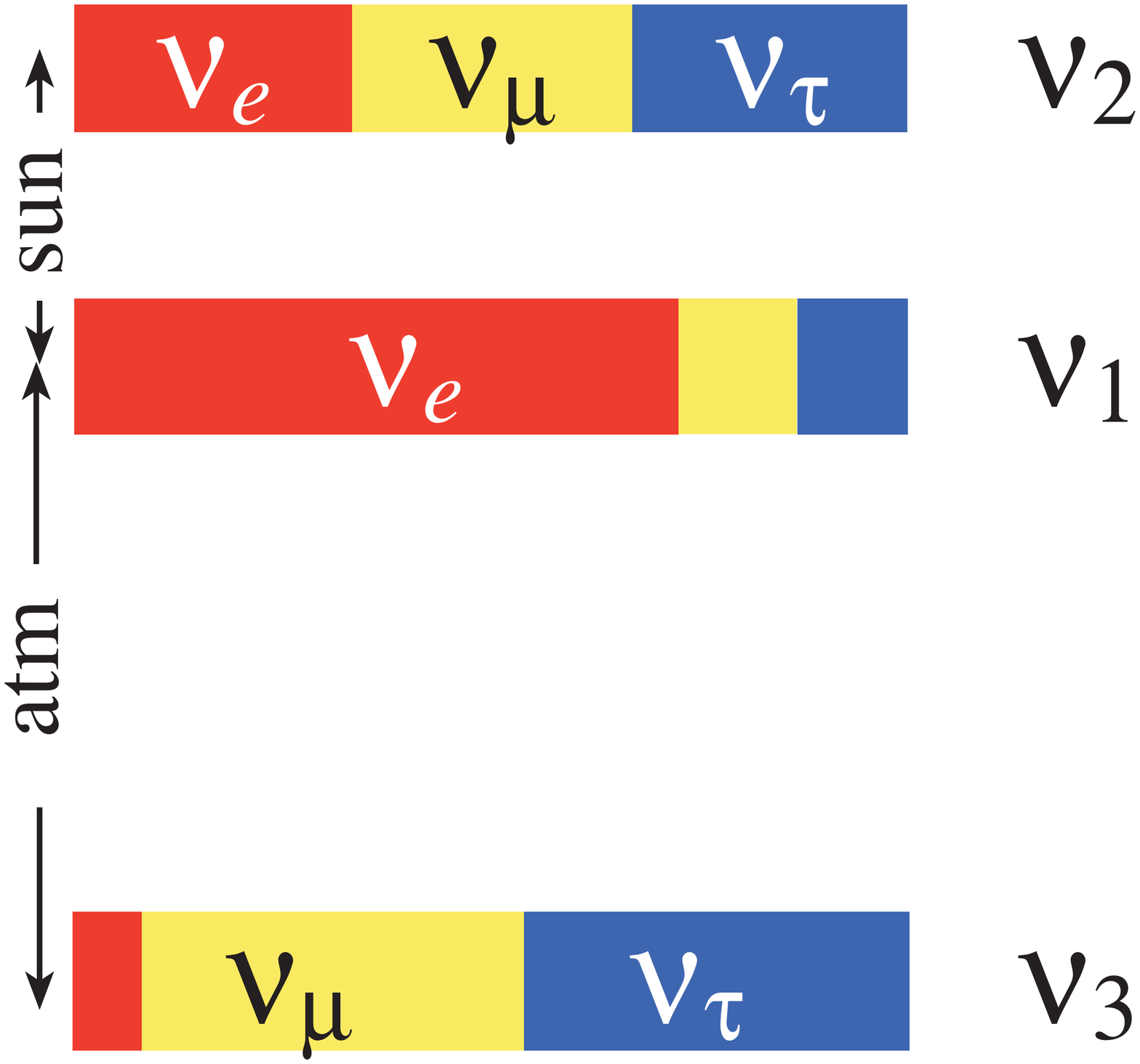}
\caption{\it Neutrino mass and flavour spectra for the
 normal (left) and inverse (right) mass hierarchies.
 The distribution of the flavours in the mass eigenstates
 corresponds to the best-fit values of mixing parameters and
 $\sin^2{\te_{13}}=0.05$.}
\label{fig:HierarchyFig}
\end{figure}

Due to the ambiguity in the sign of the atmospheric mass squared difference, neutrinos can have two mass hierarchies: they can be normally hierarchical (NH) if $m_1<m_2<m_3$ or inversely hierarchical (IH) if $m_3<m_1<m_2$. Furthermore, if the absolute mass scale is much larger than the mass squared differences then we cannot speak about hierarchy: in this case the neutrino spectrum is quasi degenerate (QD) and we speak about mass ordering. In figure \ref{fig:HierarchyFig} we display the two possible hierarchical spectra. It is common to redefine the atmospheric mass squared difference to account for the type of the hierarchy: indeed $\De m^2_{atm}$ is taken to be the mass squared difference between the heaviest and the lightest mass eigenstates and therefore
\beq
|m_3^2-m_1^2(m_2^2)| \equiv \De m^2_{atm}
\label{SM:PhysicalBasisDeltamAtm2}
\eeq
for the normal (inverse) hierarchy.

There are some weak indications in favour of normal mass hierarchy from supernova SN1987A data \cite{SN87}. However in view of small statistics and uncertainties in the original fluxes it is not possible to make a firm statement.

\begin{figure}[h!]
\centering
\vspace{-4mm}
\includegraphics[width=7.8cm]{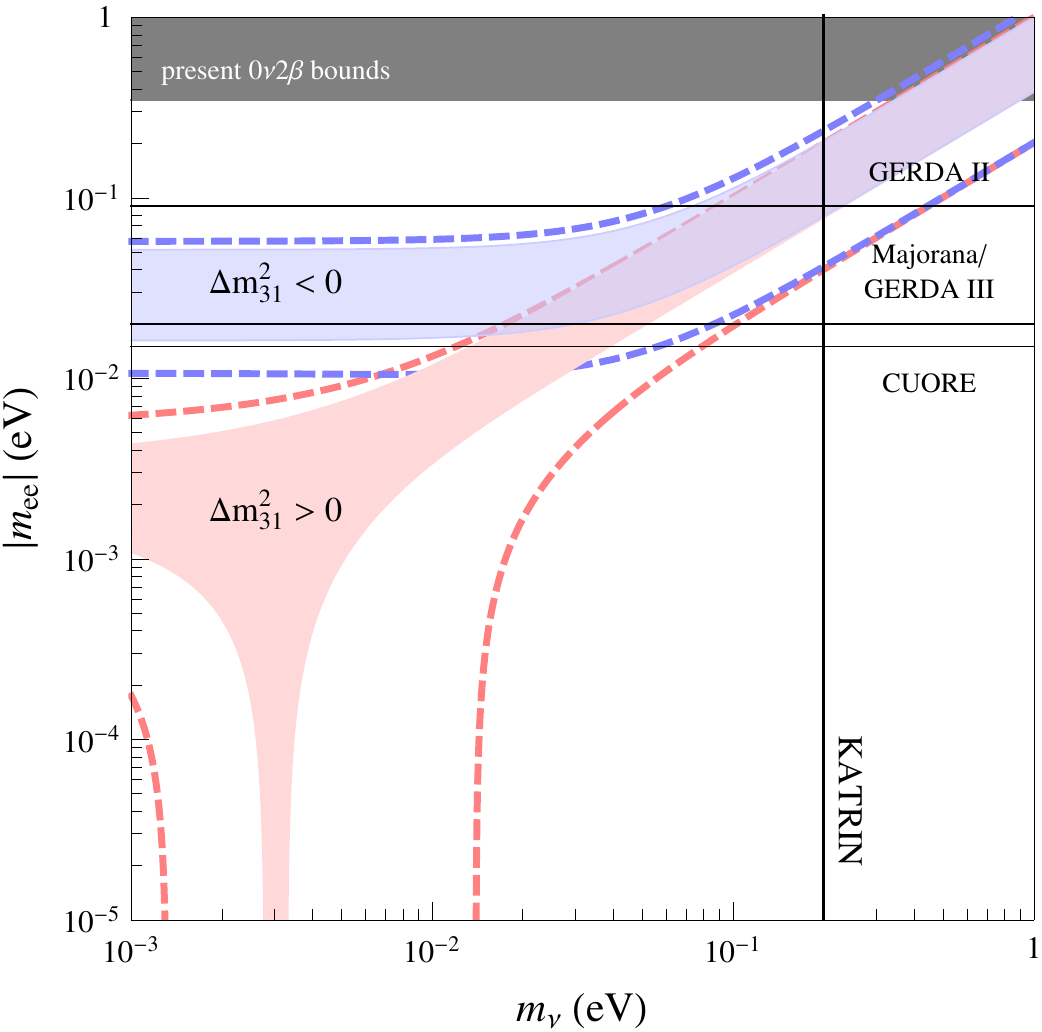}
\vspace{-4mm}
\caption{\it $|m_{ee}|$ as a function of the lightest neutrino mass for the normal ($\Delta m_{31}^2>0$) and inverse ($\Delta m_{31}^2<0$) mass hierarchies. The coloured regions show the allowed range for the best-fit values of the parameters from \cite{Fogli:Indication}. The dashed lines refer to the allowed region when the $3\sigma$ errors are considered. The black continuous lines represent future experimental sensitivities as described in the text.}
\label{fig:0nu2betaGeneral}
\end{figure}

Regarding the absolute neutrino mass scale there are several sources of information from non-oscillation experiments and from cosmological analysis. They are:
\begin{itemize}
\item $\beta$-decay experiments which measure the endpoint of the tritium decay and to good approximation probe $m_{\nu_e}^2=\sum_i|U_{ei}^2|m_i^2$. The most recent experiment is \textsc{Mainz} \cite{mainz} and it puts an upper bound at $99\%$ of CL of $m_{\nu_e}<2.1$ eV.
The \textsc{Katrin} experiment will improve the sensitivity to $m_{\nu_e}$  by one order of magnitude down to $\sim0.2$ eV \cite{katrin}.

\item the neutrinoless-double-beta ($0\nu2\beta$) decay is a viable decay for a little class of nuclei and only in the hypothesis of Majorana nature for neutrinos. Dedicated experiments could probe the $ee$ element of the neutrino Majorana mass:
\beq
m_{ee}=\sum_iU^2_{ei}m_i=\cos\theta_{13}^2(m_1\cos\theta_{12}^2e^{i\varphi_1}+ m_2\sin\theta_{12}^2e^{i\varphi_2})+m_3\sin\theta_{13}^2\;,
\eeq
where $\varphi_i$ are the Majorana phases, which will be defined in the following. Nowadays only an upper bound of $0.35$ eV on this quantity has been put by the Heidelberg-Moscow collaboration \cite{HM}, but the future experiments are expected to reach better sensitivities: $90$ meV \cite{gerda} (GERDA), $20$ meV \cite{majorana} (Majorana), $50$ meV \cite{supernemo} (SuperNEMO), $15$ meV \cite{cuore} (CUORE) and $24$ meV \cite{exo} (EXO). In figure \ref{fig:0nu2betaGeneral} we show the $0\nu2\beta$-decay effective mass as a function of the lightest neutrino mass for both the hierarchies together with the future experimental sensitivities.

\item cosmology can set an upper bound on the sum of the neutrino masses: there are typically five representative combinations of the cosmological data, which lead to increasingly stronger bounds and as a result \cite{CosmoNu} $\sum_i m_i<0.19\div2.6$ eV.
\end{itemize}

\subsubsection{The CKM and PMNS Mixing Matrices}
\setcounter{footnote}{3}

Moving to the physical basis, the unitary matrices $V_i$ and $U_i$ should enter into all the fermion interactions. As already noted, the associated transformations bring the Yukawa couplings of fermions with the Higgs boson in the diagonal form. The coupling of the $Z^0$ boson and the photon have the original diagonal form even after these rotations. It follows that there is no tree-level flavour changing neutral current mediated by the $Z^0$ boson or by the photon in the Standard Model. Most significantly, these transformations bring the charged current interactions in a non diagonal form: considering for simplicity only the negative charged current $J_{\mu}^-$, we see that
\beq
J_{\mu}^-=\overline\nu\ga_\mu e + \overline{u}\ga_\mu d\longrightarrow
\overline\nu\ga_\mu U_\nu^{\dag}U_e e +\overline{u}\ga_\mu V_u^{\dag}V_d d\;.
\eeq
The products of the diagonalising unitary matrices are defined as the mixing matrices for leptons, the Pontecorvo-Maki-Nakagawa-Sakata (PMNS) matrix \cite{PMNS}, and for quarks, the Cabibbo-Kobayashi-Maskawa (CKM) matrix \cite{Cabibbo,KM},
respectively:
\beq
U=U_e^{\dag}U_\nu\qquad\text{and}\qquad 
V=V_u^{\dag}V_d\;.
\eeq
Each one of these matrices is also unitary and has nine parameters, but only some of them are physical. It is possible to absorb the non-physical parameters through further unitary transformations on the fields and finally $V$ is described by 3 angles and 1 phase and $U$ by 3 angles and 3 phases. This difference is due to the Majorana nature of the neutrino mass term. However the meaning of the parameters are equivalent in both sectors: the angles rule the mixing between the flavour eigenstates and the phases are responsible for CP violation. If we now assume the conservation of the lepton number, neutrinos develop masses only due to the introduction of the right-handed neutrinos. In this case, as already discussed, the mass term is of the Dirac form $(m_\nu)_{ij}\nu^c_i\nu_j$ and going through the same passages as before, we note one main difference: the physical parameters of $U$ can be reduce to only 3 angles and 1 phase.\\

There are several parametrisations of the CKM matrix. We recall now the standard one, by the use of the angles $\theta_{12}$, $\theta_{13}$ and $\theta_{23}$ and of the phase $\delta$:
\beq
\begin{split}
V&=\,\left(\begin{array}{ccc}
                          1 & 0 & 0 \\
                          0 & c_{23} & s_{23} \\
                          0 & -s_{23} & c_{23} \\
                        \end{array}\right)\cdot
                    \left(\begin{array}{ccc}
                              c_{13} & 0 & s_{13}e^{-i\delta} \\
                              0 & 1 & 0 \\
                              -s_{13}e^{i\delta} & 0 & c_{13} \\
                            \end{array}\right)\cdot
                    \left(\begin{array}{ccc}
                              c_{12} & s_{12} & 0 \\
                              -s_{12} & c_{12} & 0 \\
                              0 & 0 & 1 \\
                            \end{array}\right)\\[3mm]
&=\,\left(\begin{array}{ccc}
                        c_{12}c_{13} & c_{13}s_{12} & s_{13}e^{-i\delta} \\
                        -c_{23}s_{12}-c_{12}s_{13}s_{23}e^{i\delta} &
                        c_{12}c_{23}-s_{12}s_{13}s_{23}e^{i\delta} & c_{13}s_{23} \\
                        s_{12}s_{23}-c_{12}c_{23}s_{13}e^{i\delta} &
                        -c_{12}s_{23}-c_{23}s_{12}s_{13}e^{i\delta} & c_{13}c_{23} \\
                        \end{array}
                        \right)\;,
\end{split}
\label{SM:PhysBasis:MixingMatrixVCKM}
\eeq
where $c_{ij}$ and $s_{ij}$ stand for $\cos\te_{ij}$ and $\sin\te_{ij}$ (with $0\leq\te_{ij}\leq\pi/2$), respectively, and the Dirac CP-violating phase lies in the range $0\leq\delta<2\pi$. This notation has various advantages: the rotation angles are defined and labelled in a way which is related to the mixing of two specific generations; as a result if one of these angles vanishes, so does the mixing between the two respective generations. Moreover in the limit $\theta_{23}=\theta_{13}=0$ the third generation decouples and the situation reduces to the usual Cabibbo mixing of the first two generations with $\sin\theta_{12}$ identified to the Cabibbo angle \cite{Cabibbo}.

Experimentally, the mixing matrix has well defined entries \cite{PDG08}: a fit on the data, considering the unitary conditions, gives the following results,
\beq
\begin{split}
|V|&=\,\left(
      \begin{array}{ccc}
        V_{ud} & V_{us} & V_{ub} \\
        V_{cd} & V_{cs} & V_{cb} \\
        V_{td} & V_{ts} & V_{tb} \\
      \end{array}
    \right)\\[3mm]
&=\,\left(
     \begin{array}{ccc}
       0.97419\pm0.00022 & 0.2257\pm0.0010 & (3.59\pm0.16)\times10^{-3} \\[2mm]
       0.2256\pm0.0010 & 0.97334\pm0.00023 & (41.5^{+0.0010}_{-0.0011})\times10^{-3} \\[2mm]
       (8.74^{+0.26}_{-0.37})\times10^{-3} & (40.7\pm1.0)\times10^{-3} & 0.999133^{+0.000044}_{-0.000043} \\
     \end{array}
   \right)\;.
\end{split}
\label{SM:PhysBasis:QuarkCKMmatrix}
\eeq
Making use of the standard parametrisation, it is possible to extract the values of the quark mixing angles: in terms of $\sin\theta_{ij}$ we naively have
\beq
\sin\theta_{12}\simeq 0.2243\;,\qquad
\sin\theta_{23}\simeq 0.0413\;,\qquad
\sin\theta_{13}\simeq 0.0037\;.
\label{SM:PhysBasis:QuarkAngles}
\eeq
In the same way it is possible to recover the value of the Dirac CP-violating phase:
\beq
\delta=(77^{+30}_{-32})^\circ\;.
\label{SM:PhysBasis:QuarkPhase}
\eeq
It is also useful to report the value of the Jarlskog invariant \cite{Jarlskog}, which measures the amount of the CP violation: it is defined as 
\bea
 J_{CP} & = & \frac{1}{2} \left| \im (V_{ud}^*V_{us}V_{cd}V_{cs}^*)\right| \,=\, \frac{1}{2} \left| \im (V_{ud}^*V_{ub}V_{td}V_{tb}^*)\right| \nonumber\\
 & =  &\frac{1}{2} \left| \im (V_{cs}^*V_{cb}V_{ts}V_{tb}^*)\right| \, = \, \frac{1}{2}\left|c_{12}\,c_{13}^2\, c_{23}\,\sin \delta \, s_{12}\,s_{13}\,s_{23}\right|\;.
\eea
and the result of the data fit is
\beq
J_{CP}=(3.05^{+0.19}_{-0.20})\times10^{-5}\;.
\eeq
From eq. (\ref{SM:PhysBasis:QuarkAngles}, \ref{SM:PhysBasis:QuarkPhase}), it is obvious the presence of a hierarchy between the size of the angles, $\sin\theta_{12}\gg \sin\theta_{23}\gg \sin\theta_{13}$, and an approximation to the standard parametrisation has been proposed by Wolfenstein \cite{Wolfenstein} which emphasises this feature. It is possible to use only four parameters to describe the CKM matrix: they are $\lambda$, $A$, $\rho$ and $\eta$ defined as
\beq
\la\equiv\dfrac{|V_{us}|}{\sqrt{|V_{ud}|^2+|V_{us}|^2}}\;,\qquad
A\equiv\dfrac{1}{\lambda}\left|\dfrac{V_{cb}}{V_{us}}\right|\;,\qquad
\rho+i\eta\equiv\dfrac{V_{ub}^*}{A\lambda^3}\;.
\eeq
In terms of powers of $\la$, up to $\cO(\la^4)$ we have
\beq
V=\left(\begin{array}{ccc}
          1-\la^2/2 & \la & A\la^3(\rho-i\eta) \\
          -\la & 1-\la^2/2 & A\la^2 \\
          A\la^3(1-\rho-i\eta) & -A\la^2 & 1 \\
        \end{array}\right)+\mcal{O}(\la^4)\;.
\eeq
These four parameters are experimentally determined as follows:
\beq
\ba{ll}
\lambda=0.2257^{+0.0009}_{-0.0010}\;,\quad
&A=0.814^{+0.021}_{-0.022}\;,\\[3mm]
\rho\left(1-\dfrac{\lambda^2}{2}\right)=0.135^{+0.031}_{-0.016}\;,\quad
&\eta\left(1-\dfrac{\lambda^2}{2}\right)=0.349^{+0.015}_{-0.017}\;.
\ea
\eeq
Due to the closeness of the Cabibbo angle, $\sin\te_{12}$, to the parameter $\lambda$, it is common to identify the two quantities and the error, which is potentially introduced, is of $\cO(\lambda^4)$.\\

The standard parametrisation of the lepton mixing is similar to eq. (\ref{SM:PhysBasis:MixingMatrixVCKM}): we can write the PMNS matrix as the product of four parts
\beq
\begin{split}
U&=\,R_{23}(\theta_{23})\cdot R_{13}(\theta_{13},\,\delta)\cdot R_{12}(\theta_{12})\cdot P\\[3mm]
&=\,\left(\begin{array}{ccc}
                        c_{12}c_{13} & c_{13}s_{12} & s_{13}e^{-i\delta} \\
                        -c_{23}s_{12}-c_{12}s_{13}s_{23}e^{i\delta} &
                        c_{12}c_{23}-s_{12}s_{13}s_{23}e^{i\delta} & c_{13}s_{23} \\
                        s_{12}s_{23}-c_{12}c_{23}s_{13}e^{i\delta} &
                        -c_{12}s_{23}-c_{23}s_{12}s_{13}e^{i\delta} & c_{13}c_{23} \\
                        \end{array}
                        \right)\cdot P\;,
\end{split}
\label{SM:PhysBasis:MixingMatrixUPMNS}
\eeq
where $P$ is the matrix of the Majorana phases $P=\diag(e^{i\varphi_1/2},\,e^{i\varphi_2/2},\,1)$, $c_{ij}$ and $s_{ij}$ represent $\cos\te_{ij}$ and $\sin\te_{ij}$, respectively, and $\delta$ is the dirac CP-violating phase. Angles and phases have well defined ranges: $0\leq\te_{12},\,\te_{23},\,\te_{13}\leq\dfrac{\pi}{2}$ and $0\leq\delta,\,\varphi_1,\,\varphi_2<2\pi$. It is interesting to note that the Dirac CP-violating phase is always present in the combination $s_{13}e^{\pm i\delta}$: this means that when the reactor angle is vanishing, not excluded from the experimental data in table \ref{table:OscillationData}, $\delta$ is undetermined and does not appear in the mixing matrix. A further consideration is that the Majorana phases are not present in eq. (\ref{SM:PhysBasis:MixingMatrixUPMNS}) if lepton number is assumed to be conserved and the right-handed neutrinos are responsible for the neutrino mass term.

A convenient summary of the neutrino oscillation data is given in table \ref{table:OscillationData}. The pattern of the mixings is characterised by two large angles and a small one: $\theta_{23}$ is compatible with a maximal value, but the accuracy admits relatively large deviations; $\theta_{12}$ is large, but about $5\sigma$ far from the maximal value; $\theta_{13}$ has only an upper bound. According to the type of the experiments which measured them, the mixing angle $\theta_{23}$ is called atmospheric, $\theta_{12}$ solar and $\theta_{13}$ reactor.
We underline that there is a tension among the two global fits presented in table \ref{table:OscillationData} on the central value of the reactor angle: in \cite{Fogli:Indication} we can read a suggestion for a positive value of $\sin^2\theta_{13}\simeq0.016\pm0.010$ [$1.6\sig$], while in \cite{Maltoni:Indication} a best fit value consistent with zero within less than $1\sig$ is found. Therefore we need for a direct measurement \cite{MezzettoSchwetz} by experiments like DOUBLE CHOOZ \cite{doublechooz}, Daya Bay \cite{dayabay}, MINOS \cite{minos}, RENO \cite{reno}, T2K \cite{T2K} and NOvA \cite{NOvA}.

It is interesting to note that the large lepton mixing angles contrast with the small angles of the CKM matrix. Furthermore, to compare with eq. (\ref{SM:PhysBasis:QuarkCKMmatrix}), we display the allowed ranges of the entries of the PMNS matrix \cite{PMNSEntriesOld}:
\beq
|U|=\,\left(
      \begin{array}{ccc}
        U_{e1} & U_{e2} & U_{e3} \\
        U_{\mu1} & U_{\mu2} & U_{\mu3} \\
        U_{\tau1} & U_{\tau2} & U_{\tau3} \\
      \end{array}
    \right)\;
=\, \left(
       \begin{array}{ccc}
         0.79-0.88 & 0.47-0.61 & <0.20 \\
         0.19-0.52 & 0.42-0.73 & 0.58-0.82\\
         0.20-0.53 & 0.44-0.74 & 0.56-0.81 \\
       \end{array}
     \right)\;.
\eeq\\

In analytical and numerical analysis, quark and lepton mixing matrices are not in the standard form as in eqs. (\ref{SM:PhysBasis:MixingMatrixVCKM}, \ref{SM:PhysBasis:MixingMatrixUPMNS}) but it is possible to recover the mixing angles $\te_{ij}$ and the phases $\delta$, $\varphi_1$ and $\varphi_2$ through the following procedure. Note that, when considering the CKM matrix, the Majorana phases are absent. Denoting the generic mixing matrix as $W$, the mixing angles are given by
\beq
\sin\theta_{13}=|W_{13}|\;,\qquad
\tan\theta_{12}=\left(\dfrac{|W_{12}|}{|W_{11}|}\right)\;,\qquad
\tan\theta_{23}=\left(\dfrac{|W_{23}|}{|W_{33}|}\right)\;,
\eeq
if $|W_{11}|$ ($|W_{33}|$) is non-vanishing, otherwise $\theta_{12}$ ($\theta_{23}$) is equal to $\pi/2$. For the Dirac CP-violating phase we use the relation
\beq
W_{ii}^*W_{ij}W_{ji}W_{jj}^* = c_{12}\,c_{13}^2\, c_{23}\,s_{13} \left(e^{-i\delta}\,s_{12}\,s_{23} - c_{12}\,c_{23}\,s_{13}\right)
\eeq
which holds for $i,j\in\{1,2,3\}$ and $i\ne j$. Then the phase $\delta$ is given by
\beq
\delta=-\arg\left(\dfrac{\displaystyle\dfrac{W_{ii}^*W_{ij}W_{ji}W_{jj}^*}
        {c_{12}\,c_{13}^2\,c_{23}\,s_{13}}
        +c_{12}\,c_{23}\,s_{13}}
        {s_{12}\,s_{23}}\right)
\eeq
where $i,j\in\{1,2,3\}$ and $i\ne j$. Similarly, we can write the Jarlskog invariant in terms of mixing angles and the phase $\delta$:
\beq
\begin{split}
J_\mathrm{CP} =&\, \frac{1}{2} \left| \im (W_{11}^*W_{12}W_{21}W_{22}^*)\right| \,=\, \frac{1}{2} \left| \im (W_{11}^*W_{13}W_{31}W_{33}^*)\right|\\
= &\, \frac{1}{2} \left| \im (W_{22}^*W_{23}W_{32}W_{33}^*)\right| \, = \, \frac{1}{2}\left|c_{12}\,c_{13}^2\, c_{23}\,\sin \delta \, s_{12}\,s_{13}\,s_{23}\right|\;.
\end{split}
\eeq
To conclude the Majorana phases are given by
\beq
\varphi_1=2\arg(e^{i\delta_e}\,W_{11}^*)\;,\qquad\qquad
\varphi_2=2\arg(e^{i\delta_e}\,W_{12}^*)\;,
\eeq
where $\delta_e=\arg(e^{i\delta}\,W_{13})$.

\section{Supersymmetry}
\label{Sec:SUSY}
\setcounter{footnote}{3}

In the Standard Model two Higgs parameters appear in the scalar potential: $m_H$ and $\lambda_H$, which are the mass and the quartic coupling of the Higgs boson, respectively. The Higgs VEV is linked to these parameter as
\beq
\mean{H}\equiv\dfrac{v}{\sqrt2}=\sqrt{-\dfrac{m_H^2}{2\lambda_H}}\;.
\eeq
Since $\lambda_H$ is bounded from above by various consistency conditions (such as perturbative unitarity), it follows that it should be roughly $-m^2_H\sim (100\GeV)^2$. However the mass parameter $m_H$ is expected to receive large radiative corrections, indeed it depends on the cutoff scale at which new physics is introduced: this leads to the well known ``hierarchy problem'' of particle physics.
If the cutoff scale is taken to be close to the Planck scale $M_P\approx10^{19}$ GeV, the corrections due to the fermion loops are much larger than the weak scale.

An elegant solution to the hierarchy problem is the low-energy supersymmetric theory. Supersymmetry (SUSY) can be considered an extension of the usual 4-dimensional space-time  Poincar\'e symmetry, in which new fermionic transformations, that change the spin of fields, are introduced. We consider only $N=1$ Supersymmetry, which is the simplest supersymmetric theory, in which a single set of supersymmetric generators is introduced.

It is not difficult to generalise the Standard Model description of section \ref{Sec:SM} to the supersymmetric context: each Standard Model field is considered to be a part of a superfield, $z$, and the Lagrangian of the model can be written as a sum of different terms in the following way
\beq
\LL\,=\;\int d^2\theta d^2\overline{\theta} {\cal K}(\ov{z}, e^{2 V} z)+\left[\int d^2 \theta w(z)+\hc\right]
+\frac{1}{4}\left[\int d^2\theta f(z) {\cal W W}+\hc\right]\;,
\label{leel}
\eeq
where $\cK(\ov{z},z)$ is the K\"ahler potential, a real gauge-invariant function of the chiral superfields $z$ and their conjugates, of dimensionality $(\mathrm{mass})^2$; $w(z)$ is the superpotential, an analytic gauge-invariant function of the chiral superfields, of dimensionality $(\mrm{mass})^3$; $f(z)$ is the gauge kinetic function, a dimensionless analytic gauge-invariant function; $V$ is the Lie-algebra valued vector supermultiplet, describing the gauge fields and their superpartners. Finally $\cW$ is the chiral superfield describing, together with the function $f(z)$, the kinetic terms of gauge bosons and their superpartners.

In the minimal supersymmetric Standard Model (MSSM) the field content of the Standard Model is increased by an extra Higgs $SU(2)_L$-doublet: the usual Standard Model Higgs $H$, defined in table \ref{SM:SMFermions}, is renamed to $H_d$ and it is responsible for giving mass to the down quarks, to the charged leptons and to their superpartners. The extra Higgs, $H_u$, is required to generate the Dirac mass of the up quarks (and of neutrinos if $\nu^c$ are included) and of their superpartners, as the holomorphicity requirement of the superpotential prevents the charge conjugate of $H_d$ from playing that role (in contrast to what happens with $H$ in the Standard Model).

The usual Standard Model fermions are contained in chiral superfields with their respective superpartners, bosons with spin 0 usually denoted as sfermions (squarks and sleptons). Analogously, the vector superfields contain the usual Standard Model gauge bosons and their own superpartners, spin $1/2$ fermions usually called gauginos (gluinos, photino, bino and winos). Finally, the two Higgs belong to chiral superfields with their superpartners, spin $1/2$ fermions denoted as Higgsinos. It is common to indicate with a ``$\sim$'' the component of a superfield which represents the superpartner of a Standard Model field.

The scalar potential $V\equiv V(\tilde{z},\,\tilde{z}^\dag)$ is composed of two contributions. One is usually called the $F$-term, obtained from the superpotential as $F_{i} \equiv \derp w(\tilde{z})/\derp\tilde{z}_{i}$, where $i$ is an index labelling the components of whatever representation the field has under the gauge group (for example, two components if the chiral superfield containing $\tilde{z}_{i}$ is a doublet of $SU(2)$). The other contribution is usually called the $D$-term, and is associated with the gauge group: $D^{a} \equiv g_a (M^a_{FI})^2-g_a \tilde{z}^{\dagger} T^{a} \tilde{z}$, where $a$ labels the generators $T^{a}$ of the group and $(M^a_{FI})^2$ denotes the contribution of the Fayet-Iliopoulos (FI) term, which may be non-zero only for Abelian $U(1)$ factors of the group. Assuming a canonical  K\"ahler  potential, $\cK=\ov{z}_iz_i$ and summarising the two contributions we have
\beq
V = F^\dag F + \dfrac{1}{2} D^{2} = \sum_{i} \left| \dfrac{dw(z)}{d\tilde{z}_{i}} \right|^{2} + \dfrac{1}{2} \sum_{a}\left(g_a (M^a_{FI})^2- g_{a}\tilde{z}^\dag T^{a} \tilde{z}  \right)^2\;.
\eeq

In terms of the hierarchy problem, $m_H$ receives new contributions from the Standard Model superpartners in such a way that the loop diagrams with superparticles in the loop have exactly the same value as those ones with Standard Model particles in the loop, but with opposite sign (due to the minus sign coming from the fermion loop): Supersymmetry enables the exact cancellation of the quadratic divergence, leaving only milder logarithmic divergences.

If from one side in supersymmetric theories there is a natural explanation of the hierarchy problem, dangerous gauge-invariant, renormalisable operators appear: the most general superpotential would include also terms which violate either the baryon number ($B$) or the total lepton number (L). The existence of these terms corresponds to $B$- and $L$-violating processes, which however have not been observed: a strong constraint comes from the non-observation of the proton decay. A possible way out to this problem is represented by the introduction of a new symmetry in MSSM, which allows the Yukawa terms, but suppresses $B$- and $L$-violating terms in the renormalisable superpotential. This new symmetry is called ``matter parity'' or equivalently ``$R$-parity''. The matter parity is a multiplicative conserved quantum number defined as
\beq
P_M=(-1)^{3(B-L)}
\eeq
for each particle in the theory. It is easy to check that quark and lepton supermultiplets have $P_M=-1$, while Higgs supermultiplets, gauge bosons and gauginos have $P_M=+1$. In the superpotential only terms for which $P_M=+1$ are allowed. The advantage of such a solution is that B and L are violated only due to non-renormalisable terms in the Lagrangian and therefore in tiny amounts.

It is common to use also a second definition of this symmetry: the $R$-parity refers to
\beq
P_R=(-1)^{3(B-L)+2s}\;,
\eeq
where $s$ is the spin of the particle. The two definitions are precisely equivalent, since the product of $(-1)^{2s}$ is always equal to $+1$ for the particles in a vertex that conserves angular moment. Under this symmetry all the Standard Model particles and the Higgs bosons have even $R$-parity ($P_R=+1$), while all their superpartners have odd $R$-parity ($P_R=-1$).

The phenomenological consequences of an $R$-parity conservation in a theory are extremely important: the lightest sparticle with odd $R$-parity is called lightest supersymmetric particle (LSP) and is absolutely stable; each sparticle, other then the LSP, must eventually decay in a state with an odd number of LSPs; sparticles can only be produced in even numbers, at colliders.\\

\begin{figure}[ht!]
\centering
\subfigure
{\includegraphics[width=7.8cm]{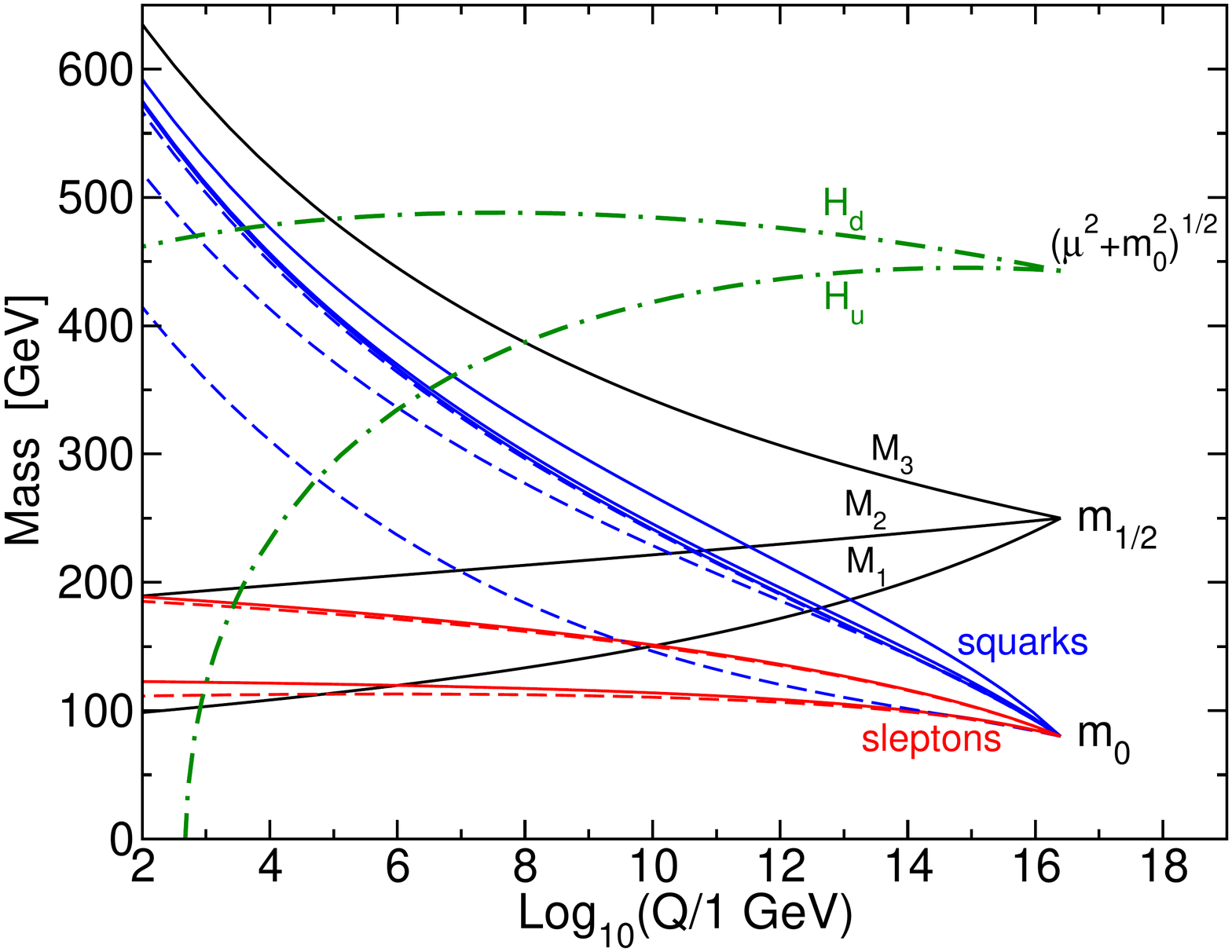}}
\subfigure
{\includegraphics[width=7.8cm]{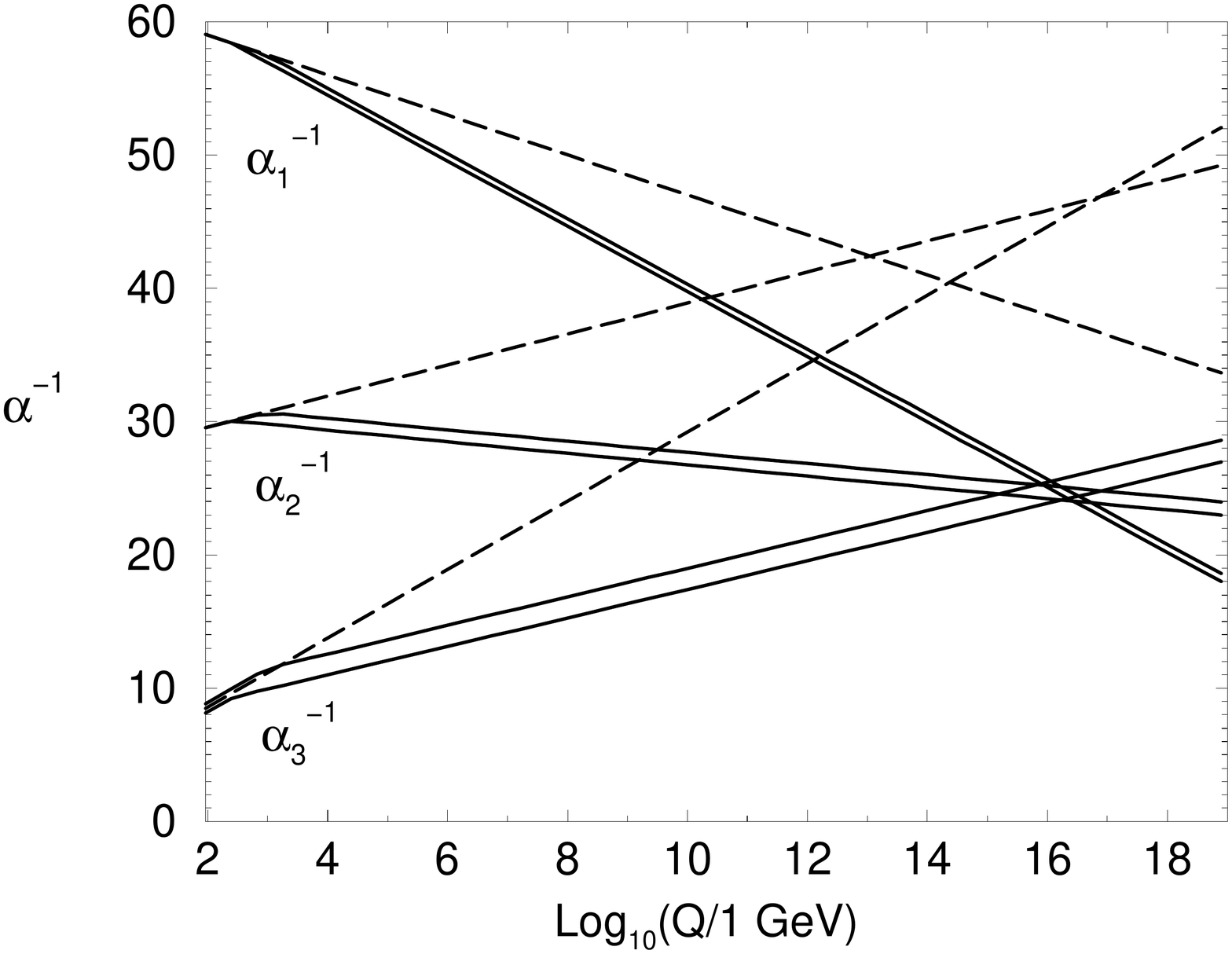}}
\caption{\it On the left, running of the superpartner masses (from \protect\cite{Martin}). On the right, running coupling constants (from \protect\cite{Martin}). The strong coupling represented by $\alpha_{3}(m_{Z})$ is varied from $0.113$ to $0.123$ and the mass thresholds are varied between $250$ and $1000$ GeV.}
\label{fig:SUSYmasses}
\end{figure}

Since any superparticle has not been observed yet, Supersymmetry must be broken at some scale higher than the electroweak scale. However, in order to solve the hierarchy problem, the breaking scale $\msusy$ has to be relatively low, not much higher than $1$ TeV.
The superparticle mass spectrum depends strongly on the Supersymmetry breaking mechanism. In figure \ref{fig:SUSYmasses} it is given an example of the evolution of superparticle masses with the energy scale $Q$, driven by radiative corrections of gauge (positive) and Yukawa (negative) contributions. Supergravity inspired boundary conditions have been implemented in the plot: common masses $m_0$ for the scalar partners and $m_{1/2}$ for the gauginos, imposed approximately at a unification scale $M_{GUT}$ of about $10^{16}$ GeV \cite{Martin}.

In figure \ref{fig:SUSYmasses}, $\mu$ represents the so-called $\mu$-term, the coupling of the two Higgs $\mu H_{u} H_{d}$. $M_{3}$, $M_{2}$ and $M_{1}$ are the gaugino masses, corresponding to the $SU(3)_{c}$, $SU(2)_{L}$ and $U(1)_{Y}$ gauge groups respectively, running from the common fermion mass $m_{1/2}$. The dashed lines refer to the third generation sfermions, and the solid lines to the other sfermions, all running from the common scalar mass $m_0$. It is interesting to note that radiative corrections due to the strong interactions dominate, driving the gluinos and the squarks considerably heavier than the other gauginos and sleptons. Furthermore the third generation sfermions are respectively lighter (particularly the stop and the sbottom) than the other two, receiving stronger Yukawa negative contributions.

From figure \ref{fig:SUSYmasses} we find another interesting feature of supersymmetric models: in the Standard Model there is no reason to have a negative $m_H^2$, but in figure \ref{fig:SUSYmasses} we can see that the mass of $H_u$ can be driven negative at low $Q$, due to the negative Yukawa contributions (largely due to the coupling to the top quark) which dominates over the gauge contributions. As a result in the supersymmetric context there is a natural explanation of the electroweak symmetry breaking.

The unification of the gauge coupling constants is strictly connected to the radiative corrections of the supersymmetric spectrum: there is no apriori reason to require this feature in a model, but in the Standard Model the gauge couplings seem to converge to a common value (dashed lines in figure \ref{fig:SUSYmasses}) and this suggests that the introduction of new physics could improve the unification. This is the case of the superparticles: considering again the supersymmetric breaking scale $\msusy$ at about $1$ TeV, the evolution is changed and the three couplings run together, as shown by the solid lines of figure \ref{fig:SUSYmasses}. Clearly, if $\msusy$ had been of a different order of magnitude, the unification would be lost. $\alpha_3$, $\alpha_2$ and $\alpha_1$ are the hyperfine constants defined as $\alpha_{a}=g_{a}^{2}/4 \pi$ and associated with $SU(3)_c$, $SU(2)_L$ and $U(1)_Y$, in the GUT normalisation, such that $g_2=g$ and $g_{1}=\sqrt{5/3} g'$.

\section{Grand Unified Theories}
\label{Sec:GUT}
\setcounter{footnote}{3}

Grand unified theories (GUTs) are the result of a common belief among many physicists that the apparent variety of interactions in Nature should eventually find a unified description, i.e. a theory with only one gauge coupling constant which is spontaneously broken at a very high energy scale. The idea is to have, at energies up to $\mgut \gg m_Z$, a simple gauge group G which is spontaneously broken down to the Standard Model gauge group:
\beq
G\stackrel{\mgut}{\longrightarrow} SU(3)_c\times SU(2)_L\times U(1)_Y\stackrel{m_Z}{\longrightarrow}SU(3)_c\times U(1)_{em}\;.
\eeq
To have complete unification (a single gauge coupling constant) and to have the Standard Model as the low-energy effective representative (Standard Model as a subgroup), the group $G$ must be simple and of rank $r\geq 4$. Furthermore, $G$ must allow for complex but anomaly-free representations in order to correctly embed the Standard Model fermions. There are only few groups which fulfill all these requirements and the simplest solutions are $SU(5)$ and $SO(10)$ of rank 4 and 5, respectively. It is relevant to note that the minimal versions of GUTs are not realistic: they suffer from serious problems such as the explanation of the correct symmetry breaking scheme, the prediction of a sufficiently long proton lifetime and the correct description of fermion masses and mixings.\\

The simplest GUT is the minimal $SU(5)$ model by Georgi and Glashow \cite{MinimalSU5}. The gauge bosons of this model belong to the adjoint $\bf24$-dimensional representation of $SU(5)$, which decomposes under the Standard Model gauge group as:
\beq
{\bf24}=({\bf8},\,{\bf1},0)+({\bf1},\,{\bf3},0)+({\bf1},\,{\bf1},0)+({\bf3},\,{\bf2},-5/6)+({\bf\overline{3}},\,{\bf2},5/6)
\eeq
where the first three terms are the usual Standard Model gauge bosons and the others are $12$ new bosons with both colour and weak isospin. Each fermion generation must be arranged in a representation of $SU(5)$ which decomposes under $SU(3)_c\times SU(2)_L$ as
\beq
2\times({\bf\ov{3},{\bf1}})+({\bf3},{\bf2})+({\bf1},{\bf2})+({\bf1},{\bf1})
\eeq
and this property is present in the reducible ${\bf\ov{5}}+{\bf 10}$ representation of $SU(5)$:
\bea
{\bf\ov{5}}&=&({\bf\ov{3},{\bf1}})+({\bf1},{\bf2})=\left(d_1^c,\,d_2^c,\,d_3^c,\,e,\,-\nu\right)^T\\[3mm]
{\bf10}&=&({\bf\ov{3},{\bf1}})+({\bf3},{\bf2})+({\bf1},{\bf1})=\left(
                                                                 \begin{array}{ccc|cc}
                                                                   0 & u_3^c & -u_2^c & -u_1 & -d_1 \\
                                                                   -u_3^c & 0 & u_1^c & -u_2 & -d_2 \\
                                                                   u_2^c & -u_1^c & 0 & -u_3 & -d_3 \\
                                                                   &&&&\\[-5mm]
                                                                   \hline\\[-5mm]
                                                                   u_1 & u_2 & u_3 & o & -e^c \\
                                                                   d_1 & d_2 & d_3 & e^c & o \\
                                                                 \end{array}
                                                               \right)\;.
\eea

The scalar content of the model contains two fields, $\phi_{24}$ and $\phi_{\ov5}$ which transform as $\bf24$ and $\bf{\ov{5}}$ of $SU(5)$, respectively. When these fields get VEVs then, $\mean{\phi_{24}}\approx\mgut$ breaks $SU(5)$ down to $SU(3)_c\times SU(2)_L\times U(1)_Y$ and after $\mean{\phi_{\ov5}}\approx m_Z$ down to $SU(3)_c\times U(1)_\mathrm{em}$. The two breakings occur at two very distinct energy scales: the first at $\mgut\sim10^{15\div16}$ GeV and the second at $m_Z$. As a result there is a large ratio between the two VEVs: this is the well-known hierarchy problem and corresponds to a fine-tuning on the parameters of about $14$ order of magnitude. Many attempts have been proposed to solve this problem, but all of them require a non-minimal extension of the model.\\

In order to go further, it is possible to extend the symmetry to $SO(10)$ \cite{SO10Original}, where there is not only gauge coupling unification, but also every fermion of one generation, including right-handed neutrinos, fits in one single fundamental representation, the $\bf16$ of $SO(10)$: it decomposes under $SU(3)_c\times SU(2)_L$ as
\beq
\begin{split}
{\bf16}&=\,({\bf3},\,{\bf2})+2\times({\bf{\ov3}},\,{\bf1})+({\bf1},\,{\bf2})+({\bf1},\,{\bf1})+({\bf1},\,{\bf1})\\[3mm]
&=\,\left(\nu,\,u_1,\,u_2,\,u_3,\,e,\,d_1,\,d_2,\,d_3,\,| \,-d_3^c,\,d_2^2,\,d_1^c,\,-e^c,\,u_3^c,\,-u_2^2,\,-u_1^c,\,\nu^c\right)^T\;.
\end{split}
\eeq
The gauge bosons belong to the $\bf45$ representation of $SO(10)$: it contains the same gauge bosons of $SU(5)$ and additional 21 new states.

In order to break $SO(10)$ down to the Standard Model gauge group it is necessary to use the spinorial representation $\bf{\ov{16}}$ and the representation $\bf126$, which is contained in the symmetric part of the $\bf{\ov{16}}\times\bf{\ov{16}}$ product (the VEV of an adjoint representation does not lower the rank of the group). In this minimal $SO(10)$ model, the VEVs of these scalar fields contain three free parameters which determine the type of the breaking: it is possible to have a superstrong breaking (directly down to $SU(3)_c\times SU(2)_L\times U(1)_Y$) at $\mgut\approx10^{15\div16}$ GeV  as well as a two-fold breaking
\beq
SO(10)\stackrel{M_X}{\longrightarrow}G'\stackrel{\mgut}{\longrightarrow} SU(3)_c\times SU(2)_L\times U(1)_Y\;,
\eeq
with $M_X>\mgut$. There are two inequivalent maximal breaking patterns: \linebreak\mbox{$SO(10)\longrightarrow SU(5)\times U(1)_X$} or $SO(10)\longrightarrow SU(4)_c\times SU(2)_L\times SU(2)_R$. The first possibility corresponds to the case discussed above, where the Standard Model gauge group is achieved through the subsequent breaking of $SU(5)$. What is relevant is that the additional $U(1)_X$ can remain unbroken up to a scale close to $m_Z$, thus giving rise to modifications of the usual neutral current phenomenology. On the other hand the second possibility corresponds to the well-known Pati-Salam (PS) GUT \cite{PatiSalam}.\\

The Pati-Salam group is one example of a partial GUT which ties quarks and leptons together: the leptons are seen as the extra ``colour'' of $SU(4)_c$. Furthermore the $SU(2)_R$ factor makes the model left-right symmetric. Although with Pati-Salam there was fermion unification at some extent, the gauge couplings remain independent parameters and for this reason it is only a partial GUT.

Each of the three families has one left-handed multiplet including left-handed quark and lepton doublets $F=(Q,\ell)$, and one right-handed multiplet $F^c=(Q^{c},\ell^{c})$ including the charge conjugates of the right-handed states that now belong to their own doublets ($Q^{c}$ and $\ell^{c}$). The explicit matrix representation of the two multiplets is given by
\beq
F\sim({\bf4},\,{\bf2},\,{\bf1})=\left(
                                  \begin{array}{cccc}
                                    u_1 & u_2 & u_3 & \nu \\
                                    d_1 & d_2 & d_3 & e \\
                                  \end{array}
                                \right)\;,\qquad
F^c\sim({\bf{\ov{4}}},\,{\bf1},\,{\bf{\ov{2}}})=\left(
                                  \begin{array}{cccc}
                                    d_1^c & d_2^c & d_3^c & e^c \\
                                    u_1^c & u_2^c & u_3^c & \nu^c \\
                                  \end{array}
                                \right)\;.
\label{GUT:PSassignment}
\eeq
From eq. (\ref{GUT:PSassignment}), we can see that the right-handed neutrinos are now naturally introduced together with the charge conjugates of the right-handed charged leptons, $e^{c}$.

The breaking to the Standard Model gauge group originates by the introduction of three different scalar fields: $\Delta_L\sim({\bf\ov{10}},\,{\bf2},\,{\bf1})$, $\Delta_R\sim({\bf\ov{10}},\,{\bf1},\,{\bf2})$ and $\phi\sim({\bf1},\,{\bf2},\,{\bf1})$, being the $\bf10$ the symmetric part of the ${\bf4}\times{\bf4}$ product, which under $SU(3)_c$ decomposes as ${\bf10}={\bf6}+{\bf3}+{\bf1}$. The VEV of the $SU(3)_c$ singlet part of $\Delta_R$ does not break $SU(3)_c$, $SU(2)_L$ and $U(1)_\mathrm{em}$, but since $\Delta_R$ is charged under $SU(4)_c$ and $SU(2)_R$, these two groups are spontaneously broken. As a result, when $\Delta_R$ develops a VEV, the first breaking step occurs:
\beq
SU(4)_c\times SU(2)_L\times SU(2)_R \stackrel{\mean{\Delta_R}}{\longrightarrow} SU(3)_c\times SU(2)_L\times U(1)_Y\;.
\eeq
The last step of the electroweak symmetry breaking is accomplished to the VEV of $\phi$.\\

A common interesting feature of GUTs is the presence of new interactions, which could have a strong phenomenological impact: among the others the proton decay is a strong test for any GUT. Considering for example the minimal $SU(5)$ model, proton decay arises from four fermion operators with a prediction for the proton lifetime smaller that the present lower bound. This represents a failure of the minimal $SU(5)$ model, which however can be overcome considering non-minimal extension of this model.

Before concluding the section, we briefly comment on the possibility of supersymmetric GUTs. In this kind of models, we can see the possibility to solve both the hierarchy problem and the proliferation of many free parameters. Taking as an example the minimal supersymmetric $SU(5)$ model \cite{MinimalSUSYSU5}, the simplest supersymmetric GUT realisation, gauge bosons and fermions are given to the same representations of the non-supersymmetric minimal $SU(5)$, but are promoted to corresponding supermultiplets. The Higgs sector is enlarged by the introduction of an additional $\phi_{5}$ which transforms as a $\bf 5$ under the gauge group. Also in the supersymmetric variant of the model it is necessary a parameter fine-tuning in order to keep small the mass of the Higgs doublets.

In this minimal supersymmetric $SU(5)$ model the hierarchy problem finds a natural solution, the gauge coupling constant unification is improved with respect to the MSSM and a solution to the proton decay can be implemented by the introduction of the $R$-parity. On the other hand the Supersymmetry breaking mechanism (whatever it is) introduces a lot of new parameters.

We have commented on different possibilities for a GUT scenario, including the interplay with Supersymmetry, but in any of these models the existence of three families and the explanation of their mass hierarchies and mixings remain an open problem which could only be solved by the extension of these theories to a non-minimal treatment. In the following chapter we directly face the problem of the flavour.

\clearpage{\pagestyle{empty}\cleardoublepage}


\newpage
\chapter{The Flavour Puzzle}
\label{Sec:FlavourPuzzle}
\setcounter{equation}{0}
\setcounter{footnote}{3}

In this chapter we focus on the flavour sector of the Standard Model, which is the origin of the majority of the free parameters. Including the neutrino masses, the complete list accounts for $26$ or $28$ low-energy free parameters, depending on the lepton number conservation or violation (or alternatively on the Dirac or Majorana nature of neutrinos). Five of these are flavour universal: the three gauge coupling constants $g,\,g',\,g_3$, one Higgs quartic coupling $\lambda_H$ and one Higgs mass squared $m_H^2$. The rest are parameters associated to the fermion masses and mixings: the masses for the six quarks, the three charged leptons and three neutrinos; three angles and one Dirac phase in the quark sector; three angles, one Dirac phase and possibly two Majorana phases in the lepton sector. The last parameter is the strong CP-violating parameter $\ov{\te}_{CP}$, which is intimately related to the quark masses.

While from the experimental point of view there is abundant information on the numerical values of (almost all) these parameters, from the theoretical side there are several fundamental open questions: why are there three generations of fermions? why are quarks and charged leptons strongly hierarchical? why do neutrinos not show the same hierarchy? why is the neutrino absolute mass scale much smaller than the charged fermion masses? why are the quark mixing angles much smaller than (at least two of) the lepton mixing ones? are the masses and the mixings free parameters? are the mixing parameters related to the masses? why is $\ov{\te}_{CP}<10^{-9}$? what is the origin of the CP violation? The lack of a fundamental understanding of all these problems is addressed as the ``flavour puzzle''.\\

Moving to more general considerations, the flavour problem is one of the key aspects in all the extensions of the Standard Model. Grand unified theories (GUTs) can help in (partially) improving the situation: besides their aesthetic appeal, GUTs reduce the number of free parameters. Apart the unification of the gauge coupling constants, several relations among fermion masses have been found: in the minimal $SU(5)$ model, at the GUT scale, this relation holds
\beq
M_d=M_e^T\;,
\eeq
which leads to the following expressions for the mass eigenvalues
\beq
m_b=m_\tau\;,\qquad m_s=m_\mu\;,\qquad m_d=m_e\;.
\label{FP:MassesMinimalSU5}
\eeq
It is easy, however, to see that this result is not in agreement with the experimental measurements. In order to test the validity of these predictions we should extrapolate the masses from the GUT threshold to the low-energy scale. We can see, however, that the last two relations of eq. (\ref{FP:MassesMinimalSU5}) turn out to be not acceptable even without going through the renormalisation group (RG) evolution: eq. (\ref{FP:MassesMinimalSU5}) implies $m_s/m_d=m_\mu/m_e$ which is independent from the running. We can directly compare these mass ratios with the observations and we find that $m_s/m_d\simeq20$ and $m_\mu/m_e\simeq200$, concluding that this relation is off by an order of magnitude.

With this simple example, we understand that a non-minimal extension of the model should be considered and, usually, this introduces new parameters which reduce the advantages of using a GUT to explain the origin of fermion masses and mixings.\\

When we consider (broken) supersymmetric theories, mainly motivated to solve the hierarchy problem, the flavour sector suffers for the introduction of many new mass and mixing parameters. Furthermore, while in the Standard Model, not including right-handed neutrinos, the flavour symmetry $[U(3)]^5$ is strongly violated only by the CKM matrix, so that there is a natural suppression of all flavour-changing and CP-violating effects, in a general supersymmetric theory, once non-universal soft breaking terms are added, new sources of flavour and CP violation are introduced and the theory is subject to very stringent constraints from the experimental measurements. A dangerous source of these effects are the off-diagonal entries of the sfermion masses in the generation space: strong constraints come from $\mu\to e\gamma$ and $b\to s\gamma$ decays, the $K^0-\ov{K}^0$ and $B^0-\ov{B}^0$ systems, electric and magnetic dipole moments, etc \ldots

It is interesting to note, however, that the minimal supersymmetric Standard Model (MSSM) respects all these constraints, when universal boundary conditions on the soft terms are assigned: indeed the sfermion masses are universal at the unification scale and the only non-universality is the one induced by the renormalisation group running down to the electroweak energy scale. However, this result is not a proper feature of MSSM, but it characterises a larger class of models in which the flavour sector is determined by the universality conditions. In the next section we briefly discuss about this general approach to soften the flavour problem.

In section \ref{Sec:FS} we illustrate a second interesting approach which consists in studying the presence of vanishing entries in the fermion mass or mixing matrices: people usually refers to this kind of studies with the name of texture zeros and are useful to understand from a theoretical point of view which flavour pattern might be a good approximation of the experimental measurements. Among these, we focus on the bimaximal and the tribimaximal mixing patterns, which received particularly attention in the last years.

In section \ref{Sec:FSym} we give a brief overview on the flavour symmetries which have been introduced to recover particular flavour structures, able to reproduce the measured fermion masses and mixings: in particular, we comment on the advantages and disadvantages of using symmetries which can be either Abelian or non-Abelian, either local or global, either continuous or discrete. We focus only on flavour symmetries which commute with the underlying gauge symmetry group. In this section we also comment on the necessary requirement to implement a (spontaneous) breaking mechanism of the flavour symmetry in order to correctly describe fermion masses and mixings.

\section{The Minimal (Lepton) Flavour Violation}
\label{Sec:MFV}
\setcounter{footnote}{3}

The Minimal Flavour Violation (MFV) models \cite{MFV} are the simplest class of extensions of the Standard Model attempting to solve the flavour problem. They are based on the introduction of the symmetry $G_f=[U(3)]^5$, acting only among the fermion generations, which corresponds to the largest group of unitary field transformations that commutes with the Standard Model gauge group, not including right-handed neutrinos. $G_f$ can be decomposed as
\beq
G_f=SU(3)_Q^3\times SU(3)_L^2\times U(1)_B\times U(1)_L\times U(1)_Y\times U(1)_{PQ}\times U(1)_{e^c}
\eeq
where
\bea
SU(3)_Q^3 &=& SU(3)_q\times SU(3)_{u^c}\times SU(3)_{d^c}\\[3mm]
SU(3)_L^2 &=& SU(3)_\ell\times SU(3)_{e^c}\;.
\eea
The $U(1)$ factors can be identified with the baryon number $U(1)_B$, the lepton number $U(1)_L$, the hypercharge $U(1)_Y$, the Peccei-Quinn symmetry $U(1)_{PQ}$ of two-Higgs-doublet models \cite{PecceiQuinnMFV} and with a rotation which affects $e^c$ only, $U(1)_{e^c}$. Under $SU(3)_Q^3\times SU(3)_L^2$ the fermions transform as
\bea
&q\sim({\bf\ov3},\,{\bf1},\,{\bf1};\,{\bf1},\,{\bf1})\;,\qquad
u^c\sim({\bf1},\,{\bf3},\,{\bf1};\,{\bf1},\,{\bf1})\;,\qquad
d^c\sim({\bf1},\,{\bf1},\,{\bf3};\,{\bf1},\,{\bf1})\;,&\nn\\[3mm]
&\ell\sim({\bf1},\,{\bf1},\,{\bf1};\,{\bf\ov3},\,{\bf1})\;,\qquad
e^c\sim({\bf1},\,{\bf1},\,{\bf1};\,{\bf1},\,{\bf3})\;.&\nn
\eea

In the Standard Model the Yukawa interactions break the symmetry group $SU(3)_Q^3\times SU(3)_L^2\times U(1)_{PQ}\times U(1)_{e^c}$, but preserve $B$, $L$ and $Y$. We can recover flavour invariance by introducing dimensionless auxiliary fields $Y_e$, $Y_d$ and $Y_u$ transforming under $SU(3)_Q^3\times SU(3)_L^2$ as
\beq
Y_e\sim ({\bf1},\,{\bf1},\,{\bf1};\,{\bf3},\,{\bf\ov3})\;,\qquad
Y_d\sim ({\bf3},\,{\bf1},\,{\bf\ov3};\,{\bf1},\,{\bf1})\;,\qquad
Y_u\sim ({\bf3},\,{\bf\ov3},\,{\bf1};\,{\bf1},\,{\bf1})\;.
\eeq
This allows to write the Lagrangian with the same appearance of the Yukawa interactions
\beq
\LL_{\rm MFV} = e^c\,Y_e\,H^\dag\ell + d^c\,Y_d\,H^\dag q + u^c\,Y_u\,\widetilde{H}^\dag q + \hc\;.
\eeq
This expression describes the most general coupling of the fields $Y_i$ to the renormalisable Standard Model operators.

The fermion masses and the quark mixing are then recovered allowing the auxiliary fields to develop a VEV. Using the flavour symmetry, it is possible to write these VEVs as
\bea
\mean{Y_e} &=& \dfrac{\sqrt2}{v}\diag(m_e,\,m_\mu,\,m_\tau)\;,\nn\\[3mm]
\mean{Y_d} &=& \dfrac{\sqrt2}{v}\diag(m_d,\,m_s,\,m_b)\;,
\label{MFV:DiagonalBasis}\\[3mm]
\mean{Y_u} &=&\dfrac{\sqrt2}{v}\diag(m_u,\,m_c,\,m_t)V\;,\nn
\eea
where $m_i$ are the fermion masses, $v/\sqrt2$ is the VEV of the neutral component of the Higgs field and $V$ is the CKM matrix.

We can now rewrite the MFV leading principle in different terms: an effective low-energy theory satisfies the criterion of MFV if all higher-dimensional operators, constructed from the Standard Model fields and the $Y_i$ auxiliary fields, are invariant under CP and under the flavour group $G_f$. In this way any flavour and CP violation contribution is completely determined by the CKM matrix.\\

In order to account for the neutrino masses it is necessary to extend such a treatment to the Minimal Lepton Flavour Violation (MLFV) context \cite{MLFV1,MLFVother}. Similarly to the discussion in section \ref{Sec:SM}, we should distinguish the case in which the source of neutrino masses is the Weinberg operator or the introduction of new fields:
\begin{description}
\item[Minimal field content.] In this case the MLFV respects two hypothesis: the breaking of the lepton number occurs at a very high energy scale $\Lambda_L$ which is unrelated to the flavour symmetry $G_f$; with respect to the MFV scenario, there is only an additional source of flavour violation in the lepton sector, which is $Y_\nu\sim({\bf1},\,{\bf1},\,{\bf1};\,{\bf6},\,{\bf1})$ defined as
\beq
\LL_{\rm MLFV} = \LL_{\rm MFV} +  \dfrac{1}{2\Lambda_L} (\widetilde{H}^\dag\ell)^T\,Y_\nu\,(\widetilde{H}^\dag\ell)+ \hc\;.
\eeq
The neutrino masses originate when $Y_\nu$ develops a VEV and, using the invariance under $G_f$ to rotate the fields in the basis of eq. (\ref{MFV:DiagonalBasis}), we can express $\mean{Y_\nu}$ in terms of neutrino masses and lepton mixings:
\beq
\mean{Y_\nu}=\dfrac{\Lambda_L}{4v^2}U^*\diag(m_1,\,m_2,\,m_3)U^\dag\;,
\eeq
where $m_i$ are the neutrino masses and $U$ is the PMNS mixing matrix.

\item[Extended field content.] In this scenario three right-handed neutrinos are considered and the flavour group enlarges to account for an additional $SU(3)$ term: $G_f\times SU(3)_{\nu^c}$. Only the right-handed neutrinos transform under the additional symmetry term, $\nu^c\sim({\bf1},\,{\bf1},\,{\bf1};\,{\bf1},\,{\bf1},\,{\bf3})$. There is a large freedom in the way of breaking this group to generate the observed masses and mixings: it is possible to introduce neutrino mass terms transforming as $({\bf1},\,{\bf1},\,{\bf1};\,{\bf6},\,{\bf1},\,{\bf1})$, $({\bf1},\,{\bf1},\,{\bf1};\,{\bf1},\,{\bf1},\,{\bf\ov6})$ and $({\bf1},\,{\bf1},\,{\bf1};\,{\bf3},\,{\bf1},\,{\bf\ov3})$. All these possibilities correspond to a Majorana mass term for the left-handed neutrinos, for the right-handed neutrinos and to a Dirac mass term, respectively. In \cite{MLFV1} a discussion of these cases is presented and some differences with respect to the previous scenario are found.
\end{description}

In \cite{MFV,MLFV1,MLFVother} a general classification of six-dimensional effective operators is presented, allowing the study of several flavour violating transitions. The conclusions are that MFV respects the strong constraints coming from LFV and FCNC processes, such as $\mu\to e\gamma$, $b\to s\gamma$, etc \ldots. However, while such a choice has the advantage that it can accommodate any pattern of fermion masses and mixing angles, it does not provide any explanation for the origin of the particular structure for the VEV of the auxiliary fields which break the flavour symmetry or, in other words, any explanation for the fermion mass hierarchies and for the specific patterns of the mixing matrices. This defect suggests to search for other flavour symmetries which, leading to correct fermion masses and mixings without introducing unwanted flavour changing phenomena, explain the origin of the observed flavour phenomenology.

\section{Flavour Structure Approach}
\label{Sec:FS}
\setcounter{footnote}{3}

In this section we review some ideas to go further with respect to the MFV approach. Instead of starting from a universality principle, the idea is to study particular flavour patterns for the fermion mass matrices which deal to realistic mixing angles and fermion hierarchies. A very simple strategy is to require that certain elements of the fermion mass matrices are negligible and, as a result, some matrix entries can be set to zero. People usually refer to these constructions as texture zeros: they are not motivated by any specific principle, but only to increase the predictivity of the model, indeed this approach allows to reduce the number of free parameters.

In the quark sector, four \cite{4ZerosQuarks}, five and six \cite{56ZerosQuarks} texture zeros have been extensively studied and some of them are able to correlate the quark masses with some elements of the CKM matrix, getting relations as the following \cite{GST_Relation,QuarkRelations}:
\beq
\left|\dfrac{V_{td}}{V_{ts}}\right| = \sqrt{\dfrac{m_d}{m_s}}\;,\qquad\qquad
\left|\dfrac{V_{ub}}{V_{cb}}\right| = \sqrt{\dfrac{m_u}{m_c}}\;.
\eeq

In the lepton sector, mass matrices with different number of zeros and with zeros in various places have been considered \cite{ZerosNeutrinos}. In particular, the charged lepton mass matrix is usually assumed to be diagonal and all the flavour information are encoded in the neutrino sector: if neutrinos are of Majorana type the mass matrix must be symmetric and, as a result, there are several textures with only two independent zeros, while schemes with a larger number of zeros appear to be excluded by experiments; on the contrary, if neutrinos are of Dirac nature textures with more than two zeros are allowed. A common problem of these studies is the stability of the textures zeros under corrections due to the renormalisation group running from the cutoff, at which the zeros are imposed, down to the electroweak scale: this effect is particularly relevant when the neutrino spectrum is quasi degenerated or inversely hierarchical, as discussed more in details in section \ref{Sec:Running}.

The approach of the textures zeros should be taken with some caution: for instance, it is not clear the origin of the zeros which appear in the mass matrices; furthermore the zeros could not be exactly zeros, partially reducing the predictivity of the model. In any case, these flavour patterns could help in understanding the correct approach to explain fermion mass hierarchies and mixings.\\

Following the philosophy of the textures zeros, which can be seen as particular flavour patterns of the mass matrices, a similar approach has been used to discuss interesting flavour structures for the mixing matrices, with particular attention to the lepton sector.

As already discussed in section \ref{Sec:SM:PhysicalBasis}, the pattern of the lepton mixings is characterised by two large angles and a small one: the atmospheric angle is compatible with a maximal value; the solar angle is large, but not maximal; the reactor angle only has an upper bound and it is well compatible with a vanishing value. Looking at this scenario, for long time people tried to reproduce models with a lepton mixing matrix characterised by a maximal angle and a vanishing one, $\theta_{23}=\pi/4$ and $\theta_{13}=0$: apart from sign convention redefinition,
\beq
U_{\mu-\tau}=\left(
         \begin{array}{ccc}
           c_{12} & s_{12} & 0 \\
           -s_{12}/\sqrt2 & c_{12}/\sqrt2 & -1/\sqrt2 \\
           -s_{12}/\sqrt2 & c_{12}/\sqrt2 & +1/\sqrt2 \\
         \end{array}
       \right)\;,
\eeq
in the basis where charged leptons are diagonal. The most general neutrino mass matrix which can be diagonalised by $U_{\mu-\tau}$ is $\mu-\tau$ symmetric, for which $(m_\nu)_{2,2}=(m_\nu)_{3,3}$ and $(m_\nu)_{1,2}=(m_\nu)_{1,3}$, and is given by \cite{MuTauSymMatrix}
\beq
m_\nu=\left(
  \begin{array}{ccc}
    x & y & y \\
    y & z & w \\
    y & w & z \\
  \end{array}
\right)\;.
\eeq
Since the reactor angle is vanishing, there is no CP violation and the only phases are of Majorana type. Collecting apart the Majorana phases, the mass matrix depends on four real parameters: the three masses and the remaining angle, the solar one, which can be written in terms of mass parameters as
\beq
\sin^2{2\theta_{12}}=\dfrac{8y^2}{(x-y-z)^2+8y^2}\;.
\label{FS:MuTauSolarAng}
\eeq
Many models have been constructed introducing a flavour symmetry in addition to the gauge group of the Standard Model in order to provide the $\mu-\tau$ symmetric pattern for the neutrino mass matrix, but in all of these the solar angle remains undetermined. It is therefore necessary some new ingredients other than the $\mu-\tau$ symmetry to describe correctly the neutrino mixings from the theoretical point of view. In what follows we review two relevant flavour structures, which can be considered an upgrade of the $\mu-\tau$ symmetry: the bimaximal (BM) and the tribimaximal (TB) patterns.

\subsection{The Bimaximal Mixing Pattern}
\label{Sec:FS:BM}
\setcounter{footnote}{3}

In the so-called bimaximal pattern \cite{BMmixing} while $\theta_{13}=0$, $\theta_{23}$ and $\theta_{12}$ are assumed to be maximal. A maximal solar angle can be alternatively written as $\sin^2{\theta_{12}}=1/2$, which, comparing with eq. (\ref{FS:MuTauSolarAng}), corresponds to a well defined relation between the mass parameters: $w=x-z$. The most general mass matrix of the BM-type can be written as
\beq
m_\nu=\left(
  \begin{array}{ccc}
    x & y & y \\
    y & z & x-z \\
    y & x-z & z \\
  \end{array}
\right)
\label{FS:BM:GeneralMassMatrix}
\eeq
and satisfies to the $\mu-\tau$ symmetry and to an additional symmetry for which $(m_\nu)_{1,1}=(m_\nu)_{2,2}+(m_\nu)_{2,3}$. The $\mu-\tau$ symmetry is responsible for $\theta_{13}=0$ and, as discussed in section \ref{Sec:SM:PhysicalBasis}, the CP-violating phase does not contribute. Apart from the Majorana phases, eq. (\ref{FS:BM:GeneralMassMatrix}) depends on only three real parameters, the masses, which can be written in terms of the mass parameters $x$, $y$ and $z$:
\beq
m_1=x+\sqrt2y\;,\qquad
m_2=x-\sqrt2y\;,\qquad
m_3=2z-x\;.
\eeq
These masses are the eigenvalues of eq. (\ref{FS:BM:GeneralMassMatrix}), while the eigenstates define the unitary transformation which diagonalises the mass matrix in such a way that $m_\nu^\diag=U_{BM}^T m_\nu U_{BM}$, where the unitary matrix is given by
\beq
U_{BM}=\left(
         \begin{array}{ccc}
           1/\sqrt2 & -1/\sqrt2 & 0 \\
           1/2 & 1/2 & -1/\sqrt2 \\
           1/2 & 1/2 & +1/\sqrt2 \\
         \end{array}
       \right)\;.
\label{FS:BM:BMmixing}
\eeq
Notice that $U_{BM}$ does not depend on the mass parameters $x$, $y$, $z$, or on the mass eigenvalues, in contrast with the quark sector, where the entries of the CKM matrix can be written in terms of the ratio of the quark masses. This feature puts the bimaximal pattern in the class of the mass-independent textures \cite{LV_Theorem}.

It is useful to express eq. (\ref{FS:BM:GeneralMassMatrix}) in terms of $m_i$ instead of $x$, $y$ and $z$:
\beq
\begin{split}
m_\nu&=\,U_{BM}\,\diag(m_1,\,m_2,\,m_3)\,U_{BM}^T\\[3mm]
&=\,\dfrac{m_1}{4}\left(\begin{array}{ccc}
                        2 & \sqrt2 & \sqrt2 \\
                        \sqrt2 & 1 & 1 \\
                        \sqrt2 & 1 & 1 \\
                        \end{array}
                    \right)
    +\dfrac{m_2}{4}\left(\begin{array}{ccc}
                        2 & -\sqrt2 & -\sqrt2 \\
                        -\sqrt2 & 1 & 1 \\
                        -\sqrt2 & 1 & 1 \\
                        \end{array}
                    \right)
    +\dfrac{m_3}{2}\left(\begin{array}{ccc}
                        0 & 0 & 0 \\
                        0 & 1 & -1 \\
                        0 & -1 & 1 \\
                        \end{array}
                    \right)\;.
\end{split}
\eeq
Clearly, all type of hierarchies among neutrino masses can be accommodated. The smallness of the ratio $r=\Delta m^2_{sun}/\Delta m^2_{atm}$
requires either $\vert xy\vert \ll \vert z^2\vert$ (normal hierarchy) or $\vert x\vert \sim \vert z\vert\ll \vert y\vert$ (inverse hierarchy) or $\vert y\vert \ll \vert x\vert \sim \vert z\vert$ (approximate degeneracy except for $x\sim 2z$).

A final comment on the agreement of this scheme with the experimental data is worth. In the bimaximal pattern the solar angle is assumed maximal, $\sin^2{\theta_{12}}= 1/2$, to be compared with the latest experimental determination: at $3\sigma$ error level, 
\mbox{$\sin^2{\theta_{12}}= 0.26-0.37$} from \cite{Fogli:Indication} or $\sin^2{\theta_{12}}= 0.25-0.37$ from \cite{Maltoni:Indication}, and the bimaximal pattern can be considered at most as a zeroth order approximation that needs large corrections.

\subsection{The Tribimaximal Mixing Pattern}
\label{Sec:FS:TB}
\setcounter{footnote}{3}

In the so-called tribimaximal or Harrison-Perkins-Scott pattern \cite{HPS} a vanishing reactor angle, a maximal atmospheric angle and $\sin^2{\theta_{12}}=1/3$ are assumed. From eq. (\ref{FS:MuTauSolarAng}) it results $w=x+y-z$ and therefore the most generic mass matrix of the TB-type is given by
\beq
m_\nu=\left(
  \begin{array}{ccc}
    x & y & y \\
    y & z & x+y-z \\
    y & x+y-z & z \\
  \end{array}
\right)\;.
\label{FS:TB:GeneralMassMatrix}
\eeq
This matrix satisfies the $\mu-\tau$ symmetry and the so-called magic symmetry, for which \mbox{$(m_\nu)_{1,1}=(m_\nu)_{2,2}+(m_\nu)_{2,3}-(m_\nu)_{1,3}$}. The $\mu-\tau$ symmetry assures that the reactor angle is vanishing and as a result the CP phase disappears. Disregarding the Majorana phases, the mass matrix depends on only three real parameters, the masses, which can be written in terms of the parameters $x$, $y$ and $z$:
\beq
m_1=x-y\;,\qquad
m_2=x+2y\;,\qquad
m_3=2z-x-y\;.
\eeq
These eigenvalues come from the diagonalisation of eq. (\ref{FS:TB:GeneralMassMatrix}) by the use of a unitary transformation in such a way that $m_\nu^\diag=U_{TB}^T m_\nu U_{TB}$, where the unitary matrix is given by
\beq
U_{TB}=\left(
         \begin{array}{ccc}
           \sqrt{2/3} & 1/\sqrt3 & 0 \\
           -1/\sqrt6 & 1/\sqrt3 & -1/\sqrt2 \\
           -1/\sqrt6 & 1/\sqrt3 & +1/\sqrt2 \\
         \end{array}
       \right)\;.
\label{FS:TB:TBmixing}
\eeq
Notice that $U_{TB}$ does not depend on the mass eigenvalues, in complete analogy to the bimaximal pattern of eq. (\ref{FS:BM:BMmixing}), and therefore it belongs to the class of mass-independent textures.

It is useful to write eq. (\ref{FS:TB:GeneralMassMatrix}) in terms of $m_i$ instead of $x$, $y$ and $z$:
\beq
\begin{split}
m_\nu&=\,U_{TB}\,\diag(m_1,\,m_2,\,m_3)\,U_{TB}^T\\[3mm]
&=\,\dfrac{m_3}{2}\left(\begin{array}{ccc}
                        0&0&0\\
                        0&1&-1\\
                        0&-1&1\end{array}\right)
    +\dfrac{m_2}{3}\left(\begin{array}{ccc}
                        1&1&1\\
                        1&1&1\\
                        1&1&1\end{array}\right)
    +\dfrac{m_1}{6}\left(\begin{array}{ccc}
                        4&-2&-2\\
                        -2&1&1\\
                        -2&1&1\end{array}\right)\;.
\end{split}
\eeq
All the type of neutrino spectra can be accommodated: $m_3>>m_2>>m_1$ defines a normal hierarchy; a degenerate model is given by choosing $m_3\approx-m_2 \approx m_1$; for $m_1 \approx - m_2$ and $m_3 \approx 0$ the inverse hierarchy case is achieved. However, stability under renormalisation group running strongly prefers opposite signs for the first and the second eigenvalue which are related to solar oscillations and have the smallest mass squared splitting (see section \ref{Sec:Running} for details).

Finally we underline that this mixing pattern is a very good approximation of the experimental data: the tribimaximal values for the atmospheric and the reactor angles are inside the $1\sigma$ error level, while that one for the solar angle is very close to the upper $1\sigma$ value.\\

The study of promising patterns for the mass matrices or for the mixing matrices are useful and predictive tools to describe the experimental data, but they suffer for the lack of an explanation of their origin and their stability under corrections. In order to improve the situation, it is necessary to search for an intimate reason of their appearance. In the next section we will examine the use of flavour symmetries in order to recover such patterns.

\section{Flavour Symmetry Overview}
\label{Sec:FSym}
\setcounter{footnote}{3}

In section \ref{Sec:MFV} we discussed the maximal size for a flavour symmetry in the Standard Model: $[U(3)]^5$ without right-handed neutrinos or otherwise $[U(3)]^6$. If we deal with a GUT model, however, the maximal size is reduced (for example in $SO(10)$ GUT, it is a single $U(3)$). Furthermore, in the current literature, there is a tendency of adopting a flavour symmetry which can be embedded into $SU(3)$ or $SO(3)$, to give the reason of some kind of rotation among the families. In any case, as we will see in a while, the symmetry which is introduced has to be broken: it is a general requirement of the Yukawa interactions and it is a necessary condition in order to be consistent with the observed fermion masses and mixings.

There is large variety of symmetries which can be used: they can be either Abelian or non-Abelian, either local or global (or even a combination of them) and finally either discrete or continuous. Historically, flavour symmetries were first used to describe the quark sector and the Abelian $U(1)$ symmetry has been shown to be able to explain the observed quark mass hierarchies and mixings. In this approach developed by Froggatt and Nielsen in 1979 \cite{FN}, there is a flavon field $S$, a gauge-invariant scalar, which acquires a vacuum expectation value (VEV) and breaks the $U(1)$ symmetry. It is possible to define a small parameter $\ep=\mean{S}/\Lambda_f$, where the cutoff $\Lambda_f$ is the scale of flavour dynamics usually associated with some heavy fermions which are integrated out. This symmetry breaking is then communicated to fermions with a non-universal mechanism, in such a way that different fermions receive different powers of $\ep$. The advantage of this mechanism is that the Yukawas can be of $\cO(1)$ and the fermion masses and mixings are explained as powers of the expansion parameter $\ep$. On the other hand, the main disadvantage consists in the lack of well-defined predictions: masses and mixings angles are only predicted up to unknown $\cO(1)$ coefficients. Furthermore, certain mixing patterns such as the bimaximal and the tribimaximal schemes cannot be achieved with an Abelian symmetry. Therefore we can conclude that the predictive power of a non-Abelian symmetry is in general larger than that of an Abelian one.

Concerning the local or the global attribute of a flavour symmetry, we have to remember that the requirement of anomaly freedom for a local symmetry can put strong constraints on the charge assignment of the fermions. Furthermore, locality preserves the symmetry from being broken by quantum gravity effects at the Planck scale.

We now discuss the advantages and disadvantages of using a continuous or a discrete group. In the case of a spontaneously broken symmetry a continuous one leads to the appearance of Goldstone or gauge bosons. On the other hand, the breaking of a discrete group is safe from such a consequence but could be affected by the problem of domain walls \cite{DomainWalls} (solvable by inflation). Furthermore, using the continuous groups such as $SO(3)$ or $SU(3)$, we have only a single non-trivial possibility to describe the three fermion families and the type of contractions is also strongly limited. On the contrary, adopting a discrete symmetry, there are several small representations which can be fairly used. Even if these disadvantages in treating continuous symmetries, they have been extensively studied in literature. A first attempt has been proposed in \cite{SymmU2} where the investigated flavour symmetry is the group $U(2)$, under which the three families transform as a ${\bf2}+{\bf1}$. This reflects
the fact that the third families are the heaviest ones: in this way it is possible to explain a relatively large mixing between the first two families, while those with the third generation are smaller. It fits well the quark sector, where the Cabibbo angle is the largest, but it does not work with the lepton sector, where two of the angles are large.

An upgraded approach has been pursued with the use of the $SO(3)$ \cite{SymmSO3,SymmSO3King} and $SU(3)$ \cite{SymmSU3} symmetries, which account for all the three generations. In the models, realistic fermion mass spectra and mixings are achieved, but with the introduction of additional heavy degrees of freedom and auxiliary symmetries, which suppress unwanted operators. It is worth to note that in all of these models it is not-trivial to explicitly realise an embedding of the flavour symmetry with an underlying GUT.\\

After this brief summary, we restrict to the context of non-Abelian discrete flavour symmetries which are in general more predictive than Abelian ones and that are safe from dangerous effects such as the appearance of Goldstone or gauge bosons. Furthermore, the particular mixing pattern in the lepton sector can be very well explained by the use of certain discrete symmetries, which are all subgroups of $SU(3)$. In rest of this section we do not enter in the details of each group, referring to \cite{GroupRepresentations} for the general group theory and to the following chapters where some of these group are treated in detail.

\subsection{Discrete Flavour Symmetries}
\label{Sec:FS:Discrete}
\setcounter{footnote}{3}

Here we give a brief overview of the main discrete flavour symmetries which has been (recently) adopted in order to describe quarks and leptons.

Of particular relevance is the $A_4$ flavour group \cite{TBA4,MR_A4,BMV_A4,AF_Extra,BH_Geometric,Ma_TBMandSUSYwithA4,AF_Modular,HKV_A4, AFL_Orbifold,He_Renorm,MPT_SO10xA4,Yin_A4,BKM_A4Zn,Altarelli_Lectures,AFH_SU5,BMPT_SU3,AG_AF,HMV_Nie,LinPredictive,CDGG_Extra,JM_A4Lepto,BFM_SO10xA4, Riazuddin,CK_Form,Lin_Lepto,BFRS_A4Lepto,AM_Lin,HMV_ILSS,Lin_LargeReactor,BBFN_Lepto,HMP_Lepto,ABMMM_Lepto}. $A_4$ has been widely used as a flavour symmetry since it is the smallest group with an irreducible three-dimensional representation. In this way, it is possible to collect all the three families in a unique representation, such as with $SO(3)$ or $SU(3)$, but with more freedom in the type of couplings. The introduction of $A_4$ as the flavour symmetry in the lepton sector can produce the tribimaximal pattern as the lepton mixing matrix and a great effort has been done in order to study the related neutrino phenomenology. However, when the quark sector is considered, it results a highly non-trivial task to get a (even non grand) unified description of leptons and quarks. We will illustrate this point in more details in section \ref{Sec:FlavourModelsTBM}. We only report here that a possible strategy to overcome the problem is to enlarge the symmetry.

Two groups have been used in order to mimic the behavior of $A_4$ in the lepton sector, but pursuing a correct description also of the quark mass hierarchy and mixings: the group $T'$ \cite{FHLM_Tp,Tp_other,CF_Tp}, whose features will be illustrated in section \ref{Sec:TpTBM}, and the group $S_4$ \cite{BM_S4,BMM_S4,BMM_SS,Lam_S4,MP_S4}, which will be discussed in section \ref{Sec:S4TBM}. It is worth to report that the group $S_4$ has been already studied in literature but with different aims in\cite{S4Old} and it has also been recently used for a revival of the bimaximal pattern in the context of the MSSM \cite{AFM_BimaxS4}, which will be the subject of section \ref{Sec:FlavourModelsBM}, and for the construction of a realistic Pati-Salam GUT \cite{ABM_PSS4}, discussed in section \ref{Sec:AFM:PS}.

The groups $T'$ and $S_4$ are the smallest groups which well fit the lepton and the quark sectors at the same time, but several other studies based on different flavour symmetries show interesting results: in particular the recently analysed symmetry group $\Delta(27)$ \cite{SymmDelta27} is worth to be mentioned.\\

Before concluding this section, it is relevant to underline a general feature of models based on (discrete) groups: it turns out that the symmetry alone is not sufficient to fully account for the fermion mass hierarchies and mixings in the majority of the cases. A first problem concerns the differences between leptons and quarks: two (of three) large lepton mixing angles with respect to three small and hierarchical quark ones; neutrinos with a much milder mass hierarchy with respect to the charged fermions. A viable solution consists in avoiding interferences among the two sectors, at least in first approximation, and to keep them separated additional groups, such as the Abelian factors $Z_n$, are implemented in the complete flavour symmetry group. A second problem refers to the use of the three-dimensional representation, which is usually adopted to describe leptons: the components of a triplet show degenerate masses, unless some breaking parameter is introduced. From this the problem of how to describe the charged lepton mass hierarchy follows and two kind of solutions have been proposed: the Froggatt-Nielsen (FN) mechanism, which consists in introducing an additional (global or local) $U(1)_{FN}$ factor under which right-handed fermions transform, is the most used; otherwise it is possible to introduce additional symmetry groups, usually very small, such as $Z_3$ or $S_3$. In the following chapters we will focus only on models where the Froggatt-Nielsen mechanism is responsible of the charged lepton mass hierarchy.

\subsection{The Flavour Symmetry Breaking}
\label{Sec:FS:SSB}
\setcounter{footnote}{3}

In this section we comment on the necessary requirement of a flavour symmetry breaking mechanism, which can occur explicitly or spontaneously. Since the explicit breaking generally introduces several additional parameters, we focus only on the mechanism of the spontaneous symmetry breaking.

The gauge group of the Standard Model prevents direct fermion mass terms and the Higgs mechanism is addressed to be responsible for them. When a flavour symmetry is implemented in a model, some new fields are needed: the simplest example is the M(L)FV which we have discussed in section \ref{Sec:MFV}, where the Yukawa couplings are promoted to gauge singlet scalars. Even in this simple approach, it is necessary that the Yukawa fields develop a VEV in order to generate mass terms for fermions. A similar requirement holds also when the flavour symmetries, which we have presented in the previous section, are introduced. We have already seen this aspect discussing the FN symmetry: a new scalar field $S$ is introduced and its VEV, communicated to the fermions, accounts for masses and mixings. People usually refer to this kind of new degrees of freedom with the name of ``flavons'': they usually are invariant under the gauge group of the underlying theory, either Standard Model or GUT, and transform only under the flavour symmetry; their masses are typically much larger then the electroweak scale and this means that they introduce a further energy scale in the model; they develop a VEV which defines the energy scale at which the flavour symmetry is broken. In order not to introduce further scales into the theory, an alternative approach has been pursued: the flavour and the electroweak symmetries are broken together due to the introduction of several copies of the Standard Model Higgs doublet which transform non-trivially under the flavour group. It is well-known that such multi-Higgs models find strong constraints by direct searches for Higgs bosons and by indirect bounds from flavour changing neutral current and lepton flavour violating processes. For this reason we confine ourselves to models in which the flavour symmetry is broken at an higher energy scale with respect to the electroweak one by the VEV of some flavon fields.\\

The requirement of a broken flavour symmetry is also the result of a well-known no-go theorem \cite{LV_Theorem,FeruglioSymBreaking}, which deals with the lepton mixing angles and in particular with the atmospheric angle. We report here the main reasoning, for which, under quite general
conditions, it can never be $\theta_{23}=\pi/4$ as a result of an exact flavour symmetry.

A general condition is that the flavour symmetry (either global or local, either continuous or discrete), is only broken by small effects. Furthermore, in the basis of canonical kinetic terms, the symmetry acts on the field content of the Standard Model, potentially considering also the right-handed neutrinos, through unitary transformations. Finally, the proof is restricted only to the limit of exact symmetry and then it is possible to neglect the symmetry breaking sector.

The lepton mass matrices can be written as the sum of two terms: the first, denoted as $M_e^0$ and $m_\nu^0$, respectively, are the dominant contributions and are the mass matrices in the exact symmetry phase; the second parts contain the subleading symmetry breaking effects. Confronting with the measured value for the charged lepton masses, it is a natural requirement that $M_e^0$ should be of rank less or equal to $1$, otherwise the differences between the two or three non-vanishing masses (of the same order in the exact symmetry limit) should be explained by large breaking effects or by fine-tunings of the parameters, which both we have excluded. On the other hand a vanishing rank is also a bad starting point since in this case the charged mixing angles are undetermined and, as we will see in a while, also $\theta_{23}$ is completely undetermined. The only remaining possibility is that the rank is equal to $1$.

By a unitary transformations it is always possible to go in the basis where
\beq
M_e^0=\left(
        \begin{array}{ccc}
        0 & 0 & 0 \\
        0 & 0 & 0 \\
        0 & 0 & m_\tau^0\\
        \end{array}
       \right)\;.
\eeq
Considering $U_\nu$ and $U_e$ the unitary matrices that diagonalise $m_\nu^0$ and $M_e^{0\dagger} M_e^0$, it will be possible to adopt the parametrisation in eq. (\ref{SM:PhysBasis:MixingMatrixUPMNS}) for $U_\nu$, putting a suffix $\nu$ on angles and phases, and
\beq
U_e=R_{12}(\theta^e_{12})\;,
\eeq
where the angle $\theta^e_{12}$ is completely undetermined.

The physical mixing matrix is defined as the product $U=U_e^\dagger U_\nu$ and it follows that
\beq
\left|\tan\theta_{23}\right|=\left|\cos\theta^e_{12} \tan\theta^\nu_{23} + \sin\theta^e_{12}\dfrac{\tan\theta^\nu_{13}}{\cos\theta^\nu_{23}} e^{-i\delta}\right|\;.
\eeq
Therefore, in general, the atmospheric mixing angle is always undetermined in the limit of the exact symmetry. Only when small breaking parameters are considered in the mass matrices, it is possible to recover $\theta_{23}=\pi/4$. This goal is provided if these breaking terms have suitable orientations in the flavour space and this is connected to the VEV (mis)alignment of the flavons: if the breaking terms are produced by a spontaneous symmetry breaking, in general two independent sectors of flavons are needed, indeed one of them communicates the breaking to charged fermions and the other one to neutrinos. It is worth to underline that the VEV (mis)alignment of the flavons is an highly non-trivial problem to solve, which could put severe constraints on the choice of the group representations and on the minimal number of new degrees of freedom. In the following chapters, we will face this problem providing for each model a suitable explanation of the specific VEV (mis)alignment of all the flavons.

\clearpage{\pagestyle{empty}\cleardoublepage}

\newpage
\chapter{Flavour Models with the Tribimaximal Mixing}
\label{Sec:FlavourModelsTBM}
\setcounter{equation}{0}
\setcounter{footnote}{3}

In this chapter we enter in the details of three flavour models which have the common prediction of the lepton mixing matrix of the tribimaximal (TB) form. As already pointed out in the previous sections, the tribimaximal pattern is a very good approximation of the measured mixings and for this reason it represents an attractive starting point to describe leptons.

There is a model based on the symmetry group $A_4$ \cite{AF_Extra,AF_Modular,AFL_Orbifold} which is extremely appealing, thanks to its simplicity and predictivity. $A_4$ is the group of the even permutations of four objects and has twelve elements and four irreducible representations, which are three singlets, $\bf1$, $\bf1'$ and $\bf1''$, and one triplet $\bf3$. This model manages in deriving the tribimaximal mixing and presents a prediction for the neutrinoless-double-beta decay parameter $|m_{ee}|$ as a function of the lightest neutrino mass: considering the Weinberg operator as the origin of the neutrino masses only the normally hierarchical spectrum is admitted and the prediction reads as
\beq
|m_{ee}|^2=\dfrac{1}{9}\left(9 m_1^2+5 \De m^2_{sol}- \De m^2_{atm}\right)\;.
\eeq
These results come from the assumption that the $A_4$ symmetry is realised at a very high energy scale $\La_f$ and that leptons transform in a non trivial way under $A_4$. Afterward the symmetry is spontaneously broken by a set of scalar multiplets $\Phi$, the flavons, whose vacuum expectation values (VEVs) receive a specific alignment. As a consequence the tribimaximal mixing is corrected by the higher-order terms by quantities of the order of $\mean{\Phi}/\La<1$ and the reactor angle is no longer vanishing and becomes proportional to $\mean{\Phi}/\La$.

The drawback of this model is the difficulty to describe correctly the quark sector. First of all the quark mixing matrix is completely different from its lepton counterpart: the first shows little angles and, on the contrary, the second presents two large angles. As a result, while the lepton mixing matrix can be fairly achieved through the $A_4$ flavour symmetry, the quark mixings seem to be better described by some continuous symmetries, like $U(2)$ \cite{SymmU2}. Indeed, according to the left- and right-handed quark representation assignments, the $A_4$ flavour symmetry tends to predict no mixing at all in the quark sector, $V_{CKM}=\unity$, or too large mixing angles. On the other hand, the results obtained by the $U(2)$-based models suggest that the use of the doublet representation in the quark sector should help in describing the quark mixing. However, this possibility is prevented in the $A_4$-based models, since there are no doublet representations. The solutions which have been proposed consist in the possibility of add several $Z_n$ symmetries \cite{BKM_A4Zn}, in order to suppress the unwanted terms, or in adopting a larger group, which manages in reproducing the lepton sector similarly as in the $A_4$-based model and possesses some doublet representations useful to describe quarks. We followed this second strategy studying two discrete symmetries: $T'$ \cite{FHLM_Tp} the double-valued group of the tetrahedral symmetry and $S_4$ \cite{BMM_SS,BMM_S4} the group of permutation of four objects. Both of these groups have $24$ elements, but they differ in the type of the representations: $T'$ contains exactly the same representations of $A_4$, i.e. three singlets, $\bf1$, $\bf1'$ and $\bf1''$, and one triplet $\bf3$, and additional three doublets, $\bf2$, $\bf2'$ and $\bf2''$; $S_4$ has only five representations, i.e. two singlets, $\bf1$, $\bf1'$, one doublet $\bf2$ and two triplets, $\bf3$ and $\bf3'$.

As we will see in the next sections, the lepton sector of $T'$ is described exactly in the same way as in the $A_4$ model, using the doublets to account for quark masses and mixings: to recover a realistic description, in particular the correct order of magnitude of the ratio $m_u/m_c$, it is however necessary a moderate fine-tuning of order $\lambda$. Apart from the predictions in the lepton sector, two relations are present in the quark sector:
\beq
\sqrt{\dfrac{m_d}{m_s}}=\left\vert V_{us}\right\vert+O(\lambda^2)\;,\qquad\qquad
\sqrt{\dfrac{m_d}{m_s}}=\left\vert\dfrac{V_{td}}{V_{ts}}\right\vert+O(\lambda^2)\;.
\eeq
These relations can be verified by the experimental data: from \cite{PDG08} we have $\sqrt{m_d/m_s}=0.213\div 0.243$, $\vert V_{us}\vert=0.2257\pm0.0010$ and $\vert V_{td}/V_{ts}\vert=0.209\pm0.001\pm0.006$. Unfortunately, the theoretical errors affecting the predictions, dominated respectively by the unknown $O(\lambda^2)$ term in $V_{us}$ and by the unknown $O(\lambda^4)$ term in
$V_{td}$, are of order $20\%$. For this reason, and for the large uncertainty on the ratio $m_d/m_s$, it is not possible to turn these predictions into precise tests of the model.

The presence of the doublet representations of $S_4$ introduce new features in the lepton sector: it is possible to describe the same neutrino mass matrix as in the $A_4$ model or alternatively a phenomenologically distinct mass texture. We investigate this second possibility and we find small, but non-negligible, parts of the parameter space in which the models are different. Apart from the lepton mixing angles, a prediction for $|m_{ee}|$ is present also in this model:
\baq
\text{NH}\qquad\quad |m_{ee}| &=& \dfrac{1}{3}\sqrt{3m_1^2+2\De m^2_{atm}-\De m^2_{sol}}\;,\\[3mm]
\text{IH}\qquad\quad |m_{ee}| &=& \dfrac{1}{3}\sqrt{3m_3^2+\De m^2_{atm}-2\De m^2_{sol}}\;.
\eaq
In the quark sector the model well explains the observed mass hierarchies, but it is necessary a moderate fine-tuning to recover the Cabibbo angle: the $(12)$ entry of the CKM matrix is the difference of two complex terms with absolute values of order $\lambda^2$ and therefore we ask a definite relation between the phases of the two terms in order to have an accidental enhancement of order $1/\lambda$.

Discussing all these models we do not consider the renormalisation group (RG) running, postponing a detailed analysis in section \ref{Sec:Running}: we only anticipate that RG corrections have a minor impact on the predictions and on the observables of the models, a part when the spectrum is inverse hierarchical and only for particular relations among the Majorana phases.

\mathversion{bold}
\section{$A_4$-Based Model}
\label{Sec:AFTBM}
\setcounter{footnote}{3}
\mathversion{normal}

We recall here the main features of the Altarelli-Feruglio (AF) model \cite{AF_Extra,AF_Modular,AFL_Orbifold}, which is based on the flavour group $G_f=A_4\times Z_3\times U(1)_{FN}$: the spontaneous breaking of $A_4$ is responsible for the tribimaximal mixing; the cyclic symmetry $Z_3$ prevents the appearance of dangerous couplings and helps keeping separated the charged lepton sector and the neutrino one; the $U(1)_{FN}$ provides a natural hierarchy among the charged lepton masses.

$A_4$ is the group of the even permutations of $4$ objects, isomorphic to the group of discrete rotations in the three-dimensional space that leave invariant a regular tetrahedron. It is generated by two elements $S$ and $T$ obeying the relations\cite{GroupRepresentations}:
\beq
S^2=(ST)^3=T^3=1\;.
\eeq
It has three independent one-dimensional representations, $\bf1$, $\bf1'$ and $\bf1''$ and one three-dimensional representation $\bf3$. We present a set of generators $S$ and $T$ for the various representations, and the relevant multiplication rules in appendix \ref{AppA:A4}. The group $A_4$ has two obvious subgroups: $G_S$, which is a reflection subgroup generated by $S$, and $G_T$, which is the group generated by $T$, isomorphic to $Z_3$. These subgroups are of interest for us because $G_S$ and $G_T$ are the relevant low-energy symmetries of the neutrino and the charged-lepton sectors at the leading order, respectively. The tribimaximal mixing is then a direct consequence of this special symmetry breaking pattern, which is achieved via the vacuum misalignment of triplet scalar fields. If $\Phi=(\Phi_1,\Phi_2,\Phi_3)$ denotes the generic scalar triplet, the VEV
\beq
\mean{\Phi}\propto (1,1,1)
\eeq
breaks $A_4$ down to $G_S$, while
\beq
\mean{\Phi}\propto (1,0,0)
\eeq
breaks $A_4$ down to $G_T$. The flavour symmetry breaking sector of the model includes the scalar fields $\varphi_T$, $\varphi_S$, $\xi$ and $\theta$. In table \ref{table:AFtransformations}, we can see the fermion and the scalar content of the model and their transformation properties under $G_f$.

\begin{table}[ht]
\begin{center}
\begin{tabular}{|c||cccc||ccccc|}
\hline
&&&&&&&&&\\[-4mm]
 & $\ell$ & $e^c$ & $\mu^c$ & $\tau^c$ &$H$ & $\theta$ & $\phit$ & $\phis$ & $\xi$ \\[2mm]
\hline
&&&&&&&&&\\[-4mm]
$A_4$ & $\bf3$ & $\bf1$ & $\bf1''$ & $\bf1'$ & $\bf1$ & $\bf1$ & $\bf3$ & $\bf3$ & $\bf1$ \\[2mm]
$Z_3$ & $\om$ & $\om^2$ & $\om^2$ & $\om^2$ & 1 & 1 & 1 & $\om$ & $\om$ \\[2mm]
$U(1)_{FN}$ & 0 & 2 & 1 & 0 & 0 & -1 & 0 & 0 & 0  \\[2mm]
\hline
\end{tabular}
\end{center}
\caption{\it The transformation properties of the fields under $A_4$, $Z_3$ and $U(1)_{FN}$.}
\label{table:AFtransformations}
\end{table}

As anticipated above, the specific breaking patter of the symmetry which leads to the tribimaximal scheme and to hierarchical masses for leptons requires that $\xi$ and $\theta$ develop a non vanishing VEV and that the following specific vacuum misalignment for the triplets occurs:
\beq
\langle\phit\rangle=(v_T,\,0,\,0)\;,\qquad
\langle\phis\rangle=(v_S,\,v_S,\,v_S)\;.
\eeq
In \cite{AF_Extra,AF_Modular} it has been shown a natural explanation of this misalignment. These VEVs can be very large, much larger than the electroweak scale. From the analysis in \cite{AF_Extra,AF_Modular}, it is reasonable to choose:
\beq
\dfrac{VEV}{\Lambda_f}\approx \lambda^2\;,
\label{AFTBM:vevratio}
\eeq
where VEV stands for the generic non-vanishing VEV of the flavons, $\Lambda_f$ the cutoff of the theory and $\lambda$ the Cabibbo angle.
Since the ratio in eq. (\ref{AFTBM:vevratio}) represents the typical expansion parameter when including higher dimensional operators, it keeps all the leading order results stable, up to correction of relative order $\lambda^2$.
A very useful parametrisation of $VEV/\Lambda_f$ is the following:
\beq
\dfrac{\langle\phit\rangle}{\Lambda_f}= (u,\,0,\,0)\;,\quad
\dfrac{\langle\phis\rangle}{\Lambda_f}=c_b(u,\,u,\,u)\;,\quad
\dfrac{\mean{\xi}}{\Lambda_f}=c_a\,u,\,\quad
\dfrac{\mean{\theta}}{\La_f}=t\;,
\label{AFTBM:vevs}
\eeq
where $c_{a,b}$ are complex numbers with absolute value of order one, while $u$ and $t$ are the small symmetry breaking parameters of the theory (they can be taken real through field redefinitions).

Once defined the transformations of all the fields under $G_f$, it is possible to write down the Yukawa interactions: at the leading order they read
\bea
\LL_e&=&\dfrac{y_e}{\Lambda_f^3} \theta^2e^c H^\dagger \left(\varphi_T \ell\right)
+\dfrac{y_\mu}{\Lambda_f^2} \theta\mu^c H^\dagger \left(\varphi_T \ell\right)'
+\dfrac{y_\tau}{\Lambda_f} \tau^c H^\dagger \left(\varphi_T \ell\right)''+h.c.
\label{AFTBM:Ll}\\
\nn\\
{\LL}_\nu&=& \dfrac{x_a}{\Lambda_f\Lambda_L} \xi ({\tilde H}^\dagger \ell {\tilde H}^\dagger \ell) + \dfrac{x_b}{\Lambda_f\Lambda_L} (\varphi_S {\tilde H}^\dagger \ell {\tilde H}^\dagger \ell)+h.c.\;,
\label{AFTBM:Lnu}
\eea
where $y_i$ and $x_i$ are complex numbers with absolute value of order one. The contractions under $SU(2)_L$ are understood and the notation $(\ldots)$, $(\ldots)'$ and $(\ldots)''$ refers to the contractions in $\bf1$, $\bf1'$ and $\bf1''$, respectively. We distinguish two different energy scales: $\Lambda_f$ refers to the energy scale of the flavour dynamics while $\Lambda_L$ to the scale at which the lepton number is violated. We assume here that $\Lambda_f\sim\Lambda_L$.

When the flavons develop VEVs in agreement with eq. (\ref{AFTBM:vevs}) and after the electroweak symmetry breaking, the leading order mass matrix of charged leptons takes the following form: in the basis of canonical kinetic terms\footnote{It has been shown in a series of papers \cite{Kahler} that the corrections, from the transformations needed to move in the basis of canonical kinetic terms, appear at most as NLO deviations.}, 
\beq
M_e=\left(
    \begin{array}{ccc}
        y_e t^2 & 0& 0\\
        0& y_\mu t& 0\\
        0& 0& y_\tau
        \end{array}
    \right)\dfrac{v\, u}{\sqrt2} \;.
\label{AFTBM:ChargedMass}
\eeq
Once in the physical basis, the entries on the diagonal are identified to the masses of the charged leptons and the relative hierarchy among them is given by the parameter $t$: when
\beq
t\approx 0.05
\eeq
then the mass hierarchy is in agreement with the experimental measurements. As we will see in the following sections, the model admits a well defined range for the parameter $u$ which can approximatively be set to
\beq
0.003\lesssim u \lesssim 0.05\;.
\label{AFTBM:RangeuSM}
\eeq

In the neutrino sector, the leading order Majorana mass matrix is given by
\beq
m_\nu=\left(
        \begin{array}{ccc}
            a+2 b/3 & -b/3 & -b/3 \\
            -b/3 & 2b/3 & a-b/3 \\
            -b/3 & a-b/3 & 2 b/3 \\
        \end{array}
        \right)\dfrac{v^2}{\La_L}\;,
\label{AFTBM:LONuMasses}
\eeq
where $a\equiv x_a\,c_a\,u$ and $b\equiv x_b\,c_b\,u$. At this order the mass matrix is diagonalised by 
\beq
U_\nu^T m_\nu U_\nu =\dfrac{v^2}{\Lambda_L}\diag(|a+b|,\,|a|,\,|-a+b|)\;,
\eeq
where $U_\nu=U_{TB}P$. The matrix $U_{TB}$ is the tribimaximal transformation of eq. (\ref{FS:TB:TBmixing}), while $P$ is the matrix of the Majorana phases, 
\beq
P=\diag(e^{i\al_1/2},\,e^{i\al_2/2},\,e^{i\al_3/2})\;,
\label{AFTBM:Pmatrix}
\eeq
with $\al_1=-\arg(a+b)$, $\al_2=-\arg(a)$ and $\al_3=-\arg(-a+b)$.

It is possible to generalise this description also to the supersymmetric context. In this case $G_f$ accounts for an additional term, a continuous $R$-symmetry $U(1)_R$, that contains the usual $R$-parity as a subgroup and simplifies the constructions of the scalar potential: under this symmetry, the matter superfields transform as $U(1)_R=1$, while the scalar ones are neutral.

It is easy to extend eqs. (\ref{AFTBM:Ll}, \ref{AFTBM:Lnu}) in the supersymmetric case: two Higgs doublets $H_{(d,u)}$, invariant under $A_4$, substitute $H$ and $\widetilde H$, respectively; the Lagrangian $\LL_e$ is identified to the leading order charge lepton superpotential $w_e$ and $\LL_\nu$ is identified to the leading order neutrino superpotential $w_\nu$. Moreover, it is necessary to introduce a further flavon $\xit$, which exactly transforms as $\xi$ but does not acquire any VEV. As a result it does not have any impact on the previous discussion and its relevance is only linked to the way in which the VEV misalignment is recovered (see \cite{AF_Modular} for further details).

While $t$ is still equal to $0.05$ in order to have a correct charged lepton mass hierarchy, the range for $u$ slightly changes:
\beq
0.007\lesssim u \lesssim 0.05\;.
\label{AFTBM:RangeuMSSM}
\eeq

\subsection{The Neutrino Mass Spectrum}
\label{Sec:AFTBM:NuSpectrum}
\setcounter{footnote}{3}

We now summarise the results for the neutrino mass spectrum. Notice that the following analysis is valid in the Standard Model as well as in its supersymmetric extension, by substituting $v$ with $v_u$ when necessary.
The neutrino masses are given by
\beq
m_1=|a+b|\dfrac{v^2}{\Lambda_L}\;,\qquad\qquad m_2=|a|\dfrac{v^2}{\Lambda_L}\;,\qquad\qquad m_3=|-a+b|\dfrac{v^2}{\Lambda_L}\;.
\eeq
They can be expressed in terms of only three independent parameters: a possible choice that simplifies the analysis consists in taking $|a|$, $\rho$ and $\Delta$, where $\rho$ and $\Delta$ are defined as
\beq
\dfrac{b}{a}=\rho\,e^{i\Delta}\;,
\label{AFTBM:RhoDeltaDef}
\eeq
with $\Delta$ in the range $[0,\,2\pi]$.
From the experimental side only the squared mass differences have been measured and as a result the spectrum is not fully determined and indeed $\Delta$ is still a free parameter: we can, however, bound $\Delta$ requiring that $|\cos\Delta|\leq1$. Before proceeding it is useful to express $\rho$ and $\cos\Delta$ as functions of some physical observables. To this purpose we calculate the following mass ratios: for both the hierarchies we have
\beq
\dfrac{m_{1(3)}^2}{m_2^2}=1\pm2 \rho\cos\Delta+\rho^2\;.
\eeq
It is then easy to express $\rho$ and $\cos\Delta$ as a function of the neutrino masses:
\beq
\rho=\sqrt{\dfrac{m_1^2-2m_2^2+m_3^2}{2m_2^2}}\;,\qquad\qquad
\cos\Delta=\dfrac{m_1^2-m_3^2}{2\sqrt2m_2\sqrt{m_1^2-2m_2^2+m_3^2}}\;.
\label{AFTBM:RoDelta}
\eeq
Using now the definitions of the mass squared differences,
\beq
\De m^2_{sol} \equiv m_2^2-m_1^2\;,\qquad\qquad
\De m^2_{atm} \equiv |m_3^2-m_1^2(m_2^2)|\;,
\label{AFTBM:DeltaMassesNu}
\eeq
it is possible to express $\cos\Delta$ as a function of only the lightest neutrino mass. Imposing the constraint $|\cos\Delta|\leq1$, it results that only the normal hierarchy is allowed and taking the most conservative case (the $3\sigma$ upper value for $\Delta m^2_\mathrm{sol}$ and the $3\sigma$ lower value for $\Delta m^2_\mathrm{atm}$ as in \cite{Fogli:Indication}) we have
\baq
m_1>14.1\;\meV\;.
\label{AFTBM:Boundm1}
\eaq
This value corresponds to $\cos\Delta=-1$ and it is the value for which the spectrum presents the strongest hierarchy: the values of the masses of the other two neutrinos are given by
\beq
m_2=16.7\;\meV\qquad\text{and}\qquad m_3=47.5\;\meV\;.
\eeq
Furthermore the sum of the neutrino masses in this case is about $78.3$ meV. When $\cos\Delta$ approached the zero, the neutrino spectrum becomes quasi degenerate.

Not only the neutrino masses can be written as a function of the lightest neutrino mass, but also the phases: since in the tribimaximal mixing the reactor angle is vanishing, the Dirac CP phase is undetermined at the leading order; on the contrary the Majorana phases are well defined and they can be expressed through $\rho$ and $\Delta$. Since we are interested in physical observables, we report only phase differences, $\alpha_{ij}\equiv(\alpha_i-\alpha_j)/2$: in terms of $\rho$ and $\Delta$ in order to keep compact the expressions,
\beq
\sin(2\alpha_{13})=\dfrac{2\rho\sin\Delta}{\sqrt{(\rho^2-1)^2+4\rho^2\sin^2\Delta}}\;,\qquad\quad
\sin(2\alpha_{23})=\dfrac{\rho\sin\Delta}{\sqrt{1-2\rho\cos\Delta+\rho^2}}\;.
\label{AFTBM:Majorana1}
\eeq
It will be useful to report also $\sin(2\alpha_{12})$:
\beq
\sin(2\alpha_{12})=-\dfrac{\rho\sin\Delta}{\sqrt{1+2\rho\cos\Delta+\rho^2}}\;.
\label{AFTBM:Majorana2}
\eeq

These results are valid only at the leading order and some deviations are expected with the introduction of the higher-order terms, that is illustrated in the following section. The corrections are expected to be of relative order $u$, whose allowed range is defined in eqs. (\ref{AFTBM:RangeuSM}, \ref{AFTBM:RangeuMSSM}). However, close to $\cos\Delta=-1$, where the bounds are saturated, the corrections to both the numerator and the denominator of eq. (\ref{AFTBM:RoDelta}) remain of relative order $u$ and as a result the lower bound on $m_1$ of eq. (\ref{AFTBM:Boundm1}) is not significantly affected. Major effects could appear when the spectrum is quasi degenerate, \mbox{$\cos\Delta\approx0$}.


\begin{figure}[ht!]
 \centering
\includegraphics[width=7.8cm]{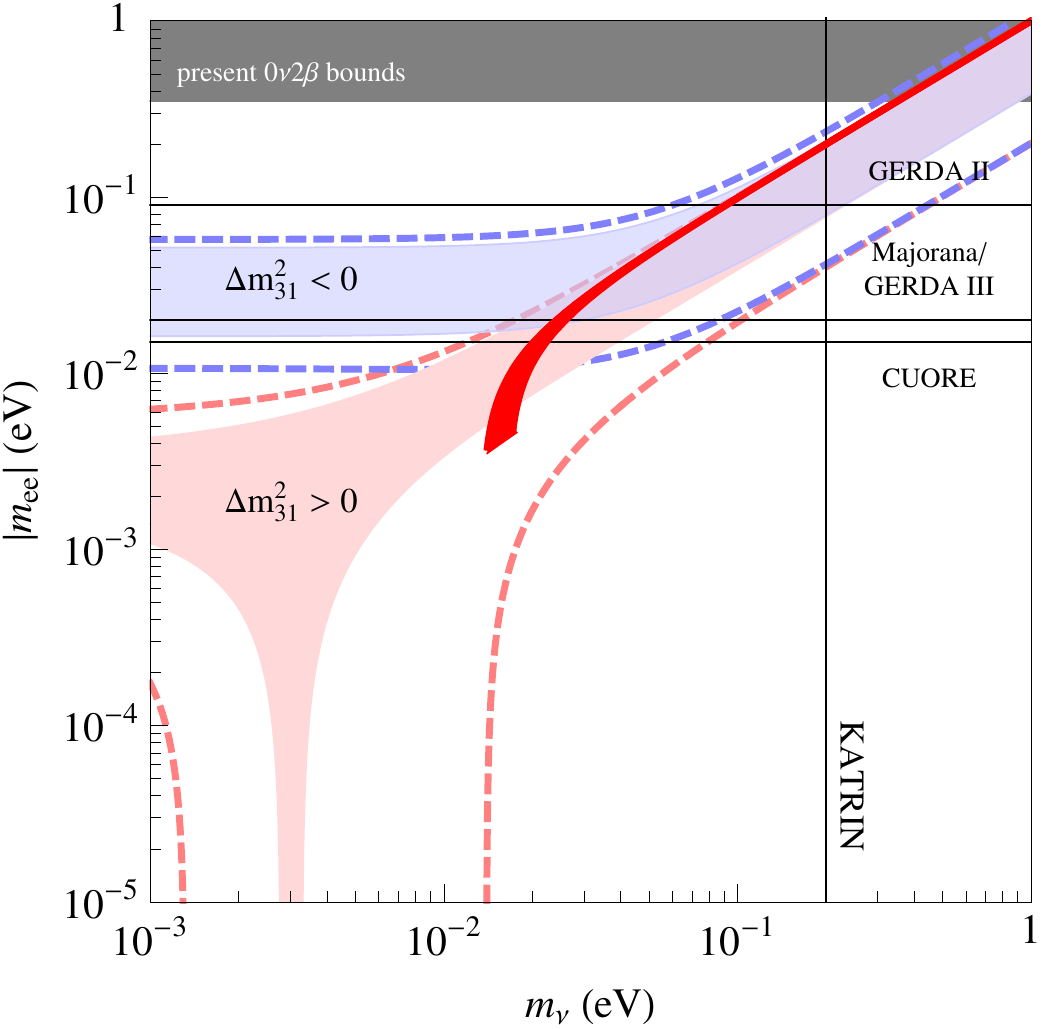}
\caption{\it $|m_{ee}|$ as a function of the lightest neutrino mass $m_1$ for the NH. The light coloured regions show the allowed range for the best-fit values of the parameters from \cite{Fogli:Indication}. The dashed lines refer to the allowed region when the $3\sigma$ errors are considered for the mixing angles as well as for the mass squared differences. The dark red area refers to the model in consideration when the $3\sigma$-error ranges have been implement for $\De m^2_{sol}$ and $\De m^2_{atm}$. The black continuous lines represent future experimental sensitivities as described in the text.}
 \label{fig:AF_0nu2beta}
\end{figure}

\subsubsection{Neutrinoless-Double-Beta Decay}

Still working in the leading order approximation, we can study the value of $|m_{ee}|$, the parameter which characterises the violation of the total lepton number in the $0\nu2\be$-decay. By using eqs. (\ref{AFTBM:RoDelta}, \ref{AFTBM:DeltaMassesNu}), $|m_{ee}|$ can be written in terms of the mass squared differences and of the lightest neutrino mass:
\beq
|m_{ee}|^2=\dfrac{1}{9}\left(9 m_1^2+5 \De m^2_{sol}- \De m^2_{atm}\right)\;.
\eeq
This expression constitutes a prediction of the model. We display the dependence of $|m_{ee}|$ as a function of the lightest neutrino mass, but similar relation can be constructed with $m_2$ or also $m_3$ \cite{AF_Modular,AF_Extra}.

For $\cos\Delta=-1$ and taking the best-fit values for $\De m^2_{sol}$ and $\De m^2_{atm}$, it results $|m_{ee}|=3.8$ meV, which is at the lower edge of the range allowed for the NH considering the best-fit values, as shown in figure \ref{fig:0nu2betaGeneral}. This value is close, but unfortunately just below, to the sensitivities of the future experiments, such as CUORE ($15$ meV) and Majorana ($20$ meV). In figure \ref{fig:AF_0nu2beta}, we show $|m_{ee}|$ as a function of the lightest neutrino $m_1$, the present upper bound from the Heidelberg-Moscow collaboration and future experimental sensitivities from GERDA, Majorana and CUORE. In the plot, the $3\sigma$-error ranges have been implemented for $\De m^2_{sol}$ and $\De m^2_{atm}$. From the figure we can conclude that the model at the leading order is in agreement with the experimental data inside the $1\sigma$ level for all the allowed range for $m_1$, apart when $m_1$ is close to its minimum in which the model slightly pass the $1\sigma$ edge.

When considering the introduction of the NLO terms, we expect that the dark red area in figure \ref{fig:AF_0nu2beta} will enlarge, but the deviations are still expected to be at most of $\cO(u)$ level.

\subsection{The Next-To-Leading Order Contributions}
\label{Sec:AFTBM:NLO}
\setcounter{footnote}{3}

Another important implication of the spontaneously broken flavour symmetry is that the leading order predictions are always subjected to corrections due to higher-dimensional operators. The latter are suppressed by additional powers of the cutoff $\Lambda_f$ and can be organised in a suitable double power expansion in $u$ and $t$.

At the NLO there are may additional terms which can be added to the Lagrangian. Since $\varphi_T$ is the only scalar field which
is neutral under the Abelian part of the flavour symmetry, all the NLO terms contain the terms already present in the leading order Lagrangian
with an additional insertion of $\varphi_T /\Lambda_f$.
In addition to these terms, there are also corrections to the leading vacuum alignment in eq. (\ref{AFTBM:vevs}):
\beq
\ba{ccl}
\dfrac{\langle\varphi_T\rangle}{\Lambda_f}&=&(u,0,0)+(c_1 u^2,\,c_2 u^2,\,c_3 u^2)\\[3mm]
\dfrac{\langle\varphi_S\rangle}{\Lambda_f}&=& c_b(u,u,u)+(c_4 u^2,\,c_5 u^2,\,c_6 u^2)\\[3mm]
\dfrac{\langle\xi\rangle}{\Lambda_f}&=&c_a u+c_7 u^2\;,
\ea
\label{AFTBM:vevsplus}
\eeq
where $c_i$ are complex numbers with absolute value of order one. Note that in the supersymmetric version, the model predicts $c_2=c_3$.
Here we will not perform a detailed analysis for NLO operators and the origin of eq. (\ref{AFTBM:vevsplus}) (see  \cite{AF_Extra,AF_Modular} for a detailed study). As a result of these NLO contributions, the quantities generally deviate from their initial values for terms of relative order $u$:
\beq
Y_e + \delta Y_e\;, \qquad m_\nu + \delta m_\nu \;.
\eeq
These corrections affect also the mixing angles and it is not difficult to see that deviations from tribimaximal are also of relative order $u$ with respect to their leading order values \cite{AF_Extra,AF_Modular}:
\beq
\sin ^2 \theta_{23} = \dfrac12 + \mathcal{O}(u), \qquad \sin^2 \theta_{12}=\dfrac 13 + \mathcal{O}(u),
\qquad \sin \theta_{13} = \mathcal{O}(u).
\eeq
Since the solar mixing angle is, at present, the most precisely known, we require that its value remains inside the $3\sigma$ range \cite{NeutrinoData}. This requirement results in an upper bound on $u$ of about $0.05$. On the other hand, from eq. (\ref{AFTBM:ChargedMass}), we have the following relations:
\beq
\ba{rll}
u&=\, \dfrac{1}{|y_\tau|} \dfrac{\sqrt{2} m_\tau}{v}\approx 0.01 \dfrac{1}{|y_\tau|}&\qquad\text{in the SM}\\[5mm]
u&\simeq\,\dfrac{\tan\beta}{|y_\tau|} \dfrac{\sqrt{2} m_\tau}{v} \approx 0.01 \dfrac{\tan\beta}{|y_\tau|}&\qquad\text{in the MSSM}
\ea
\label{AFTBM:tanb&u&yt}
\eeq
where for the $\tau$ lepton we have used its pole mass $m_\tau=(1776.84 \pm 0.17) \;\rm{MeV}$ \cite{PDG08}. Requesting $|y_\tau|<3$ we find a lower limit for $u$ of about $0.003$ in the Standard Model case; in the supersymmetric context, the same requirements provides a lower bound close to the upper bound $0.05$ for $\tan\beta=15$, whereas for $\tan\beta=2$ it is $u>0.007$. From now on, we will choose the maximal range of $u$ as
\beq
0.003\lesssim u \lesssim 0.05
\eeq
for the Standard Model context, while for the supersymmetric case we take
\beq
0.007\lesssim u \lesssim 0.05\;,
\eeq
which shrinks when $\tan\beta$ is increased from 2 to 15.

The NLO terms affect also the previous results for the neutrino phases. All the new parameters which perturb the leading order results are complex and therefore they introduce corrections to the phases of the PMNS matrix. Due to the large amount of such a parameters, we expect non-negligible deviations from the leading order values.

\subsection{Type I See-Saw Realisation}
\label{Sec:AFTBM:SeeSaw}
\setcounter{footnote}{3}

It is possible to easily modify the previous model to accommodate the (type I) See-Saw mechanism. In this part we do such an extension and analyse the differences with the effective model. Notice that this part is written considering an underlying Standard Model context, but the extension to the supersymmetric one is trivial, following the same strategy as in the effective model.

We introduce conjugate right-handed neutrino fields $\nu^c$ transforming as a triplet of $A_4$ and we modify the transformation properties of some other fields according to table \ref{table:AF+SeeSawtransformations}.

\begin{table}[ht]
\begin{center}
\begin{tabular}{|c||c|ccc|}
\hline
&&&&\\[-4mm]
 & $\nu^c$ & $\phis$ & $\xi$ & $\xit$ \\[2mm]
\hline
&&&&\\[-4mm]
$A_4$ & $\bf3$ & $\bf3$ & $\bf1$ & $\bf1$ \\[2mm]
$Z_3$ & $\om^2$ & $\om^2$ & $\om^2$ & $\om^2$ \\[2mm]
$U(1)_{FN}$ & 0 & 0 & 0 & 0  \\[2mm]
\hline
\end{tabular}
\end{center}
\caption{\it The transformation properties of $\nu^c$, $\phis$, $\xi$ and $\xit$ under $A_4\times Z_3\times U(1)_{FN}$. The rest of the fields still transform as in table \ref{table:AFtransformations}. Notice that $\xit$ is present only in the supersymmetric context.}
\label{table:AF+SeeSawtransformations}
\end{table}

The Lagrangian changes only in the neutrino sector and it is given by
\beq
\LL_\nu = y(\nu^c \widetilde H^\dag \ell)+x_a\xi(\nu^c\nu^c)+x_b(\varphi_S\nu^c\nu^c)+h.c.+\ldots\;,
\label{AFTBM:LnuSeeSaw}
\eeq
where dots stand for higher-order contributions.

The vacuum alignment of the flavons is exactly the one described in eqs. (\ref{AFTBM:vevs}, \ref{AFTBM:vevsplus}). When the flavons develop VEVs in agreement with eq. (\ref{AFTBM:vevs}) and after the electroweak symmetry breaking, the Dirac and the Majorana mass matrices, at the leading order, are given by
\beq
m_D = \dfrac{y\,v}{\sqrt2} \left(
                              \begin{array}{ccc}
                                1 & 0 & 0 \\
                                0 & 0 & 1 \\
                                0 & 1 & 0 \\
                              \end{array}
                            \right)\;,\qquad
M_R = \left(
                \begin{array}{ccc}
                a+2 b/3 & -b/3 & -b/3 \\
                -b/3 & 2b/3 & a-b/3 \\
                -b/3 & a-b/3 & 2 b/3 \\
                \end{array}
            \right)\;,
\label{AFTBM:SSMassMatrices}
\eeq
where $a\equiv 2x_a c_a u$ and $b\equiv 2x_b c_b u$. The complex symmetric matrix $M_R$ is diagonalised by the transformation
\beq
\hat{M}_R=U_R^T M_R U_R\;,
\label{AFTBM:Mhat}
\eeq
where $\hat M_R$ is a diagonal matrix with real and positive entries, given by
\beq
\hat M_R\equiv \diag(M_1,\,M_2,\,M_3)=\diag(|a+b|,|a|,|-a+b|)\;,
\label{AFTBM:RHEigenvalues}
\eeq
while the unitary matrix $U_R$ can be written as $U_R=U_{TB}P$, where $P$ is the diagonal matrix of the Majorana phases already defined in eq. (\ref{AFTBM:Pmatrix}). After the electroweak symmetry breaking, the mass matrix for the light neutrinos is recovered from the well known type I See-Saw formula
\beq
m_\nu=-m_D^T M_R^{-1}m_D=-\dfrac{y^2\,v^2}{2}M_R^{-1}\;
\eeq
where the last passage is possible considering that $M_R^{-1} m_D=m_D M_R^{-1}$. From \eq{AFTBM:Mhat}, $U_R^\dag M_R^{-1} U_R^*=\diag(M_1^{-1},\,M_2^{-1},\,M_3^{-1})$ and as a result the light neutrino mass matrix can be diagonalised by
\beq
\hat{m}_\nu=U_\nu^Tm_\nu U_\nu\;,
\eeq
where $U_\nu=iU_R^*=iU_{TB}P^*$. The diagonal matrix $\hat{m}_\nu$ has real and positive entries written as
\beq
m_i=\dfrac{v^2}{2}\dfrac{y^2}{M_i}\;,
\label{AFTBM:LightMasses}
\eeq
which explicitly give the following values
\beq
m_1=\dfrac{v^2}{2}\dfrac{y^2}{|a+b|}\;,\qquad\qquad
m_2=\dfrac{v^2}{2}\dfrac{y^2}{|a|}\;,\qquad\qquad
m_3=\dfrac{v^2}{2}\dfrac{y^2}{|-a+b|}\;.
\eeq
In these expressions we have absorbed the possible phase of $y$ inside the matrix $P$: this phase however is not observable and thus we could have assumed a positive $y$ from the beginning without loss of generality.
We can repeat the analysis presented in section \ref{Sec:AFTBM:NuSpectrum} to study the light neutrino spectrum in this case. Taking $|a|=M_2=v^2y^2/(2m_2)$, we find that both the orderings can be described and that the lightest neutrino masses span the following ranges: for the most conservative case,
\baq
\text{normal hierarchy:}\qquad&4.3\;\mathrm{meV}<m_1<6.2\;\mathrm{meV}&\\[3mm]
\text{inverse hierarchy:}\qquad&m_3>15.8\;\mathrm{meV\;.}&
\label{AFTBM:RangeMasses}
\eaq

\begin{figure}[ht!]
 \centering
\includegraphics[width=6cm]{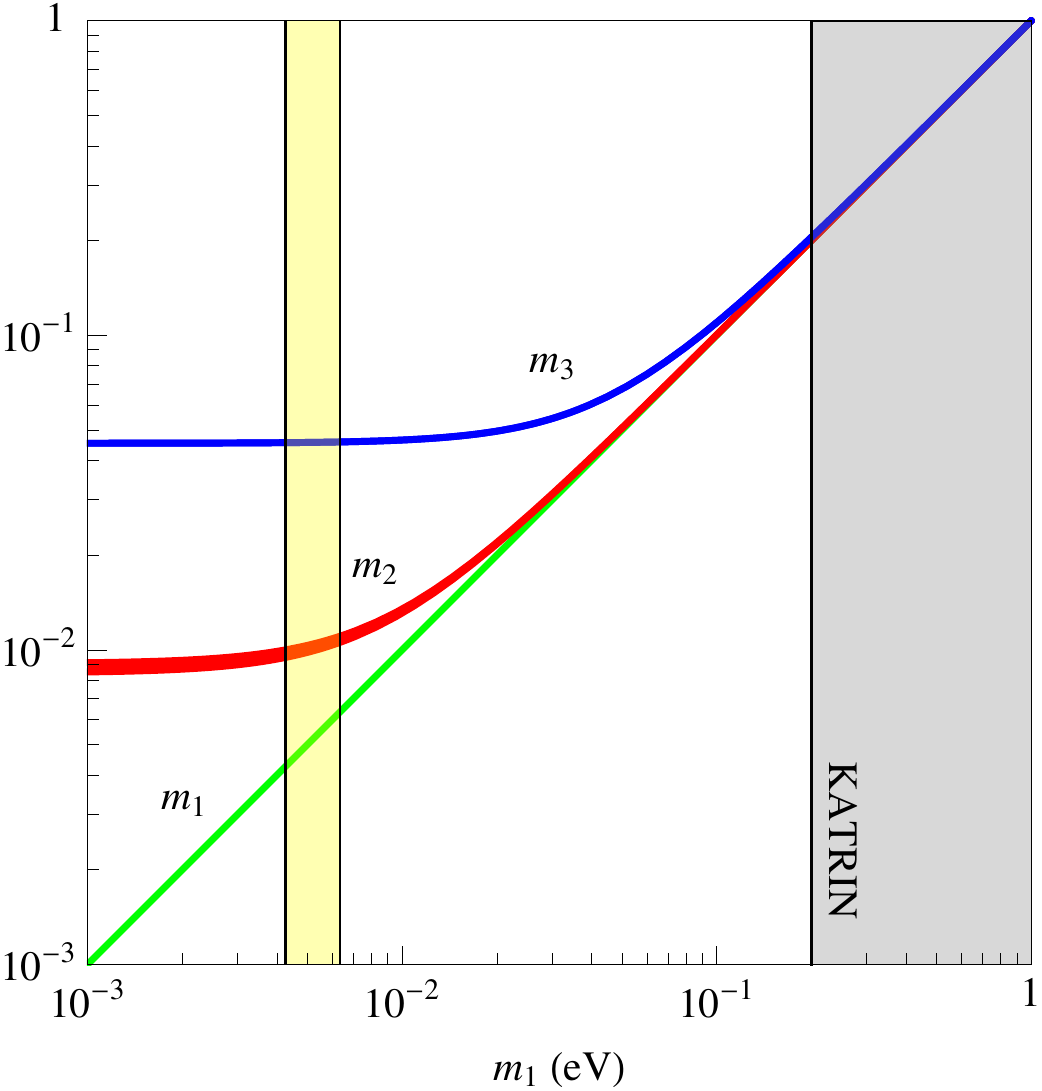}\qquad
\includegraphics[width=6cm]{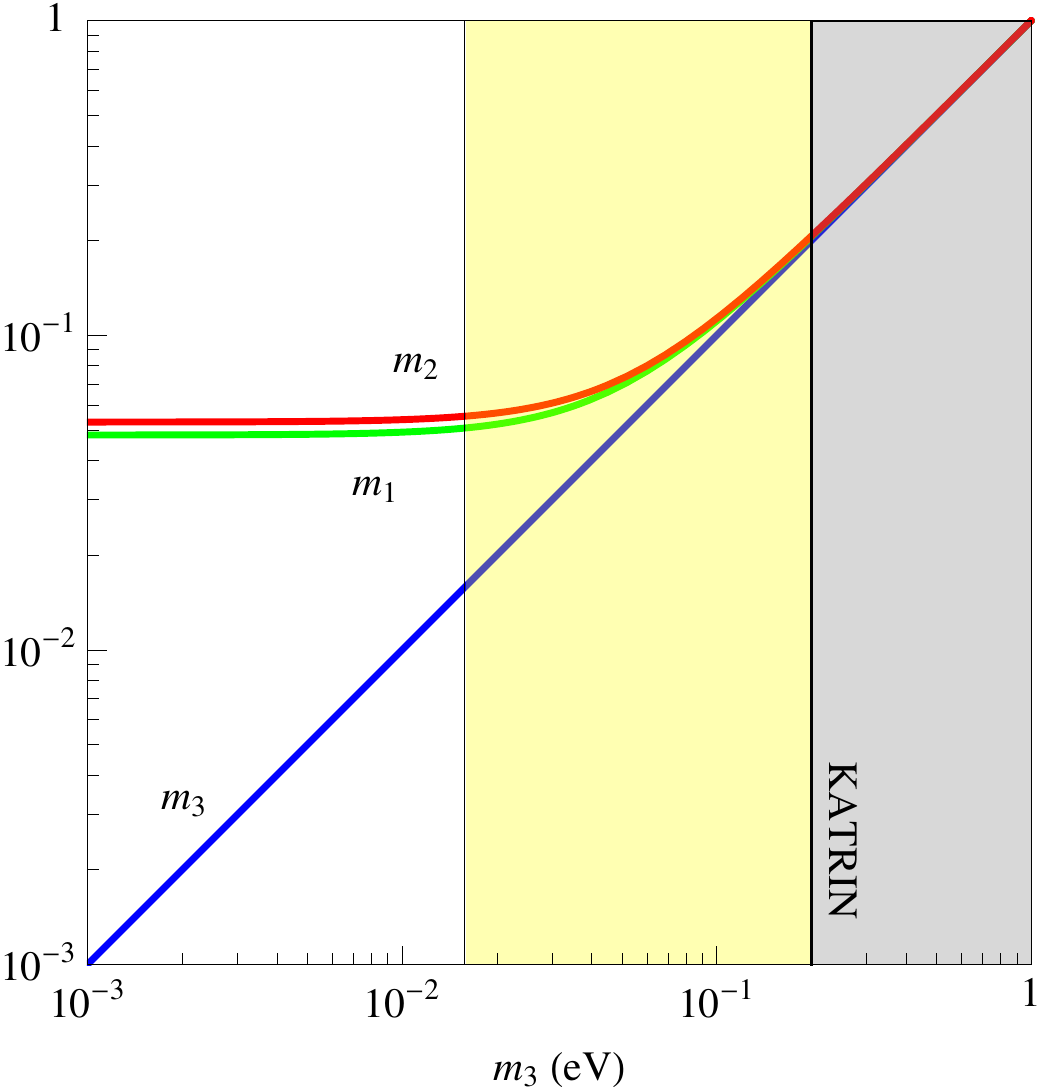}
\includegraphics[width=6cm]{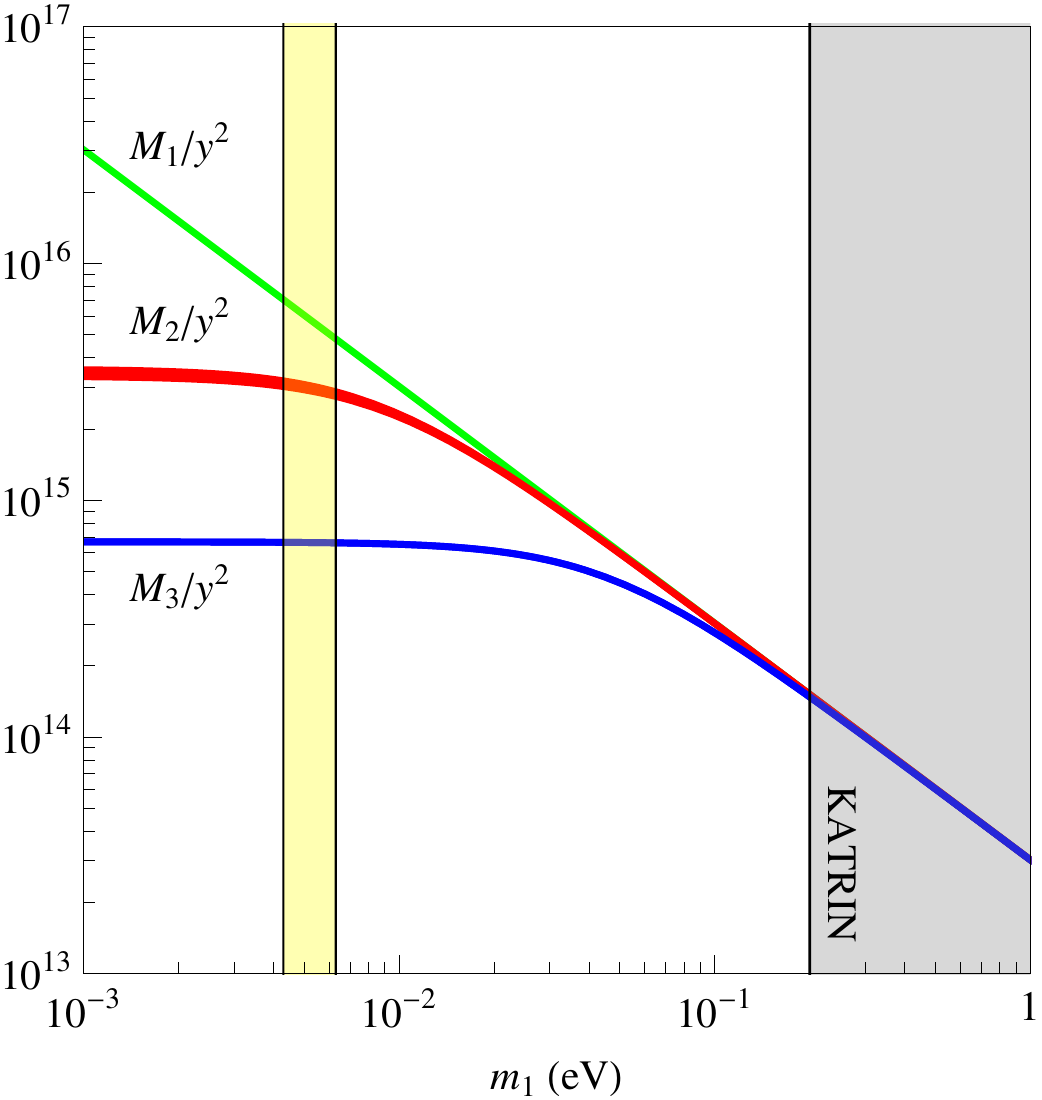}\qquad
\includegraphics[width=6cm]{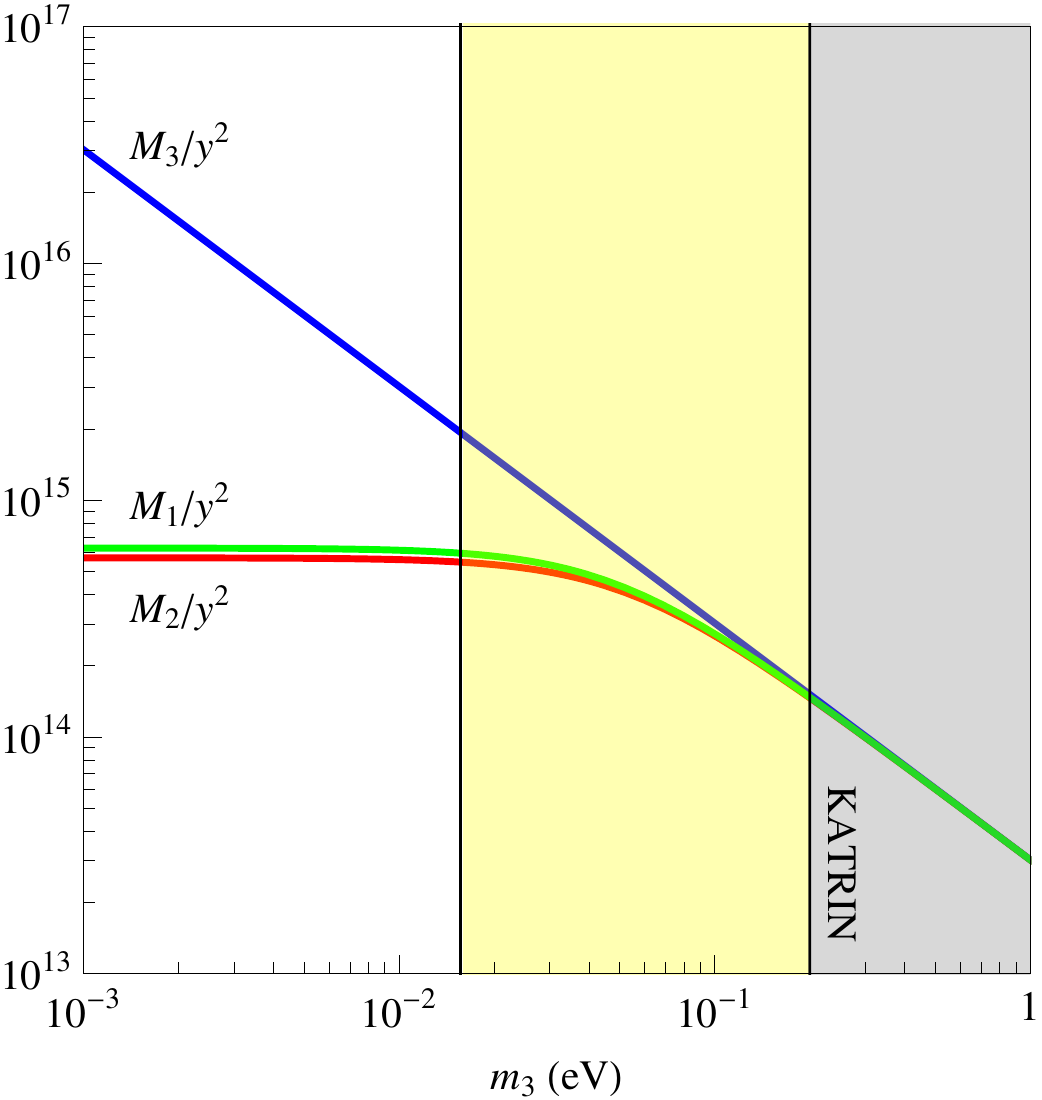}
\caption{\it Plots of the light (above) and heavy (below) neutrino masses, as a function of the lightest left-handed neutrino mass. On the left the normal hierarchy and on the right the inverse hierarchy. The yellow areas refer to the allowed range for $m_{1(3)}$ as in eq. (\ref{AFTBM:RangeMasses}). The vertical black lines correspond to the future sensitivity of KATRIN experiment.}
 \label{fig:AFSeeSaw}
\end{figure}

For the normal hierarchy, $m_1$ spans a narrow range of values, which corresponds to values of $\Delta$ close to zero. This completely determines   the neutrino masses inside a very small range and represents a prediction of the model. On the other hand, for the inverse hierarchy, $m_3$ is bounded only from below and the minimum is achieved when $\Delta$ is close to $\pm\pi$. In figure \ref{fig:AFSeeSaw} we can read off the light neutrino spectrum and its dependence with the lightest neutrino mass.

From eq. (\ref{AFTBM:LightMasses}) it is possible to describe the leading order spectrum of the right-handed neutrinos as a function of a unique parameter, which is the lightest left-handed neutrino mass. In all the allowed range for $m_{1,3}$, the order of magnitude of the right-handed neutrino masses falls between $10^{14}$ GeV and  $10^{15}$ GeV. In fig. (\ref{fig:AFSeeSaw}) we show explicitly the right-handed neutrino masses for normal hierarchy and inverse hierarchy, on the left and on the right respectively. The ratios among the masses are well defined for the NH, thanks to the narrow allowed range for $m_1$: $M_1/M_3\sim11$ and $M_2/M_3\sim5$. In the case of the IH, the ratio $M_1/M_2$ is fixed at $1$ while $M_3/M_2$ varies from about $3$ to $1$, going from the lower bound of $m_3$ up to the KATRIN sensitivity.

The analysis done for the Majorana phases in eqs. (\ref{AFTBM:Majorana1}, \ref{AFTBM:Majorana2}) is still valid here. \\

It is interesting to comment also in this context about the results for the neutrinoless-double-beta decay. The parameter $|m_{ee}|$ can be written as
\baq
\mathrm{NH:}&&|m_{ee}|=\dfrac{1}{3}\sqrt{\dfrac{9 m_1^4+2\De m^2_{atm}\De m^2_{sol}+m_1^2(10\De m^2_{atm}+\De m^2_{sol})}{m_1^2+\De m^2_{atm}}}\\[5mm]
\mathrm{IH:}&&|m_{ee}|=\dfrac{1}{3m_3}\sqrt{9 m_3^4+m_3^2(8\De m^2_{atm}-\De m^2_{sol})+\De m^2_{atm}(-\De m^2_{atm}+\De m^2_{sol})}\;.\nn
\eaq

\begin{figure}[ht!]
 \centering
\includegraphics[width=7.8cm]{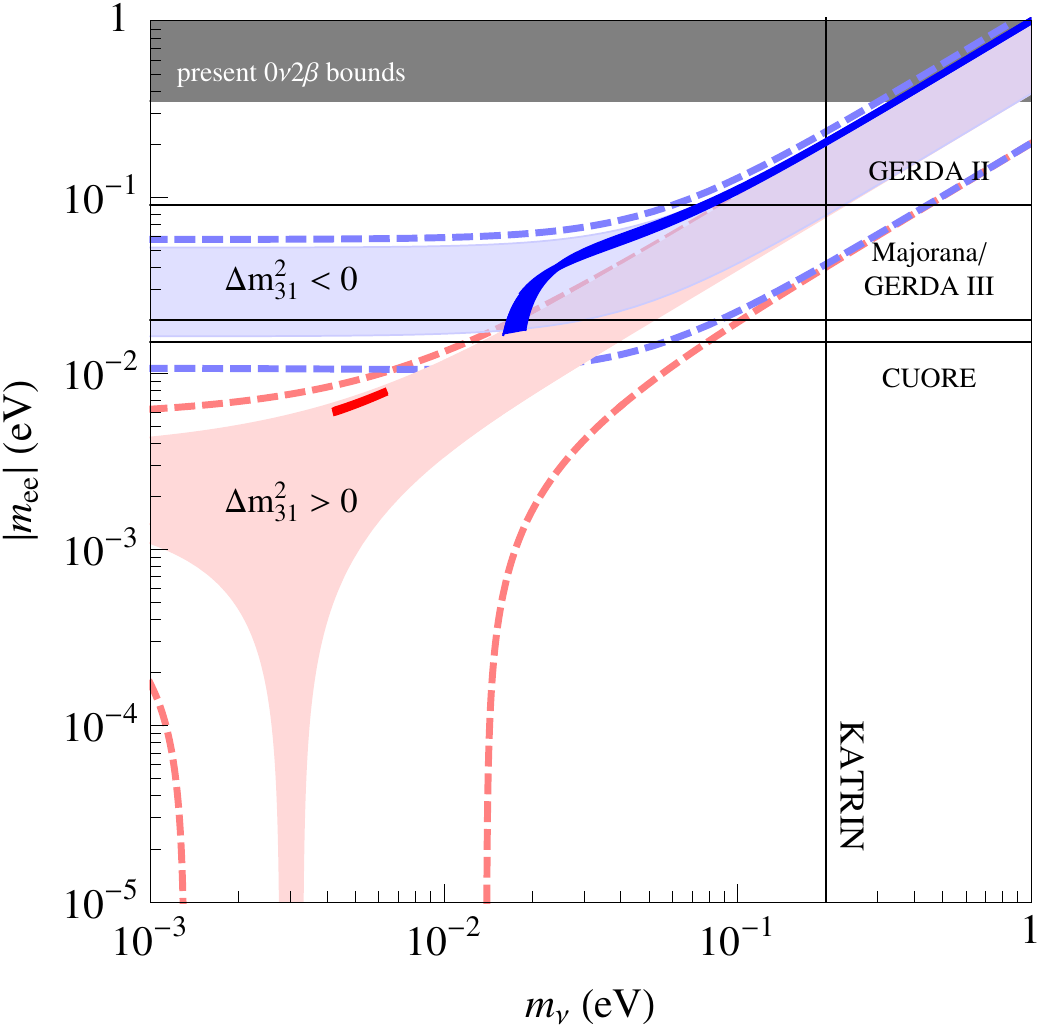}
\caption{\it $|m_{ee}|$ as a function of the lightest neutrino mass $m_{1(3)}$ for the NH (IH). See figure \ref{fig:AF_0nu2beta} for the details of the plots.}
 \label{fig:AF_0nu2betaSS}
\end{figure}

In figure \ref{fig:AF_0nu2betaSS}, we show the behaviour of $|m_{ee}|$ as a function of the lightest neutrino mass for both the mass hierarchies: in red the NH and in blue the IH. The red profile for the NH case is restricted to a really narrow range of values and we can conclude that $|m_{ee}|$ remains just below the future experiment sensitivity. On the contrary, the blue line which represents the IH case remains well above the future experiment sensitivity, except in the most pessimistic situation when the lower bound is saturated. As a concluding comment we can say that the Altarelli-Feruglio predicts that if the neutrino masses are explained by the type I See-Saw and the neutrino ordering is inverse, then a $0\nu2\be$-decay signal will be observed in the next future with very high probability.

These results are valid only at the leading order and some deviations are expected with the introduction of the higher-order terms. The result of a direct computation shows that for the NH spectrum the corrections leave approximatively unaffected eqs. (\ref{AFTBM:RangeMasses}); this is true for the IH case too, apart when the neutrino masses reach values at about $0.1$ eV for which the deviations become significant.

\subsection{Extension to Quarks}
\label{Sec:AFTBM:Quarks}
\setcounter{footnote}{3}

In this section we address the question of looking for a realistic description of quarks through the flavour group $A_4\times Z_3\times U(1)_{FN}(\times U(1)_R)$ in the context of the Standard Model (MSSM). An attractive possibility is to adopt for quarks the same representations under $A_4$ that have been used for leptons: the left-handed quark doublets $q$ transform as a triplet $\bf3$, while the right-handed quarks $(u^c,\,d^c)$, $(c^c,\,s^c)$ and $(t^c,\,b^c)$ transform as $\bf1$, $\bf1''$ and $\bf1'$, respectively. We can similarly extend to quarks the transformations of $Z_3$ (and $U(1)_R$) given for leptons. As a result the Lagrangian for the quark sector reads:
\beq
\begin{split}
\LL_q=&\phantom{+}\dfrac{y_d}{\Lambda_f}\, d^c H^\dagger \left(\varphi_T q\right)
+\dfrac{y_s}{\Lambda_f}\, s^c H^\dagger \left(\varphi_T q\right)'
+\dfrac{y_b}{\Lambda_f}\, b^c H^\dagger \left(\varphi_T q\right)''+\\
&+\dfrac{y_u}{\Lambda_f}\, u^c H^\dagger \left(\varphi_T q\right)
+\dfrac{y_c}{\Lambda_f}\, c^c H^\dagger \left(\varphi_T q\right)'
+\dfrac{y_t}{\Lambda_f}\, t^c H^\dagger \left(\varphi_T q\right)''+\hc+\ldots
\end{split}
\eeq
where dots stand for higher-order terms. When the flavour and the electroweak symmetry breakings occur this Lagrangian provide the mass matrices for quarks: it is straightforward to verify that the mass matrices are diagonal, leading to a diagonal CKM mixing matrix, that represents a good first order approximation. In order to explain the mass hierarchies, a suitable charge assignment under $U(1)_{FN}$ can be implemented.

Looking at the Lagrangian we find a disadvantage of adopting for quarks the same representations used for lepton: the top Yukawa does not arise at the renormalisable level. Furthermore, we expect that the NLO corrections introduce additional terms in the CKM in order to switch on the mixing angles and in particular the Cabibbo angle. However, we have already seen that the NLO corrections are of relative order $u\approx \lambda^2$ with respect to the starting values and therefore they are too small to describe the Cabibbo angle. To a closer sight, the NLO corrections do not come from higher-order operators in the Lagrangian, but from the new vacuum as described in eq. (\ref{AFTBM:vevsplus}). As a results the corrections are the same in the up and down sectors, apart negligible differences, and therefore they almost exactly cancel in the CKM matrix.

The conclusion is that new symmetry breaking sources are needed in order to describe quarks adopting a similar description used for leptons. Alternatively, we can try to enlarge the flavour symmetry group in such a way to reproduce similar results in the lepton sector as in the Altarelli-Feruglio model and to find a new method to correctly describe quarks. In the next two sections we follow this second approach, adopting as flavour symmetry the $T'$ and the $S_4$ discrete groups.

\mathversion{bold}
\section{$T'$-Based Model}
\label{Sec:TpTBM}
\setcounter{footnote}{3}
\mathversion{normal}

In this section we recall the main features of the flavour model based on $T'$ \cite{FHLM_Tp}, the double-valued group of the tetrahedral symmetry, which is isomorphic to $A_4$. Further synonyms of $T^{\prime}$ are Type $24/13$ and $SL_{2} (F_{3})$ \cite{CF_Tp}. The key role in our construction is played by the fact that $T'$ is the double covering of the tetrahedral group $A_4$. The relation between $T'$ and $A_4$ can be understood by thinking of $A_4$, the group of proper rotation in the three-dimensional space leaving a regular tetrahedron invariant, as a subgroup of $SO(3)$. Thus the $12$ elements of $A_4$ are in a one-to-one correspondence with $12$ sets of Euler angles. Now consider $SU(2)$, the double covering of $SO(3)$, possessing ``twice'' as many elements as $SO(3)$. There is a correspondence from $SU(2)$ to $SO(3)$ that maps two distinct elements of $SU(2)$ into the same set of Euler angles of $SO(3)$. The group $T'$ can be defined as the inverse image under this map of the group $A_4$.

The group $T'$ has $24$ elements and has two kinds of representations. It contains the representations of $A_4$: one triplet $\bf3$ and three singlets $\bf1$, $\bf1'$ and $\bf1''$. When working with these representations there is no way to distinguish the group $T'$ from the group $A_4$. In particular, in these representations, the elements of $T'$ coincide two by two and can be described by the same matrices that represent the elements in $A_4$. The other representations are three doublets $\bf2$, $\bf2'$ and $\bf2''$. Note that $A_{4}$ is not a subgroup of $T'$, since the two-dimensional representations cannot be decomposed into representations of $A_{4}$. 

It is generated by two elements $S$ and $T$ fulfilling the relations:
\begin{equation}
S^{2}=\mathbb{R}\;, \;\; T^{3}=\mathbb{1}\;, \;\; (S T)^{3}=\mathbb{1}\;, \;\; \mathbb{R}^{2}=\mathbb{1}\;,
\end{equation}
where $\mathbb{R}=\mathbb{1}$ in case of the odd-dimensional representation and $\mathbb{R}=-\mathbb{1}$ for $\bf2$, $\bf2'$ and $\bf2''$ such that $\mathbb{R}$ commutes with all elements of the group. Beyond the center of the group, generated by the elements $\unity$ and $\mathbb{R}$, there are other Abelian subgroups: $Z_3$, $Z_4$ and $Z_6$. In particular, there is a $Z_4$ subgroup here denoted by $G_S$, generated by the element $TST^2$ and a $Z_3$ subgroup here called $G_T$, generated by the element $T$.
As we will see $G_S$ and $G_T$ are of great importance for the structure of our model. Realisations of $S$ and $T$ for all the representations can be found in the appendix \ref{AppA:Tp}.

The multiplication rules of the representations are as follows:
\beq
\begin{array}{l}
{\bf1}^a \times {\bf r}^b = {\bf r}^b\times {\bf1}^a={\bf r}^{a+b}\qquad\qquad\text{for}\;{\bf r}={\bf1},\,{\bf2}\\
{\bf1}^a \times {\bf3} = {\bf3} + {\bf1}^a = {\bf3}\\
{\bf2}^a \times {\bf2}^b = {\bf3} + {\bf1}^{a+b}\\
{\bf2}^a \times {\bf3} = {\bf3}\times {\bf2}^a = {\bf2} + {\bf2'} + {\bf2''}\\
{\bf3} \times {\bf3} = {\bf3} + {\bf3} + {\bf1} + {\bf1'} + {\bf1''}
\end{array}
\label{TpTBM:mult}
\eeq
where $a,\,b=0,\pm1$ and we have denoted ${\bf1}^0\equiv{\bf1}$, ${\bf1}^{1}\equiv{\bf1'}$, ${\bf1}^{-1}\equiv{\bf1''}$ and similarly for the doublet representations. On the right-hand side the sum $a+b$ is modulo 3. The Clebsch-Gordan coefficients for the decomposition of product representations are shown in the appendix \ref{AppA:Tp}.

\subsection{Outline of the Model}
\label{Sec:TpTBM:Outline}
\setcounter{footnote}{3}

In this section we introduce our model and we illustrate its main features. We choose the model to be supersymmetric, which help us when discussing the vacuum selection and the symmetry breaking pattern of  $T'$. The model is required to be invariant under a flavour symmetry group $G_f=T' \times Z_3 \times U(1)_{FN} \times U(1)_R$. The group factor $T'$ is the one responsible for the tribimaximal lepton mixing. The group $T'$ is unable to produce the necessary mass suppressions for all the fermions. These suppressions originate in part from a spontaneously broken $U(1)_{FN}$, according to the original Froggatt-Nielsen mechanism. The $Z_3$ factor helps
in keeping separate the contributions to neutrino masses and to charged fermion masses, and it is an important ingredient in the vacuum alignment analysis. Finally, the $U(1)_R$ contains the usual $R$-parity as a subgroup and simplifies the constructions of the scalar potential. The fields of the model, together with their transformation properties under the flavour group, are listed in table \ref{table:Tptransformations}.

\begin{table}[!ht]
\hspace{-.5cm}
\begin{tabular}{|c||c|c|c|c||c|c|c|c|c|c||c|c|c|c|c|c|c|}
\hline
&&&&&&&&&&&&&&&&&\\[-4mm]
& $\ell$ & $e^c$ & $\mu^c$ & $\tau^c$ &  $D_q$ & $D_u^c$ & $D_d^c$ & $q_3$ & $t^c$ & $b^c$  &  $H_{u,d}$ & $\theta$ & $\phit$ & $\phis$ & $\xi$, $\xit$ & $\eta$ & $\xipp$ \\[2mm]
\hline
&&&&&&&&&&&&&&&&&\\[-4mm]
$T'$ & {\bf3} & {\bf1} & $\bf1''$ & $\bf1'$ & $\bf2''$ & $\bf2''$ & $\bf2''$ & {\bf1} & {\bf1} & {\bf1} & {\bf1} & {\bf1} & {\bf3} & {\bf3} & {\bf1} &  $\bf2'$ & $\bf1''$ \\[2mm]
$Z_3$ & $\om$ & $\om^2$ & $\om^2$ & $\om^2$ & $\om$ & $\om^2$ & $\om^2$ & $\om$ & $\om^2$ & $\om^2$ & $1$ & $1$ & $1$ & $\om$ & $\om$  & 1 & 1 \\[2mm]
$U(1)_{FN}$ & 0 & 2 & 1 & 0 & 0 & 1 & 1 & 0 & 0 & 1 & 0 & $-1$ & 0 & 0 & 0 & 0 & 0 \\[2mm]
$U(1)_R$ & 1 & 1 & 1 & 1 & 1 & 1 & 1 & 1 & 1 & 1 & 0 & 0 & 0 & 0 & 0 & 0 & 0 \\[2mm]
\hline
\end{tabular}
\caption{\it The transformation rules of the fields under the symmetries associated to the groups $T'$, $Z_3$, $U(1)_{FN}$ and $U(1)_R$.  We denote $D_q=(q_1,q_2)^t$ where $q_1=(u,d)^t$ and $q_2=(c,s)^t$ are the electroweak $SU(2)_L$-doublets of the first two generations, $D_u^c=(u^c,c^c)^t$ and $D_d^c=(d^c,s^c)^t$.  $D_q$, $D_u^c$ and $D_d^c$ are doublets of $T'$. $q_3=(t,b)^t$ is the electroweak $SU(2)_L$-doublet of the third generation. $q_3$, $t^c$ and $b^c$ are all singlets under $T'$.}
\label{table:Tptransformations}
\end{table}

The most important feature of our model is the pattern of symmetry breaking of the flavour group $T'$. We will see that, at the leading order, $T'$ is broken down to the subgroup $G_S$, generated by the element $TST^2$, in the neutrino sector and to the subgroup $G_T$, generated by $T$, in the charged fermion sector. This pattern of symmetry breaking is achieved dynamically and corresponds to a local minimum of the scalar potential of the model. This result is already sufficient to understand the predicted pattern of fermion mixing angles.
Indeed, given the $T'$ assignment of the matter fields displayed in table \ref{table:Tptransformations} and the explicit expressions
of the generators $S$ and $T$ for the various representations (see appendix \ref{AppA:Tp}), specific mass textures are obtained from the requirement of invariance under $TST^2$ or $T$. For instance, neutrinos are in a triplet of $T'$ and the element $TST^2$ in the triplet representations is given by:
\beq
TST^2=\dfrac{1}{3}\left(\begin{array}{ccc}
                        -1 & 2 & 2 \\
                        2 & -1 & 2 \\
                        2 & 2 & -1 \\
                    \end{array}\right)\;.
\eeq
The most general mass matrix for neutrinos invariant under $G_S$, in arbitrary units, is given by:
\beq
m_\nu=\left(
            \begin{array}{ccc}
            a+c & -b/3-c+d & -b/3 \\
            -b/3-c+d & c & a-b/3 \\
            -b/3 & a-b/3 & d
            \end{array}
        \right)
\label{TpTBM:t1}
\eeq
where $a$, $b$, $c$ and $d$ are arbitrary parameters. Similarly, the most general mass matrices for charged fermions invariant under $G_T$ have the following structure:
\beq
M_e=\left(
        \begin{array}{ccc}
        \times & 0 & 0 \\
        0 & \times & 0 \\
        0 & 0 & \times \\
        \end{array}
    \right)\;,\qquad
M_{u,d}=\left(
                \begin{array}{ccc}
                0 & 0 & 0 \\
                0 & \times & \times \\
                0 & \times & \times \\
                \end{array}
            \right)
\label{TpTBM:t2}
\eeq
where a cross denotes a non-vanishing entry.

The lepton mixing originates completely from $m_\nu$ and, with an additional requirement, reproduces the tribimaximal scheme. This additional requirement is the condition $c=d$, which is not generically implied by the invariance under $G_S$. In our model the fields that break $T'$ along the $G_S$ direction are a triplet $\varphi_S$ and an invariant singlet $\xi$. There are no further scalar singlets, transforming as $\bf1'$ or $\bf1''$ that couple to the neutrino sector. We will see in a moment that due to this restriction our model gives rise to a particular version of the neutrino mass matrix in eq (\ref{TpTBM:t1}), where $c=d=2b/3$, which implies directly a tribimaximal mixing. This feature holds also in the Altarelli-Feruglio model and it works exactly in the same way.

It is interesting to note that, while the requirement of $G_T$ invariance implies a diagonal mass matrix in the charged lepton sector, this is not the case for the quark sector, due to the different $T'$ assignment. At the leading order, in both up and down sectors, we get mass matrices with vanishing first row and column, eq. (\ref{TpTBM:t2}). Moreover, the element $(33)$ of both mass matrices is larger than the other elements, since it is invariant under the full $T'$ group, not only the $G_T$ subgroup. The other non-vanishing elements carry a suppression factor originating from the breaking of $T'$ down to $G_T$. This pattern of quark mass matrices, while not yet fully realistic, is however encouraging, since it reproduces correctly masses and mixing angle of the second and third generations. As we will see, the textures in eqs. (\ref{TpTBM:t1}, \ref{TpTBM:t2}, \ref{TpTBM:t2}) are modified by subleading effects. These effects are sufficiently small to keep the good features of the leading order approximation, and large enough to provide a realistic description of the quark sector.

Fermion masses are generated by the superpotential $w$:
\beq
w=w_\ell+w_q+w_d
\label{TpTBM:fullw}
\eeq
where $w_\ell$ is the term responsible for the Yukawa interactions in the lepton sector, $w_q$ is the analogous term for quarks and $w_d$ is the term responsible for the vacuum alignment. We will consider the expansion of $w$ in inverse powers of the cutoff scale $\Lambda_f$ and we will write down only the first non-trivial terms of this expansion. This will provide a leading order approximation, here analysed in detail. Corrections to this approximation are produced by higher dimensional operators contributing to $w$. As described in section \ref{AppB:Tp}, at the leading order,
the scalar components of the supermultiplets $\varphi_T$, $\varphi_S$, $\xi$, $\tilde{\xi}$, $\eta$ and $\xi''$ develop VEVs according to
\bea
\mean{\varphi_S}=(v_S,\,v_S,\,v_S)\;,&\qquad
\mean{\xi}=v_\xi\;,&\qquad
\mean{\xit}=0\;,
\label{TpTBM:love1}\\[3mm]
\mean{\varphi_T}=(v_T,\,0,\,0)\;,&\qquad
\mean{\eta}=(v_1,0)\;,&\qquad
\mean{\xi''}=0\;.
\label{TpTBM:love2}
\eea
Notice that the VEVs of $\varphi_T$, $\varphi_S$, $\xi$ and $\tilde{\xi}$ correspond to those in the Altarelli-Feruglio model and, as in that realisation, it is reasonable to choose:
\beq
\dfrac{VEV}{\Lambda_f}\approx \lambda^2\;.
\label{TpTBM:vevratio}
\eeq
Furthermore, there is a neat misalignment in flavour space between $\langle\varphi_T\rangle$, $\langle\eta\rangle$ and $\langle\varphi_S\rangle$:
$\langle\varphi_T\rangle=(v_T,0,0)$, $\langle\eta\rangle=(v_1,0)$ and $\langle\xi''\rangle=0$ break $T'$ down to the subgroup $G_T$, while $\langle\varphi_S\rangle=(v_S,v_S,v_S)$ breaks $T'$ down to the subgroup $G_S$. It is precisely this misalignment the origin of the mass textures in eqs. (\ref{TpTBM:t1}, \ref{TpTBM:t2}, \ref{TpTBM:t2}).

A certain freedom is present in our formalism and this can lead to models that are physically equivalent though different at a superficial level, when comparing VEVs or mass matrices. One source of freedom is related to the possibility of working with different basis for the generators $S$ and $T$.  Another source of freedom is related to the fact that vacua that break $T'$ are degenerate and lie in orbits of the flavour group. For instance, when we say that the set of VEVs in eq. (\ref{TpTBM:love2}) breaks $T'$ leaving invariant the $Z_3$ subgroup generated by $T$, VEVs obtained from this set by acting with elements of $T'$ are degenerate and they preserve other $Z_3$ subgroups of $T'$. Both these sources of freedom can lead to mass matrices different from those explicitly shown in eqs. (\ref{TpTBM:t1}, \ref{TpTBM:t2}, \ref{TpTBM:t2}). It is however easy to show that the different ``pictures'' are related by field redefinitions and the physical properties of the system, such as the mass eigenvalues and the physical mixing angles, are always the same. Thus it is not restrictive to work in a particular basis and to choose a single representative VEV configuration, as we will do in the following.

\subsection{Fermion Masses and Mixings at the Leading Order}
\label{Sec:TpTBM:MassesMixingsLO}
\setcounter{footnote}{3}

Lepton masses are described by $w_\ell$, given by:
\beq
\begin{split}
w_\ell=&\phantom{+}\dfrac{y_e}{\Lambda_f^3} \theta^2e^c H_d \left(\varphi_T \ell\right) +\dfrac{y_\mu}{\Lambda_f^2} \theta\mu^c H_d \left(\varphi_T \ell\right)' +\dfrac{y_\tau}{\Lambda_f} \tau^c H_d \left(\varphi_T \ell\right)''+\\[3mm]
&+\dfrac{x_a}{\Lambda_f\Lambda_L} \xi (\ell\ell)H_uH_u + \dfrac{x_b}{\Lambda_f\Lambda_L} (\varphi_S\ell\ell)H_uH_u+\ldots\;,
\end{split}
\label{TpTBM:wlplus}
\eeq
where dots here and in the following formulae stand for higher dimensional operators. The contractions under $SU(2)_L$ are understood and the notation $(\ldots)$, $(\ldots)'$ and $(\ldots)''$ refers to the contractions in $\bf1$, $\bf1'$ and $\bf1''$, respectively. This superpotential corresponds to the Lagrangian in eq. (\ref{AFTBM:Ll}, \ref{AFTBM:Lnu}) when the supersymmetric context is considered and therefore all the results listed in the Altarelli-Feruglio model exactly hold in this model based on $T'$. Just for simplicity, we recall here the mass matrices for the charged leptons and for neutrinos: here and in the following we make the notation more compact indicating with $t$ the ratio of the VEV of the flavon $\theta$ over the cutoff of the theory, similarly as in the Altarelli-Feruglio model, and we get
\beq
M_e=\left(
      \begin{array}{ccc}
        y_e t^2 & 0 & 0 \\
        0 & y_\mu t & 0 \\
        0 & 0 & y_\tau \\
      \end{array}
    \right)\dfrac{v_d\,v_T}{\sqrt2\La_f}\;,\qquad
m_\nu=\left(
        \begin{array}{ccc}
            a+2 b/3 & -b/3 & -b/3 \\
            -b/3 & 2b/3 & a-b/3 \\
            -b/3 & a-b/3 & 2 b/3 \\
        \end{array}
        \right)\dfrac{v_u^2}{\La_L}\;,
\label{TpTBM:LOMasses}
\eeq
where $a\equiv x_a\,v_\xi/\La_f$ and $b\equiv x_b\,v_S/\La_f$.\\

The contribution to the superpotential in the quark sector is given by
\beq
\begin{split}
w_q =&\phantom{+}y_t\left(t^c q_3\right)H_u+\dfrac{y_b}{\Lambda_f}\theta\left(b^c q_3\right)H_d+\\[3mm]
    &+\dfrac{y_1}{\Lambda_f^2}\theta(\phit D_u^c D_q) H_u + \dfrac{y_5}{\La_f^2}\theta(\phit D_d^c D_q) H_d+\\[3mm]
    &+\dfrac{y_2}{\La_f^2}\theta\xipp(D_u^c D_q)' H_u + \dfrac{y_6}{\La_f^2}\theta\xipp(D_d^c D_q)' H_d+\\[3mm]
    &+\dfrac{1}{\La_f}\left[y_3\, t^c (\eta D_q) + \dfrac{y_4}{\La_f}\theta(D_u^c \eta)q_3\right] H_u+\\[3mm]
    &+\dfrac{1}{\Lambda_f^2}\theta\Big[y_7\, b^c (\eta D_q) + y_8\, (D_d^c \eta) q_3 \Big] H_d+\ldots
\end{split}
\label{TpTBM:wmq}
\eeq
Observe that the supermultiplets $\varphi_S$, $\xi$ and $\tilde{\xi}$, which control the neutrino mass matrices, do not couple to the quark sector, at the leading order. Conversely, the supermultiplets $\varphi_T$, $\eta$ and $\xi''$, which give masses to the charged fermions, do not couple to neutrinos at the leading order. This separation is partly due to the discrete $Z_3$ symmetry, described in table \ref{table:Tptransformations}. By recalling the VEVs of eq. (\ref{TpTBM:love1}, \ref{TpTBM:love2}), we can write down the mass matrices for the up and down quarks: at the leading order we have
\beq
M_u=\left(\begin{array}{ccc}
                    0 & 0 & 0 \\[2mm]
                    0 & y_1\,t\, \dfrac{v_T}{\La_f} & y_4\,t\, \dfrac{v_1}{\La_f} \\[2mm]
                    0 & y_3\, \dfrac{v_1}{\La_f} & y_t \\[2mm]
            \end{array}\right)\dfrac{v_u}{\sqrt2}\;,\qquad
M_d=\left(\begin{array}{ccc}
                    0 & 0 & 0 \\[2mm]
                    0 & y_5\, \dfrac{v_T}{\La_f} & y_8\, \dfrac{v_1}{\La_f} \\[2mm]
                    0 & y_7\, \dfrac{v_1}{\La_f} & y_b \\[2mm]
            \end{array}\right)\dfrac{v_d\,t}{\sqrt2}\;.
\label{TpTBM:qmassm}
\eeq
These mass matrices are the most general ones that are invariant under $G_T$, see eq. (\ref{TpTBM:t2}). The following absolute values for quark masses and mixing angles are predicted, at the leading order:
\beq
\ba{ccc}
m_u=0\;, & \qquad m_c\approx y_1 \dfrac{v_u\, v_T\, t}{\sqrt2\,\La_f}\;, &\qquad  m_t\approx y_t\, \dfrac{v_u}{\sqrt2}\\[3mm]
m_d=0\;, & \qquad m_s\approx y_5 \dfrac{v_d\, v_T\, t}{\sqrt2\,\La_f}\;, &\qquad  m_b\approx y_b\, \dfrac{v_d\, t}{\sqrt2}\\[3mm]
V_{us}=0\;, & \qquad V_{ub}=0\;, & \qquad V_{cb}\approx \left(\dfrac{y_7}{y_b}-\dfrac{y_3}{y_t}\right)\dfrac{v_1}{\La_f}\;.
\ea
\eeq
The mass of the top quark is expected to be of the order of the VEV \mbox{$v_u\approx\cO(100\GeV)$}. The mass of the bottom quark is suppressed compared with $m_t$ by the Froggatt-Nielsen mechanism so that it is of the same order of $m_\tau$. For values of order one of the dimensionless
coefficients $y_b$ and $y_5$, the ratio $m_s/m_b$ is correctly reproduced since it is approximately given by $VEV/\Lambda_f$, which we already chose of order $\lambda^2$, see eq. (\ref{TpTBM:vevratio}). The mass of the charm quark is $m_c\approx \lambda^4v_u$ and therefore the ratio $m_c/m_t\approx\lambda^4$ holds. Finally the element $V_{cb}$ is of order $v_1/\La_f\approx\lambda^2$, in agreement with the experiments. Masses and mixing angles are still unrealistic, since $m_u/m_c$, $m_d/m_s$, $V_{ub}$ and $V_{us}$ are vanishing, at this level.
We will see that all these parameters can be generated by higher-order corrections, in particular those affecting the VEVs in eqs. (\ref{TpTBM:love1}, \ref{TpTBM:love2}).

\subsection{Higher-Order Corrections}
\label{Sec:TpTBM:NLO}
\setcounter{footnote}{3}

The inclusion of higher-order corrections is essential in our model. First of all, from these corrections we hope to achieve a realistic mass spectrum in the quark sector. The leading order result is encouraging, but quarks of the first generations are still massless at this level and there is no mixing allowing communication between the first generations and the other ones. Moreover we should check that the higher-order corrections do not spoil the leading order results. At the leading order there is a neat separation between the scalar fields giving masses to the neutrino sector and those giving masses to the charged fermion sector. As a result the $T'$ flavour symmetry is broken down in two different directions in the two sectors: neutrino mass terms are invariant under the subgroup $G_S$, while the charged fermion mass terms are invariant under the subgroup $G_T$. It is precisely this misalignment the source of the tribimaximal lepton mixing. Such a sharp separation is not expected to survive when higher dimensional operators are included and this will cause the breaking of the subgroup $G_S$ $(G_T)$ in the neutrino (charged fermion) sector. It is important to check that this further breaking does not modify too much the misalignment achieved at the leading order and that the tribimaximal mixing remains stable.

The corrections are induced by higher dimensional operators, compatible with all the symmetries of our model, that can be included in the superpotential $w$, thus providing the next terms in a $1/\Lambda_f$ expansion. Here we do not enter into the details of the analysis already presented in \cite{FHLM_Tp} and we simply give the results. 

When looking at the corrections to the flavon sector (see appendix \ref{AppB:Tp}), we can parametrise the new VEVs as
\beq
\begin{array}{c}
\mean{\varphi_T}=(v_T+\delta v_{T1},\delta v_{T2},\delta v_{T3})\;,\qquad
\mean{\varphi_S}=(v_S+\delta v_{S1},v_S+\delta v_{S2},v_S+\delta v_{S3})\;\\[3mm]
\mean{\xi}=v_\xi\;,\qquad
\mean{\tilde{\xi}}=\delta v_{\tilde{\xi}} \;,\qquad
\mean{\eta}=(v_1+\delta v_1,\delta v_2)\;,\qquad\langle\xi''\rangle=\delta v_{\xi''}\;.
\end{array}
\eeq
This modification will affect also the fermion mass matrices, since new flavour structures are expected to appear when these new VEVs are introduced in the superpotential $w$ of eqs. (\ref{TpTBM:wlplus}, \ref{TpTBM:wmq}).

Lepton masses and mixing angles are modified by terms of relative order $\lambda^2$, exactly in the same way as in the Altarelli-Feruglio model . This correction is close to the $3\sigma$ experimental error for $\theta_{12}$ and largely within the current uncertainties of $\theta_{23}$ and $\theta_{13}$. From the experimental view point, a small non-vanishing value $\theta_{13}\approx \lambda^2$ and a deviation from $\pi/4$ of order $\lambda^2$ of $\theta_{23}$, are both close to the reach of the next generation of neutrino experiments and will provide a valuable test of this model.

In the quark sector all the subleading effects contribute to give these new quark mass matrices:
\begin{gather}
M_u=\left(\begin{array}{ccc}
i\, y_1\,t\,\dfrac{\delta v_{T2}}{\Lambda_f} +...& (1-i)\,y_1\,t\,\dfrac{\delta v_{T3}}{2\Lambda_f}\,+\,y_2\,t\,\dfrac{\delta v_{\xi''}}{\La_f}& -\,y_4\,t\,\frac{\delta v_2}{\La_f} \\[2mm]
(1-i) \,y_1\,t\,\dfrac{\delta v_{T3}}{2\Lambda_f}-\,y_2\,t\,\dfrac{\delta v_{\xi''}}{\La_f} & \,y_1\,t\, \dfrac{v_T}{\La_f} & \,y_4\,t\, \dfrac{v_1}{\La_f} \\[2mm]
-\,y_3\dfrac{\delta v_2}{\La_f} & \,y_3 \dfrac{v_1}{\La_f} & \,y_t \\[2mm]
\end{array}\right)\frac{v_u}{\sqrt2}\\[3mm]
M_d=\left(\begin{array}{ccc}
i \,y_5\dfrac{\delta v_{T2}}{\La_f}+... & (1-i) \,y_5\dfrac{\delta v_{T3}}{2\Lambda_f}+\,y_6\dfrac{\delta v_{\xi''}}{\La_f}& -\,y_8 \frac{\delta v_2}{\La_f} \\[2mm]
(1-i) \,y_5\dfrac{\delta v_{T3}}{2\Lambda_f}-\,y_6\dfrac{\delta v_{\xi''}}{\La_f} & \,y_5 \dfrac{v_T}{\La_f} & \,y_8 \dfrac{v_1}{\La_f} \\[2mm]
- \,y_7 \dfrac{\delta v_2}{\La_f} & \,y_7 \dfrac{v_1}{\La_f} & \,y_b \\[2mm]
\end{array}\right)\dfrac{v_d\,t}{\sqrt2}\;,
\label{hoqm}
\end{gather}
where the dots in the $(11)$ entry of $m_u$ and $m_d$ stand for additional contributions from higher dimensional operators.
Not all the available parameter space is suitable to correctly reproduce the masses and the mixing angles of the first generation quarks. To explain this point we rewrite the quark mass matrices indicating only the order of magnitudes in terms of $\lambda$ for each entry:
\beq
M_u\sim\left(
      \begin{array}{ccc}
        \la^6 & \la^6 & \la^6 \\
        \la^6 & \la^4 & \la^4 \\
        \la^4 & \la^2 & 1 \\
      \end{array}
    \right)
\frac{v_u}{\sqrt2}\;,\qquad
M_d\sim\left(
    \begin{array}{ccc}
        \la^4 & \la^4 & \la^4 \\
        \la^4 & \la^2 & \la^2 \\
        \la^4 & \la^2 & 1 \\
    \end{array}
  \right)
\frac{v_d\,\la^2}{\sqrt2}\;.
\eeq
It is now easy to see that, up to small corrections,
\beq
\dfrac{m_u}{m_c}= \dfrac{m_d}{m_s}= \dfrac{\delta v_{T2}}{v_T}\approx \lambda^2\;,
\eeq
which is not correct in the up sector. To overcome this difficulty we assume that the correction $\delta v_{T2}$ is somewhat smaller than its natural value:
\beq
\dfrac{\delta v_{T2}}{v_T}\approx \lambda^4\;.
\label{TpTBM:ass1}
\eeq
This brings the up quark mass in the correct range but depletes too much the down quark mass. To get the appropriate mass for the down quark
we assume that the dimensionless coefficient $y_6$ is larger than one by a factor $1/\lambda$:
\beq
y_6\approx\dfrac{1}{\lambda}\;.
\label{TpTBM:ass2}
\eeq
We cannot justify the two assumptions in eqs. (\ref{TpTBM:ass1}, \ref{TpTBM:ass2}) within our approach, where, in the absence of a theory for the higher-order terms, we have allowed for the most general higher-order corrections. From our effective lagrangian approach, they should be seen as two moderate tunings that we need in order to get up and down quark masses. To summarise, in our parameter space we naturally have all dimensionless parameters of order one, with the exception of $y_6$. Concerning the VEVs, we can naturally accommodate $VEV/\Lambda_f\approx\delta VEV/VEV\approx \lambda^2$, with the exception of $\delta v_{T2}$. Looking at the equations in appendix \ref{AppB:Tp}, it is easy to see that it has no consequences on the other shifts of the VEVs. Within the restricted region of the parameter space where the two relations in eqs. (\ref{TpTBM:ass1}, \ref{TpTBM:ass2}) are approximately valid, the quark mass matrices have the following structures:
\beq
M_u=\left(\begin{array}{ccc}
                \lambda^8 & \lambda^6 & \lambda^6 \\
                \lambda^6 & \lambda^4 & \lambda^4 \\
                \lambda^4 & \lambda^2 & 1 \\
        \end{array}\right)\dfrac{v_u}{\sqrt2}\;,\qquad
M_d=\left(\begin{array}{ccc}
                \lambda^6 & \lambda^3 & \lambda^4 \\
                \lambda^3 & \lambda^2 & \lambda^2 \\
                \lambda^4 & \lambda^2 & 1 \\
        \end{array}\right)\dfrac{v_d\lambda^2}{\sqrt2}\;.
\label{qmtextures}
\eeq
By diagonalising the matrices in eq. (\ref{hoqm}) with standard perturbative techniques we obtain:
\beq
\begin{array}{ll}
m_u\approx \left\vert y_1\dfrac{v_u\,t}{\sqrt2} \left\{i\dd\frac{\delta v_{T2}}{\Lambda_f}-\left[\left(\dd\frac{1-i}{2}\right)^2 \dd\frac{\delta v_{T3}^2}{v_T\Lambda_f}-\dd\frac{y_2^2}{y_1^2} \dd\frac{\delta v_{\xi''}^2}{v_T\Lambda_f}\right]\right\}+...\right\vert\;,&
m_d\approx \left\vert \dfrac{v_d\,t}{\sqrt2} \dd\frac{y_6^2}{y_5}\dd\frac{\delta v_{\xi''}^2}{v_T\Lambda_f}\right\vert\;,\\[5mm]
m_c\approx \left\vert y_1 \dfrac{v_u\,t}{\sqrt2} \dd\frac{v_T}{\Lambda_f}\right\vert+O(\lambda^6)\;,&
m_s\approx \left\vert y_5 \dfrac{v_d\,t}{\sqrt2} \dd\frac{v_T}{\Lambda_f}\right\vert+O(\lambda^6)\;,\\[5mm]
m_t\approx \left\vert y_t \dfrac{v_u\,t}{\sqrt2}\right\vert+O(\lambda^4)\;, &
m_b\approx \left\vert y_b \dfrac{v_d\,t}{\sqrt2}\right\vert+O(\lambda^6)\;.
\end{array}
\eeq
For the mixing angles, we get:
\beq
\ba{l}
V_{ud}\approx V_{cs}\approx 1+O(\lambda^2)\;,\qquad\qquad V_{tb}\approx 1\;,\\[5mm]
V_{us}^*\approx -V_{cd}\approx-\dd\frac{y_6}{y_5}\dd\frac{\delta v_{\xi''}}{v_T}-
\left[\left(\dd\frac{1-i}{2}\right)\dd\frac{\delta v_{T3}}{v_T}-\dd\frac{y_2}{y_1}\dd\frac{\delta v_{\xi''}}{v_T}\right]+O(\lambda^3)\;,\\[5mm]
V_{ub}^*\approx-\left(\dd\frac{y_7}{y_b}-\dd\frac{y_3}{y_t}\right) \left\{\dd\frac{\delta v_2}{\Lambda_f}+\dd\frac{v_1}{v_T}
\left[\left(\dd\frac{1-i}{2}\right)\dd\frac{\delta v_{T3}}{\Lambda_f}-\dd\frac{y_2}{y_1}\dd\frac{\delta v_{\xi''}}{\Lambda_f}\right]     \right\}\;,
\ea
\eeq
\beq
\ba{l}
V_{cb}^*\approx-V_{ts}\approx\left(\dd\frac{y_7}{y_b}-\dd\frac{y_3}{y_t}\right) \dd\frac{v_1}{\Lambda_f}\;,\\[5mm]
V_{td}\approx-\dd\frac{y_6}{y_5} \left(\dd\frac{y_7}{y_b}-\dd\frac{y_3}{y_t}\right) \dd\frac{v_1\delta v_{\xi''}}{v_T \Lambda_f}+
\left(\dd\frac{y_7}{y_b}-\dd\frac{y_3}{y_t}\right) \dd\frac{\delta v_2}{\Lambda_f}\;,
\ea
\eeq
where, when not explicitly indicated, the relations include all terms up to $O(\lambda^4)$. In the previous expressions, where all the quantities are generically complex, is possible to remove all phases except the one carried by the combination $(y_7/y_b-y_3/y_t)\delta v_2/\Lambda_f$ which enters $V_{ub}$ and $V_{td}$ at the order $\lambda^4$. Notice that in our model $V_{ub}$ is of order $\lambda^4$ whereas $V_{td}$ is of order $\lambda^3$. In the Wolfenstein parametrisation of the mixing matrix, this corresponds to a combination $\rho+i\eta$ of order $\lambda$, which is phenomenologically viable. Notice that quark masses and mixing angles are all determined within their correct order of magnitudes and enough parameters are present to fit the data. Moreover, despite the large number of parameters controlling the quark sector, our model contains a well-known \cite{GST_Relation} non-trivial relation between masses and mixing angles:
\beq
\sqrt{\dfrac{m_d}{m_s}}=\left\vert V_{us}\right\vert+O(\lambda^2)\;.
\label{TpTBM:p1}
\eeq
Due to the approximate unitarity relation $V_{td}+V_{us}^* V_{ts}+V_{ub}^*=0$ and due to the fact that $V_{ub}$ is of order $\lambda^4$
in our model, from the relation (\ref{TpTBM:p1}) we also get:
\beq
\sqrt{\dfrac{m_d}{m_s}}=\left\vert\dfrac{V_{td}}{V_{ts}}\right\vert+O(\lambda^2)\;.
\label{TpTBM:p2}
\eeq
These relations well compare with the data: from \cite{PDG08} we have $\sqrt{m_d/m_s}=0.213\div 0.243$, $\vert V_{us}\vert=0.2257\pm0.0010$
and $\vert V_{td}/V_{ts}\vert=0.209\pm0.001\pm0.006$. Unfortunately, the theoretical errors affecting eqs. (\ref{TpTBM:p1}) and (\ref{TpTBM:p2}), dominated respectively by the unknown $O(\lambda^2)$ term in $V_{us}$ and by the unknown $O(\lambda^4)$ term in
$V_{td}$, are of order $20\%$. For this reason, and for the large uncertainty on the ratio $m_d/m_s$, it is not possible to turn these predictions into precise tests of the model.
It is interesting to compare our predictions with those of early models of quark masses based on $U(2)$ or $T'$ flavour symmetries \cite{4ZerosQuarks,56ZerosQuarks,QuarkRelations}.
They also predict eq. (\ref{TpTBM:p2}), with a smaller theoretical error of order $\lambda^3$.
Moreover, due to the characteristic two zero textures, in their early versions they predict $\sqrt{m_u/m_c}=\vert V_{ub}/V_{cb}\vert$,
which is off by approximately a factor two. In our model the mass of the up quark depends on additional free parameters,
that modify this wrong relation by a relative factor of order one.

\mathversion{bold}
\section{$S_4$-Based Model}
\label{Sec:S4TBM}
\setcounter{footnote}{3}
\mathversion{normal}

In the previous section we used the symmetry group $T'$ in order to describe the quark sector, keeping almost unchanged the lepton sector description of the Altarelli-Feruglio model. In this section we illustrate an alternative way to describe both leptons and quarks in a unified context based on the discrete group $S_4$. It is the group of the permutations of four objects and is composed by $24$ elements. It can be defined by two generators $S$ and $T$ that satisfy
\begin{equation}
S^4 = T^3 =  (ST^2)^2 = 1 \;.
\label{S4TBM:rel}
\end{equation}
The  three relations reported above directly indicate which are the discrete Abelian subgroups of $S_4$: $Z_4$, $Z_3$ and $Z_2$ respectively. Furthermore, $S_4$  presents $5$ irreducible representations \footnote{It has the same number of elements of $T'$, but the representations are different.}: two singlets, $\bf1$, $\bf1'$, one doublet, $\bf2$, and two triplets, $\bf3$ and $\bf3'$. All the technical details are reported in appendix \ref{AppA:S4}.

The presence of the two-dimensional representations of $S_4$ allows for new patterns of the neutrino mass matrix, eventually different from the one in the Altarelli-Feruglio model. Indeed, the most general neutrino mass matrix which can be diagonalised by the tribimaximal mixing can be written as
\beq
m_\nu\sim\left(
            \begin{array}{ccc}
            a+2c& b-c& b-c\\
            b-c& b+2c& a-c\\
            b-c& a-c& b+2c
            \end{array}
            \right)\;,
\label{S4TBM:massaNuTB}
\eeq
in arbitrary units. It is easy to recognise in eq. \eqref{S4TBM:massaNuTB} the $\mu-\tau$ and the magic symmetries; furthermore, this description is equivalent to eq. \eqref{FS:TB:GeneralMassMatrix}. Usually this pattern can be obtained constructing the Lagrangian in such a way that the usual Weinberg operator, $\ell\ell$ (understanding the presence of the Higgses), is forbidden at the leading order, but appears only at higher-orders with additional flavons. The previous models based on $A_4$ and on $T'$ are characterised by $b=0$. The factors $a$ in eq. (\ref{S4TBM:massaNuTB}) come from the term $\ell\ell F_1$ and the factors $c$ from $\ell\ell F_3$, where $F_1$ and $F_3$ are flavons transforming respectively as a singlet $\bf1$ and as a triplet $\bf3$ of $A_4$.

The presence of the doublet representation of $S_4$ introduces a new feature in the neutrino mass matrix: indeed the terms which contribute to $m_\nu$ are $\ell\ell F_1$, $\ell\ell F_3$ and the new $\ell\ell F_2$, where $F_2$ represents a flavon transforming as a doublet $\bf2$. In eq. (\ref{S4TBM:massaNuTB}), this last contribution is represented by the term $b$. However the presence of three parameters in order to describe three masses prevents any predictions on the neutrino masses. For this reason the $S_4$-based model in which a singlet $F_1$, a doublet $F_2$ and also a triplet $F_3$ couple to $\ell\ell$ is not phenomenologically interesting. It is not restrictive to construct a model in which only a singlet and a doublet contribute to the neutrino mass matrix, but in this case $m_1=m_3$ and it is ruled out by the experimental observations. Moreover it is possible to think about a model in which only a singlet and a triplet contribute to the neutrino mass matrix: we have verified that such a model can be built, with a natural vacuum alignment. This model provides exactly the neutrino mass matrix with $b=0$ and therefore it has the same predictions in the lepton sector as in the $A_4$-based models. For this reason in this section we study the case in which only a doublet and a triplet couple to the term $\ell\ell$ and as a result we get an unusual neutrino mass matrix
\beq
m_\nu\sim\left(
            \begin{array}{ccc}
            2c & b-c & b-c \\
            b-c & b+2c & -c \\
            b-c & -c & b+2c \\
            \end{array}
            \right)
\label{S4TBM:massaNuS4General}
\eeq
which can still be diagonalised by the tribimaximal mixing. This new pattern provides different predictions for the $0\nu2\be$-decay and thus we expect to be able to distinguish this realisation from the other two which predict the tribimaximal mixing just looking at some observables related to the neutrino oscillations.

Moving to the quark sector, the idea is to use the doublet representations in order to describe masses and mixings. Since the Clebsch-Gordan coefficients are different with respect to those of the $T'$ model, we expect different predictions.

\subsection{The Lepton Sector}
\label{Sec:S4TBM:Leptons}
\setcounter{footnote}{3}

In this part we illustrate the model in the lepton sector, predicting an exact tribimaximal mixing at the leading order and a realistic charged lepton mass hierarchy, by the use of a flavour group $G_f$ in addition to the gauge group of the Standard Model. The complete flavour group is $G_f=S_4\times Z_5\times U(1)_{FN}$, where the three factors play different roles: the spontaneous breaking of $S_4$ down to its subgroup $Z_2\times Z_2$ in the neutrino sector is directly responsible for the tribimaximal mixing\footnote{This breaking is extremely unusual, indeed the common preserved subgroup is $Z_2$. Here $Z_2\times Z_2$ provides the same flavour structure for the neutrino mass matrix as $Z_2$ in the $A_4$-based models and it is associated to one element of the class $\mcal{C}_2$ and one of the class $\mcal{C}_4$.}; the $Z_5$ factor keeps separated the different sectors of the theory, quarks from leptons and furthermore neutrinos from charged leptons; moreover $Z_5$ plays a similar role as the baryon on the total lepton numbers, avoiding some dangerous terms, and, together to the $U(1)_{FN}$, is responsible for the hierarchy among the charged fermion masses. In table \ref{table:S4lepton_transformation}, we can see the lepton sector fields of the model and their transformation properties under $G_f$.

\begin{table}[h]
\begin{center}
\begin{tabular}{|c||c|c|c|c|c||c||c|c||c|c||c|}
  \hline
  &&&&&&&&&&& \\[-0,3cm]
  & $\ell$ & $e^c$ & $\mu^c$ & $\tau^c$ & $H_{u,d}$ & $\theta$ & $\psi$ & $\eta$ & $\Upsilon$ & $\varphi$ & $\xi'$ \\
  &&&&&&&&&&& \\[-0,3cm]
  \hline
  &&&&&&&&&&& \\[-0,3cm]
  $S_4$ & $\bf3$ & $\bf1'$ & $\bf1'$ & $\bf1$ & $\bf1$ & $\bf1$ & $\bf3$ & $\bf2$ & $\bf3$ & $\bf2$ & $\bf1'$  \\
  &&&&&&&&&&& \\[-0,3cm]
  $Z_5$ & $\om$ & $\om^3$ & 1 & $\om^2$ & 1 & 1 & $\om^2$ & $\om^2$ & $\om^3$ & $\om^3$ & 1  \\
  &&&&&&&&&&& \\[-0,3cm]
  $U(1)_{FN}$ & 0 & 1 & 0 & 0 & 0 & -1 & 0 & 0 & 0 & 0 & 0  \\
  \hline
  \end{tabular}
\end{center}
\caption{\it Transformation properties of the matter fields in the lepton sector and of all the flavons of the model. We distinguish the flavon fields on their role and thus we can consider $\psi$ and $\eta$ mainly connected to the charged lepton sector and $\Upsilon$ and $\varphi$ to the neutrino sector. All these fields together to $\xi'$ are present in the quark sector.}
\label{table:S4lepton_transformation}
\end{table}

We treat the model in a supersymmetric scenario, because the minimisation of the scalar potential is simplified, but it is not compulsory for the construction of the model itself. For this purpose an additional continuous $R$-parity is introduced, under which fermion fields are singly charged while scalars are uncharged.

The superpotential for leptons can be written as an expansion in inverse powers of the cutoff of the theory $\Lambda_f$:
\bea
w_e&=&\sum_{i=1}^{4}\dfrac{\theta}{\La_f}\dfrac{y_{e,i}}{\La_f^3}e^c(\ell X_i)'H_d+\dfrac{y_\mu}{\La_f^2}\mu^c(\ell\psi\eta)'H_d+\dfrac{y_\tau}{\La_f}\tau^c(\ell\psi)H_d
\label{S4TBM:eq:wd:leptons}\\[0.3cm]
w_\nu&=&\dfrac{x_d}{\La_f\La_L}(\ell H_u\ell H_u\varphi)+\dfrac{x_t}{\La_f\La_L}(\ell H_u\ell H_u\Upsilon)
\label{S4TBM:eq:wd:neutrinos}
\eea
where
\beq
X=\left\{\psi\psi\eta,\;\psi\eta\eta,\;\Upsilon\Upsilon\xi',\;\Upsilon\varphi\xi'\right\}
\eeq
using $(\ldots)$ to refer to the contraction in $\bf1$ and $(\ldots)'$ to the contraction in $\bf1'$. We indicate as usual the scale of the lepton number violation by the symbol $\La_L$, which we assume to be of the same order of $\La_f$. It is interesting to underline that the first contributions containing $e^c$ should be
\beq
\dfrac{\theta}{\La_f}\dfrac{y'_{e,1}}{\La_f^2}e^c(\ell\Upsilon\Upsilon)'H_d+\dfrac{\theta}{\La_f}\dfrac{y'_{e,2}}{\La_f^2}e^c(\ell\Upsilon\varphi)'H_d\;,
\eeq
which would dominate with respect to the terms in eq. (\ref{S4TBM:eq:wd:leptons}). However an explicit computation will show that these two terms are vanishing, once we assume that the flavons get this specific VEV alignment:
\bac{rclrcl}
\mean{\psi}&=&(0,\,v_\psi,\,0)\;,&\qquad
\mean{\eta}&=&(0,\,v_\eta)\;,\\[3mm]
\mean{\Upsilon}&=&(v_\Upsilon,\,v_\Upsilon,\,v_\Upsilon)\;,&\qquad
\mean{\varphi}&=&(v_\varphi,\,v_\varphi)\;,\\[3mm]
\mean{\xi'}&=&v_{\xi'}\;,&\qquad
\mean{\theta}&=&v_\theta\;.
\label{S4TBM:vev:allleptons}
\eac
We will demonstrate that this particular VEV alignment is a natural solution of the scalar potential in appendix \ref{AppB:S4}; moreover we will see that all the VEVs are of the same order of magnitude and for this reason we will parametrise the ratio $VEV/\La_f$ by the parameter $u$. The only VEV which originates with a different mechanism with respect to the others is $v_\theta$ and we indicate the ratio $v_\theta/\La_f$ by the parameter $t$.\\
With this setting, the mass matrix for the charged leptons is
\beq
M_e=\left(
         \begin{array}{ccc}
           y_e^{(1)} u^2t & y_e^{(2)} u^2t & y_e^{(3)} u^2t \\
           0 & y_\mu u & 0 \\
           0 & 0 & y_\tau \\
         \end{array}
       \right)\dfrac{u\,v_d}{\sqrt2}
\eeq
where the $y_e^{(i)}$ contains all the different contributions $y_{e,i}$. This matrix is already in a almost diagonal form and therefore $U_e$ and $U_{e^c}$, its diagonalising unitary matrices, are the unity matrix, apart from negligible corrections of the order of $u\,t$:
\beq
U_{e^c}^\dag M_e U_e=\diag\left(y_e\, u^2\,t,\;y_\mu\, u,\;y_\tau\right)u\,v_d/\sqrt2\;.
\eeq
For the neutrinos we get the following mass matrix
\beq
m_\nu=\left(
              \begin{array}{ccc}
                2c & b-c & b-c \\
                b-c & b+2c & -c \\
                b-c & -c & b+2c \\
              \end{array}
            \right)\dfrac{v_u^2}{\Lambda_f}\;,
\label{S4TBM:massaNuS4}
\eeq
where $b=x_d\,v_\varphi/\La_f$ and $c=x_t\,v_\Upsilon/\La_f$. Notice that it is $\mu-\tau$ symmetric and presents the magic symmetry, which assure that the matrix is of the TB-type.

At this level of approximation, the PMNS matrix is given by
\beq
U\equiv U_e^\dag\, U_{TB} = U_{TB}\;.
\eeq
When we introduce the higher-order terms in the Lagrangian, we expect corrections into the tribimaximal mixing of relative order u \cite{BMM_S4}. Comparing with the experimental values, the maximal deviation from the tribimaximal pattern is $0.05$ and therefore we can put an upper bound on $u$ of the same order of magnitude. For $u>0.05$ the model provides a $\theta_{12}$ angle which is not in agreement at $3\sig$ error with respect to the experimental data in table \ref{table:OscillationData}. The mass hierarchy of the charged leptons is a consequence of the symmetry of the model and it is possible to further constrain the expansion parameters, $u$ and $t$: indeed, looking at the mass of the $\tau$ we have
\beq
u \simeq\,\dfrac{\tan\beta}{|y_\tau|} \dfrac{\sqrt{2} m_\tau}{v} \approx 0.01 \dfrac{\tan\beta}{|y_\tau|}\;,
\eeq
where for the $\tau$ lepton we have used its pole mass $m_\tau=(1776.84 \pm 0.17) \;\rm{MeV}$ \cite{PDG08}. Requesting $|y_\tau|<3$ we find a lower bound for $u$ close to the upper bound $0.05$ for $\tan\beta=15$, whereas for $\tan\beta=2$ it is $u>0.007$. From the requirement that also $y_\mu$ remains in the perturbative regime, the lower bound on $u$ of $0.007$ is slightly raised and we fix it at $0.01$. From now on, we will choose the maximal range of $u$ as $0.01 \lesssim u \lesssim 0.05$, which shrinks when $\tan\beta$ is increased from $2$ to $15$. In order to explain the ratio $m_e/m_\mu$, we get a range of values for the parameter $t$ which is similar to that one for $u$. Finally we can write
\beq
0.01 \lesssim u,\,t \lesssim 0.05\,.
\label{S4TBM:vev:uet}
\eeq

\subsubsection{Phenomenological Analysis}
In this section we perform a phenomenological analysis on the neutrino spectrum, along the same lines as section \ref{Sec:AFTBM:NuSpectrum}.

The neutrino mass matrix in eq. (\ref{S4TBM:massaNuS4}) is diagonalised by the tribimaximal mixing and the diagonal neutrino mass matrix is given by
\beq
U_\nu^T m_\nu U_\nu =\dfrac{v^2}{\Lambda_L}\diag(|3c-b|,\,|2b|,\,|3c+b|)\;,
\eeq
where $U_\nu=U_{TB}P$ and $P$ contains the Majorana phases $\al_1=-\arg(3c-b)$, $\al_2=-\arg(2b)$, $\al_3=-\arg(3c+b)$. We parametrise the mass eigenvalues in terms of $|b|=m_2/2$, $\rho$ and $\Delta$, where $\rho$ and $\Delta$ are defined as
\beq
\dfrac{c}{b}=\rho\,e^{i\Delta}\;,
\label{S4TBM:RhoDeltaDef}
\eeq
with $\Delta$ is in the range $[0,\,2\pi]$. Imposing the constraint $|\cos\Delta|\leq1$, it results that the model can accommodate both the mass orderings: taking the most conservative case, we have for the normal and the inverse hierarchies, respectively,
\beq
m_1>25.2\;\meV\;,\qquad\qquad m_3>0.68\;\meV\;.
\label{S4TBM:Boundm1}
\eeq
These values correspond to $\cos\Delta=\pm1$ and they are the values for which the spectrum presents the strongest hierarchy: the values of the masses of the other two neutrinos are given by
\beq
\ba{rclcrcl}
\text{NH:}\qquad\quad m_2&=&26.7\;\meV\qquad&\text{and}\qquad m_3&=&51.9\;\meV\;,\\[3mm]
\text{IH:}\qquad\quad m_1&=&52.3\;\meV\qquad&\text{and}\qquad m_2&=&53.0\;\meV\;.
\ea
\eeq
Furthermore the sum of the neutrino masses is about $103.8$ meV for the normal hierarchy and $106.0$ meV for the inverse hierarchy. When $\cos\Delta$ approaches the zero, the neutrino spectrum becomes quasi degenerate.

It is also interesting to study the $0\nu2\be$ parameter $|m_{ee}|$ as a function of  the lightest neutrino mass:
\beq
|m_{ee}|=m_2\,\rho\;,
\eeq
and specifying the type of mass hierarchy we get
\baq
\text{NH:}\qquad\quad |m_{ee}| &=& \dfrac{1}{3}\sqrt{3m_1^2+2\De m^2_{atm}-\De m^2_{sol}}\;,\\[3mm]
\text{IH:}\qquad\quad |m_{ee}| &=& \dfrac{1}{3}\sqrt{3m_3^2+\De m^2_{atm}-2\De m^2_{sol}}\;.
\eaq

\begin{figure}[h!]
\begin{center}
\includegraphics[width=7.8cm]{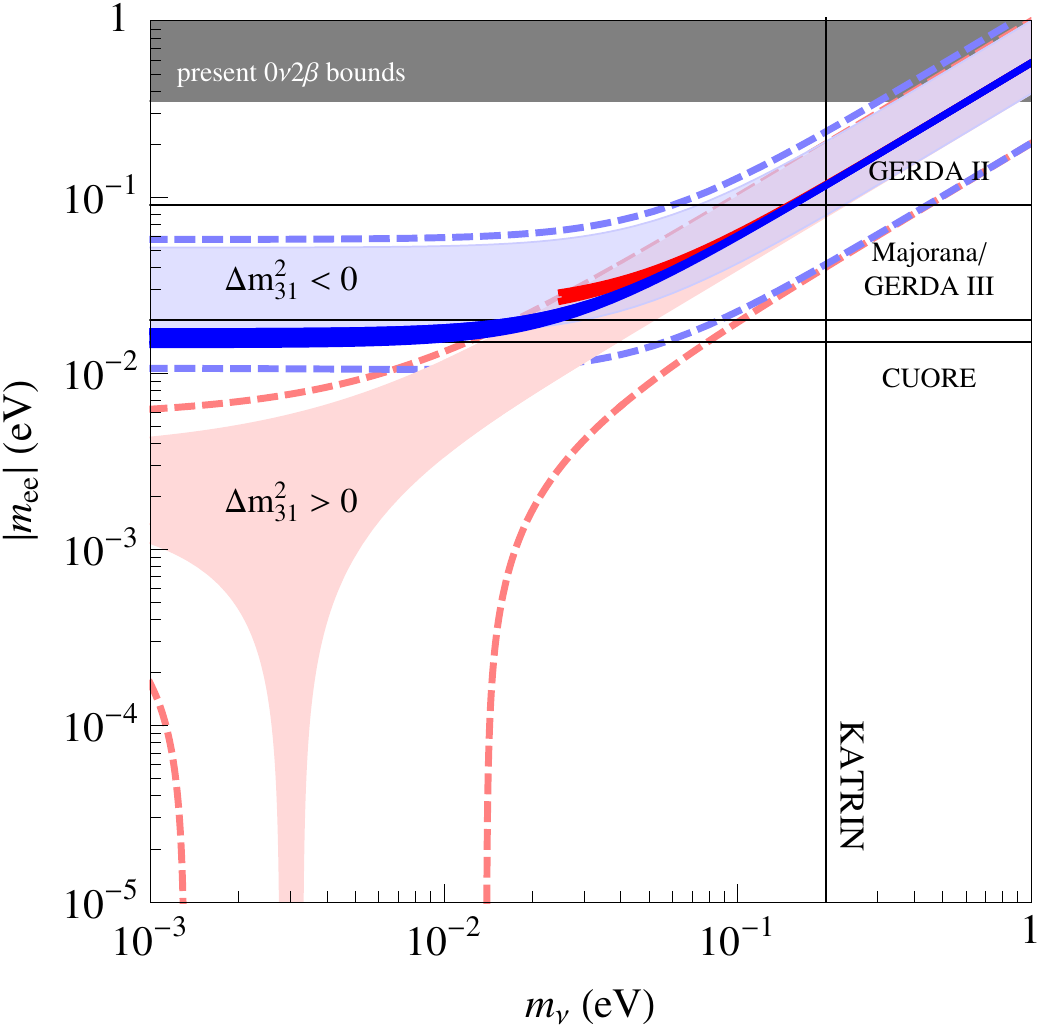}
\includegraphics[width=7.8cm]{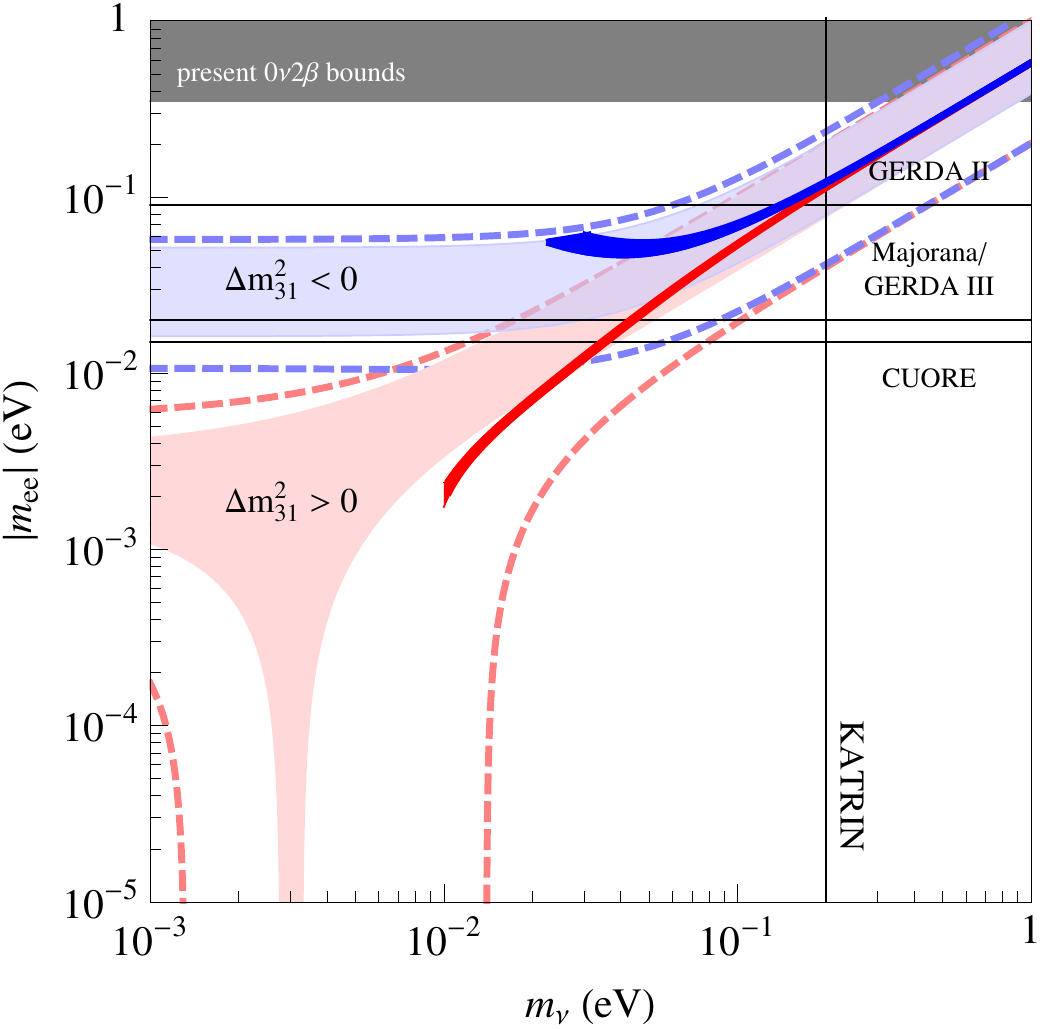}
\caption{\it $|m_{ee}|$ as a function of the lightest neutrino mass, $m_1$ in the case of the normal hierarchy and $m_3$ in the case of the inverse hierarchy. On the left the effective model, while on the right the type I See-Saw version. See figure \ref{fig:AF_0nu2beta} for the description of the plot.}
\label{fig:S4first}
\end{center}
\end{figure}

In fig.(\ref{fig:S4first}), we plot $|m_{ee}|$ as a function of the lightest neutrino mass eigenstate, $m_1$ in the normal hierarchy case and $m_3$ in the inverse hierarchy one. Looking at the dark red area we can see that it falls in the quasi degenerate spectrum band and therefore we speak about normal ordering and not about normal hierarchy. When $|\cos\Delta|=1$, $|m_{ee}|$ reaches the minima which are given by:
\beq
|m_{ee}|\geq25.7\meV\qquad\text{and}\qquad|m_{ee}|\geq15.5\meV\;,
\eeq
in the normal ordering and inverse hierarchy, respectively. We observe that, considering only the leading order approximation level, the present model can be distinguished from the Altarelli-Feruglio model of section \ref{Sec:AFTBM} and the model based on $T'$ of section \ref{Sec:TpTBM}, just looking to the lower bound on $|m_{ee}|$: in fact those models predict a lower bound for $|m_{ee}|$, which is about $5$ meV, while in the present proposal it is larger. The next future experiments should be able, in principle, to test the present model, since the lower bound is close to the future experimental sensitivities, which are expected to reach the values of $90$ meV \cite{gerda} (GERDA), $20$ meV \cite{majorana} (Majorana), $50$ meV \cite{supernemo} (SuperNEMO), $15$ meV \cite{cuore} (CUORE) and $24$ meV \cite{exo} (EXO).\\

These results are valid only at the leading order and some deviations are expected with the introduction of the higher-order terms, that is illustrated in the following sections. The corrections are expected to be of relative order $u$ or $t$, whose allowed range is defined in eqs. (\ref{S4TBM:vev:uet}). However, close to $\cos\Delta=-1$, where the bounds are saturated, the corrections remain of relative order $u$ or $t$ and as a result the lower bounds on $m_1$ and $m_{3}$ of eqs. (\ref{S4TBM:Boundm1}) are not significantly affected. Major effects could appear when the spectrum is quasi degenerate, \mbox{$\cos\Delta\approx0$}.

\subsection{The Quark Sector}
\label{Sec:S4TBM:Quark}
\setcounter{footnote}{3}

In this part we illustrate the model in the quark sector, getting a good approximation of the experimental quark mixing matrix. In table \ref{table:S4quark_transformation}, we can see the quark sector fields of the model and their transformation properties under $S_4\times Z_5\times U(1)_{FN}$.

The superpotential in the quark sector can be written as
\bac{rl}
w_q\;=&y_tt^cq_3H_u+\dfrac{y_b}{\Lambda_f}b^cq_3\xi'H_d+\\[0.2cm]
&+\sum_{i=1}^{2}\dfrac{y_{tc,i}}{\Lambda_f^2}t^c\left(D_qX^{(1)}_i\right)H_u+ \sum_{i=1}^{2}\dfrac{y_{bs,i}}{\Lambda_f^2}b^c\left(D_qX^{(1)}_i\right)'H_d+\\[0.3cm]
&+\sum_{i=1}^{6}\dfrac{y_{tu,i}}{\Lambda_f^3}t^c\left(D_qX^{(2)}_i\right)H_u+ \sum_{i=1}^{6}\dfrac{y_{bd,i}}{\Lambda_f^3}b^c\left(D_qX^{(2)}_i\right)'H_d+\\[0.3cm]
&+\sum_{i=1}^{2}\dfrac{y_{c,i}}{\Lambda_f^2}c^c\left(D_qX^{(1)}_i\right)'H_u+
\sum_{i=1}^{2}\dfrac{y_{s,i}}{\Lambda_f^2}s^c\left(D_qX^{(1)}_i\right)'H_d+\\[0.3cm]
&+\dfrac{y_{ct}}{\Lambda_f}c^cq_3\xi'H_u+
\sum_{i=1}^{3}\dfrac{y_{sb,i}}{\Lambda_f^2}s^cq_3\left(\eta\varphi\right)'H_d+\\[0.3cm]
&+\sum_{i=1}^{6}\dfrac{y_{cu,i}}{\Lambda_f^3}c^c\left(D_qX^{(2)}_i\right)'H_u+
\sum_{i=1}^{6}\dfrac{y_{sd,i}}{\Lambda_f^3}s^c\left(D_qX^{(2)}_i\right)'H_d+\\[0.3cm]
&+\sum_{i=1}^{2}\dfrac{y_{u,i}}{\Lambda_f^2}\dfrac{\theta^2}{\Lambda_f^2}u^c\left(D_qX^{(4)}_i\right)H_u+ \sum_{i=1}^{2}\dfrac{y_{d,i}}{\Lambda_f^2}\dfrac{\theta}{\Lambda_f}d^c\left(D_qX^{(4)}_i\right)H_d\\[0.3cm]
&+\sum_{i=1}^{4}\dfrac{y_{ut,i}}{\Lambda_f^3}\dfrac{\theta^2}{\Lambda_f^2}u^c\left(q_3X^{(5)}_i\right)H_u+ \sum_{i=1}^{4}\dfrac{y_{db,i}}{\Lambda_f^3}\dfrac{\theta}{\Lambda_f}d^c\left(q_3X^{(5)}_i\right)H_d
\eac
where
\baq
&&X^{(1)}=\left\{\eta\eta+\psi\psi\right\}\nn\\[0.3cm]
&&X^{(2)}=\left\{\eta\eta\xi',\;\psi\psi\xi',\;\Upsilon\Upsilon\Upsilon,\; \Upsilon\Upsilon\varphi,\;\Upsilon\varphi\varphi,\;\varphi\varphi\varphi\right\}\nn\\[0.3cm]
&&X^{(3)}=\left\{\psi\Upsilon,\;\eta\varphi,\;\xi'\xi'\right\}\nn\\[0.3cm]
&&X^{(4)}=\left\{\varphi\varphi,\;\Upsilon\Upsilon\right\}\nn\\[0.3cm]
&&X^{(5)}=\left\{\psi\psi\Upsilon,\;\psi\psi\varphi,\;\psi\eta\Upsilon,\;\eta\eta\varphi\right\}\nn\;.
\eaq
Since the quantum numbers of $b^c$ and $s^c$ are exactly the same, there are no fundamental distinctions between $b^c$ and $s^c$. We can therefore define $b^c$ as the field which couples to $q_3\xi'H_d$ in the superpotential $w_q$.

\begin{table}[h!]
\begin{center}
\begin{tabular}{|c||c|c|c|c|c|c|c|c||c||c|c||c|c||c|}
  \hline
  &&&&&&&&&&&&&& \\[-0,3cm]
  & $D_q$ & $q_3$ & $u^c$ & $d^c$ & $c^c$ & $s^c$ & $t^c$ & $b^c$ & $\theta$ & $\psi$ & $\eta$ & $\Upsilon$ & $\varphi$ & $\xi'$ \\
  &&&&&&&&&&&&&& \\[-0,3cm]
  \hline
  &&&&&&&&&&&&&& \\[-0,3cm]
  $S_4$ & $\bf2$ & $\bf1$ & $\bf1'$ & $\bf1'$ & $\bf1'$ & $\bf1'$ & $\bf1$ & $\bf1'$ & $\bf1$ & $\bf3$ & $\bf2$ & $\bf3$ & $\bf2$ & $\bf1'$  \\
  &&&&&&&&&&&&&& \\[-0,3cm]
  $Z_5$ & $\om^4$ & $\om^3$ & $1$ & $1$ & $\om^2$ & $\om^2$ & $\om^2$ & $\om^2$ & 1 & $\om^2$ & $\om^2$ & $\om^3$ & $\om^3$ & 1 \\
  &&&&&&&&&&&&&& \\[-0,3cm]
  $U(1)_{FN}$ & 0 & 0 & 2 & 1 & 0 & 0 & 0 & 0 & -1 & 0 & 0 & 0 & 0 & 0  \\
  \hline
  \end{tabular}
\end{center}
\caption{\it Transformation properties of all the fields in the quark sector.}
\label{table:S4quark_transformation}
\end{table}

Notice that all the flavons are present in the quark sector and as a result the $S_4$ symmetry is completely broken in the quark sector. When the flavour and the electroweak symmetries are broken, the mass matrices for the up- and down-quarks are
\beq
m_u=\left(
         \begin{array}{ccc}
           y_u u^2t^2 & y_u u^2t^2 & y_{ut}u^3t^2 \\
           y_{cu} u^3 & y_cu^2 & y_{ct}u \\
           y_{tu} u^3 & y_{tc}u^2 & y_t \\
         \end{array}
       \right)\dfrac{v_u}{\sqrt2}\;,\qquad
m_d=\left(
      \begin{array}{ccc}
        y_dut & y_dut & y_{db}u^2t \\
        y_{sd}u^2 & y_su & y_{sb}u \\
        y_{bd}u^2 & y_{bs}u & y_b \\
      \end{array}
    \right)\dfrac{u\,v_d}{\sqrt2}\;,
\eeq
where the Yukawas are the sum of all the different terms, which appear in the superpotential.\\
These mass matrices can be diagonalised by the following transformations:
\beq
V_{u^c}^\dag m_u V_u=(y_uu^2t^2,\;y_cu^2,\;y_t)\dfrac{v_u}{\sqrt2}\;,\qquad\qquad
V_{d^c}^\dag m_d V_d=(y_dut,\;y_su,\;y_b)\dfrac{u\,v_d}{\sqrt2}
\eeq
where the unitary matrices can be written in terms of order of magnitude of $u$ and $t$ as
\bac{rlrl}
V_u=&\left(
      \begin{array}{ccc}
        1 & O(u) & O(u^3) \\
        -O(u) & 1 & O(u^2) \\
        -O(u^3) & -O(u^2) & 1 \\
      \end{array}
    \right)\;,&
V_d=&\left(
      \begin{array}{ccc}
        1 & O(u) & O(u^2) \\
        -O(u) & 1 & O(u) \\
        -O(u^2) & -O(u) & 1 \\
      \end{array}
    \right)\;,\\[0.8cm]
V_{u^c}=&\left(
      \begin{array}{ccc}
        1 & O(t^2) & O(u3t^2) \\
        -O(t^2) & 1 & O(u) \\
        -O(ut^2) & -O(u) & 1 \\
      \end{array}
    \right)\;,\quad&
V_{d^c}=&\left(
      \begin{array}{ccc}
        1 & O(t) & O(u^2t) \\
        -O(t) & 1 & O(u) \\
        -O(ut) & -O(u) & 1 \\
      \end{array}
    \right)\;.
\eac
While the mass of the top quark is expected to be of the order of $v_u\approx\cO(100\GeV)$, the mass of the bottom quark is suppressed compared with $m_t$ by the factor $u\,v_d/v_u$ so that it is of the same order of $m_\tau$. The other measured quark masses can be accommodated thanks to the $Z_5$ and the $U(1)_{FN}$ suppressions. The resulting quark mixing matrix is
\beq
V\equiv V_u^\dag V_d\simeq\left(
                                  \begin{array}{ccc}
                                    1 & \left(\dfrac{y_{sd}}{y_s}-\dfrac{y_{cu}}{y_c}\right)u & \left(\dfrac{y_{bd}y_{c}-y_{bs}y_{cu}}{y_by_c}\right)u^2 \\[0.3cm]
                                    -\left(\dfrac{y_{cu}}{y_c}-\dfrac{y_{sd}}{y_s}\right)u & 1 & \dfrac{y_{bs}}{y_b}u \\[0.3cm]
                                    \left(y_{bs}y_{sd}-\dfrac{y_{bd}y_{s}}{y_by_s}\right)u^2 & -\dfrac{y_{bs}}{y_b}u & 1 \\
                                  \end{array}
                                \right)\;.
\eeq
In order to fit the experimental values of the mixing angles we need to invoke a moderate fine-tuning in some parameters. The (23) entry of $V$ has to be of order $\lambda^2\simeq0.05$ and therefore suggests for $u$ a value close to its upper bound. However this is not a strict constraint because this value can be well explained for the entire range of $u$ considering the Yukawas. On the other hand, the entry (12) requires an accidental enhancement of the combination $\left(\dfrac{y_{sd}}{y_s}-\dfrac{y_{cu}}{y_c}\right)$ of order $1/\lambda\sim4$ in order to describe the correct Cabibbo angle. It is possible to explain such an enhancement considering particular values of the relative phase, $\Delta_q$, between $\dfrac{y_{sd}}{y_s}$ and $\dfrac{y_{cu}}{y_c}$, which is connected to the CP violating phase: if $\Delta_q=\pi$, then the two factors sum up and the required values are easily explained.

\subsection{Higher Order Corrections}
\label{Sec:S4TBM:NLO}
\setcounter{footnote}{3}

We now discuss the deviations to the leading order results. A detailed analysis is presented in the original paper \cite{BMM_S4}, while here we summarise only the results.

Looking at the flavon sector, we find that the VEV alignment in eq. (\ref{S4TBM:vev:allleptons}) is stable under the NLO corrections and the deviations are of relative order $u$ with respect to the leading order results (see appendix \ref{AppB:S4}). This affects the fermion mass matrices, when these new VEVs are introduced in the superpotential $w$.

In the lepton sector, all the corrections from the higher-order terms introduce deviations to the lepton mixing matrix of relative order $u$ with respect to the leading order. This is in line to what happens in the Altarelli-Feruglio model, where the tribimaximal values of the atmospheric and the solar angles are corrected by $\cO(u)$ terms and the reactor angle deviates from zero by $\cO(u)$ contributions.

The analysis for the up and down quark mass matrices point out tha  the corrections coming from the NLO operators of the superpotential and from the deviations to the VEVs introduce new terms of relative order $u$ in each entry of the mass matrices. As a result the quark mixing angles receive deviations of relative order $u$ with respect the initial values, which do not spoil the leading order results.

\subsection{See-Saw Extensions}
\label{Sec:S4TBM:SeeSaws}
\setcounter{footnote}{3}

In this section we deal with the study of a possible explanation of the Weinberg operator used in the previous sections in order to describe the smallness of neutrino masses. We focus only on the neutrino sector, indeed the quark and the charged lepton sectors are described by the same superpotential as in the effective model and the flavon content is not modified at all, in particular the VEV misalignment mechanism works exactly in the same way.

The simplest approach is the type I See-Saw mechanism, extending the matter content of the model by adding three right-handed neutrinos $\nu^c$ which transform as a triplet of $S_4$. Here we summarise the phenomenological results of the analysis presented in \cite{BMM_SS} and we discuss the differences with the effective model. 

In the See-Saw model, it is possible to describe both the neutrino mass hierarchies and it is interesting to find the allowed ranges for the lightest neutrinos in both the cases. For the most conservative case and for the normal and inverse hierarchies, respectively, we get 
\beq
m_1>10.4\;\meV\;,\qquad\qquad m_3>25.9\;\meV\;.
\label{S4TBM:Boundm1SS}
\eeq
These value corresponds to the condition for which the spectrum present the strongest hierarchy: the values of the masses of the other two neutrinos are given by
\beq
\ba{rclcrcl}
\text{NH:}\qquad\quad m_2&=&13.4\;\meV\qquad&\text{and}\qquad m_3&=&46.6\;\meV\;,\\[3mm]
\text{IH:}\qquad\quad m_1&=&51.5\;\meV\qquad&\text{and}\qquad m_2&=&52.3\;\meV\;.
\ea
\eeq
Furthermore the sum of the neutrino masses in this case is about $70.4$ meV for the normal hierarchy and $129.7$ meV for the inverse case. When $m_1$ and $m_3$ increase their values, the spectrum becomes degenerate.

The $0\nu2\be$ parameter $|m_{ee}|$ is different with respect to the effective model:
\baq
\text{NH:}\qquad\quad |m_{ee}| &=& \dfrac{1}{3}\sqrt{-\left(m_1^2+\De m^2_{sol}\right)+2m_1^2\left(\dfrac{m_1^2+\De m^2_{sol}}{m_1^2+\De m^2_{atm}}+1\right)}\;,\\[5mm]
\text{IH:}\qquad\quad |m_{ee}| &=& \dfrac{1}{3m_3}\sqrt{3m_3^4+m_3^2\left(5\De m^2_{atm}-4\De m^2_{sol}\right)+2\De m^2_{atm}\left(\De m^2_{atm}-\De m^2_{sol}\right)}\;.\nn
\eaq

In fig.(\ref{fig:S4first}), we plot $|m_{ee}|$ as a function of the lightest neutrino mass eigenstate, $m_1$ for the normal hierarchy and $m_3$ for the inverse hierarchy. It is interesting to note that, while the dark blue area referring to the inverse hierarchy falls well inside the $1\sigma$ error band, the dark red area goes out of the $1\sigma$ error band. This small discrepancy can be easily motivated by a shift of $1\sig$ level on the lepton mixing angles. However, unfortunately the future experimental sensitivities will not reach such small values and therefore in the next experiments it will not be possible to test these deviations in the mixing angles, as predicted by the model.

When $m_1$ and $m_3$ reaches the values as in eqs. (\ref{S4TBM:Boundm1SS}), $|m_{ee}|$ reaches the minimum which is given by:
\beq
|m_{ee}|\geq2.47\meV\qquad\text{and}\qquad|m_{ee}|\geq51.8\meV\;,
\eeq
in the normal and inverse hierarchies, respectively.

The See-Saw model presents different features with respect to the effective model, as it easy to understand looking at figures \ref{fig:S4first}. While in the See-Saw model the inverse hierarchy is almost restricted in the quasi degenerate spectrum area, this happens to the normal hierarchy in the effective model. Furthermore, an absence of any signal linked to some lepton number violating processes in the next experiments would fix an experimental upper bound on $|m_{ee}|$ which would completely rule out the inverse hierarchy allowing only the normal hierarchy.

Considering now the comparison with the Altarelli-Feruglio model or the model based on $T'$ and only the leading order terms, it seems possible, but difficult, to distinguish among the different realisations. However, the introduction of the higher-order corrections makes the predictions overlap in a large part of the parameter space, and it will be hard to test these little differences in the future.

It is interesting to note that the Altarelli-Feruglio and the $T'$ models present exactly the same behaviour in the quasi degenerate region, which is different from all the $S_4$ proposals: in the first case the profile of $|m_{ee}|$ is at the upper edge of the allowed region, while in the second case the profile follows approximately the central line. This behavior is due to a constraint on the Majorana phases, which are determined as a function of the neutrino masses: in the $A_4$- and $T'$-based models the constraint forces the different terms in $|m_{ee}|$ to sum up, while in the $S_4$-based model it requires a partial cancellation (see \cite{BMM_SS} for details).

In \cite{BMM_SS} we have analysed also the other possibilities of the type II and type III See-Saw mechanisms. Here we report only the phenomenological conclusions of these approaches, referring to the original paper for the details.

Introducing a scalar triplet or three fermion triplets as explained in section \ref{Sec:SM:SeeSaw}, it is possible to choose a suitable charge assignment in order to find the tribimaximal pattern as the lepton mixing matrix. Moreover, the phenomenology linked to these two alternatives is identical to the one of the effective model and of the type I See-Saw: indeed the light neutrino mass matrix in the type II See-Saw coincides with the one of the effective model, once the VEV of the triplet is identified with the VEV of $H_u$; the mass matrices in the type III See-Saw correspond to those ones of the type I, identifying $M_R$ with the mass matrix of the fermion triplets.
As a result, it is clear that it is not possible to discriminate among type I, II and III See-Saw models based on the $S_4$ symmetry group only relying on the neutrino phenomenology. New observables are needed, such as the mixing of the fermion triplets with the charged leptons or the decays of the scalar triplet.

\section{Conclusions of the Chapter}
\label{Sec:TBM:Conclusions}
\setcounter{footnote}{3}

In this chapter we focussed on three flavour models which reproduce a lepton mixing matrix of the tribimaximal type. The model based on the $A_4$ discrete symmetry is the simplest realisation in which the tribimaximal pattern naturally arises: its simplicity relies on the small dimension of the flavour group and on the small number of flavons which break the symmetry; its naturalness refers to the presence of only a moderate fine-tuning to accommodate the parameter $r\equiv \De m_{sol}^2/\De m_{atm}^2$, which translates in a small range of values for $\Delta$, the phase between the two parameters which describe the neutrino mass matrix. In the model, neutrino masses belong to a limited range and in particular there is always a lower bound for the lightest neutrino mass. Furthermore there is also a prediction for the neutrinoless-double-beta decay as a function of the neutrino masses.

The $A_4$-based model applies only to the lepton sector, indeed it fails when extending the description to the quark sector. Two possible solutions are the larger groups $T'$ and $S_4$: the first one is the double covering of $A_4$, while the latter contains $A_4$ as a subgroup. The $T'$ model reproduces exactly the same results as in the $A_4$ model and describes quarks using the doublet representations, similarly to $U(2)$-based models developed long ago. As a result, realistic mass hierarchies and mixings are achieved, introducing only a small fine-tuning to the parameters of the order of $\lambda$ in order to reproduce the measured ratio $m_u/m_c$.

The $S_4$ discrete group has the same number of elements as $T'$, but different representations. This enables to describe neutrinos with a different mass matrix, still diagonalised by the tribimaximal mixing. This leads to a completely new neutrino phenomenology. Considering only the leading order terms, it seems possible, even if difficult, to distinguish among the different realisations; unfortunately, the introduction of the higher-order corrections makes the predictions overlap in all the parameter space, apart from very small areas, which will be difficult to test in the near future. In the quark sector, the mass hierarchies are well explained through the Froggatt-Nielsen mechanism, while the mixings necessitate of a small fine-tuning of the order of $\lambda$: the Cabibbo angle is described as the difference of two complex terms of order of $\lambda^2$ and the fine-tuning corresponds to the requirement of having a coherent sum of the two terms.

In all these models, a fundamental aspect is the vacuum misalignment of the flavons, which break the discrete group down to certain subgroups which represent the effective low-energy flavour structures: it is this mechanism that assures the tribimaximal mixing for the neutrinos in the basis of diagonal charged leptons. Furthermore, the main discrete group is usually extended by additional terms, usually $Z_n$, which help keeping separated each sector from the others, at least at leading order. The symmetry breaking can be complete when speaking about quarks. The introduction of the higher-order corrections have a deep impact on the neutrino phenomenology, introducing non-negligible deviations from the tribimaximal values of the mixing angles: as a result, the solar angle can reach its measured best-fit value, the atmospheric one can slightly deviate from the maximal value, and finally the reactor angle could be non-vanishing, but still very close to zero. We expect $\theta_{13}$ to be less than or equal to approximately $0.05$ not far from the reach of the future high-intensity neutrino beam facilities. 

\clearpage{\pagestyle{empty}\cleardoublepage}


\newpage
\chapter{A Flavour Model with the Bimaximal Mixing}
\label{Sec:FlavourModelsBM}
\setcounter{equation}{0}
\setcounter{footnote}{3}

In the previous sections we investigated on a series of models which reproduce al leading order the tribimaximal pattern for the lepton mixing. When considering the next-to-the-leading order corrections, all the three mixing angles receive corrections of the same order of magnitude. This is a very general feature of tribimaximal models, in the absence of specific dynamical tricks (see \cite{Lin_LargeReactor} for a model in which such a trick is implemented). Since the experimentally allowed departures of $\theta_{12}$ from the tribimaximal value are small, at most of $\mathcal{O}(\lambda^2)$, it follows that both $\theta_{13}$ and the deviation of $\theta_{23}$ from the maximal value are expected in these models to also be at most of $\mathcal{O}(\lambda^2)$. A value of $\theta_{13} \sim \mathcal{O}(\lambda^2)$ is within the sensitivity of the experiments which are now in preparation and will take data in the near future. However, the present data do not exclude a larger value for $\theta_{13}$ with $\theta_{13} \sim \mathcal{O}(\lambda)$, as suggested in \cite{Fogli:Indication}. If experimentally it is found that $\theta_{13}$ is close to its present upper bound, this could be interpreted as an indication that the agreement with the tribimaximal mixing is accidental. Then a scheme where instead the bimaximal (BM) mixing is the correct first approximation modified by terms of $\mathcal{O}(\lambda)$ could be relevant. This is in line with the well known empirical observation that $\theta_{12}+\lambda\sim \pi/4$, a relation known as ``quark-lepton complementarity'' \cite{SymmSO3King,Complementarity}, or similarly $\theta_{12}+\sqrt{m_\mu/m_\tau} \sim \pi/4$, since $\sqrt{m_\mu/m_\tau}\sim\lambda$. No compelling model leading without parameter fixing to the exact complementary relation has been produced so far. Probably the exact complementarity relation is to be replaced with something like $\theta_{12}+\mathcal{O}(\lambda)\sim \pi/4$, which we could call ``weak'' complementarity.

In the following we construct a model based on the permutation group $S_4$ which naturally leads to the bimaximal mixing at the leading order. We adopt a supersymmetric formulation of the model in 4 space-time dimensions. The complete flavour group is $S_4\times Z_4 \times U(1)_{FN}$. In the lowest approximation, the charged leptons are diagonal and hierarchical and the light neutrino mass matrix, after See-Saw, leads to the exact bimaximal mixing. The model is built in such a way that the dominant corrections to the bimaximal mixing, from higher dimensional operators in the superpotential, only arise from the charged lepton sector at the NLO and naturally inherit $\lambda$ as the relevant expansion parameter. As a result the mixing angles deviate from the bimaximal values by terms of $\mathcal{O}(\lambda)$ (at most), and weak complementarity holds. A crucial feature of the model is that only $\theta_{12}$ and $\theta_{13}$ are corrected by terms of $\mathcal{O}(\lambda)$ while $\theta_{23}$ is unchanged at this order (which is essential to make the model agree with the present data).

\section{The Structure of the Model}
\label{Sec:AFM:Structure}
\setcounter{footnote}{3}

We discuss here the general properties of our model. We have already introduced in the previous chapter the $S_4$ discrete group in order to perform a tribimaximal model. That description, however, does not fit our purpose and we need to define a new set of generators in order to achieve the bimaximal mixing in a clever way: the two new operators $S$ and $T$ satisfy to
\beq
T^4=S^2=(ST)^3=(TS)^3=1
\eeq
and their explicit forms in each of the irreducible representations and the Clebsch-Gordan coefficients in our basis are collected in the appendix \ref{AppA:S4BM}. Notice that the multiplication rules are the same as in the previous chapter. 

This description is suitable in our case, because, by requiring invariance under the action of $S$
\beq
m_\nu= S m_\nu S\;,
\label{AFM:invS}
\eeq
the resulting effective neutrino mass matrix is given by 
\beq
m_\nu=\left(
  \begin{array}{ccc}
    x & y & y \\
    y & z & x-z \\
    y & x-z & z \\
  \end{array}
\right)\;,
\label{AFM:GeneralMassMatrix}
\eeq
which can be diagonalised by the bimaximal mixing, as stated in section \ref{Sec:FS:BM}.

We formulate our model in the framework of the See-Saw mechanism (even though it would also be possible to build a version where light neutrino masses are directly described by the Weinberg operator). For this we assign the $3$ generations of left-handed lepton doublets $\ell$ and of right-handed neutrinos $\nu^c$ to two triplets $\bf3$, while the right-handed charged leptons $e^c$, $\mu^c$ and $\tau^c$ transform as $\bf1$, $\bf1'$ and $\bf1$, respectively. The $S_4$ symmetry is then broken by suitable triplet flavons.  Additional symmetries are needed, in general, to prevent unwanted couplings and to obtain a natural hierarchy among $m_e$, $m_\mu$ and $m_\tau$. In our model, the complete flavour symmetry is $S_4\times Z_4\times U(1)_{FN}$. A flavon $\theta$, carrying a negative unit of the $U(1)_{FN}$ charge, acquires a VEV and breaks $U(1)_{FN}$. In view of a possible GUT extension of the model at a later stage, we adopt a supersymmetric context, so that two Higgs doublets $H_{u,d}$, invariant under $S_4$, are present in the model. The usual continuous $U(1)_R$ symmetry, related to $R$-parity and the presence of driving fields in the flavon superpotential, is implemented in the model. Supersymmetry also helps producing and maintaining the hierarchy $\langle H_{u,d}\rangle=v_{u,d}\ll \Lambda_f$ where $\Lambda_f$ is the cutoff scale of the theory.

\begin{table}[h]
\begin{center}
\begin{tabular}{|c||c|c|c|c|c|c||c||c|c|c|c||c|c|c|c|}
  \hline
  &&&&&&&&&&&&&&&\\[-0,3cm]
  & $\ell$ & $e^c$ & $\mu^c$ & $\tau^c$ & $\nu^c$ & $H_{u,d}$ & $\theta$ & $\varphi_\ell$ & $\chi_\ell$ & $\psi_\ell^0$ & $\chi_\ell^0$ & $\xi_\nu$ &$\varphi_\nu$ & $\xi_\nu^0$ & $\varphi_\nu^0$ \\
  &&&&&&&&&&&&&&&\\[-0,3cm]
  \hline
  &&&&&&&&&&&&&&&\\[-0,3cm]
  $S_4$ & $\bf3$ & $\bf1$ & $\bf1^\prime$ & $\bf1$ & $\bf3$ & $\bf1$ & $\bf1$ & $\bf3$ & $\bf3^\prime$ & $\bf2$ & $\bf3'$ & $\bf1$ & $\bf3$ & $\bf1$ & $\bf3$  \\
  &&&&&&&&&&&&&&&\\[-0,3cm]
  $Z_4$ & 1 & -1 & -i & -i & 1 & 1 & 1 & i & i & -1 & -1 & 1 & 1 & 1 & 1 \\
  &&&&&&&&&&&&&&&\\[-0,3cm]
  $U(1)_{FN}$ & 0 & 2 & 1 & 0 & 0 & 0 & -1 & 0 & 0 & 0 & 0 & 0 & 0 & 0 & 0  \\
  &&&&&&&&&&&&&&&\\[-0,3cm]
  $U(1)_R$ & 1 & 1 & 1 & 1 & 1 & 1 & 0 & 0 & 0 & 2 & 2 & 0 & 0 & 2 & 2  \\
  \hline
  \end{tabular}
\end{center}
\caption{\it Transformation properties of all the fields.}
\label{table:TransformationsS4BM}
\end{table}

The fields in the model and their classification under the symmetry are summarised in Table \ref{table:TransformationsS4BM}. The complete superpotential can be written as
\beq
w=w_e+w_\nu+w_d\;,
\eeq
where $w_d$ is responsible for the flavon VEV alignment as discussed in appendix \ref{AppB:S42}, while $w_e$ and $w_\nu$ refer to the charged lepton and neutrino sectors and can be written as 
\beq
\ba{l}
\begin{split}
w_e\;=&\;\frac{y_e^{(1)}}{\La_f^2}\frac{\theta^2}{\La_f^2}e^c(\ell\varphi_\ell\varphi_\ell)H_d+ \frac{y_e^{(2)}}{\La_f^2}\frac{\theta^2}{\La_f^2}e^c(\ell\chi_\ell\chi_\ell)H_d+ \frac{y_e^{(3)}}{\La_f^2}\frac{\theta^2}{\La_f^2}e^c(\ell\varphi_\ell\chi_\ell)H_d+\\
&+\frac{y_\mu}{\La_f}\frac{\theta}{\La_f}\mu^c(\ell\chi_\ell)^\prime H_d+\frac{y_\tau}{\La_f}\tau^c(\ell\varphi_\ell)H_d+\dots
\end{split}
\label{AFM:wl}\\[1mm]
w_\nu\;=\;y(\nu^c\ell)H_u+M \Lambda_f (\nu^c\nu^c)+a(\nu^c\nu^c\xi_\nu)+b(\nu^c\nu^c\varphi_\nu)+\dots
\ea
\eeq
indicating with $(\ldots)$ the singlet $\bf{1}$, with $(\ldots)^\prime$ the singlet ${\bf1^\prime}$ and with $(\ldots)_R$ the representation R ($R={\bf2},\,{\bf3},\,{\bf3'}$). Note that the parameter $M$ defined above is dimensionless. In the above expression for the superpotential $w$, only the lowest order operators in an expansion in powers of $1/\Lambda_f$ are explicitly shown. Dots stand for higher
dimensional operators that will be discussed later on. The stated symmetries ensure that, for the leading terms, the flavons that appear in $w_e$ cannot contribute to $w_\nu$ and viceversa.

We will show in appendix \ref{AppB:S42} that the potential corresponding to $w_d$ possesses an isolated minimum for the following VEV configuration:
\bea
&\dfrac{\mean{\varphi_\ell}}{\La_f}=\left(
                     \begin{array}{c}
                       0 \\
                       1 \\
                       0 \\
                     \end{array}
                   \right)A\;,\qquad\qquad
&\dfrac{\mean{\chi_\ell}}{\La_f}=\left(
                     \begin{array}{c}
                       0 \\
                       0 \\
                       1 \\
                     \end{array}
                   \right)B\;,
\label{AFM:vev:charged:best}\\
&\dfrac{\mean{\varphi_\nu}}{\La_f}=\left(
                     \begin{array}{c}
                       0 \\
                       1 \\
                       -1 \\
                     \end{array}
                   \right)C\;,\qquad\quad
&\dfrac{\mean{\xi_\nu}}{\La_f}=D\;,
\label{AFM:vev:neutrinos}
\eea
where the factors  $A$, $B$, $C$, $D$ should obey to the relations:
\bea
&\sqrt{3}f_1A^2+\sqrt{3}f_2B^2+f_3AB=0\;,
\label{AFM:AB}\\[1mm]
&D=-\dfrac{M_\varphi}{g_2}\;,\qquad\qquad
C^2=\dfrac{g_2^2M_\xi^2+g_3M_\varphi^2-g_2M_\varphi M'_\xi}{2 g_2^2g_4}
\label{AFM:CD}\;.
\eea

In the discrete component $S_4\times Z_4$ of the full flavour group  we can choose generators $(S,T,i)$,
where the imaginary unit $i$ denotes the generator of the $Z_4$ factor.
The flavons $\xi_\nu$ and $\varphi_\nu$ are invariant under $Z_4$ and their VEVs are eigenvectors of the generator $S$ corresponding to the eigenvalue 1,
so that the corresponding breaking of $S_4\times Z_4$ preserves the subgroup $G_{\nu}$ generated by $(S,i)$.
In the charged lepton sector $S_4\times Z_4$ is broken down to the subgroup $G_\ell$, generated by the product $i T$.
Indeed the generator $iT$ is given by $\pm\diag(-i ,1 ,-1)$, with the plus (minus) sign for the $\bf3$ $({\bf3'})$ $S_4$ representation.
Such a generator, acting in the appropriate representation on the VEVs of eq. (\ref{AFM:vev:charged:best}), leaves them invariant.
It is precisely the mismatch, present at the leading order, between the subgroups $G_{\nu}$ and $G_\ell$
preserved in the neutrino and charged lepton sectors, respectively, that produces the bimaximal lepton mixing, as we will explicitly see in this section.

Similarly, the Froggatt-Nielsen flavon $\theta$ gets a VEV, determined by the $D$-term associated to the local $U(1)_{FN}$ symmetry (as in the previous models), and it is denoted by $\mean{\theta}/\La_f= t$.

With this VEVs configuration, the charged lepton mass matrix is diagonal
\beq
M_e=\left(
         \begin{array}{ccc}
           (y_e^{(1)}A^2-y_e^{(2)}B^2+y_e^{(3)}AB)t^2 & 0 & 0 \\
           0 & y_\mu Bt & 0 \\
           0 & 0 & y_\tau A \\
         \end{array}
       \right)\dfrac{v_d}{\sqrt2}
\eeq
so that at the leading order there is no contribution to the lepton mixing matrix from the diagonalisation of charged lepton masses.
In the neutrino sector for the Dirac and right-handed Majorana matrices we have
\beq
m_D=\left(
          \begin{array}{ccc}
            1 & 0 & 0 \\
            0 & 0 & 1 \\
            0 & 1 & 0 \\
          \end{array}
        \right)\dfrac{y\,v_u}{\sqrt2}\;,\qquad\quad
M_R=\left(
              \begin{array}{ccc}
                2M+2aD & -2bC & -2bC \\
                -2bC & 0 & 2M+2aD \\
                -2bC & 2M+2aD & 0 \\
              \end{array}
            \right)\Lambda_f\;.
\label{AFM:Feq:RHnu:masses}
\eeq
The matrix $M_R$ can be diagonalised by the bimaximal mixing matrix $U_{BM}$, which represents the full lepton mixing at the leading order, while the Majorana phases are absorbed by the introduction of the diagonal matrix $P$, defined as
\beq
P=\diag(e^{i\al_1/2},\,e^{i\al_2/2},\,e^{i\al_3/2})\;,
\label{AFM:Pmatrix}
\eeq
with $\al_1=-\arg(M+aD-\sqrt{2}bC)$, $\al_2=-\arg(M+aD+\sqrt{2}bC)$, $\al_3=-\arg(M+aD)$. As a result, defining $U_R=U_{BM}P$, the eigenvalues are given by
\beq
U_R^T M_R U_R\,=\,\diag(M_1,\,M_2,\,M_2)\qquad\text{with}\quad
                                        \left\{\begin{array}{rcl}
                                          M_1 &=& 2|M+aD-\sqrt{2}bC|\Lambda_f\;, \\[3mm]
                                          M_2 &=& 2|M+aD+\sqrt{2}bC|\Lambda_f\;, \\[3mm]
                                          M_3 &=& 2|M+aD|\Lambda_f\;.
                                        \end{array}\right.
\eeq
After See-Saw, since the Dirac neutrino mass matrix commutes with $M_R$ and its square is a matrix proportional to unity,
the light neutrino Majorana mass matrix, given by the See-Saw relation \mbox{$m_\nu=-m_D^TM_R^{-1}m_D$}, is also diagonalised by the bimaximal mixing matrix and, defining $U_\nu=U_{BM}P^*$, the eigenvalues are
\beq
U_\nu^T m_\nu U_\nu\,=\,\diag(m_1,\,m_2,\,m_2)\qquad\text{with}\quad m_i\equiv\dfrac{|y^2|v_u^2}{2M_i}\;.
\label{AFM:spec}
\eeq
Notice that since the neutrino sector is not charged under the $Z_4$ symmetry, we have operators of dimension 5 which contribute to the neutrino masses and may correspond to some heavy exchange other than the right-handed neutrinos $\nu^c$. The contribution from these terms is
\beq
m_\nu^{eff}=\left(
              \begin{array}{ccc}
                2M'+2a'D & -2b'C & -2b'C \\
                -2b'C & 0 & 2M'+2a'D \\
                -2b'C & 2M'+2a'D & 0 \\
              \end{array}
            \right)\dfrac{v_u^2}{2\Lambda_f}\;,
\label{AFM:meff}
\eeq
where the choice of the labels $M'$, $a'$ and $b'$ is not accidental but reflects the fact that $m_\nu^{eff}$ is similar to $M_R$. When considering the interesting domain of parameters, we find that this effective contribution is subdominant and for this reason we will discuss its effects when dealing with the NLO terms.

At the leading order, the light neutrino mass matrix depends on only 2 effective parameters indeed the terms $M$ and $aD$ enter the mass matrix in the combination $F\equiv M+a D$. The coefficients $y_e^{(i)}$, $y_\mu$, $y_\tau$, $y$, $a$ and $b$ are all expected to be of $\mathcal{O}(1)$. A priori $M$ could be of $\mathcal{O}(1)$, corresponding to a right-handed neutrino Majorana mass of $\mathcal{O}(\Lambda_f)$, but, actually, we will see that it must be of the same order as $C$ and $D$.

We expect a common order of magnitude for the VEVs (scaled by the cutoff $\Lambda_f$):
\beq
A \sim B \sim v\;,\qquad \qquad C \sim D \sim v'\;.
\eeq
However, due to the different minimisation conditions that determine $(A,B)$ and $(C,D)$, we may tolerate a moderate hierarchy
between $v$ and $v'$. Similarly the order of magnitude of $t$ is in principle unrelated to those of $v$ and $v'$.
It is possible to estimate the values of $v$ and $t$ by looking at the mass ratios of charged leptons:
\beq
\dfrac{m_\mu}{m_\tau} \sim t\;, \qquad\qquad \dfrac{m_e}{m_\mu} \sim vt\;.
\eeq
In order to fit these relations with the data, we must have approximately $t \sim 0.06$ and $v \sim 0.08$ (modulo coefficients of $\mathcal{O}(1)$).\\

To summarise, at the leading order we have diagonal and hierarchical charged leptons together with the exact bimaximal mixing for neutrinos. It is clear that substantial NLO corrections are needed to bring the model to agree with the data on $\theta_{12}$. A crucial feature of our model is that the neutrino sector flavons  $\varphi_\nu$ and $\xi_\nu$ are invariant under $Z_4$ which is not the case for the charged lepton sector flavons $\varphi_\ell$ and $\chi_\ell$. The consequence is that $\varphi_\nu$ and $\xi_\nu$  can contribute at the NLO to the corrections in the charged lepton sector, while at the NLO $\varphi_\ell$ and $\chi_\ell$ cannot modify the neutrino sector couplings. As a results the dominant genuine corrections to the bimaximal mixing only occur  at the NLO through the diagonalisation of the charged leptons. We will discuss the NLO corrections in section \ref{Sec:AFM:NLO} after having proven that the necessary VEV alignment is in fact realised at the leading order.

\subsection{The Light Neutrino Spectrum and the Value of $r$}
\label{Sec:AFM:LightSpectrum}
\setcounter{footnote}{3}

We now discuss the constraints on the parameters of the neutrino mass matrix in order to get the correct value for the ratio $r$.  Like for the previous models, also in this case some fine-tuning is needed to accommodate the value of $r$. In fact the triplet assignment for left-handed lepton doublets and for right-handed neutrinos tends to lead to $r\sim1$. We find useful to begin the presentation by analysing the leading order terms, even though a more complete phenomenological discussion with the inclusion of the NLO contributions will be illustrated in section \ref{Sec:AFM:Phem}.

We redefine the parameters in eqs. (\ref{AFM:spec}) as $F=M+aD$ and $Y=-\sqrt{2}bC$ so that $Y\sim v'$, while $F$, like $M$, a priori could be larger, of $\mathcal{O}(1)$. We make the phases of $F$ and $Y$ explicit by setting
\beq
F\rightarrow Fe^{i\phi_F}\qquad\qquad Y\rightarrow Ye^{i\phi_Y}\;,
\eeq
where now $F$ and $Y$ are real and positive parameters. Defining the phase difference $\Delta\equiv\phi_Y-\phi_F$ we can explicitly write the absolute values of the neutrino masses as
\bea
m_1&=&\frac{1}{\sqrt{F^2+Y^2+2FY\cos{\Delta}}}\frac{|y^2|v_u^2}{4\Lambda_f}\nn \\[3mm]
m_2&=&\frac{1}{\sqrt{F^2+Y^2-2FY\cos{\Delta}}}\frac{|y^2|v_u^2}{4\Lambda_f}
\label{AFM:nuMasses}\\[3mm]
m_3&=&\frac{1}{F}\frac{|y^2|v_u^2}{4\Lambda_f}\nn\;.
\eea
Note that the phase $\Delta$ is related by a non-trivial relation to the Majorana CP phase $2 \al_{12}$, which, by definition, is
the phase difference between the complex eigenvalues $m_1$ and $m_2$ of the neutrino mass matrix.
Furthermore $\cos\Delta$ must be positive in order to guarantee $m_2>m_1$. By defining $F/Y\equiv\rho$, we can write the expression of the ratio $r$:
\beq
r=\dfrac{4\rho^3\cos{\Delta}}{(\rho^2+1+2\rho\cos{\Delta})(1-2\rho\cos{\Delta})}\;.
\label{AFM:eq:rf}
\eeq

In order to have $r$ small either we take $\rho$ small or $\cos{\Delta}$ small (or both). If $F\sim \mathcal{O}(1)$ then $\rho\sim \mathcal{O}(1/v')$, $r\sim 4\rho\cos{\Delta}$ and $\cos{\Delta}$ must be extremely small: $\cos{\Delta}\sim v' r/4 \sim 10^{-3}$. We prefer to take $\rho$ small, such that
\beq
r\sim4\rho^3\cos\Delta\;.
\eeq
If so, in order to accommodate the value of $r$, we only need, for example,  $4\cos\Delta\sim1$ and $\rho\sim 1/3$. In conclusion, we have to take $F\sim M\sim \mathcal{O}(v')$ and $\rho=F/Y$ moderately small.
We interpret the relation $M\sim F \sim v'$, necessary to reproduce the value of $r$, as related to the fact that the right-handed neutrino Majorana M mass must empirically be smaller than the cutoff $\Lambda_f$ of the theory.  In the context of a grand unified theory this corresponds to
the requirement that $M$ is of $\mathcal{O}(M_{GUT})$ rather than of $\mathcal{O}(M_{Planck})$.

With these positions, the neutrino spectrum shows a moderate normal hierarchy, with
\beq
m_{1,2}\sim\dfrac{1}{Y}\;\sim\;\mathcal{O}\left(\frac{1}{v'}\right)\;,\qquad\qquad
m_3\sim\mathcal{O}\left(\dfrac{1}{\rho Y}\right)\;,
\eeq
in units of $|y^2|v_u^2/4\Lambda_f$. At the leading order an inverse ordering of the neutrino masses is forbidden, as
we can see from eq. (\ref{AFM:nuMasses}), which, for $m_2>m_1$, always implies $m_3>m_1$.\\

At the leading order the lightest neutrino mass $m_1$ has a lower bound. Indeed, the only possible way to decrease $m_1$
is to take $Y$ as large as possible. By expanding eqs. (\ref{AFM:nuMasses}) in powers of $\rho$, we have
\beq
m_1\approx \rho m_3\ge \rho \sqrt{\Delta m^2_{atm}}\ge 10\, {\rm meV}\;.
\eeq
As we will see in section \ref{Sec:AFM:Phem}, this lower bound can be evaded by including NLO corrections, but values of $m_1$ much smaller than $10$ meV would require an additional tuning of the parameters.

\section{The Next-To-Leading Order Corrections}
\label{Sec:AFM:NLO}
\setcounter{footnote}{3}

We now summarise the set of subleading corrections to the superpotential that are essential to bring the model in agreement with the data. The detailed analysis can be found in the original paper \cite{AFM_BimaxS4}. 

Classifying the corrections according to an expansion in inverse powers of $\Lambda_f$, the new flavon VEV configuration can be written as 
\beq
\langle \Phi \rangle=\langle \Phi \rangle_{LO}+\delta \Phi
\eeq
where $\Phi=(\xi_\nu,~\varphi_\nu, ~\varphi_\ell, ~\chi_\ell)$ and  $\langle \Phi \rangle_{LO}$ are given by eqs. (\ref{AFM:vev:charged:best}) and (\ref{AFM:vev:neutrinos}). As illustrated in the appendix \ref{AppB:S42}, in the neutrino sector the shifts $\delta \xi_\nu,~\delta \varphi_\nu$ turn out to be proportional to the leading order VEVs $\langle \Phi \rangle_{LO}$ and can be absorbed in a redefinition of the parameters $C$ and $D$. Instead, in the charged lepton sector, the shifts $\delta \varphi_\ell, ~\delta  \chi_\ell$ have a non trivial structure, so that the leading order texture is modified:
\beq
\mean{\varphi_\ell}=\left(
                     \begin{array}{c}
                       {\delta \varphi_\ell}_1\\
                       A' \La_f \\
                       0\\
                     \end{array}
                   \right)\qquad
\qquad\mean{\chi_\ell}=\left(
                     \begin{array}{c}
                       {\delta \chi_\ell}_1 \\
                       0 \\
                       B' \La_f \\
                     \end{array}
                   \right)
\label{AFM:vev:charged:nlo}
\eeq
where $A'$ and $B'$ satisfy a relation similar to that in eq. (\ref{AFM:AB}) and the shifts ${\delta \varphi_\ell}_1$ and ${\delta \chi_\ell}_1$ are proportional to $v'v\La_f$, that are, in other words, suppressed by a factor $v'$ with respect to the leading order entries $A\La_f$ and $B\La_f$, respectively. 

The fermion mass matrices receive corrections due to higher-order operators in the respective superpotential and due to the new VEV configurations.
The NLO operators contributing to the lepton masses can be obtained by inserting in all possible ways $\xi_\nu$ or $\varphi_\nu$ in the leading order operators and by extracting the $S_4\times Z_4\times U(1)_{FN}$ invariants. Insertions of one power of the flavons $\varphi_\ell$ or $\chi_\ell$
are forbidden by the invariance under the $Z_4$ component of the flavour symmetry group. We find that the corrected mass matrix for the charged leptons at this order has the (23) and (32) elements still vanishing. By omitting all order one coefficients, the charged lepton mass matrix and the unitary matrix that realises the transformation to the physical basis, where the product $(M_e^\dag M_e)$ is diagonal at the NLO, are given by
\beq
M_e=\left(
         \begin{array}{ccc}
           vt^2 & vv't^2 & vv't^2 \\
           v't & t & 0 \\
           v' & 0 & 1 \\
         \end{array}
       \right)\dfrac{v\,v_d}{\sqrt2}\;,\qquad
U_e=\left(
         \begin{array}{ccc}
           1 & V_{12} v' & V_{13} v' \\
           -V_{12} v' & 1 & 0 \\
           -V_{13} v' & 0 & 1 \\
         \end{array}
       \right)\;,
\eeq
where the coefficients $V_{ij}$ are of $\mathcal{O}(1)$.

Moving to the neutrino sector, we have already stated that the structure of the leading order VEVs of the neutrino flavons is unchanged and therefore the only corrections come from the higher-order terms in $w_\nu$. However, it is straightforward to verify that, even after the inclusion of the NLO corrections, both $M_R$ and $m_\nu$ can be exactly diagonalised by the bimaximal mixing, which therefore represents the total contribution to lepton mixing of the neutrino sector. These corrections introduce also some terms in the mass eigenvalues of relative order $v'$ with respect to the leading order results, but they do not affect the type of the spectrum.

Since the neutrino mass matrix is diagonalised by $U_{BM}$, the PMNS matrix can be written as
\beq
U=U_e^\dag U_{BM}P^{'*}\;,
\eeq
where $P'$ is the diagonal matrix of the Majorana phases which differs from the original $P$ only due to the NLO contributions. The corrections from $U_e$ affect the neutrino mixing angles at the NLO according to
\beq
\sin^2\theta_{12}=\dfrac{1}{2}-\frac{1}{\sqrt{2}}(V_{12}+V_{13})v'\;,\qquad
\sin^2\theta_{23}=\dfrac{1}{2}\;,\qquad
\sin\theta_{13}=\dfrac{1}{\sqrt{2}}(V_{12}-V_{13})v'\;.
\label{AFM:sinNLO}
\eeq
By comparing these expressions with the current experimental values of the mixing angles in table \ref{table:OscillationData}, we see that, to correctly reproduce $\theta_{12}$ we need a parameter $v'$ of the order of the Cabibbo angle $\lambda$. Moreover, barring cancellations of/among some the $V_{ij}$ coefficients, also the reactor angle is corrected by a similar amount.

\begin{figure}[h!]
\centering
{\includegraphics[width=11cm]{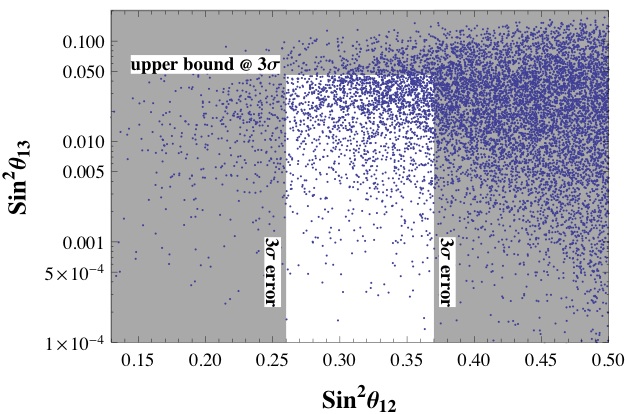}}
\caption{\it $\sin^2\theta_{13}$ as a function of $\sin^2\theta_{12}$ is plotted, following eqs. (\ref{AFM:sinNLO}). The plot is symmetric with respect $\sin^2\theta_{12}=0.5$ and we report here only the left-part. The parameters $V_{ij}$ are treated as random complex numbers of absolute value between 0 and 2, while $|v'|$ has been fixed at the indicative value of 0.15. The gray bands represent the regions excluded by the experimental data \cite{Fogli:Indication}: the horizontal one corresponds to the $3\sigma$-upper bound for $\sin^2\theta_{13}$ of 0.46 and the vertical ones to the region outside the $3\sigma$ error range $[0.26 - 0.37]$ for $\sin^2\theta_{12}$.}
\label{fig:AFMfigure1213}
\end{figure}

Any quantitative estimates are clearly affected by large uncertainties due to the presence of unknown parameters of order one, as we can see in figure \ref{fig:AFMfigure1213}, but in our model a value of $\theta_{13}$ much smaller than the present upper bound would be unnatural.

All this discussion works at the NLO, but we expect that at the NNLO the value of $\theta_{23}$ will also be modified with deviations of about $\mathcal{O}(\lambda^2)$  at most. The next generation of experiments, in particular those exploiting a high intensity neutrino beam, will probably reduce the experimental error on $\theta_{23}$ and the sensitivity on $\theta_{13}$ to few degrees. If no significant deviations from zero of $\theta_{13}$ will be detected, our construction will be ruled out.

A salient feature of our model is that, at the NLO accuracy, the large corrections of $\mathcal{O}(\lambda)$ only apply to $\theta_{12}$ and $\theta_{13}$ while $\theta_{23}$ is unchanged at this order. As a correction of $\mathcal{O}(\lambda)$ to $\theta_{23}$ is hardly compatible with the present data this feature is very crucial for the phenomenological success of our model. It is easy to see that this essential property depends on the selection in the neutrino sector of flavons $\xi_\nu$ and $\varphi_\nu$ that transform as $\bf1$ and $\bf3$ of $S_4$, respectively. We have checked that if, for example, the singlet $\xi_\nu$ is replaced by a doublet $\psi_\nu$ (and correspondingly the singlet driving field  $\xi_\nu^0$ is replaced by a doublet $\psi_\nu^0$), all other quantum numbers being the same, one can construct a variant of the model along similar lines, but in this case all the 3 mixing angles are corrected by terms of the same order. This confirms that a  particular set of $S_4$ breaking flavons is needed in order to preserve $\theta_{23}$ from taking as large corrections as the other two mixing angles.

\section{Phenomenological Implications}
\label{Sec:AFM:Phem}
\setcounter{footnote}{3}

We now develop a number of important physical consequences of our model and derive some predictions. We consider the predicted spectrum and the effective mass $m_{ee}$ for neutrinoless-double-beta decay. The light neutrino mass matrix, including the NLO corrections, is given by:
\beq
m_\nu=-m_D^TM_R^{-1}m_D+m_\nu^{eff}\;,
\eeq
where $m_\nu$ and $M_R$  are the mass matrices at the NLO and $m_\nu^{eff}$ is given in eq. (\ref{AFM:meff}). It is diagonalised by the bimaximal unitary transformation and its complex eigenvalues are given by:
\bea
m_1&=&\left[(F'+Y')-\dfrac{(y'+Y_2)^2}{4(F+Y+a_2 C^2)}\right]\dfrac{v_u^2}{\Lambda_f}\nonumber \\
m_2&=&\left[(F'-Y')-\dfrac{(y'-Y_2)^2}{4(F-Y+a_2 C^2)}\right]\dfrac{v_u^2}{\Lambda_f}\\
m_3&=&\left[-F'+\dfrac{y'^2}{4(F-2 a_2 C^2)}\right]\dfrac{v_u^2}{\Lambda_f}\;,\nonumber
\label{AFM:numNLO}
\eea
where
\beq 
F'=M'+a' D\;,\qquad Y'=-\sqrt{2} b' C\;,\qquad Y_2=-\sqrt{2} y_2 C\;,\qquad  y'=y+y_1 D\;,
\eeq
with $y_{1,2}$ and $a_2$ NLO parameters (see \cite{AFM_BimaxS4} for details). We see that the leading order expressions are recovered in the limit $F'=Y'=y_{1,2}=a_2=0$. By exploiting eqs. (\ref{AFM:numNLO}) we can study some observables like the effective $0\nu2\beta$-decay mass, $|m_{ee}|$, the lightest neutrino mass, $m_1$, and the sum of the neutrino masses directly from the experimental data. We perform a numerical analysis, by treating all the leading order, NLO and effective parameters as random complex numbers with absolute value between 0 and 3.

\begin{figure}[h!]
\centering
\subfigure[$|m_{ee}|$ vs. $m_1$]
{\includegraphics[width=7cm]{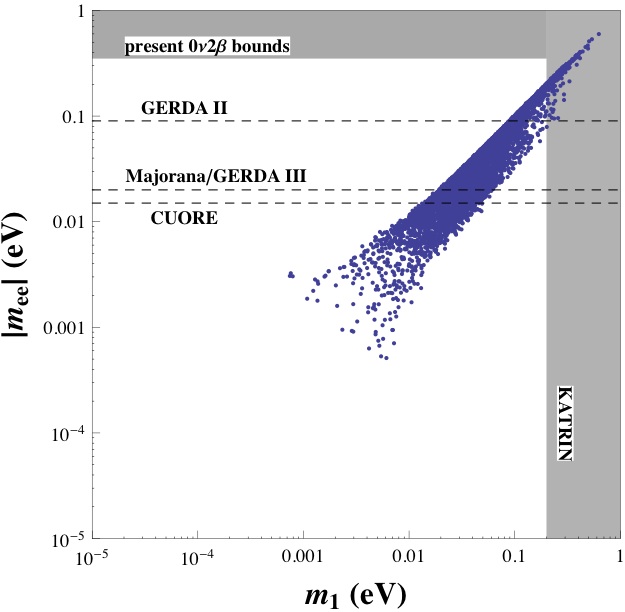}}
\subfigure[$\sum m_i$ vs. $m_1$]
{\includegraphics[width=7cm]{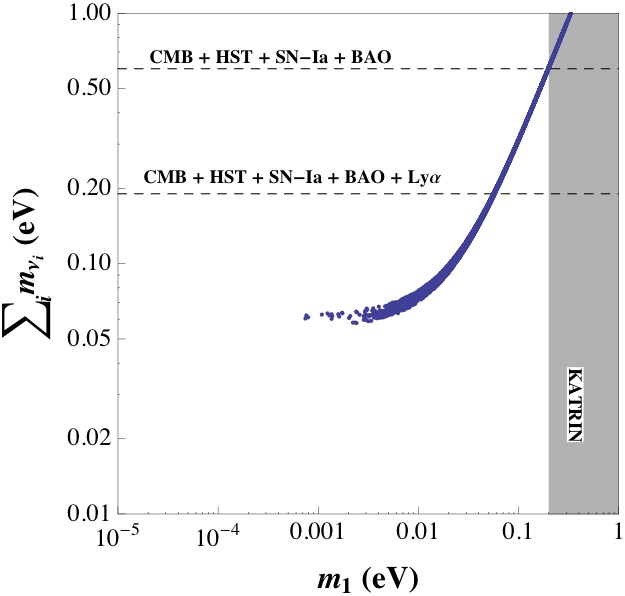}}
\caption{\it In figure (a), $|m_{ee}|$ as a function of the lightest neutrino mass, $m_1$, is plotted. The constraints which have been imposed in the plot are the experimental values at $3\sigma$ for $\De m^2_{atm}$, $\De m^2_{sol}$ and the mixing angles. All the parameters of the model are treated as random complex parameters. The experimental bounds are equal as in figure \ref{fig:0nu2betaGeneral}. In figure (b) we show the sum of the neutrino masses as a function of the lightest neutrino mass.}
\label{fig:AFMfigure2}
\end{figure}

In figure \ref{fig:AFMfigure2}(a), we plot $|m_{ee}|$ as a function of the lightest neutrino mass. The points correspond to the case of normal ordering of the neutrino masses, with a moderate hierarchy or a quasi degenerate spectrum.
However, at variance with the results of the leading order, some solutions of our numerical simulation also reproduce an inverse hierarchical spectrum. The plot displays only the points corresponding to choices of the parameters reproducing
$\De m^2_{atm}$, $\De m^2_{sol}$ and the mixing angles within a $3\sig$ interval. The figure suggests that a lower bound of about $0.1$ meV holds for the lightest neutrino mass. Indeed, with the inclusion of the NLO corrections, from eq. (\ref{AFM:numNLO}) we see that $m_1$ can vanish if a cancellation between the NLO and leading order contributions takes place. This however requires an additional fine-tuning of the parameters which has been reproduced in our numerical analysis only partially and by very few points. Similarly the scatter plot indicates a lower bound for $|m_{ee}|$ of about $0.1$ meV.

In figure \ref{fig:AFMfigure2}(b), we plot the sum of the neutrino masses as a function of the lightest neutrino mass, $m_1$. The vertical band refers to the future sensitivity of KATRIN experiment, while the horizontal ones to the cosmological bounds \cite{CosmoNu}. There are typically five representative combinations of the cosmological data, which lead to increasingly stronger upper bounds on the sum of the neutrino masses: we are showing the two strongest ones, at $0.60$ eV and $0.19$ eV. This plot is typical for normal hierarchy or quasi degenerate spectrum. The only special feature is the lower bound on $m_1$, which, as explained above, relies on a naturalness assumption.

\section{Extension to Quarks: a GUT Realisation}
\label{Sec:AFM:PS}
\setcounter{footnote}{3}

In this section we are interested in the extension to the quark sector. A first attempt is to adopt for quarks the same representations under $S_4$ that have been used for leptons: the left-handed quark doublets $q$ transform as a triplet $\bf3$, while the right-handed quarks $(u^c,\,d^c)$, $(c^c,\,s^c)$ and $(t^c,\,b^c)$ transform as $\bf1$, $\bf1'$ and $\bf1$, respectively. We can similarly extend to quarks the transformations of $Z_4$ (and $U(1)_R$) given for leptons. As a result, it is easy to see that the quark mass matrices are diagonal, at the leading order on the expansion parameters, exactly as for the charged leptons and to account for the correct mass hierarchies the $U(1)_{FN}$ has to be suitably implemented. At this level the CKM matrix is the unity matrix and to get realistic mixings, the higher-order corrections should switch on off-diagonal entries with a well-defined pattern: $(12)\sim\lambda$, $(23)\sim\lambda^2$ and $(13)\sim\lambda$. By an explicit computation we find the following result for the quark mass matrices: in terms of order of magnitude
\beq
M_d=\left(
          \begin{array}{ccc}
            v\,t^2 & v\,v'\,t^2 & v\,v'\,t^2 \\
            v'\,t & t & v^{\prime\,2}\,t \\
            v' & v^{\prime\,2} & 1 \\
          \end{array}
        \right)\dfrac{v\,v_d}{\sqrt2}\;,\qquad
M_u=\left(
          \begin{array}{ccc}
            v\,t^3 & v\,v'\,t^3 & v\,v'\,t^3 \\
            v'\,t^2 & t^2 & v^{\prime\,2}\,t^2 \\
            v' & v^{\prime\,2} & 1 \\
          \end{array}
        \right)\dfrac{v\,v_u}{\sqrt2}\;.
\eeq
Calculating now the unitary matrices which diagonalise $M_d^\dag M_d$ and $M_u^\dag M_u$ we find
\beq
V_d\sim V_u=\left(
          \begin{array}{ccc}
            1 & v' & v' \\
            -v' & 1 & v^{\prime\,2} \\
            -v' & -v^{\prime\,2} & 1 \\
          \end{array}
        \right)\;.
\eeq
At a first sight, barring cancellations among the single entries, we can see that the CKM matrix $V=V_u^\dag V_d$ should be similar to $V_u$ and $V_d$ and as a consequence it cannot correctly describe the quark mixings: while the entries $(12)$ and $(23)$ well reproduce the measured values, the large values in the $(13)$ and $(31)$ entries would require a large fine-tuning of order $\lambda^2$. To a closer look at the superpotential which generate such a result, we see that the higher-order corrections have two independent sources: new higher-order operators calculated with the VEVs at the leading order and the original superpotential calculated with the NLO VEVs. As a result we do not expect any cancellation between $V_u$ and $V_d$ when multiplied to give the CKM matrix, contrary to what happens in the Altarelli-Feruglio model in section \ref{Sec:AFTBM:Quarks}.\\

An alternative possibility is to further investigate on the complementarity relations:
\baq
\theta_{12}+\lambda &\simeq& \pi/4\;,\\[3mm]
\theta_{23}+\lambda^2 &\simeq& - \pi/4\;.
\label{ABM:anglescomplement}
\eaq
These equations suggest that the angles in the CKM and PMNS matrices may have a common origin which can be motivated for example in Pati-Salam models, where the following relation holds, 
\beq
U_e\sim V_d\;.
\label{ABM:Comple}
\eeq
We can use this to write the CKM and PMNS matrices as
\beq
\ba{l}
U\;=\;R_{23}\left(-\dfrac{\pi}{4}\right) R_{13}(\lambda) R_{12}\left(\dfrac{\pi}{4} - \lambda\right)
\;=\;\Big(\underbrace{R_{23}\left(\dfrac{\pi}{4}\right) R_{13}(\lambda) R_{12}(\lambda)}_{U_e}\Big)^\dag \underbrace{R_{12}\left(\dfrac{\pi}{4}\right)}_{U_\nu}\;\\[3mm]
V\;=\; R_{12}(\lambda)
\;=\;\Big(\underbrace{R_{23}\left(\dfrac{\pi}{4}\right) R_{13}(\lambda) R_{12}(\lambda)}_{V_u}\Big)^\dagger \underbrace{R_{23}\left(\dfrac{\pi}{4}\right) R_{13}(\lambda) R_{12}(\lambda))}_{V_d}.
\ea
\label{ABM:CKMrots}
\eeq
Here $R_{ij}(\alpha)$ stand for rotations in the $(ij)$ plane of the angle $\alpha$ (apart from coefficients $\cO(1)$ in front of each angles). The coefficients of the angles in the rotations in $V_u^\dag$ and $V_d$ should be such that the rotations cancel each other in the $(13)$ sector, but not in the $(12)$ sector. Thus we should introduce terms which distinguish between the up- and down-quark sectors, indeed possible within the Pati-Salam context.

Moving to the explicit form of the mass matrices, the generic Majorana neutrino mass matrix $m_\nu$ which is diagonalised by $U_\nu=R_{12}\left(\dfrac{\pi}{4}\right)$,
\beq
m_\nu^{diag}\;=\;R_{12}\left(\dfrac{\pi}{4}\right)^T\,m_\nu \,R_{12}\left(\dfrac{\pi}{4}\right)\,,
\eeq
is given by
\beq
m_\nu \sim\left(
                \begin{array}{ccc}
                a & b & 0 \\
                b & a & 0 \\
                0 & 0 & c \\
                \end{array}
        \right)\,.
\label{ABM:Mnu1}
\eeq
Considering the charged lepton mass matrix $M_e$, the product $M_e^\dag\,M_e$ should be diagonalised by the action of $V_d$ as in eq. \eqref{ABM:CKMrots},
\beq
R_{12}(-\lambda) R_{13}(-\lambda) R_{23}\left(-\dfrac{\pi}{4}\right) \, M_e^\dag \, M_e \, R_{23}\left(\dfrac{\pi}{4}\right) R_{13}(\lambda) R_{12}(\lambda)\,.
\eeq
In the limit $m_e\to 0$ we find the generic structure for the product $M_e^\dag\,M_e$:
\beq
M_e^\dag\,M_e \sim \dfrac{m_\tau^2}{2} \left(
                                            \begin{array}{ccc}
                                            0 & \lambda & \lambda \\
                                            \lambda & 1 & 1 \\
                                            \lambda & 1 & 1 \\
                                            \end{array}
                                        \right) +
\dfrac{m_\mu^2}{2} \left(
                    \begin{array}{ccc}
                    0 & \lambda & -\lambda \\
                    \lambda & 1 & -1 \\
                    -\lambda & -1 & 1 \\
                    \end{array}
                    \right) +\mathcal{O}(\lambda^2) \;,
\label{ABM:Mlsq}
\eeq
that can be obtained if $M_e$ is given by
\beq
M_e  \sim \dfrac{m_\tau}{\sqrt{2}} \left(
                                        \begin{array}{ccc}
                                        0 & 0 & 0 \\
                                        0 & 0 & 0 \\
                                        \lambda & 1 & 1 \\
                                        \end{array}
                                    \right) +
\dfrac{m_\mu}{\sqrt{2}} \left(
                        \begin{array}{ccc}
                        0 & 0 & 0 \\
                        \lambda & 1 & -1 \\
                        0 & 0 & 0
                        \end{array}
                        \right) +\mathcal{O}(\lambda^2) \;.
\label{ABM:Ml1}
\eeq
It is interesting to note that, moving to the basis of diagonal charged leptons and considering only the leading order terms, the neutrino mass matrix results to be of the classical bimaximal type.

From eq. \eqref{ABM:Comple}, the relation $M_e\sim M_d$ follows and therefore the down-quark matrix has a similar structure as in eq. \eqref{ABM:Ml1}. Looking at eq. (\ref{ABM:CKMrots}) we find that also $M_u$ should have a similar structure. We find that we can satisfy the constraint on the $(12)$ and $(13)$ rotations if the third columns of $M_u$ and $M_d$ are proportional to each other, but the second columns are not.\\

So far we have not given yet any explanation of the origin of these mass matrices and mixings and we leave this analysis to \cite{ABM_PSS4}. Here we only say that such a construction is possible in a Pati-Salam realisation where the flavour symmetry is $S_4\times Z_4\times U(1)_{FN}$: as for the non-GUT model described in the rest of the chapter, key points are a suitable choice of the group representations for the flavons and the particular VEV misalignment whose effects are a reactor angle and a deviation from $\pi/4$ of the solar angle of the order of $\lambda$, while introducing small deviations of the order $\lambda^2$ from the maximal value of the atmospheric angle.

A difficulty with respect the non-GUT model refers to the study of the gauge coupling running and of the Higgs potential: while in a general Pati-Salam model, in particular without any flavour symmetry implementation, it is possible to reproduce a realistic sequential symmetry breaking chain, with the usual Standard Model or MSSM Higgs fields at the electroweak scale, the introduction of a flavour symmetry puts strong constraints. Indeed we need additional scalars which transform under the gauge group and the effect is to sandwich the energy scales of the different symmetry breakings, lowering to $10^{14}$ GeV the energy scale of the (almost) unification.

In \cite{ABM_PSS4} we stress out that a combined study of the flavour and Higgs sectors should be compulsory, due to the strong interplay between the two. This is in open contrast with a general attitude: usually people focus only on a single aspect, either the flavour sector or the gauge/Higgs one.

\section{Conclusions of the Chapter}
\label{Sec:AFM:Conclusions}
\setcounter{footnote}{3}

In this part we have illustrated a See-Saw model based on the flavour symmetry $S_4\times Z_4 \times U(1)_{FN}$ where the bimaximal mixing is realised at the leading order in a natural way.  The hierarchy of charged lepton masses is obtained as a combined effect of the $U(1)_{FN}$ symmetry breaking measured by the parameter $t$ and of the $S_4\times Z_4$ breaking induced by $v$, proportional to the VEVs of $\varphi_\ell$ and $\chi_\ell$. We have $m_\mu/m_\tau =t \sim 0.06$ and  $m_e m_\tau/m_\mu^2 =v \sim 0.08$.

Since exact bimaximal mixing implies a value of $\tan{\theta_{12}}$ which is excluded by the data, large corrections are needed. The dominant corrections to the bimaximal mixing arise at the NLO only through the diagonalisation of the charged lepton mass matrix. The shifts of the quantities $\sin^2{\theta_{12}}$ and $\sin{\theta_{13}}$ from the bimaximal values are linear in the parameter $v'$, proportional to the VEVs of $\varphi_\nu$ and $\xi_\nu$, which is expected to be of the same order as $v$, but not necessarily too close, as $v$ and $v'$ are determined by two different sets of minimisation equations. From the experimental value $\tan^2{\theta_{12}}= 0.45\pm 0.04$, which is sizeably different than the bimaximal value $\tan^2{\theta_{12}}= 1$, we need $v'\sim \mathcal{O}(\lambda)$.
As in most models where the bimaximal mixing is only corrected by the effect of charged lepton diagonalisation, one also expects $\theta_{13}\sim \mathcal{O}(\lambda)$. A value of $\theta_{13}$ near the present bound would be a strong indication in favour of this mechanism and a hint that the closeness of the measured values of the mixing angles to the tribimaximal values may be purely an accident. In addition, a very important feature of our model is that the shift of $\sin^2{\theta_{23}}$ from the maximal mixing value of 1/2 vanishes at the NLO and is expected to be of $\mathcal{O}(\lambda^2)$ at most. In our $S_4$ model, this property is obtained by only allowing  the breaking of $S_4$ in the neutrino sector via flavons transforming as $\bf1$ and $\bf3$ (in particular with no doublets).

In order to reproduce the experimental value of the small parameter $r=\Delta m^2_{sol}/\Delta m^2_{atm}$ we need some amount of fine-tuning. For instance, the right-handed neutrino Majorana mass $M$ should be below the cutoff $\Lambda_f$ (this is reminiscent of the fact that empirically $M \sim M_{GUT}$ rather than $M \sim M_{Planck}$). The neutrino spectrum is mainly of the normal hierarchy type (or moderately degenerate), the smallest light neutrino mass and the $0\nu \beta \beta$-parameter $|m_{ee}|$ are expected to be larger than about $0.1$ meV.

When extending such a flavour treatment to the quark sector we face several problems: simply adopting the same representations used for leptons, the model does not account for a realistic CKM matrix, due to large contributions to the $(13)$ and $(31)$ entries of $V$. An improvement is possible moving to a grand unified context, where a Pati-Salam model has been studied: the flavour symmetry is still $S_4\times Z_4\times U(1)_{FN}$ and realistic mass matrices and mixings are found. The disadvantage of this choice manifests in the presence of strong constraints on the scalar content of the model, which deeply affect the gauge coupling running.

\clearpage{\pagestyle{empty}\cleardoublepage}


\newpage
\chapter{Running Effects on Flavour Models}
\label{Sec:Running}
\setcounter{equation}{0}
\setcounter{footnote}{3}

In all the flavour models described in the previous sections, mass matrices and mixings are evaluated at a very high energy scale. On the other hand, for a comparison with the experimental results, it is necessary to evolve the observables to low energies through the renormalisation group (RG) running. In general the deviations from high energy values due to the running consist in minor corrections which cannot be measured in the future neutrino experiments, apart from some special case in which these deviations undergo a large enhancement.

In the present section we will discuss the effects of the renormalisation group running on the lepton sector when masses and mixings are the result of an underlying flavour symmetry, dealing with both the Standard Model and the MSSM contexts.

When the lightness of the neutrino masses is explained through the five-dimensional Weinberg operator, it is a general result \cite{RunningNoSeeSaw} that the running corrections become relevant only when the neutrino spectrum is almost degenerate or inversely hierarchical (and only for particular values of the Majorana phases) or when in the supersymmetric context $\tan\beta$ is large. Similar results have been found when particular flavour structures for the neutrino masses are invoked, such as the tribimaximal \cite{Running+TB} and the bimaximal \cite{Running+BM} patterns.

When we consider models in which the type I See-Saw mechanism in implemented, few studies have been proposed in literature \cite{RunningSeeSaw} and only regarding general, in particular non-flavour, models. For this reason we focus \cite{LMP_RGE} our attention only on flavour models in which the type I See-Saw is responsible for the light neutrino masses.

We first describe, in a very general context, two kinds of interesting constraints on the Dirac neutrino Yukawa $Y_\nu$ from flavour symmetries and then analyse their impact on running effects. We start considering flavour models in which $Y_\nu$ is proportional to a unitary matrix as it is the case, for example, when the right-handed singlet neutrinos or the charged leptons are in an irreducible representation of the flavour group $G_f$. Then we extend this constraint to a more general class of flavour models in which the mixing textures are independent from the mass eigenstates: as a general result, we find that in this class of models, the effect of the running through the See-Saw thresholds can always be absorbed by a small shift on neutrino mass eigenvalues and the mixing angles remain unchanged. This conclusion is, in particular, independent both from the specific mixing pattern implied by the flavour symmetry and from the basis in which we are working.

Mass-independent mixing textures usually exhibit an underlying discrete symmetry nature: the tribimaximal and the bimaximal patterns, the golden-ratio mixing \cite{golden_ratio} and some (but not all) cases of the trimaximal mixing \cite{Trimaximal} belong to the category of mass-independent mixing textures.

In a second moment, as an explicit example, we describe in detail the running effects on the tribimaximal mixing texture in the Altarelli-Feruglio model described in section \ref{Sec:AFTBM}.

\mathversion{bold}
\section{Running Effects on Neutrino Mass Operator $m_\nu$}
\label{Sec:LMP:Running}
\setcounter{footnote}{3}
\mathversion{normal}

In this section we begin to analyse, in a general context, the renormalisation group equations (RGEs) for neutrino masses below and above the See-Saw threshold, both in the Standard Model and in the MSSM extended with three right-handed neutrinos. We consider the Lagrangian in the lepton sector of the type I See-Saw already defined in eqs. (\ref{SM:LagrangianY}, \ref{SM:LagrangianTypeI}):
\beq
\LL=e^c Y_e H^\dag \ell+ \nu^c Y_\nu \widetilde H^\dag \ell + \nu^c M_R \nu^c +h.c.
\eeq
where the supersymmetric case is easily derived considering two Higgs doublets, all the fields as supermultiplets and identifying the holomorphyc part of $\LL$ with a superpotential. In what follows we concentrate only on the Standard Model particles and for this reason in our notation a chiral superfield and its $R$-parity even component are denoted by the same letter.

Given the heavy Majorana and the Dirac neutrino mass matrices, $M_R$ and $m_D=Y_\nu v/\sqrt2$ respectively, the light $m_\nu$ is obtained from block-diagonalising the complete $6\times6$ neutrino mass matrix,
\beq
m_\nu=-\dfrac{v^2}{2}Y_\nu^TM_R^{-1}Y_\nu\;,
\label{LMP:EqSee-Saw}
\eeq
The matrix $m_\nu$ is modified by quantum corrections according to the RGEs widely studied in the literature \cite{RunningSeeSaw}. For completeness, in appendix \ref{AppendixC}, we report the full RGEs for all the interested quantities in the running. 
In order to analytically study the change of $m_\nu(\mu)$ from high to low-energy, it is useful to work in the basis in which the Majorana neutrino mass is diagonal and real, $\hat{M}_R = \diag(M_S, M_M, M_L)$. The mass eigenvalues can be ordered as $M_S < M_M < M_L$. Furthermore, we can divide the running effects in three distinct energy ranges: from the cutoff $\Lambda$ of the theory down to $M_L$, the mass of the heaviest right-handed neutrino; from $M_L$ down to $M_S$, the mass of the lightest right-handed neutrino; below $M_S$ down to $\varrho$, which can be either $m_Z$, considered as the electroweak scale, or $m_{SUSY}$, the average energy scale for the supersymmetric particles.

\begin{description}
\item[$\mathbf{\La_f\longrightarrow M_L}$.]
Above the highest See-Saw scale the three right-handed neutrinos are all active and the dependence of the effective light neutrino mass matrix from the renormalisation scale $\mu$ is given by mean of the $\mu-$dependence of $Y_\nu$ and $M_R$:
\begin{equation}
m_\nu(\mu)\, =\, -\dfrac{v^2}{2} \, Y_\nu^T(\mu)\, M_R^{-1}(\mu) \,Y_\nu(\mu) \;.
\label{LMP:effnumass1}
\end{equation}
Then from the RGEs in eqs. (\ref{LMP:EqRGEMSSM}, \ref{LMP:EqRGESM}), it is not difficult to see that the evolution of the effective mass matrix $m_\nu$ is given by:
\begin{equation}
16\pi^2 \, \frac{\D m_\nu}{\D t} =\Big(C_eY_e^\dagger Y_e + C_\nu Y_\nu^\dagger Y_\nu\Big)^T \, m_\nu +m_\nu \, \Big(C_e Y_e^\dagger Y_e + C_\nu Y_\nu^\dagger Y_\nu\Big) + \bar\alpha \, m_\nu
\label{LMP:betanu1}
\end{equation}
with
\beq
\ba{ll}
C_e=\,-\dfrac{3}{2}\;,\qquad C_\nu=\dfrac{1}{2}&\qquad\text{in the SM}\\[5mm]
C_e=\,C_\nu=1&\qquad\text{in the MSSM}
\ea
\eeq
and
\beq
\ba{ccl}
\bar\alpha_{SM}&=&2\Tr\left[ 3Y_u^\dagger Y_u + 3Y_d^\dagger Y_d + Y_\nu^\dagger Y_\nu + Y_e^\dagger Y_e \right] - \dfrac{9}{10} g_1^2 - \dfrac{9}{2} g_2^2\\[5mm]
\bar\alpha_{MSSM}&=&2\Tr\Big[ 3Y_u^\dagger Y_u + Y^{\dagger}_\nu Y_\nu \Big] - \dfrac{6}{5} g_1^2 - 6 g_2^2\;.
\ea
\eeq

\item[$\mathbf{M_L\longrightarrow M_S}$.]
The effective neutrino mass matrix $m_\nu$ below the highest See-Saw scale
can be obtained by sequentially integrating out $\nu^c_n$ with $n=L,M,S$:
\begin{equation}\label{LMP:effnumass2}
m_\nu \,=\,-\dfrac{v^2}{4} \left(\,\accentset{(n)}{\kappa}+ 2 \accentset{(n)}{Y}_\nu^T\accentset{(n)}{M_R}^{-1}\accentset{(n)}{Y}_\nu\right)
\end{equation}
where $\accentset{(n)}{\kappa}$ is the coefficient of the effective neutrino mass operator $ (\widetilde H^\dag \ell)^T(\widetilde H^\dag\ell)$. From the (tree-level) matching condition, it is given by
\begin{equation}
\accentset{(n)}{\kappa}_{ij}\,=\,2(Y_\nu^T)_{in}\,M^{-1}_n\,(Y_\nu)_{nj}\;,
\label{LMP:kn}
\end{equation}
which is imposed at $\mu=M_n$.
At $M_L$, the $2\times3$ Yukawa matrix $\accentset{(L)}{Y}_\nu$ is obtained by simply removing the
$L$-th row of $Y_\nu$ and the $2\times2$ mass matrix $\accentset{(L)}{M_R}$ is found from $M_R$ by removing the $L$-th row and $L$-th column. Further decreasing the energy scale down to $M_M$, $\accentset{(M)}{Y}_\nu$ is a single-row matrix, obtained by removing the $M$-th row from $\accentset{(L)}{Y}_\nu$, and $\accentset{(M)}{M_R}$ consists of a single parameter, found by removing the $M$-th row and $M$-th column from $\accentset{(L)}{M}_R$. Finally at $M_S$, $\accentset{(S)}{Y}_\nu$ and $\accentset{(S)}{M}_R$ are vanishing.

In the Standard Model, the two parts which define $m_\nu$ in eq. (\ref{LMP:effnumass2}) evolve in different ways. We can summarise the corresponding RGEs as follows:
\beq
16\pi^2 \, \frac{\D \accentset{(n)}{X}}{\D t} = \left( \dfrac{1}{2}\accentset{(n)}{Y}_\nu^\dagger \accentset{(n)}{Y}_\nu - \dfrac{3}{2}Y_e^\dagger Y_e \right)^T \accentset{(n)}{X} + \accentset{(n)}{X} \left( \dfrac{1}{2}\accentset{(n)}{Y}_\nu^\dagger \accentset{(n)}{Y}_\nu - \dfrac{3}{2}Y_e^\dagger Y_e \right) +  \accentset{(n)}{\bar\alpha}_X \accentset{(n)}{X}
\eeq
where
\beq
\ba{ccl}
\accentset{(n)}{\bar\alpha}_\kappa &=& 2\Tr\left[ 3Y_u^\dagger Y_u + 3Y_d^\dagger Y_d + \accentset{(n)}{Y}_\nu^\dagger \accentset{(n)}{Y}_\nu + Y_e^\dagger Y_e \right] -3 g_2^2 + \lambda_H\\[5mm]
\accentset{(n)}{\bar\alpha}_{Y_\nu^TM_R^{-1}Y_\nu} &=& 2\Tr\left[ 3Y_u^\dagger Y_u + 3Y_d^\dagger Y_d + \accentset{(n)}{Y}_\nu^\dagger \accentset{(n)}{Y}_\nu + Y_e^\dagger Y_e \right]-\dfrac{9}{10} g_1^2 - \dfrac{9}{2} g_2^2\;,
\ea
\eeq
with $\lambda_H$ the Higgs self-coupling.\footnote{We use the convention that the Higgs self-interaction term in the
Lagrangian is $-\lambda_H (H^\dagger H)^2/4$.}

In MSSM the running of $\accentset{(n)}{\kappa}$ and of $\accentset{(n)}{Y}_\nu^T\accentset{(n)}{M_R}^{-1}\accentset{(n)}{Y}_\nu$ is the same and therefore we can write
\beq
16 \pi^2 \,
\frac{\D m_\nu} {\D t}\,=\,\left(Y_e^\dagger Y_e +\accentset{(n)}{Y}_\nu^\dagger\accentset{(n)}{Y}_\nu\right)^T m_\nu + m_\nu \left( Y_e^\dagger Y_e +  \accentset{(n)}{Y}_\nu^\dagger \accentset{(n)}{Y}_\nu\right) +  \accentset{(n)}{\bar\alpha}m_\nu \; ,
\label{LMP:betanu2}
\eeq
where
\beq
\accentset{(n)}{\bar\alpha} = 2\Tr\left[ 3Y_u^\dagger Y_u + \accentset{(n)}{Y}_\nu^\dagger \accentset{(n)}{Y}_\nu \right] -\dfrac{6}{5} g_1^2 - 6 g_2^2\;.
\eeq

\item[$\mathbf{M_S\longrightarrow\la}$.] For energy range below the mass scale of the lightest right-handed neutrino, all the $\nu^c_n$ are integrated out and $\accentset{(S)}{Y}_\nu$ and $\accentset{(S)}{M}_R$ vanish. In the right-hand side of eq. (\ref{LMP:effnumass2}) only the term $\accentset{(S)}{\kappa}$ is not vanishing and in this case the effective mass matrix $m_\nu$ evolves as:
\begin{equation}
16\pi^2 \, \dfrac{\D m_\nu}{\D t} = \Big(C_eY_e^\dagger Y_e\Big)^T \, m_\nu + m_\nu \, \Big(C_eY_e^\dagger Y_e\Big) + \accentset{(S)}{\bar\alpha} \, m_\nu
\label{LMP:betanu3}
\end{equation}
with
\begin{equation}
\ba{ccl}
\accentset{(S)}{\bar\alpha}_{SM}&=& 2\Tr\left[ 3Y_u^\dagger Y_u + 3Y_d^\dagger Y_d + Y_e^\dagger Y_e \right] -3 g_2^2 + \lambda_H\\[5mm]
\accentset{(S)}{\bar\alpha}_{MSSM}&=& 6\Tr\Big[Y_u^\dagger Y_u \Big] - \dfrac{6}{5} g_1^2 - 6 g_2^2\;.
\ea
\end{equation}
\end{description}

\mathversion{bold}
\subsection{Analytical Approximation to the Running Evolution of $m_\nu$}
\label{Sec:LMP:RunningApprox}
\setcounter{footnote}{3}
\mathversion{normal}

Now we analytically solve the RGEs for $m_\nu$ in the leading Log approximation. All the Yukawa couplings $Y_i^\dagger Y_i$ for $i=\nu,e,u,d$ are evaluated at their initial value at the cutoff $\Lambda$. Furthermore we will keep only the leading contributions from each $Y_i^\dagger Y_i$ term, for $i=e,u,d$, i.e. $|y_\tau|^2$, $|y_t|^2$ and $|y_b|^2$ respectively. The corrections to the leading order $Y_i^\dag Y_i$ come from their running evolution  as well as from their subleading terms and they contribute to the final result as subleading effects and we can safely neglect them in our analytical estimate.

In the MSSM context, the general solution to eqs. (\ref{LMP:betanu1}, \ref{LMP:betanu2}, \ref{LMP:betanu3}) have all the same structure, which is approximately given by
\beq
m_{\nu\,\text{(lower Energy)}} \approx I_U J_e^T J_\nu^Tm_{\nu\,\text{(higher Energy)}} J_\nu J_e
\label{LMP:generalsol}
\eeq
where $I_U$, $J_e$ and $J_\nu$ are all exponentials of integrals containing loop suppressing factors and as a result they are close to $\unity$. Note that $I_U$ is a universal contribution defined as
\beq
I_U=\exp\left[-\dfrac{1} {16\pi^2}\int\accentset{(n)}{\bar\alpha}~\D t \right]
\label{LMP:I}
\eeq
where the integral runs between two subsequent energy scales and we have extended the definition of $\accentset{(n)}{\bar\alpha}$ by identifying $\accentset{(\Lambda)}{\bar\alpha} \equiv  \bar\alpha$ in order to include the range from $\Lambda$ down to $M_L$. $J_e$ is the contribution from charged lepton Yukawa couplings which is always flavour-dependent, while $J_\nu$ is the contribution from the neutrino Yukawa coupling: they are given by 
\beq
J_{e}=\exp\left[-\dfrac{1}{16\pi^2}\int Y_e^\dagger Y_e\,\D t \right]\;,\qquad
J_{\nu}=\exp\left[-\dfrac{1}{16\pi^2}\int\accentset{(n)}{Y}_{\nu}^\dagger \accentset{(n)}{Y}_{\nu}\,\D t \right]\;,
\label{LMP:EqJeGeneral}
\eeq
where also here we have extended the definition of $\accentset{(n)}{Y}_{\nu}$ by identifying $\accentset{(\La)}{Y}_{\nu}$ with $Y_\nu$ in order to include the range between $\Lambda$ and $M_L$. \footnote{In eq. (\ref{LMP:EqJeGeneral}), the combination $Y_e^\dagger Y_e$ should enter with $\accentset{(n)}{Y}_e$ instead of $Y_e$, as one can see from the RGEs in appendix \ref{AppendixC}. In our approximation, however, they coincide.}
Differently from $J_e$, $J_\nu$ can be flavour-dependent or not.

In the Standard Model context, the running effects do not factorise, due to the different evolution of $\accentset{(n)}{\kappa}$ and $\accentset{(n)}{Y}_\nu^T \accentset{(n)}{M}_R^{-1} \accentset{(n)}{Y}_\nu$ between the See-Saw mass thresholds. However eq. (\ref{LMP:generalsol}) applies also to the Standard Model context when $m_\nu$ is a result of a flavour symmetry: in this case, by a suitable redefinition of the parameters which define te mass eigenvalues, the sum $\accentset{(n)}{\kappa}+\accentset{(n)}{Y}_\nu^T \accentset{(n)}{M}_R^{-1} \accentset{(n)}{Y}_\nu$ after the running evolution has exactly the same flavour structure of $m_{\nu\,\text{(higher Energy)}}$. For the purposes of the present discussion we simply assume that eq. (\ref{LMP:generalsol}) is valid also in the Standard Model context and an explicit example will be proposed in section \ref{Sec:LMP:RGcoefficients}.

Expanding $J_e$ and $J_\nu$ in Taylor series and summing up eq. (\ref{LMP:generalsol}) on several energy ranges one can approximately calculate the neutrino mass at low-energy as
\beq
m_{\nu(\varrho)}\simeq I_U\left(m_{\nu(\Lambda)}+
\Delta m^{(J_e)}_\nu + \Delta m^{(J_\nu)}_\nu\right)\;,
\label{LMP:generalsol2}
\eeq
where the low-energy scale $\varrho$ is $m_Z$ in the case of Standard Model and $m_{\rm SUSY}$ for MSSM. The explicit form of the universal part $I_U$ is given by:
\beq
\begin{split}
\hspace{-7mm}
I_U^{\rm SM}\;=&\;\unity\;\times\;\exp\Bigg[-\dfrac{1}{16\pi^2}\Bigg[\left(-\dfrac{9}{10}g_1^2-\dfrac{9}{2}g_2^2+ 6|y_t|^2\right)\ln\dfrac{\La_f}{m_Z}+ \left(\dfrac{9}{10}g_1^2+\dfrac{3}{2}g_2^2+\lambda_H\right)\ln\dfrac{M_S}{m_Z}+\\[3mm]
&\hspace{3cm}+y^2\left(2\ln\dfrac{M_M}{M_S}+ 4\ln\dfrac{M_L}{M_M}+7\ln\dfrac{\La_f}{M_L}\right)\Bigg]\Bigg]\;,
\end{split}
\eeq
\beq
\begin{split}
\hspace{-7mm}
I_U^{\rm MSSM}\;=&\;\unity\;\times\;\exp\Bigg[-\dfrac{1}{16\pi^2}\Bigg[\left(-\dfrac{6}{5}g_1^2-6g_2^2+ 6|y_t|^2\right)\ln\dfrac{\La_f}{m_{SUSY}}+\\[3mm]
&\hspace{3cm}+y^2\left(2\ln\dfrac{M_M}{M_S}+ 4\ln\dfrac{M_L}{M_M}+8\ln\dfrac{\La_f}{M_L}\right)\Bigg]\Bigg]\;.
\end{split}
\eeq
$\Delta m^{(J_e)}_\nu$ is the the contribution from $J_e$ and can easily be calculated as:
\beq
\Delta m^{(J_e)}_\nu = m_{\nu (\Lambda)} ~ \diag (0, 0, \Delta_\tau)+\diag (0, 0, \Delta_\tau) m_{\nu (\Lambda)}
\label{LMP:Je}
\eeq
where the small parameter $\Delta_\tau$ is given by
\beq
\ba{ccll}
\Delta_\tau&\equiv&-\dfrac{3m^2_\tau}{16\pi^2 v^2}\ln\dfrac{\Lambda}{m_Z}&\quad\text{in the SM}\\[3mm]
\Delta_\tau&\equiv&\dfrac{m^2_\tau}{8\pi^2 v^2} (1+ \tan^2 \beta) \ln\dfrac{\Lambda}{m_{SUSY}}&\quad\text{in the MSSM}\;
\ea
\label{LMP:DeltaTau}
\eeq
with $\tan\beta$ the usual ratio between the VEVs of the neutral spin zero components of $H_u$ and $H_d$, the two doublets responsible for electroweak symmetry breaking in the MSSM. On the other hand, the contribution from $J_\nu$, $\Delta m^{(J_\nu)}_\nu$, non trivially depends on the neutrino Yukawa coupling $Y_\nu$ which cannot be determined by low-energy observables without additional ingredients. In section \ref{Sec:LMP:FlavourSym_RGE}, we will analyse strong impacts of the flavour symmetries on $J_\nu$, but before proceeding, we comment on the hierarchy among the various running contributions to the neutrino mass. Indeed, assuming that the flavour symmetries have no effects on $Y_\nu$, we expect that  
\beq
Y_\nu^\dagger Y_\nu \sim \accentset{(n)}{Y}_\nu^\dagger\accentset{(n)}{Y}_\nu = \mathcal{O}(\unity)
\eeq
and therefore we conclude that the contribution from $J_\nu$ always dominates. In \cite{LMP_RGE} we explicitly show that this conclusion holds both in the Standard Model and in the MSSM even for large $\tan\beta$ (we consider $\tan\beta=60$ as the maximal value). One should expect that a similar observation holds also for the lepton mixing angles, but quite frequently flavour symmetries imply a $J_\nu$ which is flavour-independent or has no effects on mixing angles, as we will see in a moment.

\section{Flavour Symmetries and Running Effects}
\label{Sec:LMP:FlavourSym_RGE}
\setcounter{footnote}{3}

In the present section, we will apply the general results of the running evolution of the neutrino mass operator $m_\nu$ to models beyond the Standard Model, where a flavour symmetry is added to the gauge group of the Standard Model. The main task is to track some interesting connections between the running effects and the realisation of the flavour symmetry.

In a given basis, $Y_e^\dagger Y_e$ and $m_\nu$ can be diagonalised by unitary matrices, $U_e$ and $U_\nu$, respectively. The lepton mixing matrix is given by $U = U_e^\dagger U_\nu$. In a flavour model, the charged lepton Yukawa, the neutrino mass matrix and therefore the PMNS matrix are dictated by the flavour symmetry $G_f$. We have already discussed in section \ref{Sec:FS:SSB} that $G_f$ must be spontaneously broken in order to naturally describe fermion masses and mixings: here, we simply assume that $G_f$ is spontaneously broken by a set of flavon fields $\Phi$ at a very high scale. Suppose that, at the leading order, the neutrino mixing matrix is given by $U_0$ which differs from $U$ by subleading contributions $\sim \langle \Phi \rangle / \Lambda_f$ where $\Lambda_f$ is the cutoff scale of the flavour symmetry $G_f$. We will begin with some general assumptions on $U_0$ without however specifying its form. Then we will move to specialise in a concrete case in which $U_0$ is given by the tribimaximal mixing pattern. Similar studies can be done considering other mass-independent textures, such as the bimaximal, the golden-ratio and (sometimes) the trimaximal schemes.

\subsection{Running Effects on Neutrino Mixing Patterns}
\label{Sec:LMP:RunningPatterns}
\setcounter{footnote}{3}

As described in section (\ref{Sec:LMP:Running}) the relevant running effects on $m_\nu$ are encoded in the combinations $Y^\dagger_e Y_e$ and $Y^\dagger_\nu Y_\nu$. Furthermore, we observe that a relevant contribution to the running of $Y^\dagger_e Y_e$ is encoded by $Y^\dagger_\nu Y_\nu$.

We perform the analysis in the basis in which the charged leptons are diagonal, then at high energy we have
\beq
Y^\dagger_e Y_e = \diag(m^2_e, m^2_\mu, m^2_\tau)\dfrac{2}{v^2}\;.
\eeq
From now on, we will use $v$ in the notation of the Standard Model and in order to convert similar expressions to the MSSM, it is sufficient to substitute $v$ with $v_{u,d}$, when dealing with neutrinos or charged leptons, respectively. This simple form changes when evolving down to low energies. This running effect of  $Y^\dagger_e Y_e$ on $m_\nu$ is of second order and we can safely forget it. However it can generate a non trivial $U_e$ and consequently introduces additional corrections to the PMNS matrix $U$. We will return to this effect in section \ref{Sec:LMP:Chargedsector}.

Since flavour symmetries impose constraints on $Y_\nu$, they should have some impacts also on running effects. In this section we are interested in two classes of constraints. The first class is characterised by $Y_\nu$ proportional to a unitary matrix: $Y_\nu^\dagger Y_\nu \sim \unity$ or $Y_\nu Y_\nu ^\dagger \sim \unity$ is frequent in the presence of a flavour symmetry, since it is, for example, a consequence of the first Schur's lemma when $\ell$ or $\nu^c$ transforms in a irreducible representation of the group $G_f$ \cite{BBFN_Lepto}. In the second class, we assume that $m_\nu$ can be exactly diagonalised by $U_0$ according to
\beq
\hat{m}_\nu = U^T_0 m_\nu U_0
\label{LMP:Diag_nu}
\eeq
where $\hat{m}_\nu = \diag (m_1, m_2, m_3)$ with $m_i$ positive and $U_0$ is a mass-independent mixing pattern enforced by the flavour symmetry $G_f$. Independently from the way in which $G_f$ is broken, it is straightforward to see that the neutrino Yukawa coupling in the basis of diagonal right-handed Majorana neutrinos, which we indicate as $\hat{Y}_\nu$, has the following simple form
\beq
\hat{Y}_\nu = i D\,U^\dagger_0
\label{LMP:General_hatY}
\eeq
where $D=\diag (\pm \sqrt{2m_1M_1},\pm \sqrt{2m_2M_2},\pm \sqrt{2m_3M_3})/v$. Notice that $\hat{Y}_\nu$ becomes unitary if $D\sim\unity$. However, the present case is not strictly a generalisation of the previous one since a unitary $Y_\nu$ does not necessarily imply a mass-independent mixing pattern.

In \cite{LMP_RGE} we show that $m_\nu$ does not change its flavour structure under $J_\nu$ if $Y_\nu$ belongs to one of these classes: the running effects from $J_\nu$ correct only the neutrino mass eigenvalues but not the mixing angles. Therefore, the only flavour-dependent running contribution to $m_\nu$ is encoded in $J_e$.

\mathversion{bold}
\subsubsection{A Special Case $U_0 = i U_{\rm TB} P^*$ and $D \propto \mathrm{\bf diag} (1,1,-1)$}
\setcounter{footnote}{3}
\mathversion{normal}

In this part we consider a special case of $\hat{Y}_\nu =i D\, U^\dagger_0$ in which the expression of $U_0$ is enforced by the flavour symmetry group $A_4$ in the context of the Altarelli-Feruglio model described in section \ref{Sec:AFTBM}. A more detailed analysis of the running effects will be discussed in the next section. Here we only comment on the constraints on the mixing matrix $U_0 = i U_{\rm TB} P^*$ and the neutrino Yukawa coupling in the hatted basis:
\beq
\hat{Y}_\nu\equiv yPU_{TB}^TO_{23}=y P\left(\begin{array}{ccc}
                                        \sqrt{2/3}& -1/\sqrt{6} & -1/\sqrt{6}\\
                                        1/\sqrt{3} & +1/\sqrt{3}& +1/\sqrt{3} \\
                                        0 & +1/\sqrt{2} & -1/\sqrt{2}\\
                                        \end{array}\right)
\label{LMP:Ynu}
\eeq
where $y$ is a positive parameter of order $\mathcal{O}(1)$, $P$ is the usual diagonal matrix of the Majorana phases and $O_{23}$ is defined as
\beq
O_{23}=\left(
         \begin{array}{ccc}
           1 & 0 & 0 \\
           0 & 0 & 1 \\
           0 & 1 & 0 \\
         \end{array}
       \right)\;.
\eeq
In order to confront eq. (\ref{LMP:Ynu}) with the general expression $\hat{Y}_\nu =i D\, U^\dagger_0$
we observe that
\beq
\hat{Y}_\nu = yPU_{TB}^TO_{23} U_{TB} U_{TB}^T = \diag (y,y,-y) P U^T_{TB}\;.
\eeq
Then we conclude that (\ref{LMP:Ynu}) corresponds to the special case in which $D= \diag (y,y,-y)$. Furthermore, in the Altarelli-Feruglio model considered in this section, there is a very simple relation between $m_i$ and $M_i$ given by $m_i= v_u^2 y^2 /2M_i$.

Now we explicitly calculate the renormalisation group running from $\La_f$ down to $\varrho$ for this special case using the approximate analytical expressions given in section \ref{Sec:LMP:RunningApprox}. In the physical basis, it is useful to define the light neutrino mass matrix eq. (\ref{LMP:EqSee-Saw}) at the initial energy scale $\La_f$: by imposing the condition $m_{\nu(\Lambda)}=U_0^*\hat{m}_\nu U_0^\dag$, we have
\beq
\begin{split}
m_\nu^{TB}&=-U_{TB}\,P\,\hat{m}_\nu\,P\,U_{TB}^T\\[3mm]
&=-\left[\dfrac{\tilde m_3}{2}\left(\begin{array}{ccc}
                        0&0&0\\
                        0&1&-1\\
                        0&-1&1\end{array}\right)
    +\dfrac{\tilde m_2}{3}\left(\begin{array}{ccc}
                        1&1&1\\
                        1&1&1\\
                        1&1&1\end{array}\right)
    +\dfrac{\tilde m_1}{6}\left(\begin{array}{ccc}
                        4&-2&-2\\
                        -2&1&1\\
                        -2&1&1\end{array}\right)\right]\;,
\end{split}
\label{LMP:EqMnuTBMmasses}
\eeq
where $\tilde m_i=m_ie^{i\al_i}$. It is necessary to specify the kind of neutrino mass spectrum: in the normal hierarchy the light neutrinos are ordered as $m_1<m_2<m_3$ and the heavy ones as $M_3<M_2<M_1$; while in the inverse hierarchy they are arranged as $m_3<m_1\lesssim m_2$ and $M_2\lesssim M_1<M_3$.

The general result of the running effects on $m_\nu$ is given by eq.~(\ref{LMP:generalsol2}) which in our case becomes
\beq
m_{\nu(\varrho)}=I_U\left(m_\nu^{TB}+\Delta m^{(J_e)}_\nu + \Delta m^{(J_\nu)}_\nu \right)\;.
\label{LMP:mTBM}
\eeq
The analytical result for both $I_U$ and $\Delta m^{(J_e)}_\nu$ (see section \ref{Sec:LMP:Running}) does not depend on the type of the neutrino spectrum, it is sufficient to identify $M_S,M_M,M_L$ with the correct hierarchy between $M_1, M_2, M_3$\;. In particular, for the tribimaximal mixing pattern, the contribution from $J_e$ is given by
\beq
\begin{split}
\Delta m^{(J_e)}_\nu &=m_\nu^{TB} ~ \diag (0,\, 0,\, \Delta_\tau)+ \diag (0,\, 0,\, \Delta_\tau) m_\nu^{TB}\\
&=-\left(\begin{array}{ccc}
     0 & 0 & \dfrac{\tilde m_1}{3}-\dfrac{\tilde m_2}{3} \\[3mm]
     0 & 0 & -\dfrac{\tilde m_1}{6}-\dfrac{\tilde m_2}{3}+\dfrac{\tilde m_3}{2}\\[3mm]
     \dfrac{\tilde m_1}{3}-\dfrac{\tilde m_2}{3} & -\dfrac{\tilde m_1}{6}-\dfrac{\tilde m_2}{3}+\dfrac{\tilde m_3}{2} & -\dfrac{\tilde m_1}{3}-\dfrac{2\tilde m_2}{3}-\tilde m_3 \\
         \end{array}
    \right)\Delta_\tau\;.
\end{split}
\eeq
Naturally, the contribution from $J_\nu$ depends on the type of the neutrino spectrum,
however it can be written in the same form for both the spectra:
\beq
\Delta m^{(J_\nu)}_\nu=-\left[\dfrac{\tilde m'_1}{6}\left(
                                                  \begin{array}{ccc}
                                                    4 & -2 & -2 \\
                                                    -2 & 1 & 1 \\
                                                    -2 & 1 & 1 \\
                                                  \end{array}
                                                    \right)
                                +\dfrac{2 \tilde m'_2}{3}\left(
                                                  \begin{array}{ccc}
                                                    1 & 1 & 1 \\
                                                    1 & 1 & 1 \\
                                                    1 & 1 & 1 \\
                                                  \end{array}
                                                    \right)
                                +\tilde m'_3\left(
                                                  \begin{array}{ccc}
                                                    0 & 0 & 0 \\
                                                    0 & 1 & -1 \\
                                                    0 & -1 & 1 \\
                                                  \end{array}
                                                \right)\right]
                                                \label{LMP:Deltam_Jnu}
\eeq
where $\tilde m'_i$ are redefinitions of the light neutrino masses:
\begin{description}
\item[\textbf{Normal Hierarchy:}]
\beq
\ba{llll}
\tilde m'_1=\tilde m_1(p+q)\;,& \tilde m'_2=\tilde m_2(x+q)\;,& \tilde m'_3=\tilde m_3(x+z)&\text{in the SM}\\[3mm]
\tilde m'_1=0\;,& \tilde m'_2=2 \tilde m_2x\;,& \tilde m'_3=2 \tilde m_3(x+z)&\text{in the MSSM}
\ea
\label{LMP:mpNH}
\eeq
with
\beq
\ba{rcl}
p&=&-\dfrac{1}{16\pi^2}(-3g_2^2+\la+\dfrac{9}{10}g_1^2+\dfrac{9}{2}g_2^2)\ln\dfrac{M_1}{M_2}\\[3mm]
q&=&-\dfrac{1}{16\pi^2}(-3g_2^2+\la+\dfrac{9}{10}g_1^2+\dfrac{9}{2}g_2^2)\ln\dfrac{M_2}{M_3}\\[3mm]
x&=&-\dfrac{y^2}{32\pi^2}\ln\dfrac{M_1}{M_2}\\[3mm]
z&=&-\dfrac{y^2}{32\pi^2}\ln\dfrac{M_2}{M_3}\;;
\ea
\eeq
\item[\textbf{Inverse Hierarchy:}]
\beq
\ba{llll}
\tilde m'_1=\tilde m_1(x+q)\;,& \tilde m'_2=\tilde m_2 (x+z)\;,& \tilde m'_3=\tilde m_3(p+q)&\text{in the SM}\\[3mm]
\tilde m'_1=2\tilde m_1 x\;,& \tilde m'_2= 2\tilde m_2 (x+z)\;,& \tilde m'_3=0&\text{in the MSSM}
\ea
\label{LMP:mpIH}
\eeq
with
\beq
\ba{rcl}
p&=&-\dfrac{1}{16\pi^2}(-3g_2^2+\la+\dfrac{9}{10}g_1^2+\dfrac{9}{2}g_2^2)\ln\dfrac{M_3}{M_1}\\[3mm]
q&=&-\dfrac{1}{16\pi^2}(-3g_2^2+\la+\dfrac{9}{10}g_1^2+\dfrac{9}{2}g_2^2)\ln\dfrac{M_1}{M_2}\\[3mm]
x&=&-\dfrac{y^2}{32\pi^2}\ln\dfrac{M_3}{M_1}\\[3mm]
z&=&-\dfrac{y^2}{32\pi^2}\ln\dfrac{M_1}{M_2}\;.
\ea
\eeq
\end{description}

Comparing $m_\nu^{TB}$ of eq. (\ref{LMP:EqMnuTBMmasses}) with the perturbations $\Delta m_\nu$ of eqs. (\ref{LMP:Deltam_Jnu}), we note the presence of the same flavour structure for several matrices and in particular, by redefining $\tilde m_i$ to absorb the terms $\tilde m'_i$ it is possible to account for the See-Saw contributions from the renormalisation group running into $m_\nu^{TB}$. As a consequence the leading order predictions for the tribimaximal angles receive corrections only from the terms proportional to $\Delta_\tau$. This result explicitly confirms what we outlined in the previous section.

\subsection{Running Effects in the Charged Lepton Sector}
\label{Sec:LMP:Chargedsector}
\setcounter{footnote}{3}

The presence of a term proportional to $\hat{Y}^\dagger_\nu \hat{Y}_\nu$ in the RG equation for $Y_e$ can switch on off-diagonal entries in the charged lepton Yukawa matrix $Y_e$. When rotated away, this additional contribution introduces a non-trivial $U_e$ and consequently corrects the lepton mixing matrix $U$. For a unitary $\hat Y_\nu$, this correction appears only between the See-Saw mass scales while in the general case it appears already from the cutoff $\Lambda_f$.

In close analogy with the running effects on neutrino mass matrix in eq. (\ref{LMP:mTBM}), the full result of the running for charged lepton mass matrix can conventionally
be written as
\beq
(Y^\dagger_e Y_e)_ {(\varrho)}=I_e\left[ (Y^\dagger_e Y_e)_{(\Lambda_f)}
+\Delta (Y^\dagger_e Y_e) \right]\;,
\label{LMP:CorrectedYeYe}
\eeq
where $I_e$ is an irrelevant global coefficient which can be absorbed by, for example, $y_\tau$. Now we move to the case of tribimaximal mixing pattern. In this case, the flavour-dependent corrections can be explicitly calculated:
\begin{description}
\item[\textbf{NH case:}]
\beq
\Delta (Y^\dagger_e Y_e)\simeq y_\tau^2 \left[ a_e \left(
                                                  \begin{array}{ccc}
                                                    0 & 0 & 1 \\
                                                    0 & 0 & -1/2 \\
                                                    1 & -1/2 & 5 \\
                                                  \end{array}
                                                    \right)
                                		+b_e \left(
                                                  \begin{array}{ccc}
                                                    0 & 0 & 0 \\
                                                    0 & 0 & -1 \\
                                                    0 & -1 & 2 \\
                                                  \end{array}
                                                    \right)
                                		+c_e \left(
                                                  \begin{array}{ccc}
                                                    0 & 0 & 0 \\
                                                    0 & 0 & 0 \\
                                                    0 & 0 & 2 \\
                                                  \end{array}
						    \right)\right]\;,	
\label{LMP:EqDeltaYeDagYeNH}
\eeq
\item[\textbf{IH case:}]
\beq
\Delta (Y^\dagger_e Y_e)\simeq y_\tau^2 \left[a'_e  \left(
                                                  \begin{array}{ccc}
                                                    0 & 0 & 0 \\
                                                    0 & 0 & 1 \\
                                                    0 & 1 & 2 \\
                                                  \end{array}
                                                    \right)
                                		+b'_e \left(
                                                  \begin{array}{ccc}
                                                    0 & 0 & 1\\
                                                    0 & 0 & 1 \\
                                                    1 & 1 & 2 \\
                                                  \end{array}
                                                    \right)
                                		+c'_e \left(
                                                  \begin{array}{ccc}
                                                    0 & 0 & 0 \\
                                                    0 & 0 & 0 \\
                                                    0 & 0 & 2 \\
                                                  \end{array}
                                                \right)\right]\;,
\label{LMP:EqDeltaYeDagYeIH}
\eeq
\end{description}
where the coefficients are
\beq
\ba{lr}
a_e=b'_e=-\dfrac{C'_\nu}{16 \pi^2} \dfrac{y^2}{3} \ln\dfrac{M_1}{M_2},&\qquad b_e=-\dfrac{C'_\nu}{16 \pi^2} \dfrac{y^2}{2} \ln\dfrac{M_2}{M_3}\;, \\[5mm]
c_e=c'_e=-\dfrac{3 C'_e y_\tau^2}{16 \pi^2}\ln\dfrac{\Lambda_f}{m_{SUSY}(m_Z)}\;,&\qquad a'_e= -\dfrac{C'_\nu}{16 \pi^2} \dfrac{y^2}{2} \ln\dfrac{M_3}{M_1}\;,
\label{LMP:coeff}
\ea
\eeq
and $C'_\nu=-3/2 \;(1)$, $C'_e=3/2 \;(3)$  in the Standard Model (MSSM). Here we observe that the off-diagonal contributions to $Y^\dagger_e Y_e$ are encoded in $a_e$, $b_e$, $a'_e$ and $b'_e$ which depend only on the See-Saw scales $M_i$. As a result, as we will show in the
next section, $c_e$ and $c'_e$ do not affect the lepton mixing angles.

\subsection{Full Running Effects on the Tribimaximal Mixing Pattern}
\label{Sec:LMP:RGcoefficients}
\setcounter{footnote}{3}

In this section, we combine various contributions discussed in previous sections into the observable matrix $U$ from which we extract angles and phases at low-energy. Since we are interested in physical quantities, we eliminate one of the phases of $P$ and in particular we express each result as a function of $\alpha_{ij}\equiv(\alpha_i-\alpha_j)/2$, removing $\alpha_3$. The corrected mixing angles can be written as
\beq
\theta_{ij(\varrho)} = \theta^{TB}_{ij}+k_{ij}+\ldots
\eeq
where $\theta^{TB}_{13} = 0$,  $\theta^{TB}_{12} = \arcsin \sqrt{1/3}$, $\theta^{TB}_{23} = -\pi/4$, dots stand for subleading corrections and $k_{ij}$ are defined by
\bea
k_{12}&=&\dfrac{1}{3\sqrt2}\left(\dfrac{|\tilde m_1+\tilde m_2|^2}{m_2^2-m_1^2} \Delta_\tau - 3 a_e\right)\nn\\[3mm]
k_{23}&=&\left\{
          \begin{array}{ll}
            \dfrac{1}{6}\left[\left(\dfrac{|\tilde m_1+\tilde m_3|^2}{m_3^2-m_1^2}+2\dfrac{|\tilde m_2+\tilde m_3|^2}{m_3^2-m_2^2}\right) \Delta_\tau -3 a_e -6 b_e \right] &\qquad\hbox{for NH} \\[3mm]
            \dfrac{1}{6}\left[\left(\dfrac{|\tilde m_1+\tilde m_3|^2}{m_3^2-m_1^2}+2\dfrac{|\tilde m_2+\tilde m_3|^2}{m_3^2-m_2^2}\right) \Delta_\tau +3 a_e +3 a'_e \right] &\qquad \hbox{for IH}
          \end{array}
        \right.
\eea
\beq
k_{13}=\dfrac{1}{3\sqrt2}\sqrt{4m_3^2 \Delta_\tau^2\left(\dfrac{m_1\sin\alpha_{13}}{m_1^2-m_3^2}-\dfrac{m_2\sin\alpha_{23}}{m_2^2-m_3^2}\right)^2+ \left[\left(\dfrac{|\tilde m_1+\tilde m_3|^2}{m_1^2-m_3^2}-\dfrac{|\tilde m_2+\tilde m_3|^2}{\tilde m_2^2-\tilde m_3^2}\right)\Delta_\tau-3 a_e\right]^2}\;.\nn
\eeq
In the previous expressions we can clearly distinguish the contributions coming from the diagonalisation of the corrected tribimaximal neutrino mass matrix (\ref{LMP:mTBM}) and those from the diagonalisation of (\ref{LMP:CorrectedYeYe}). As it is clear from (\ref{LMP:coeff}), the corrections to the tribimaximal mixing from the charged lepton sector is important only for hierarchical right-handed neutrinos and will approach to zero as soon as the spectrum becomes degenerate. On the other hand, the corrections from the neutrino sector should be enhanced if the light neutrinos are quasi-degenerate and if the $\tan \beta$ is large, in the MSSM case.\\

The physical Majorana phases are also corrected due to the running and we found the following results:
\beq
\alpha_{ij(\varrho)}\simeq\alpha_{ij}+\delta\alpha_{ij}\Delta_\tau+\ldots
\eeq
where $\alpha_{ij}$ are the starting values at $\La_f$ and
\beq
\delta\alpha_{13}=\dfrac{2}{3}\dfrac{m_1m_2\sin(\alpha_{13}-\alpha_{23})}{m_2^2-m_1^2}\;,\qquad\qquad
\delta\alpha_{23}=\dfrac{4}{3}\dfrac{m_1m_2\sin(\alpha_{13}-\alpha_{23})}{m_2^2-m_1^2}\;.
\label{LMP:Deltaalpha} 
\eeq
At $\La_f$, $\sin{\theta^{TB}_{13}}$ is vanishing and as a result the Dirac CP-violating phase is undetermined. An alternative is to study the Jarlskog invariants which are well-defined at each energy scale. At $\La_f$, $J_{CP}$ is vanishing, while after the renormalisation group running it is given by
\beq
J_{CP}=\dfrac{1}{18}\left|m_3\left(\dfrac{m_1\sin\alpha_{13}}{m_1^2-m_3^2}- \dfrac{m_2\sin\alpha_{23}}{m_2^2-m_3^2}\right)\right|\Delta_\tau\;.
\label{LMP:Jarlskog}
\eeq
Two comments are worth. First of all, in the expression for $k_{13}$, it is easy to recover the resulting expression for $J_{CP}$ as the first term under the square root, apart global coefficients. This means that the running procedure introduces a mixing between the expression of the reactor angle and of the Dirac CP-phase. Moreover we can recover the value of the Dirac CP-phase directly from eq. (\ref{LMP:Jarlskog}) and we get the following expression:
\beq
\begin{split}
\cot\delta=&-\dfrac{m_1(m_2^2-m_3^2)\cos\alpha_{13}-m_2(m_1^2-m_3^2) \cos\alpha_{23}-m_3(m_1^2-m_2^2)}{m_1(m_2^2-m_3^2)\sin\alpha_{13}-m_2(m_1^2-m_3^2)\sin\alpha_{23}}+\\[3mm]
&-\dfrac{3 a_e (m_2^2-m_3^2)(m_1^2-m_3^2)}{2 m_3 \left[m_1(m_2^2-m_3^2)\sin\alpha_{13}-m_2(m_1^2-m_3^2)\sin\alpha_{23}\right] \Delta_\tau}  \;.
\end{split}
\eeq
In the neutrino sector, the running contributions from the See-Saw terms are present only in the resulting mass eigenvalues:
\beq
m_{i(\la)}\simeq m_i(1+\delta m_i)+\ldots
\eeq
where $m_i$ are the starting values at $\La_f$ and $\delta m_i$, in both the Standard Model and the MSSM and in both the normally and inversely hierarchical spectra, are given by
\beq
\delta m_1=\dfrac{m'_1}{m_1}-\dfrac{\Delta_\tau}{3}\;,\qquad
\delta m_2=2\dfrac{m'_2}{m_2}-\dfrac{2\Delta_\tau}{3}\;,\qquad
\delta m_3=2\dfrac{m'_3}{m_3}-\Delta_\tau\;,
\eeq
with $m'_i\equiv|\tilde m'_i|$, given as in eqs. (\ref{LMP:mpNH}, \ref{LMP:mpIH}).

\section{Running Effects in the Altarelli-Feruglio Model}
\label{Sec:LMP:AFmodel_RG}
\setcounter{footnote}{3}

In this section we will apply the analysis of renormalisation group running effects on the lepton mixing angles to the Altarelli-Feruglio model, already introduced in section \ref{Sec:AFTBM}. In order to perform such a study, it is important to verify the initial assumptions made in section \ref{Sec:LMP:RGcoefficients}, in particular, we see that eq. (\ref{LMP:Ynu}) exactly corresponds to the one implied by the Altarelli-Feruglio model, when moving to the physical basis (the phase of $y$ can be absorbed in the definition of $P$). On the other side, the presence of flavon fields has a relevant impact on the results of the analysis. In the unbroken phase, flavons are active fields and should modify the RGEs. Since the only source of the $A_4$ breaking is the VEVs of the flavons, any flavour structure is preserved above the corresponding energy scale, whatever interactions are present. In particular, the lagrangian (\ref{AFTBM:LnuSeeSaw}) contains all possible leading order terms, given the group assignments, and its invariance under $A_4$ is maintained moving downward to the scale $\mean{\varphi}$, where significant changes in the flavour structure can appear. From eqs. (\ref{AFTBM:SSMassMatrices}) and (\ref{AFTBM:RHEigenvalues}), we deduce that $\mean{\varphi}\sim M_i$ and as a result in the Altarelli-Feruglio model $\Delta_\tau$ must be proportional to $\ln(\mean{\varphi}/\varrho)$ and not to $\ln(\La_f/\varrho)$.

Furthermore, it is relevant for the subsequent discussion to recall the level of degeneracy of the neutrino masses in the allowed space of parameters. The ratios between the right-handed neutrinos are well defined for the normal hierarchy, $M_1/M_3\sim11$ and $M_2/M_3\sim5$, while in the case of the inverse hierarchy, the ratio $M_1/M_2$ is fixed at $1$ while $M_3/M_2$ varies from about $3$ to $1$, going from the lower bound of $m_3$ up to the KATRIN sensitivity.

We will separately discuss the evolution of angles and phases for both type of hierarchy. In the following, the results will be shown for the Standard Model and for the MSSM with $\tan\beta =15$ in the absence of other explicit indications. Without loss of generality, we choose $y=1$ for our numerical analysis. We also set $\mean{\varphi}=10^{15}$. The spectrum spans the range obtained in (\ref{AFTBM:RangeMasses}).

\subsection{Running of the Angles}
\label{Sec:LMP:RGAngles}
\setcounter{footnote}{3}

\begin{figure}[ht!]
 \centering
\includegraphics[width=7.8cm]{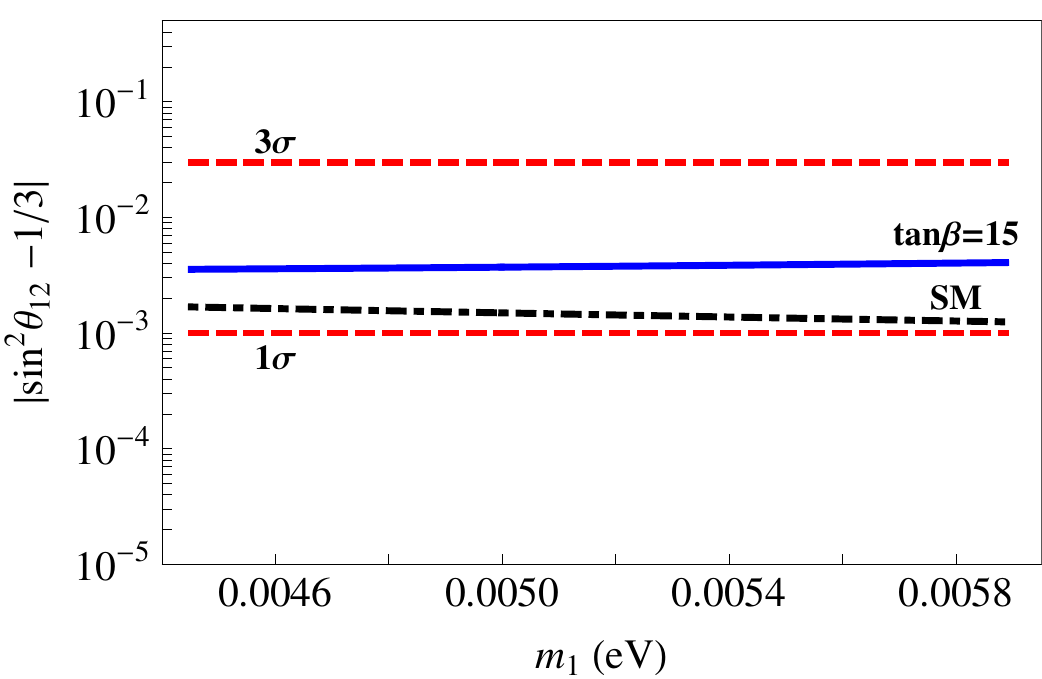}
\includegraphics[width=7.8cm]{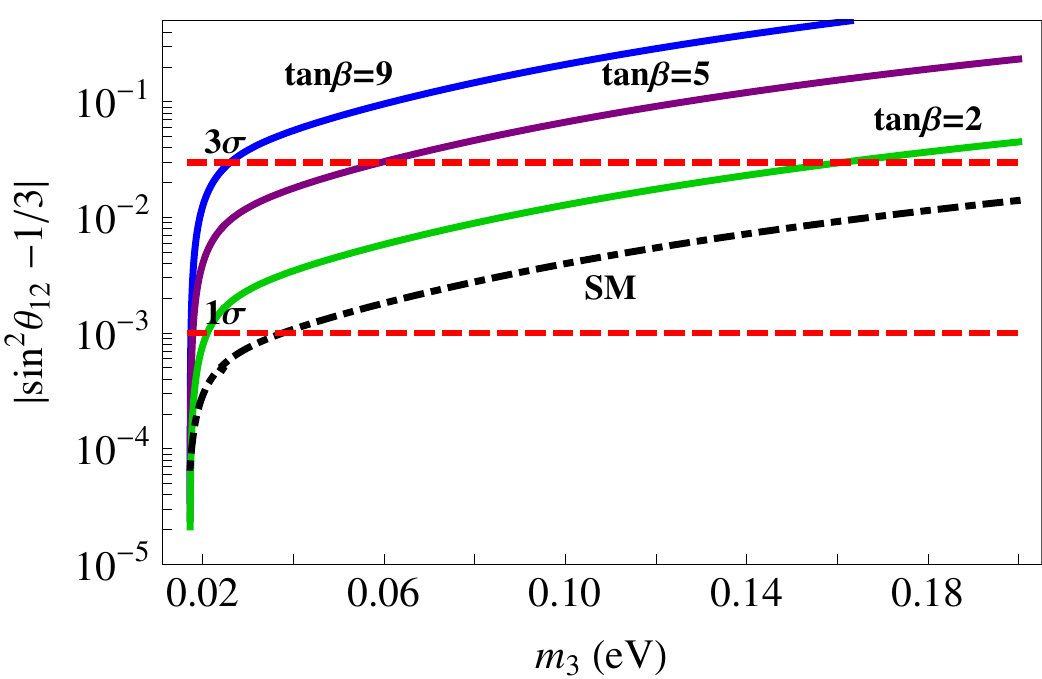}
\includegraphics[width=7.8cm]{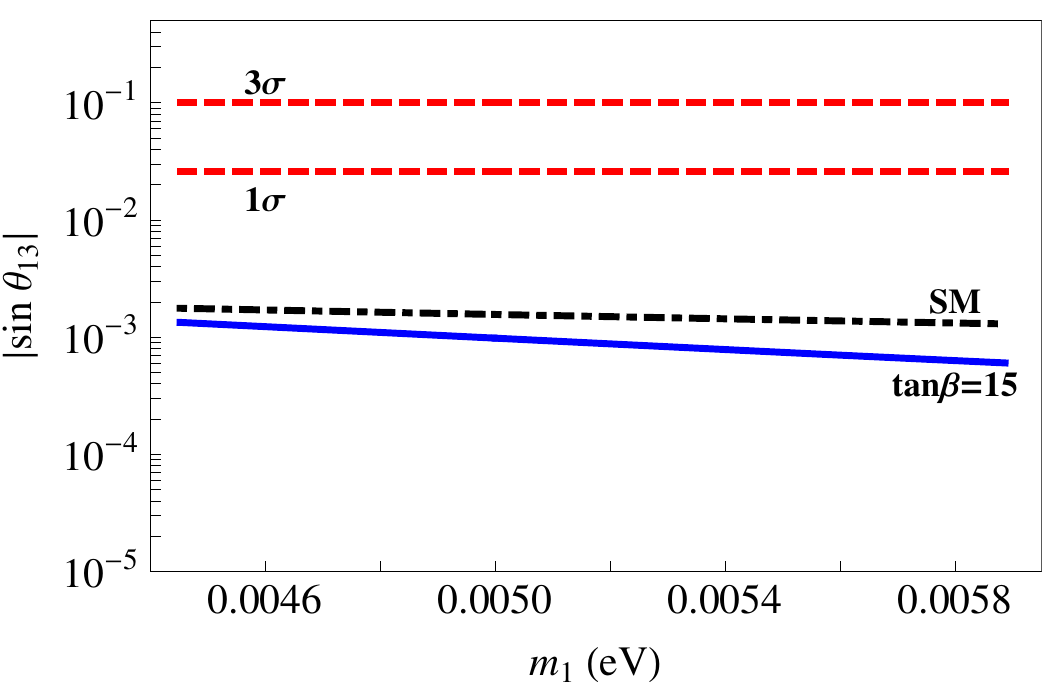}
\includegraphics[width=7.8cm]{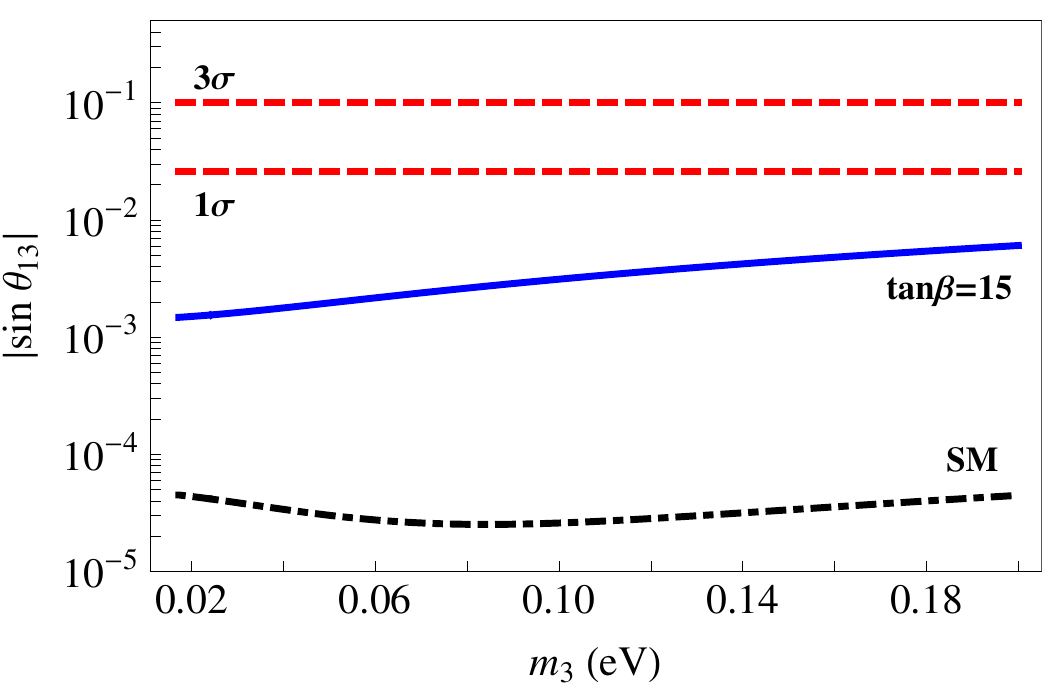}
\includegraphics[width=7.8cm]{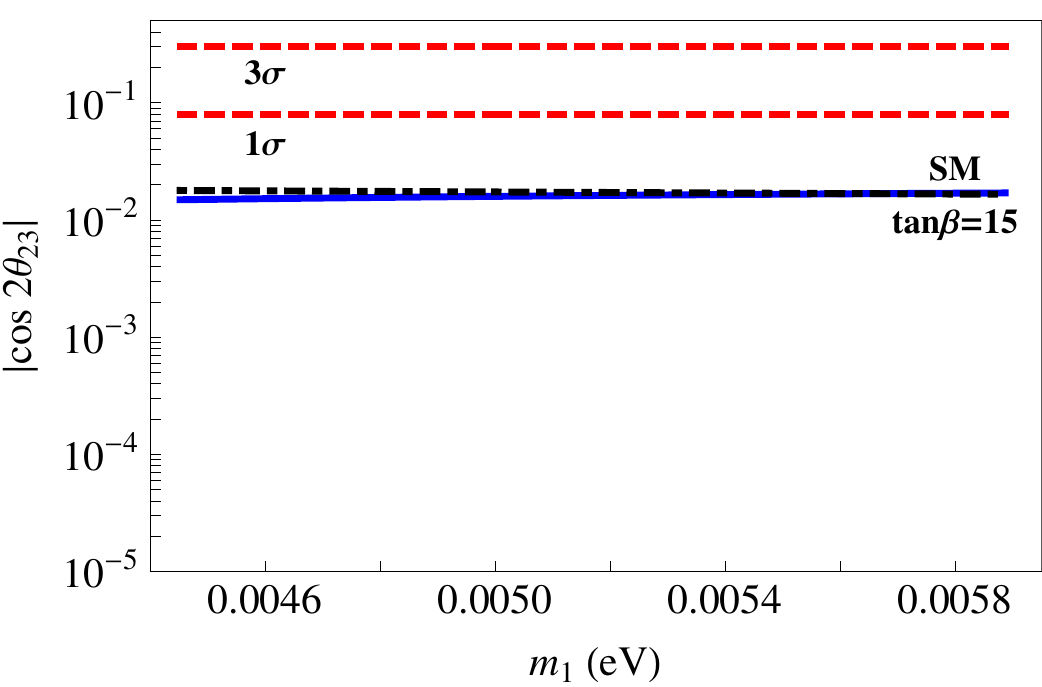}
\includegraphics[width=7.8cm]{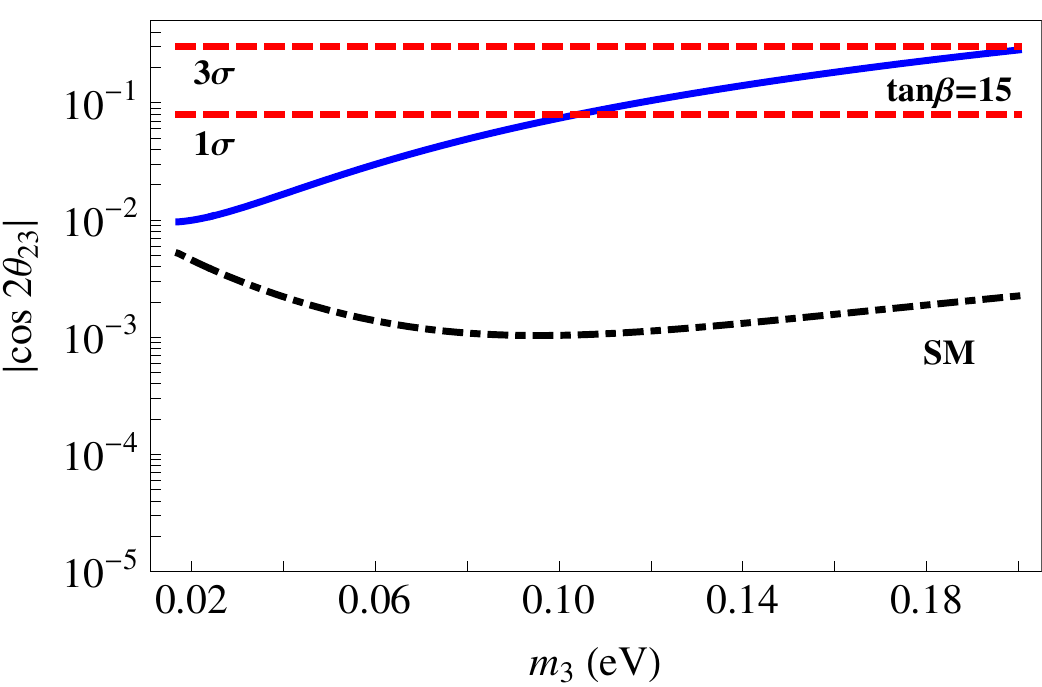}
\vspace{-0.2cm}
\caption{\it Corrections to the tribimaximal mixing angles as functions of the lightest neutrino masses, for the normal hierarchy on the left and for the inverse hierarchy on the right. The plots show the MSSM case with $\tan\beta =15$ (solid blue) and the Standard Model case (black dashed), compared to the current $1\sigma$ and $3\sigma$ limits (dashed red).  $m_{1,3}$ are restricted in a range which is given by eq. (\ref{AFTBM:RangeMasses}) or by the KATRIN bound.}
\vspace{-1cm}
\label{fig:LPM:ANG}
\end{figure}

Since we are interested in deviations of the corrected mixing angles from the tribimaximal predictions and in comparing them with experimental values, it is convenient to relate the coefficients $k_{ij}$ defined in section \ref{Sec:LMP:RGcoefficients} with physical observables. Keeping in mind that $\vert k_{ij}\vert \ll 1$ and that we start from a tribimaximal mixing matrix, it follows that
\beq
\sin\theta_{13}\simeq k_{13}\;,\qquad\cos2 \theta_{23}\simeq 2 k_{23}\;,\qquad \sin^2\theta_{12}-\dfrac{1}{3} \simeq \dfrac{2 \sqrt{2}}{3} k_{12}\;.
\eeq
The corrections to the tribimaximal mixing angles as functions of $m_{1,3}$ in the normal and inverse hierarchies are shown in figure \ref{fig:LPM:ANG}.

We begin with the case of the normal hierarchy. Since the dependence of the corrected mixing angles from $\Delta_\tau$ is the same, Standard Model corrections are generally expected to be smaller than those in MSSM. However, from figure \ref{fig:LPM:ANG} we see that, in normal hierarchy, there is not a large split between the two curves for Standard Model and MSSM. This fact suggests a dominant contribution coming from the charged lepton sector as discussed in section \ref{Sec:LMP:RGcoefficients}. For the atmospheric and reactor angles, the deviations from the tribimaximal predictions lie roughly one order of magnitude below the $1\sigma$ limit. In particular, running effects on $\sin\theta_{13}$ are even smaller than the NLO contributions analysed in section \ref{Sec:AFTBM:NLO} which are of $\mathcal{O}(u)$, without cancellations. On the other hand, since the experimental value of the solar angle is better measured than the other two, the running effects become more important in this case. Indeed, the running corrections to the tribimaximal solar angle evade the $1\sigma$ limit as it can be clearly seen in figure \ref{fig:LPM:ANG}. Anyway, we observe that for both the atmospheric and solar angles, the running contribution is of the same order as the contribution from NLO operators.

Now we move to analyse the case of the inverse hierarchy. In this case, since the neutrino spectrum predicted by the Altarelli-Feruglio model is almost degenerate and in particular $m_2/m_1\sim1$, the contribution from the charged lepton sector in eqs. (\ref{LMP:EqDeltaYeDagYeIH}) is subdominant. As a consequence the information which distinguishes the Standard Model case from the MSSM one is mainly dictated by $\Delta_\tau$ defined in eq. (\ref{LMP:DeltaTau}). As a result the running effects in the MSSM are always larger than in the Standard Model and for large $\tan \beta$ they are potentially dangerous. The curves corresponding to the atmospheric and reactor angles do not go above the $3\sigma$ and $1\sigma$ windows respectively. However, the deviation from $\theta^{TB}_{12}$ presents a more interesting situation. For example, for $\tan\beta \gtrsim 10$, the running effects push the value of the solar angle beyond the $3\sigma$ limit for the entire spectrum. For lower values of $\tan\beta$, the model is within the $3\sigma$ limit only for a (small) part of the spectrum where the neutrinos are less degenerate. Comparing with the running effects, in the inverse hierarchy, the contribution from NLO operators in the Altarelli-Feruglio model is under control.

\subsection{Running of the Phases}
\label{Sec:LMP:RGPhases}
\setcounter{footnote}{3}

Majorana phases are affected by renormalisation group running effects too. Since there is no experimental information on Majorana phases available at this moment we will simply show their values at low-energy, comparing them with the predictions in the Altarelli-Feruglio model. We stress again that they are completely determined by only one parameter, the mass of the lightest neutrino, $m_1$ for the normal hierarchy and $m_3$ for the inverse hierarchy.

\begin{figure}[ht!]
 \centering
\includegraphics[width=7.6cm]{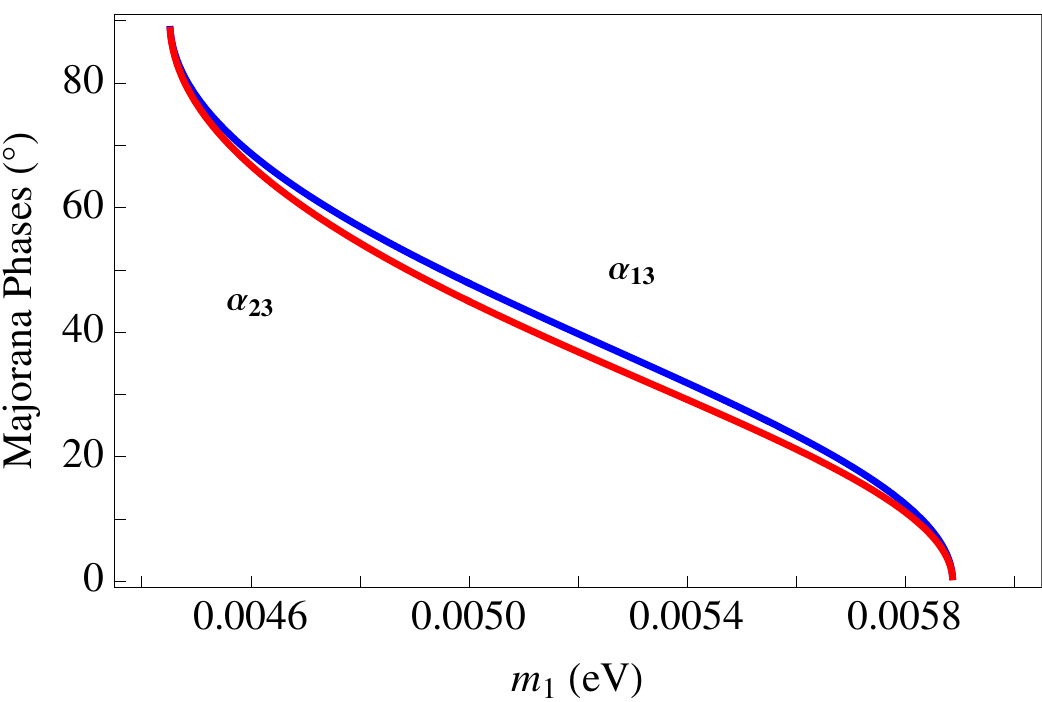}\quad
\includegraphics[width=7.6cm]{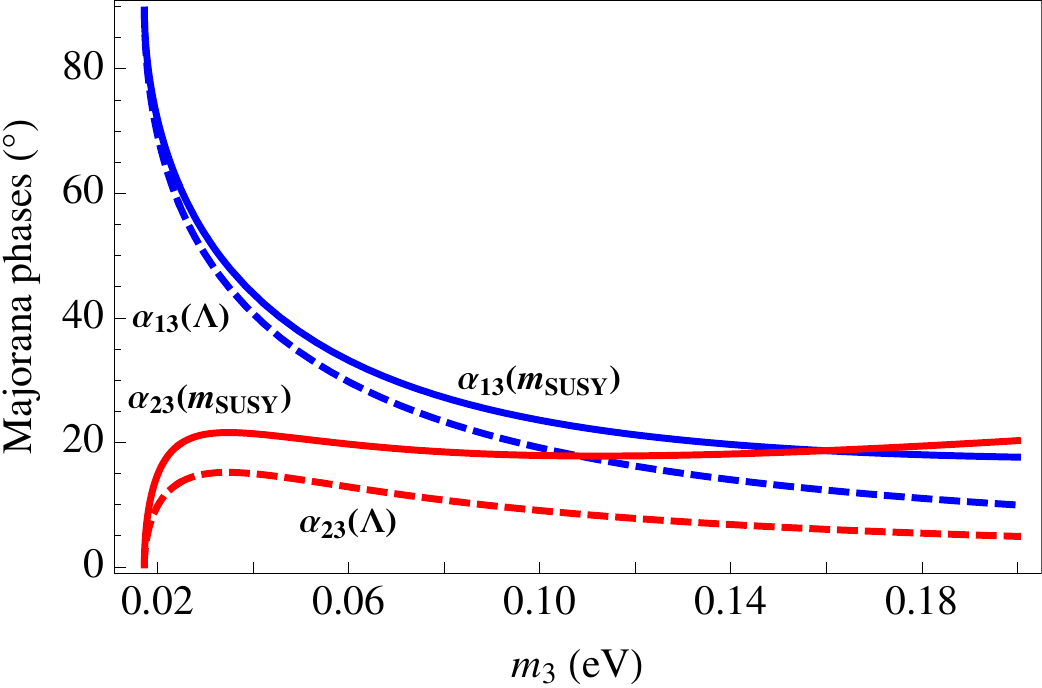}
\caption{\it Majorana phases $\alpha_{13}$ and $\alpha_{23}$ as functions of the lightest left-handed neutrino masses. For the normal hierarchy (left panel) the corresponding curves at low and high energies are undistinguishable. For the inverse hierarchy (right panel) the curves refer to low-energy values in MSSM with $\tan\beta =15$ (solid blue or red) and the Altarelli-Feruglio predictions at $\Lambda_f$ (dashed blue or red).}
 \label{fig:LPM:MAJ}
\end{figure}

In the case of normal hierarchy, Majorana phases are essentially not corrected by running effects. This feature is due to the fact that $\delta \alpha_{13}$ and $\delta \alpha_{23}$ of eqs. (\ref{LMP:Deltaalpha}) are proportional to $\sin(\alpha_{13}-\alpha_{23})$ which is close to zero, as we can see looking at the left panel of figure \ref{fig:LPM:MAJ}. In the case of inverse hierarchy, MSSM running effects always increase the values of phases when moving from high energy to low-energy and they are maximised for $\tan\beta=15$, especially when the neutrino spectrum becomes degenerate. On the contrary, in the Standard Model context, the low-energy curves cannot be distinguished from the high energy ones.

As described in section (\ref{Sec:LMP:RGcoefficients}), a definite Dirac CP violating phase $\delta$ arises from running effects even if, in the presence of a tribimaximal mixing pattern, it is undetermined in the beginning. Although the final Dirac phase can be large, Jarlskog invariant, which measures an observable CP violation, remains small because of the smallness of $\theta_{13}$ (see \cite{LMP_RGE} for details). These results are valid both for the Standard Model and for MSSM.

\section{Conclusions of the Chapter}
\label{Sec:LMP:Conclusions}
\setcounter{footnote}{3}

In this section we have studied the running effects on neutrino mixing patterns. In See-Saw models, the running contribution from the neutrino Yukawa coupling $Y_\nu$, encoded in $J_\nu$, is generally dominant at energies above the See-Saw threshold. However, this effect, which in general introduces appreciable deviations from the leading order mixing patterns, does not affect the mixing angles, under specific conditions: in the first part of the section, we have analysed two classes of models in which this indeed happens.
The first class is characterised by a $Y_\nu$ proportional to a unitary matrix. It is the case, for example, when the right-handed singlet neutrinos or the charged leptons belong to an irreducible representation of the flavour group. The second class is the mass-independent mixing pattern, in which, in particular, the effects of $J_\nu$ can be absorbed by a small shift of neutrino mass eigenvalues leaving mixing angles unchanged.
The widely studied tribimaximal mixing pattern belongs, for example, to this second class of models.

Subsequently, we focused on the Altarelli-Feruglio model. The aim is to analyse the running effects on the tribimaximal mixing pattern in addition to the NLO corrections already present in this model due to the spontaneous breaking of the symmetry and to confront them with experimental values. The analysis has been performed both in the Standard Model and MSSM. We found that for the normal hierarchy light neutrinos, the dominant running contribution comes from the charged lepton sector which weakly depends on both $\tan\beta$ and mass degeneracy. As a result, for this type of spectrum, the tribimaximal prediction is stable under running evolution. Moreover, the running contribution is of the same order or smaller with respect to the contribution from NLO operators. On the other hand, in the case of the inverse hierarchy, the deviation of the solar angle from its tribimaximal value can be larger than the NLO contribution and, in particular in MSSM, for $\tan\beta \gtrsim 10$ an inversely hierarchical spectrum is strongly disfavored. In the end, we observe that for both spectra, the reactor angle $\theta_{13}$ does not receive appreciable deviations from zero.

\clearpage{\pagestyle{empty}\cleardoublepage}

\newpage
\chapter{Flavour Violation}
\label{Sec:FlavourViolation}
\setcounter{equation}{0}
\setcounter{footnote}{3}

In the previous sections we illustrated a series of flavour models in which the fermion masses and mixings can be explained by the introduction of an additional symmetry group. All these models can accommodate the present measured fermion mass hierarchies and manifest particular mixing patterns for leptons and for quarks, which all can fit the experimental data (thanks also to the large number of parameters present in the models). Beyond these features, the study of the neutrinoless-double-beta decay, of the type of hierarchy, of the bounds on the smallest neutrino mass and on the sum of the neutrino masses could represent viable ways in order to confirm or disfavour each of these realisations. However, comparing the phenomenological predictions of the various models, we cannot see a clear distinction. In order to find new ways to characterise each model, it would be highly desirable to investigate other types of observables, not directly related to neutrino properties.

In this chapter we analyse the predictions of a class of flavour models concerning processes which violate the individual lepton number, such as $\mu\to e \gamma$, $\tau\to e\gamma$ and $\tau\to\mu\gamma$. We will focus on a flavour symmetry that contains the discrete group $A_4$, which is particularly successful in reproducing the lepton mixing angles observed in neutrino oscillations. In particular we consider the flavour group $G_f=A_4\times Z_3\times U(1)_{FN}$ and the flavon content which characterise the Alterelli-Feruglio model already discussed in section \ref{Sec:AFTBM}. The total lepton number is generally broken at a large scale $\Lambda_L$ (possibly related to the VEV of the flavons, $\langle\varphi\rangle$) and light neutrinos are assumed to be of Majorana type. Depending on the specific way the total lepton number is violated (either by higher dimensional operators or via the see-saw mechanism), the neutrino mass spectrum can have different properties that can be tested in future experiments. Other types of observables, not directly related to neutrino properties, naturally arise if there is new physics at a much lower energy scale $M$, around $1\div 10$ TeV. Indeed we have several indications, both from the experimental and from the theory side, that this can be the case. For instance, the observed discrepancy in the anomalous magnetic moment of the muon, the overwhelming evidence of dark matter, the evolution of the gauge coupling constants towards a common high-energy value and the solution of the hierarchy problem can all benefit from the existence of new particles around the TeV scale. In this chapter we assume that such a new scale exists and that the associated degrees of freedom do not provide new sources of baryon and/or lepton number violation.

At least four different scales are present in our approach: the scale of flavour dynamics $\Lambda_f$, the lepton number breaking scale $\Lambda_L$, the scale introduced by the VEVs of the flavon fields $\langle\varphi\rangle$ and the new physics scale $M$. A generic hierarchy among the scales is $M \ll \langle \varphi \rangle \ll \Lambda_f$ with $\Lambda_L$ expected to be comparable to or smaller than $\Lambda_f$.

We will first adopt an effective field theory approach, where the dominant physical effects of the new particles at low energies can be described by local six-dimensional operators, suppressed by two powers of the new mass scale $M$ and explicitly conserving baryon and lepton numbers. They will contribute to physical effects like the anomalous magnetic moments (MDMs) of the charged leptons, their electric dipole moments (EDMs) and lepton flavour violating (LFV) transitions like $\mu\to e \gamma$, $\tau\to e \gamma$ and $\tau\to\mu\gamma$. We will separately treat the general case where no further requirement is enforced and the supersymmetric case, where additional constraints are present. In this second context a cancellation is expected to take place, considerably changing the conclusion that can be derived from the existing bound on $\mu\to e \gamma$. It is then important to perform a direct computation of the branching ratios within an explicit supersymmetric model incorporating the flavour symmetry $A_4\times Z_3\times U(1)_{FN}$.

In the second part of the chapter we consider the Altarelli-Feruglio model by adding a full set of Supersymmetry breaking terms consistent with the flavour symmetry. We assume that the breaking of Supersymmetry occurs at a scale higher than or comparable to the flavour scale, simulated in our effective Lagrangian by a cutoff, so that at energies close to the cutoff scale we have non-universal boundary conditions for the soft Supersymmetry breaking terms, dictated by the flavour symmetry. Depending on the assumed mechanism of Supersymmetry breaking we may have boundary conditions different from these, possibly enforced at a smaller energy scale. For this reason, our approach maximises the possible effects on LFV processes.

We perform a detailed calculation of the slepton mass matrices in the physical basis and evaluate the branching ratios for the mentioned LFV decays in the mass insertion (MI) approximation.  We find a behaviour for these processes which is different from what expected in the supersymmetric variant of the effective Lagrangian approach: in particular we identify an additional contribution to the right-left (RL) block of the slepton mass matrix which  survives to the cancellation.  We then enumerate the conditions under which such a contribution is absent and the original behavior is recovered:  we could not find a dynamical explanation to justify the realisation of these conditions in our model, even if some of them are naturally realised in the context of supergravity (SUGRA) models. 

Finally we comment on the agreement of our results with the experimental measurements and bounds. In particular, assuming a SUGRA framework with a common mass scale $m_{SUSY}$ for soft sfermion and Higgs masses and a common mass $m_{1/2}$ for gauginos at high energies, we numerically study the normalised branching ratios of $\ell_i\to \ell_j\gamma$ using full one-loop expressions and explore the parameter space of the model. We find that the branching ratios for $\mu\to e \gamma$, $\tau\to \mu\gamma$ and $\tau \to\ e \gamma$ are all of the same order of magnitude. Therefore, applying the present MEGA bound on $BR(\mu\to e \gamma)$, this implies that $\tau\to \mu\gamma$ and $\tau \to\ e \gamma$ have rates much smaller than the present (and near future) sensitivity. Moreover, still considering the MEGA limit, we find that small values of the symmetry breaking parameter and of $\tan\beta$ are favoured for $m_{SUSY}$ and $m_{1/2}$ below $1000$ GeV, i.e. in the range of a possible detection of sparticles at LHC. Furthermore, it turns out to be rather unnatural to reconcile the values of superparticle masses necessary to account for the measured deviation in the muon anomalous magnetic moment from the Standard Model value with the present bound on the branching ration of $\mu\to e\gamma$. In our model values of such deviation smaller than $100 \times 10^{-11}$ are favoured.

\section{Low-Energy Effective Approach}
\label{Sec:LFV:Effective}
\setcounter{footnote}{3}

In this section we adopt an effective field theory approach, where the dominant physical effects of the new particles at low energies
can be described by local dimension six operators, suppressed by two powers of the new mass scale $M$ and explicitly conserving baryon and lepton numbers. We can account for the flavour breaking effects by requiring invariance of these operators under the flavour symmetry and by encoding the symmetry breaking effects in the flavon fields. This approach is strictly related to, and indeed inspired by, that of minimal flavour violation (MFV)\cite{MFV,MLFV1,MLFVother} already described in section \ref{Sec:MFV}. While such a minimal symmetry choice has the advantage that it can accommodate any pattern of lepton masses and mixing angles, it does not provide any clue about the origin of the approximate tribimaximal pattern observed in the lepton mixing matrix. For this reason, here we choose a flavour group that includes the discrete factor $A_4$: $G_f=A_4\times Z_3\times U(1)_{FN}$, a sort of minimal choice to properly account for both, the neutrino and the charged lepton mass spectra.

Following section \ref{Sec:AFTBM}, neutrinos can get their small masses through the low-energy Weinberg operator and the tribimaximal pattern, in the basis of diagonal charged leptons, is achieved via a specific vacuum misalignment of the flavons. The Lagrangian can be written as an expansion in powers of $\varphi/\Lambda_f$. When flavons develop a VEV, $\mean{\varphi}$, all the flavour information are enclosed in the Yukawa couplings: $Y_f=Y_f\left(\mean{\varphi}\right)$.
Under the assumption that $\vert\langle\varphi\rangle\vert\ll 1$, the functions $Y_f$ can be expanded in powers of $\langle\varphi\rangle$ and only a limited number of terms gives a non-negligible contribution. Finally, in the language of effective field theories, the leading terms of the relevant effective Lagrangian are given by
\beq
\LL_{eff}=\LL_K+e^{c\,T} H^\dagger Y_e \ell +\LL_\nu+ i\dfrac{e}{M^2}{e^c}^T H^\dagger\sigma^{\mu\nu}F_{\mu\nu}{\cal M}\ell +\hc+[\text{4-fermion operators}]
\label{LFV:Eff:leff}
\eeq
where $\LL_K$ stands for the kinetic terms, $\LL_\nu$ contains the terms describing the neutrino masses and $e$ is the electric charge. The complex $3\times3$ matrix $\cM$, with indices in the flavour space, is a function of $\langle\varphi\rangle$: $\cM=\cM\left(\langle\varphi\rangle\right)$. Invariance under $SU(2)_L\times U(1)_Y$ gauge transformations is guaranteed if $F_{\mu\nu}$ is any arbitrary combination of $B_{\mu\nu}$, the field strength of $U(1)_Y$, and $\sigma^a W^a_{\mu\nu}$, the non-Abelian field strength of $SU(2)_L$. Here we are interested in the component of this combination along the direction of the unbroken $U(1)_{em}$ and we identify $F_{\mu\nu}$ with the electromagnetic field strength.
We can imagine that we derive such an effective Lagrangian from a fundamental theory by integrating out two different sets of modes. In a first step we can integrate out the flavon fields and the possible degrees of freedom associated with the violation of $B-L$, thus obtaining a complete set of mass terms for all the light particles including neutrinos, charged fermions and the quanta with masses around the TeV scale. Subsequently, around the scale $M=1\div 10$ TeV, we integrate out these additional quanta and we generate the other operators listed in eq. (\ref{LFV:Eff:leff}). The latter are still invariant under $G_f$, once we treat the symmetry breaking parameters as spurions.

In this way we can fully consider all flavour symmetry breaking effects and keep a high degree of predictability since the expansion in the small symmetry breaking parameters can be truncated after few terms. Thereby, the same symmetry breaking parameters that control lepton masses and mixing angles also control the flavour pattern of the other operators in $\LL_{eff}$. Moreover the effects described by these operators are suppressed by $1/M^2$ and not by inverse powers of the larger scales $\Lambda_f$ and $\Lambda_L$ and this opens up the possibility that they might be observable in the future.

In a field basis where the kinetic terms are canonical and the charged lepton mass matrix is diagonal, where we will denote vectors and matrices with a hat, the real and imaginary parts of the matrix elements $\hat{\cM}_{ii}$ are proportional to the MDMs $a_i$ and to the EDMs $d_i$ of charged leptons, respectively \cite{Raidal,Brignole,Ciuchini,MLFV1,MLFVother}:
\beq
a_i=2 m_i \dfrac{v}{\sqrt{2} M^2}\Re\hat{\cM}_{ii}\;,\qquad\qquad d_i=e \dfrac{v}{\sqrt{2} M^2} \Im \hat{\cM}_{ii}\qquad\qquad(i=e,\mu,\tau)\;,
\eeq
\vskip 0.2cm
The off-diagonal elements $\hat{\cM}_{ij}$ describe the amplitudes for the LFV transitions $\mu\to e \gamma$, $\tau\to\mu\gamma$ and $\tau\to e \gamma$ \cite{ArgandaHerrero,Raidal,Ciuchini,MLFV1,MLFVother,Brignole,BorzumatiMasiero,HisanoFukuyama}:
\beq
R_{ij}\equiv\dfrac{BR(\ell_i\to \ell_j\gamma)}{BR(\ell_i\to \ell_j\nu_i{\ov{\nu}_j})}=\dfrac{12\sqrt{2}\pi^3 \alpha}{G_F^3 m_i^2 M^4}\left(\vert\hat{\cM}_{ij}\vert^2+\vert\hat{\cM}_{ji}\vert^2\right)\;,
\eeq
where $\alpha$ is the fine structure constant, $G_F$ is the Fermi constant and $m_i$ is the mass of the lepton $\ell_i$. Finally the four-fermion operators describe other flavour violating processes like $\tau^-\to \mu^+ e^- e^-$ and $\tau^-\to e^+ \mu^- \mu^-$. In this section our focus will be mainly on the processes $\mu\to e \gamma$, $\tau\to\mu\gamma$ and $\tau\to e\gamma$.

\subsection{The Dipole Matrix}
\label{Sec:LFV:EFF:DipoleMatrix}
\setcounter{footnote}{3}

In this section we present the analytical results for the dipole matrix, which is the main ingredient in order to evaluate MDMs, EDMs and LFV transitions. Before proceeding few comments are worth. We assume that the small symmetry breaking parameters of the Altarelli-Feruglio model, $u$ and $t$, are real parameters, encoding all the complex factors in the coupling of the Lagrangian. We further assume, without loss of generality, that the allowed range of values for $u$ is $0.001<u<0.05$, while $t$ should be around $0.05$, both in the general and in the supersymmetric contexts. The general Lagrangian which describes charged leptons and neutrinos for the class of models with the flavour symmetry $G_f$ and with the same flavon content of the Altarelli-Feruglio model is illustrated in eq. (\ref{AFTBM:Ll}, \ref{AFTBM:Lnu}) and here we only recall the elements which are useful for the present analysis: in particular we will concentrate only on the charged lepton sector. When calculating the kinetic terms, we find that the non-canonical contributions are small (at most at order $\cO(u)$ and $\cO(t^2)$). They should be taken into account when calculating lepton masses and the dipole transitions induced by the matrix $\hat{\cM}$, but an explicit calculation shows that at the leading order in our expansion parameters the charged lepton masses and the elements of $\hat{\cM}$ are not affected by these non-canonical kinetic terms \cite{Kahler}. For this reason we can safely assume canonical kinetic terms from the beginning (for a detailed discussion see the original paper \cite{FHLM_Efficace}).

Looking at eq. (\ref{LFV:Eff:leff}), we observe that the Lagrangian of the dipole moments have the same structure in flavour space of the Lagrangian of the charged leptons. As a consequence the entries of the charged lepton mass matrix and of the dipole matrix are exactly of
the same order of magnitude and only the coefficients are nominally different. After the breaking of the flavour and the electroweak symmetries, we find the following matrices:
\beq
Y_e \sim \cM = \left(
    \begin{array}{ccc}
        \cO(t^2 u) & \cO(t^2 u^2) &  \cO(t^2 u^2) \\
        \cO(t u^2) & \cO(t u) &  \cO(t u^2) \\
        \cO(u^2) & \cO(u^2) & \cO(u) \\
    \end{array}
\right)\;.
\label{LFV:Eff:yl&M_subleading}
\eeq
Notice that the off-diagonal elements of $Y_e$ and of $\cM$ originate either from the subleading contributions to the VEV of the $\varphi_T$ multiplet, or from a double flavon insertion of the type $\xi^\dagger \varphi_S$ or $\xi \varphi_S^\dagger$. In a generic case we expect that both these contributions are equally important and contribute at the same order to a given off-diagonal dipole transition. There is however a special case where the double flavon insertions $\xi^\dagger \varphi_S$ and its conjugate are suppressed compared to the subleading corrections to $\varphi_T$. This happens when the underlying theory is supersymmetric and Supersymmetry is softly broken, under the general assumption that the only sources of chirality flip are either fermion masses or sfermion masses of left-right type. Both of them, up to the order $u^2$, are described by the insertion of $\varphi_T$ or $\varphi_T^2$ in the relevant operators. The origin of the suppression of  $\xi^\dagger \varphi_S$ and  $\xi \varphi_S^\dagger$ is the holomorphycity of the superpotential in supersymmetric theories and we refer to the original paper \cite{FHLM_Efficace} for a complete discussion.
The impact of such a suppression is strong: indeed, moving into the physical basis in which the charged leptons are diagonal, with respect to what happens in the general case, an overall depletion in some elements of the dipole matrix takes place.

\subsection{Dipole Transitions}
\label{Sec:LFV:EFF:DipoleTransitions}
\setcounter{footnote}{3}

Here we give the results for the dipole moments when all the transformations to move into the physical basis have been carried out. Furthermore we compare them to the experimental values. In order to move to the physical basis it is only necessary to rotate the charged lepton into the diagonal form and we arrive at the following leading order result for the dipole matrix that, in this basis, will be denoted by $\mathcal{\hat M}$: we have
\beq
\mathcal{\hat M} = \left(
    \begin{array}{ccc}
        \cO(t^2 u) & \cO(t^2 u^2) &  \cO(t^2 u^2) \\
        \cO(t u^2) & \cO(t u) &  \cO(t u^2) \\
        \cO(u^2) & \cO(u^2) & \cO(u) \\
    \end{array}
\right)\;,\qquad
\mathcal{\hat M} = \left(
    \begin{array}{ccc}
        \cO(t^2\,u) & \cO(t^2\,u^2) &  \cO(t^2\,u^2) \\
        \cO(t\,u^3) & \cO(t\,u) &  \cO(t\,u^2) \\
        \cO(u^3) & \cO(u^3) & \cO(u) \\
    \end{array}
\right)\;,
\label{LFV:Eff:Mhat}
\eeq
in the general and in the supersymmetric cases, respectively. The main difference between the general and the supersymmetric approaches is the suppression of the elements below the diagonal: $\sim\cO(u^2)$ in the first case and $\sim\cO(u^3)$ in the second. This additional suppressing factor has a relevant consequence for the analysis on the LFV transitions. Furthermore, as reported in section \ref{Sec:AFTBM:NLO}, when the supersymmetric case is considered the VEV of $\varphi_T$ at the NLO has the second and third components equal ($c_2=c_3$), and as a result the elements ${\cal \hat{M}}_{\tau e}$ and ${\cal \hat{M}}_{\tau \mu}$ become equal and not only of the same order.

The MDMs and EDMs for the general and the supersymmetric cases are of the same order of magnitude: we display only the leading terms and they arise at first order in the $u$ parameter:
\beq
\ba{rcl}
a_e &=& 2 m_e\dfrac{v}{\sqrt{2} M^2}\, \cO(t^2 \, u) \\[3mm]
a_\mu &=& 2 m_\mu\dfrac{v}{\sqrt{2} M^2}\,\cO(t \, u) \\[3mm]
a_\tau &=& 2 m_\tau\dfrac{v}{\sqrt{2} M^2}\,\cO(u)
\ea\;,\qquad\qquad
\ba{rcl}
d_e&=&e \dfrac{v}{\sqrt{2} M^2}\, \cO(t^2 \, u) \\[3mm]
d_\mu&=&e \dfrac{v}{\sqrt{2} M^2}\, \cO(t \, u) \\[3mm]
d_\tau&=&e \dfrac{v}{\sqrt{2} M^2}\, \cO(u)\;.
\ea
\label{LFV:Eff:MDMEDM}
\eeq
Considering the explicit form of the charged lepton masses, we can write these expression as follows:
\beq
a_i=\cO\left(2\dfrac{m_i^2}{M^2}\right)\;,\qquad\qquad d_i=\cO\left(e\dfrac{m_i}{M^2}\right)\;.
\label{LFV:Eff:oom}
\eeq
We can derive a bound on the scale $M$, by considering the existing limits on MDMs and EDMs and by using eqs. (\ref{LFV:Eff:oom}) as exact equalities to fix the ambiguity of the unknown coefficients. We find the results shown in table \ref{table:LFV:Eff:BoundsMDMEDM}. We see that, in order to accept values of $M$ in the range $1\div 10$ TeV, we should invoke a cancellation in the imaginary part of $\hat{\cal{M}}_{ee}$, which can be either accidental or due to CP-conservation in the considered sector of the theory.

\begin{table}[!ht]
\centering
\begin{math}
\begin{array}{|c|c|}
    \hline
    & \\[-9pt]
    d_e<1.6\times 10^{-27}\;e\,\cm&M>80\,\TeV\\[3pt]
    \hline
    &\\[-9pt]
    d_\mu<2.8\times 10^{-19}\;e\,\cm&M>80\,\GeV\\[3pt]
    \hline
    &\\[-9pt]
    \delta a_e<3.8\times 10^{-12}&M>350\,\GeV\\[3pt]
    \hline
    &\\[-9pt]
    \delta a_\mu\approx 30\times 10^{-10}&M\approx 2.7\,\TeV\\[3pt]
    \hline
\end{array}
\end{math}
\caption{\it Experimental limits on lepton MDMs and EDMs\cite{ExperimentalBoundsMDMmu,ExperimentalBoundsMDMe,PDG08} and corresponding bounds on the scale $M$, derived from eqs. (\ref{LFV:Eff:oom}). The data on the $\tau$ lepton have not been reported since they are much less constraining. For the anomalous magnetic moment of the muon, $\delta a_\mu$ stands for the deviation of the experimental central value from the Standard Model expectation.}
\vspace{-0.5cm}
\label{table:LFV:Eff:BoundsMDMEDM}
\end{table}

Concerning the flavour violating dipole transitions for the general case, using eq. ({\ref{LFV:Eff:Mhat}) we see that the rate for $\ell_i\to \ell_j\gamma$ is dominated by the contribution ${\cal\hat M}_{ij}$, since ${\cal\hat M}_{ji}$ is suppressed by a relative factor of $\cO(t)$ for $\mu\to e\gamma $ and $\tau\to\mu\gamma$ and of $\cO(t^2)$ for $\tau\to e\gamma$. We get:
\beq
R_{ij}=\dfrac{48\pi^3 \alpha}{G_F^2 M^4}\vert w_{ij} ~u\vert^2
\label{LFV:Eff:LFV}
\eeq
where $w_{ij}$ are coefficients of $\cO(1)$ which contain two contributions, one coming from the off-diagonal element $(ij)$ of the original dipole matrix ${\cal M}$ and one coming from the effect of diagonalising the charged lepton mass matrix. Both contribute at the same order in $u$. Since the quantities $w_{ij}$ are all expected to be of order one, we can conclude that in the class of models considered here the branching ratios for the three transitions $\mu\to e\gamma $, $\tau\to\mu\gamma$ and $\tau\to e\gamma$ should all be of the same order:
\beq
BR(\mu\to e \gamma)\approx BR(\tau\to\mu\gamma)\approx BR(\tau\to e \gamma)\;.
\eeq
This is a distinctive feature of our class of models, since in most of the other existing models there is a substantial difference between the branching ratios\cite{MFV,MLFV1,MLFVother,SUSYLFV+symmetries,SUSYFP+GUTs}. In particular it often occurs that $BR(\mu\to e \gamma)<BR(\tau\to\mu\gamma)$. Given the present experimental bound $BR(\mu\to e \gamma)<1.2\times 10^{-11}$\cite{MEGA}, our result implies that $\tau\to\mu\gamma$ and $\tau\to e \gamma$ have rates much below the present and expected future sensitivity \cite{ExperimentalBoundsDecaysTau}. Moreover, from the current (future) experimental limit on $BR(\mu\to e \gamma)$\cite{MEGA,MEG} and assuming $\vert w_{\mu e}\vert=1$, we derive the following bounds
\beq
\ba{ll}
u = 0.001 &\longrightarrow\quad M>10~(30)\,\TeV\\[3mm]
u = 0.05 &\longrightarrow\quad M>70~(200)\,\TeV\;.
\ea
\eeq
This pushes the scale $M$ considerably above the range we were initially interested in. In particular $M$ is shifted above the region of
interest for $(g-2)_\mu$ and probably for LHC.

When considering the supersymmetric case we get
\beq
R_{ij}= \dfrac{48\pi^3 \alpha}{G_F^2 M^4}\left[\vert w^{(1)}_{ij} u^2\vert^2+\dfrac{m_j^2}{m_i^2} \vert w^{(2)}_{ij} u\vert^2\right]
\label{LFV:Eff:LFVsusy}
\eeq
where $w^{(1,2)}_{ij}$ are coefficients of $\cO(1)$ depending on the coefficients coming from the original dipole matrix and from the charged lepton mass matrix.

\begin{figure}[ht!]
\centering
\includegraphics[width=12cm]{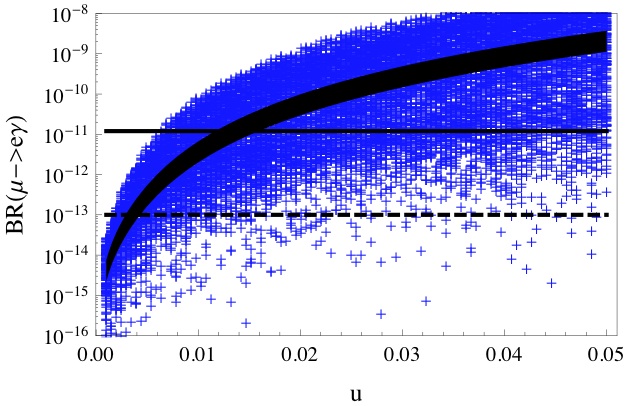}
\vspace{-0.4cm}
\caption{\it The branching ratio of $\mu\to e \gamma$ as a function of $u$, eq. (75). The deviation of the anomalous magnetic moment of the muon from the Standard Model expectation is kept fixed to its experimental range\cite{ExperimentalBoundsMDMmu}. The unknown coefficients $\tilde{w}^{(1,2)}_{\mu e}$ are equal to 1 (black region) or are random complex numbers with absolute values between zero and two (blue points). The continuous (dashed) horizontal line corresponds to the present (future expected) experimental bound on $BR(\mu\to e\gamma)$\cite{MEGA,MEG}.}
\label{fig:LFV:Eff:BR}
\vspace{-0.5cm}
\end{figure}

Notice that now the first contribution on the right-hand side of eq. (\ref{LFV:Eff:LFVsusy}) is suppressed by a factor of $u$ compared to the non-supersymmetric case. In most of the allowed range of $u$, the branching ratios of $\mu\to e \gamma$ and $\tau \to\mu \gamma$ are similar and larger than the branching ratio of $\tau\to e \gamma$. Assuming $\vert w^{(1,2)}_{\mu e}\vert=1$, the present (future) experimental limit on $BR(\mu\to e \gamma)$ implies the following bounds
\beq
\ba{ll}
u = 0.001 &\longrightarrow\quad M>0.7~(2)\,\TeV\\[3mm]
u = 0.05 &\longrightarrow\quad M>14~(48)\,\TeV\;.
\ea
\eeq
We see that at variance with the non-supersymmetric case there is a range of permitted values of the parameter $u$ for which the scale $M$ can be sufficiently small to allow an explanation of the observed discrepancy in $a_\mu$, without conflicting with the present bound on $BR(\mu\to e \gamma)$. We can eliminate the dependence on the unknown scale $M$ by combining eqs. (\ref{LFV:Eff:oom}) and (\ref{LFV:Eff:LFVsusy}). For $\mu\to e \gamma$ we get
\begin{equation}
R_{e\mu}= \dfrac{12\pi^3 \alpha}{G_F^2 m_\mu^4}\left(\delta a_\mu\right)^2
\left[\vert \tilde{w}^{(1)}_{\mu e}\vert^2 \vert u\vert^4+\dfrac{m_e^2}{m_\mu^2} \vert \tilde{w}^{(2)}_{\mu e}\vert^2\vert u\vert^2\right]
\label{LFV:Eff:muegamma}
\end{equation}
where $\tilde{w}^{(1,2)}_{\mu e}$ are unknown, order one coefficients. We plot $BR(\mu\to e\gamma)$ versus $u$ in figure \ref{fig:LFV:Eff:BR}, where the coefficients $\tilde{w}^{(1,2)}_{\mu e}$ are kept fixed to 1 (black region) or are random complex numbers with absolute values between zero and two (blue points). The deviation of the anomalous magnetic moment of the muon from the Standard Model prediction is in the interval of the experimentally allowed values, about three sigma away from zero. Even if the ignorance about the coefficients $\tilde{w}^{(1,2)}_{\mu e}$ does not allow us to derive a sharp limit on $u$, we see that the present limit on $BR(\mu\to e \gamma)$ disfavors values of $u$ larger than few percents. We recall that in this model the magnitudes of $u$ and $\theta_{13}$ are comparable.

\subsection{Comparison with Minimal Flavour Violation }
\label{Sec:LFV:EFF:MFV}
\setcounter{footnote}{3}

It is instructive to compare the previous results with those of MFV \cite{MFV,MLFV1,MLFVother}. We have already introduced the main features of MFV in section \ref{Sec:MFV} and we recall here only that, by going to a basis where the charged leptons are diagonal, the Yukawa matrices for charged leptons and neutrinos can be written as 
\beq
\hat{Y}_e=\dfrac{\sqrt{2}}{v} \diag(m_e,\,m_\mu,\,m_\tau)\;,\qquad\qquad \hat{Y}_\nu=\dfrac{\Lambda_L}{4v^2} U^* \diag(m_1,\,m_2,\,m_3) U^\dagger\;,
\eeq
where $U$ is the lepton mixing matrix. The diagonal elements of the matrix $\mathcal{\hat{M}}$ evaluated in MFV are analogous to those of the previous class of models and similar bounds on the scale $M$ are derived from the existing data on MDMs and EDMs. The off-diagonal elements are given by:
\beq
\mathcal{\hat{M}}_{ij}\;=\;\beta (\hat{Y}_e \hat{Y}_\nu^\dagger \hat{Y}_\nu)_{ij}
\;=\;\sqrt{2}\beta\dfrac{m_i}{v}\dfrac{\Lambda_L^2}{16v^4}\left[\Delta m^2_{sol} U_{i2} U^*_{j2}\pm\Delta m^2_{atm} U_{i3} U^*_{j3}\right]
\eeq
where $\beta$ is an overall coefficient of order one and the plus (minus) sign refers to the case of normal (inverted) hierarchy.
We see that, due to the presence of the ratio $\Lambda_L^2/v^2$ the overall scale of these matrix elements is much less constrained than in the previous case. This is due to the fact that MFV does not restrict the overall strength of the coupling constants $Y_\nu$, apart from the requirement
that they remain in the perturbative regime. Very small or relatively large (but smaller than one) $Y_\nu$ can be accommodated by adjusting the scale $\Lambda_L$. On the contrary this is not allowed in the case previously discussed where the size of the symmetry breaking effects is restricted to the small window ($0.001<u<0.05$) and the scale $\Lambda_L$ is determined within a factor of about fifty. The conclusion is that in MFV the non-observation of $\ell_i\to \ell_j\gamma$ could be justified by choosing a small $\Lambda_L$, while a positive signal in $\mu\to e \gamma$ with a branching ratio in the range $1.2\times 10^{-11}\div 10^{-13}$ could also be fitted by an appropriate $\Lambda_L$, apart from a small region of the $\theta_{13}$ angle, around $\theta_{13}\approx0.02$ where a cancellation can take place.

The dependence on the scale $\Lambda_L$ is eliminated by considering ratios of branching ratios:
\beq
\dfrac{BR(\mu\to e\gamma)}{BR(\mu\to e\nu_\mu{\bar \nu_e})}\frac{BR(\tau\to \mu\nu_\tau{\bar \nu_\mu})}{BR(\tau\to \mu\gamma)}=
\left\vert\frac{2\Delta m^2_{sol}}{3\Delta m^2_{atm}}\pm \sqrt{2}\sin\theta_{13} e^{i\delta}\right\vert^2<1\;,
\label{mfv}
\eeq
where we took the tribimaximal ansatz to fix $\theta_{12}$ and $\theta_{23}$. We see that $BR(\mu\to e\gamma)<BR(\tau\to \mu\gamma)$ always in MFV. Moreover, for $\theta_{13}$ above approximately $0.07$, $BR(\mu\to e\gamma)<1.2\times 10^{-11}$ implies $BR(\tau\to \mu\gamma)<10^{-9}$. For $\theta_{13}$ below $0.07$, apart possibly from a small region around $\theta_{13}\approx0.02$, both the transitions $\mu\to e \gamma$ and $\tau\to\mu\gamma$ might be above the sensitivity of the future experiments.

We also observe that in MFV the only difference between the general case and the supersymmetric one is the presence of two doublets in the low-energy Lagrangian. In MFV a chirality flip in leptonic operators necessarily requires the insertion of the matrix $\hat{Y}_e$, both in the general and in the supersymmetric case and, apart from the possibility of $\tan\beta$ enhanced contributions, similar predictions for the LFV processes are expected in the two cases.

\mathversion{bold}
\section{Explicit Supersymmetric $A_4$ Model}
\label{Sec:LFV:FullSUSYModel}
\setcounter{footnote}{3}
\mathversion{normal}

In this section we move to consider an explicit supersymmetric model incorporating the flavour symmetry $A_4\times Z_3\times U(1)_{FN}$. The Lagrangian of the model accounts for three distinct terms:
\beq
\begin{split}
\LL=&\;\int \D^2\theta_{SUSY} \D^2\ov{\theta}_{SUSY} \cK (\ov{z}, e^{2 V} z)+\left[\int \D^2 \theta_{SUSY} w(z)+\hc\right]\\[3mm]
&+\frac{1}{4}\left[\int \D^2\theta_{SUSY} f(z) {\cal W W}+\hc\right]\;,
\end{split}
\label{LFV:leel}
\eeq
where ${\cal K}(\ov{z},z)$ is the K\"ahler potential,  $w(z)$ is the superpotential, $f(z)$ is the gauge kinetic function, $V$ is the Lie-algebra valued vector supermultiplet, describing the gauge fields and their superpartners. \footnote{In our notation a chiral superfield and its $R$-parity even component are denoted by the same letter. The $R$-parity odd component is indicated by a tilde in the following and the conjugate (anti-chiral) superfield is denoted by a bar.} Finally ${\cal W}$ is the chiral superfield describing, together with the function $f(z)$, the kinetic terms of gauge bosons and their superpartners. Each of the terms on the right-hand side can be written in an expansion in powers of the flavon fields. The flavour symmetry of the Lagrangian $\LL$ is spontaneously broken by the VEVs of the flavons, which in the supersymmetric context are aligned as
\beq
\ba{ccl}
\dfrac{\langle\varphi_T\rangle}{\Lambda_f}&=&(u,0,0)+(c' u^2,c u^2,c u^2)+\cO(u^3)\\[3mm]
\dfrac{\langle\varphi_S\rangle}{\Lambda_f}&=& c_b(u,u,u)+\cO(u^2)\\[3mm]
\dfrac{\langle\xi\rangle}{\Lambda_f}&=&c_a u+\cO(u^2)\;,
\ea
\label{LFV:vevs}
\eeq
where $c$, $c'$, $c_{a,b}$ are complex numbers with absolute value of order one and $u$ is one of the two small symmetry breaking parameters in the theory. With respect to eq. (\ref{AFTBM:vevsplus}), we specify the subleading corrections only for $\mean{\varphi_T}$, since those of $\mean{\varphi_S}$ and $\mean{\xi}$ do not affect the following analysis. Moreover we simplify the notation using $c'$ instead of $c_1$ and $c$ instead of $c_2$ and $c_3$, which are equal at this level of approximation. The second symmetry breaking parameter is the VEV of the Froggatt-Nielsen $U(1)_{FN}$ symmetry, $t$. It is useful to briefly recall the mechanism which assures this specific vacuum misalignment. A set of driving fields, two $A_4$-triplets $\varphi_T^0$ and $\varphi_S^0$ plus an $A_4$-singlet $\xi^0$, is introduce in the model and the following driving superpotential can be written down:
\beq
\begin{split}
w_d\;=&\;M_T (\varphi_T^0 \varphi_T)+ g (\varphi_T^0 \varphi_T\varphi_T)+\\
&+g_1 (\varphi_S^0 \varphi_S\varphi_S)+ g_2 \tilde{\xi} (\varphi_S^0 \varphi_S)+ g_3 \xi^0 (\varphi_S\varphi_S)+ g_4 \xi^0 \xi^2+ g_5 \xi^0 \xi \tilde{\xi}+ g_6 \xi^0 \tilde{\xi}^2+\ldots\;,
\end{split}
\label{LFV:wd}
\eeq
where dots denote subleading non-renormalisable corrections. In the limit of unbroken Supersymmetry all $F$-terms vanish when the vacuum in eq. (\ref{LFV:vevs}) is considered. In \cite{AF_Modular} it is shown that this setting is an isolated minimum of the scalar potential, achieved in a completely non-tuned way. We remark that in the supersymmetric limit the VEVs of the driving fields $\varphi_T^0$, $\varphi_S^0$ and $\xi^0$ are zero. This is however in general no longer true, if we include soft Supersymmetry breaking terms into the flavon potential, as has been discussed in \cite{FHM_VEV}. The origin of the VEV for $\theta$ can be find in appendix \ref{AppB:Tp}. We recall the range of values in which $u$ and $t$ can run in the supersymmetric context:
\beq
0.007\lesssim u\lesssim0.05\;,\qquad\qquad t\approx0.05\;.
\label{LFV:Bounds}
\eeq
Since we have two independent symmetry breaking parameters, we consider a double expansion of $\LL$ in powers of $u$ and $t$. In this expansion we keep terms up to the second order in $u$, i.e. terms quadratic in the fields $\varphi_{S,T}$ and $\xi$. The expansion in the parameter $t$ is stopped at the first non-trivial order, that is by allowing as many powers of the field $\theta$ as necessary in order to obtain non-vanishing values for all entries of the matrices describing lepton masses as well as for the entries of the matrices describing kinetic terms and slepton masses. \footnote{Concerning the K\"ahler potential we observe that we can additionally write down operators involving the total invariant $\overline{\theta}\theta=|\theta|^{2}$. These contribute to the diagonal elements of the kinetic terms and the slepton masses. In the K\"ahler potential for the left-handed fields they can be safely neglected, since the leading order correction is of $O(u)$. In the right-handed sector, they contribute at the same order as the terms arising through a double flavon insertion.} Finally, second order corrections in $u$ also arise from the subleading terms of the VEV $\langle\varphi_T\rangle$ and are included in our estimates.

The soft Supersymmetry breaking terms are generated from the supersymmetric Lagrangian by promoting all coupling constants, such as Yukawa couplings, couplings in the flavon superpotential and couplings in the K\"ahler potential, to superfields with constant $\theta_{SUSY}^2$ and $\theta_{SUSY}^2\ov{\theta}_{SUSY}^2$ components \cite{LutyReview}. Through this we derive subsequently the soft masses $(m_{(e,\nu)LL}^2)_K$ and $(m_{e RR}^2)_K$ from the K\"ahler potential. One contribution to $m_{eRL}^2$, which we call $(m_{e RL}^2)_1$ in the following, arises from the Yukawa couplings present in the superpotential $w$.

Important contributions to slepton masses originate from the modification of the VEVs of flavons and driving fields due to Supersymmetry breaking effects. A detailed study of the VEVs of these fields and their dependence on the soft Supersymmetry breaking parameters is presented in \cite{FHM_VEV} and we summarise the main results here. When soft Supersymmetry breaking terms are included into the flavon potential, the VEVs in eq. (\ref{LFV:vevs}) receive additional contributions of order $m_{SUSY}$, completely negligible compared to $\Lambda_f\,u$. At the same time,
the driving fields $\varphi_T^0$, $\varphi_S^0$ and $\xi^0$ develop a VEV of the size of the soft Supersymmetry breaking scale $m_{SUSY}$.
An equivalent statement is that the auxiliary components of the flavons acquire a VEV at the leading order of the size of $m_{SUSY} \times u \, \La_f$.
Especially, for the auxiliary part on the flavon supermultiplet $\varphi_T$ we have \cite{FHM_VEV}:
\beq
\dfrac{1}{\La_f} \left\langle \dfrac{\partial w}{\partial \varphi_T} \right\rangle = \zeta \, m_{SUSY} \, \left\{ (u,0,0)
+ (c_F^\prime u^2, c_F u^2, c_F u^2) \right\}
\label{LFV:VEVsauxphiT}
\eeq
where $\zeta$, $c_F^\prime$ and $c_F$ are in general complex numbers with absolute value of order one. The parameter $\zeta$ vanishes in the special case of universal soft mass terms in the flavon potential. When different from zero, the VEVs of the auxiliary components of the flavon supermultiplet $\varphi_T$ generate another contribution to the soft masses of RL-type, which we denote as $(m_{e RL}^2)_2$. This contribution is analogous to the one which has been found before in the supergravity context and which can have a considerable effect on the size of the branching ratio of radiative leptonic decays, as shown in \cite{FtermSUGRA}. Indeed, as we shall see below, in the global supersymmetric model under consideration the leading dependence of the normalised branching ratios $R_{ij}$ on $u$ is dominated by $(m_{e RL}^2)_2$. We remark that the VEVs in eq. (\ref{LFV:VEVsauxphiT}) and those of the corresponding flavon field $\varphi_T$ in eq. (\ref{LFV:vevs}) have a similar structure but they are not proportional, in general. This is due to the different coefficients $c$, $c^\prime$ and $c_F$, $c_F^\prime$, which can be qualitatively understood as follows: the coefficients $c$, $c^\prime$ mainly depend on a set of parameters that remain in the supersymmetric limit and receive completely negligible corrections from the Supersymmetry breaking terms. On the contrary $\langle \partial w/\partial \varphi_T\rangle$ vanishes in the supersymmetric limit, to all orders in $u$, and $c_F$, $c_F^\prime$ crucially depend on the set of parameters describing the Supersymmetry breaking. We will see that, if $c$ and $c_F$ accidentally coincide (up to complex conjugation), a cancellation in the leading behaviour of $R_{ij}$ takes place.

Similarly, the VEV of the Froggatt-Nielsen field $\theta$ becomes
shifted, when soft Supersymmetry breaking terms are included into the potential, so that:
\beq
\frac{M_{FI}^2}{g_{FN}} - |\langle\theta\rangle|^2 = c_{\theta} \, m_{SUSY}^2 \; ,
\eeq
with $c_{\theta}$ being an order one number, holds. This will lead to a contribution $(m_{e RR}^2)_{D,FN}$  to the soft masses of RR-type,
since only the right-handed charged leptons $e^c$ and $\mu^c$ are charged under $U(1)_{FN}$. Apart from these there are supersymmetric contributions to $m_{(e,\nu) LL}^2$ and $m_{e RR}^2$ from $F$ and $D$-terms, $(m_{(e,\nu) LL}^2)_{F (D)}$ and $(m_{e RR}^2)_{F (D)}$,
as well as a contribution to $m_{e RL}^2$ coming from the $F$-term of $H_d$, called $(m_{e RL}^2)_{3}$ in the following.

The detailed derivation of the kinetic terms as well as of the mass matrices for fermion and sfermions can be found in the original paper \cite{FHLM_LFV}.

\section{Slepton Masses in the Physical Basis}
\label{Sec:LFV:Phys}
\setcounter{footnote}{3}

In this section we discuss the results for the slepton masses in the physical basis and comment on results found in the literature. To derive the physical masses and the unitary transformations that enter our computation, we have to go into a basis in which kinetic terms are canonical, for both, slepton and lepton, fields. Subsequently, we diagonalise the mass matrix of the charged leptons via a biunitary transformation. To avoid flavour-violating gaugino-lepton-slepton vertices in this intermediate step, we perform the same transformation on both fermion and scalar components of the involved chiral superfields. This procedure gives us the physical slepton mass matrices ${\hat m}^2_{(e,\nu)LL}$, ${\hat m}^2_{eRR}$ and ${\hat m}^2_{eRL}$. The results shown here are obtained under the assumption that all the parameters of the model are real. The analytical expressions for the slepton mass matrices in the physical basis contain the first non-vanishing order in each of the matrix elements. We start with the left-left (LL) block. The contribution from the soft breaking terms is common to charged sleptons and sneutrinos and reads:
\begin{equation}
\begin{array}{lcl}
(\hat{m}_{eLL}^2)_K &=& (\hat{m}_{\nu LL}^2)_K\\[3mm]
&&\hspace{-2cm}=\left(
                    \begin{array}{ccc}
                              n_0 + 2 \, \hat{n}_1 \, u
                            & (\hat{n}_4 + (3 \, \hat{n}_1 + \hat{n}_2) \, c) \, u^2
                            & (\hat{n}_5 + (3 \, \hat{n}_1 - \hat{n}_2) \, c) \, u^2 \\
                              (\hat{n}_4 + (3 \, \hat{n}_1 + \hat{n}_2) \, c) \, u^2
                            & n_0 - (\hat{n}_1 + \hat{n}_2) \, u
                            & (\hat{n}_6 - 2 \, \hat{n}_2 \, c) \, u^2 \\
                              (\hat{n}_5 + (3 \, \hat{n}_1 - \hat{n}_2) \, c) \, u^2
                            & (\hat{n}_6 - 2 \, \hat{n}_2 \, c) \, u^2
                            & n_0 - (\hat{n}_1 - \hat{n}_2) \, u

                    \end{array}
                \right) \, m_{SUSY}^2
\end{array}
\label{LFV:eq:m_LL_hat}
\end{equation}
where $\hat{n}_i$ are complex parameters with modulus of order 1. The supersymmetric $F$ and $D$-term contributions are given by:
\begin{equation}
(\hat{m}_{eLL}^2)_F=\hat{M}_e^T \hat{M}_e\;,\qquad\qquad(\hat{m}_{\nu LL}^2)_F\simeq0
\end{equation}
and
\begin{equation}
(\hat{m}^2_{eLL})_D=\left(-\dfrac{1}{2}+\sin^2\theta_W \right) \cos 2\beta ~m_Z^2 \times \unity\;,\qquad
(\hat{m}^2_{\nu LL})_D=\left(+\dfrac{1}{2} \right) \cos 2\beta~m_Z^2 \times \unity\;,
\end{equation}
with $\hat{M}_e$ being the mass matrix for the charged leptons in the same basis, i.e. diagonal and with canonically normalised kinetic terms. The supersymmetric $D$-term contributions are proportional to the unity matrix. Notice that in the physical basis all supersymmetric contributions are diagonal in flavour space. Both the $F$ and the $D$-term contributions are small compared to that coming from the K\"ahler potential. The relative suppression is of order $\left(\hat{M}_e^T \hat{M}_e\right)/m_{SUSY}^2$ and $m_Z^2/m_{SUSY}^2$, respectively, which do not exceed the per cent level for typical values of $m_{SUSY}$ around 1 TeV. Note also that the supersymmetric part is the only one that distinguishes between charged sleptons and sneutrinos. 

For $\hat{m}_{eRR}^2$ we find that $(\hat{m}_{eRR}^2)_K$ is given by:
\begin{equation}
\label{LFV:eq:m_RR_hat}
(\hat{m}_{eRR}^2)_K = \left( \begin{array}{ccc}
                 n_1^c
                                & 2 \, c \,  (n_1^c - n_2^c) \, \dfrac{m_e}{m_\mu} u
                                & 2 \, c \, (n_1^c - n_3^c) \, \dfrac{m_e}{m_\tau} u\\[0.1in]
                2 \, c  \, (n_1^c - n_2^c) \, \dfrac{m_e}{m_\mu} u
                                & n_2^c
                                & 2 \, c \, (n_2^c - n_3^c) \, \dfrac{m_\mu}{m_\tau} u\\[0.1in]
                2 \, c  \, (n_1^c - n_3^c) \, \dfrac{m_e}{m_\tau} u
                                & 2 \, c  \, (n_2^c - n_3^c) \,\dd \frac{m_\mu}{m_\tau} u
                                & n_3^c
\end{array}
\right) \, m_{SUSY}^2\;.
\end{equation}
The supersymmetric terms are:
\begin{equation}
(\hat{m}_{eRR}^2)_F=\hat{M}_e \hat{M}_e^T\qquad\mbox{and}\qquad
(\hat{m}^2_{eRR})_D=-\sin^2\theta_W \cos 2\beta\,m_Z^2 \times \unity\;.
\end{equation}
Also in this case the supersymmetric contributions are diagonal and numerically negligible in most of our parameter space. The dominant contribution is thus $(\hat{m}_{eRR}^2)_K$. 

Finally, coming to the RL block of the mass matrix for charged sleptons, we find:
\begin{equation}
\label{LFV:eq:m_RL_hat}
(\hat{m}_{eRL}^2)_{1} = \left( \begin{array}{ccc}
                \dfrac{z_e}{y_e} \, m_e
                                & 2 c  \, \dfrac{(z_e y_\mu - z_\mu y_e)} {y_e y_\mu}\, m_e u
                                & 2 c\, \dfrac{(z_e y_\tau - z_\tau y_e)}{y_e y_\tau}  \, m_e u\\[0.1in]
                c \, \dfrac{(z_\mu y_\mu^\prime - z_\mu^\prime y_\mu)}{y_\mu^2} \, m_\mu u^2
                                & \dfrac{z_\mu}{y_\mu} m_\mu
                                & 2 c  \,\dfrac{(z_\mu y_\tau -  z_\tau y_\mu)}{y_\mu y_\tau} m_\mu u\\[0.1in]
                c  \, \dfrac{(z_\tau y_\tau ^\prime - z_\tau ^\prime y_\tau)}{y_\tau^2} \, m_\tau u^2
                                & c \, \dfrac{(z_\tau y_\tau ^\prime - z_\tau ^\prime y_\tau)}{y_\tau^2} \, m_\tau u^2
                                & \dfrac{z_\tau}{y_\tau} \, m_\tau
\end{array}
\right) \, m_{SUSY} \; ,
\end{equation}
\begin{equation}
\label{LFV:eq:m_RL_hat_2}
(\hat{m}_{eRL}^2)_{2} = \zeta \left( \begin{array}{ccc}
                m_e
                                & (c_F -c) \, m_e u
                                & (c_F -c) \, m_e u\\[0.1in]
               (c_F -c) \, m_\mu u
                                & m_\mu
                                & (c_F -c) \, m_\mu u\\[0.1in]
               (c_F -c) \, m_\tau u
                                & (c_F -c) \, m_\tau u
                                & m_\tau
\end{array}
\right) \, m_{SUSY} \; ,
\end{equation}
and
\beq \label{LFV:eq:m_RL_hat_3}
(\hat{m}^2_{eRL})_3=-\mu \tan\beta\,\hat{M}_e\;.
\eeq
The matrix $\hat{m}_{eRL}^2$ is the sum of these three contributions. An important feature of $(\hat{m}_{eRL}^2)_1$ is that the elements below the diagonal are suppressed by a factor $u$ compared to the corresponding elements of $(\hat{m}_{eRL}^2)_2$. Nevertheless there are cases in which this second contribution can be suppressed. In the first case the VEVs of the auxiliary fields contained in the supermultiplet $\varphi_T$ vanish, i.e. the parameter $\zeta$ is zero, due to the fact that the soft Supersymmetry breaking terms in the flavon potential are (assumed to be) universal, that is equal to the terms of the superpotential $w_d$ up to an overall proportionality constant \cite{FHM_VEV}. The second possibility arises, if the VEVs of the auxiliary fields can be completely aligned with those of the flavon $\varphi_T$ at the leading order as well as NLO, such that $c_F$ becomes equal to $c$. In both cases the off-diagonal elements of $(\hat{m}_{eRL}^2)_{2}$ are further suppressed than shown in eq. (\ref{LFV:eq:m_RL_hat_2}). We emphasise this fact here, since it turns out that the suppression of the off-diagonal elements below the diagonal as it occurs in the case of $(\hat{m}_{eRL}^2)_{1}$ is relevant for the actual size of the leading behaviour of the normalised branching ratios $R_{ij}$ with respect to the expansion in $u$. As we shall see in section \ref{Sec:LFV:MI_AnaliticResults}, in a general case $R_{ij} \propto u^2$ holds, whereas, if the contribution in eq. (\ref{LFV:eq:m_RL_hat_2}) vanishes or is also suppressed, $R_{ij}$ is proportional to $u^4$. The contribution $(\hat{m}^2_{eRL})_3$ is diagonal in flavour space. Concerning the possible size of this contribution, note that $|\mu| \tan \beta/m_{SUSY}$ is the relative magnitude of the non-vanishing elements of $(\hat{m}_{eRL}^2)_3$ with respect to the corresponding ones in $(\hat{m}_{eRL}^2)_{1,2}$. Notice finally that the (31) and (32) element of $\hat{m}_{eRL}^2$ coincide.

In \cite{FHLM_LFV} we present an analysis of the renormalisation group effects on the slepton masses in the leading Log approximation and we see that these effects can be neglected or absorbed into our parametrisation of the soft mass terms.

\section{Results in the Mass Insertion Approximation}
\label{Sec:LFV:MI}
\setcounter{footnote}{3}

We can now evaluate the normalised branching ratios $R_{ij}$ for the LFV transitions $\mu\to e \gamma$, $\tau\to\mu\gamma$ and $\tau\to e \gamma$. In this section we establish the leading dependence of the quantities $R_{ij}$ on the symmetry breaking parameter $u$. We then compare the results with the conclusions of the effective approach already illustrated in section \ref{Sec:LFV:Effective}. Performing this comparison, it is important to keep in mind that when $R_{\mu e}$ is dominated by a one-loop amplitude with virtual particles of mass $m_{SUSY}$, $M$ and $m_{SUSY}$ are roughly related by $M=(4\pi/g) m_{SUSY}$ and a given lower bound on $M$ corresponds to a lower bound on $m_{SUSY}$ one order of magnitude smaller.

\subsection{Analytic results}
\label{Sec:LFV:MI_AnaliticResults}
\setcounter{footnote}{3}

It is useful to first analyse the predictions in the so-called mass insertion (MI) approximation, where we have a full control of the results in its analytic form. A more complete discussion based on one-loop results can be found in section \ref{Sec:LFV:Numerical}. For the case at hand, the MI approximation consists in expanding the amplitudes in powers of the off-diagonal elements of the slepton mass matrices, normalised to their average mass. From the expression of the mass matrices of the previous section we see that in our case such an expansion amounts to an expansion in the parameters $u$ and $t$, which we can directly compare with eq. (\ref{LFV:Eff:LFVsusy}). A common value in the diagonal entries of both LL and  RR blocks is assumed and we consequently set $n_0=n_1^c=n_2^c=n_3^c=1$ and also $\hat{n}_1=\hat{n}_2=0$ in this section, so that the average mass becomes $m_{SUSY}$. On the contrary, no assumptions have been made for the trilinear soft terms, which keep the expression as in eqs. (\ref{LFV:eq:m_RL_hat}-\ref{LFV:eq:m_RL_hat_3}). Concerning chargino and neutralino mass matrices, they carry a dependence on the vector boson masses $m_{W,Z}$ through off-diagonal matrix elements. Such a dependence is not neglected in this approximation, but only the leading order term of an expansion in $m_{W,Z}$ over the relevant supersymmetric mass combination is kept. At the same time, to be consistent, we have to neglect the supersymmetric contributions of $\hat{m}^2_{\nu LL}$ and $\hat{m}^2_{e LL}$ and therefore $\hat{m}^2_{\nu LL}$ and $\hat{m}^2_{e LL}$ coincide. Using these simplifications, the ratios $R_{ij}$ can be expressed as:
\beq
R_{ij}= \frac{48\pi^3 \alpha}{G_F^2 m_{SUSY}^4}\left(\vert A_L^{ij} \vert^2+\vert A_R^{ij} \vert^2 \right)\;.
\label{LFV:rij}
\eeq
At the leading order, the amplitudes $A_L^{ij}$ and $A_R^{ij}$ are given by:
\bea
A_L^{ij}&=&a_{LL} (\delta_{ij})_{LL} + a_{RL} \frac{m_{SUSY}}{m_i} (\delta_{ij})_{RL}\nn\\
A_R^{ij}&=&a_{RR} (\delta_{ij})_{RR} + a_{LR} \frac{m_{SUSY}}{m_i} (\delta_{ij})_{LR}
\label{LFV:ALAR}
\eea
where $a_{CC'}$ $(C,C'=L,R)$ are dimensionless functions of the ratios $M_{1,2}/m_{SUSY}$, $\mu/m_{SUSY}$ and of $\tan\theta_W$ and can be found in appendix \ref{AppE:MI} . Their typical size is one tenth of $g^2/(16\pi^2)$, $g$ being the $SU(2)_L$ gauge coupling constant. 

\begin{table}[!ht]
\centering
    \begin{math}
    \begin{array}{|c|c|c|c|c|}
        \hline
        &&&& \\[-9pt]
        ij &w^{LL}_{ij}  &w^{RL}_{ij} &w^{RR}_{ij} &w^{LR}_{ij}\\[10pt]
        \hline
        &&&&\\[-9pt]
        \mu e &\hat{n}_4 &\zeta (c_F-c)& 0 &2   \, \dfrac{(z_e y_\mu - z_\mu y_e)} {y_e y_\mu}\,~c +\zeta (c_F-c)\\[3pt]
        \hline
        &&&&\\[-9pt]
        \tau e &\hat{n}_5 &\zeta (c_F-c)&  0 &2 \, \dfrac{(z_e y_\tau - z_\tau y_e)}{y_e y_\tau} ~c+\zeta (c_F-c)\ \\[3pt]
        \hline
        &&&&\\[-9pt]
        \tau \mu &\hat{n}_6 &\zeta (c_F-c)& 0 &2  \,\dfrac{(z_\mu y_\tau -  z_\tau y_\mu)}{y_\mu y_\tau}~c+\zeta (c_F-c) \\[3pt]
        \hline
    \end{array}
    \end{math}
\caption{\it Coefficients $w^{CC'}_{ij}$ characterising the transition amplitudes for $\mu\to e \gamma$, $\tau\to e \gamma$ and $\tau\to \mu\gamma$, in the MI approximation in which $n_0$ and $n_i^c$ are set to one and $\hat{n}_{1,2}$ to zero so that $w^{RR}_{ij}$ vanish.}
\label{table:coefficientsLFV}
\end{table}

Notice that $a_{CC'}$ do neither depend on $u$ nor on the fermion masses $m_{i,j}$. Finally, $(\delta_{ij})_{CC'}$ parametrise the MIs and are defined as:
\beq
(\delta_{ij})_{CC'}=\frac{(\hat{m}^2_{eCC'})_{ij}}{m^2_{SUSY}}\;.
\eeq
From the mass matrices of the previous section, we find ($j<i$):
\beq
\ba{ll}
(\delta_{ij})_{LL}=w^{LL}_{ij} u^2\;,&\quad
(\delta_{ij})_{RL}=\dfrac{m_i}{m_{SUSY}} \left( w^{RL}_{ij} u +w^{'RL}_{ij}  u^2\right)\\[3mm]
(\delta_{ij})_{RR}=w^{RR}_{ij} \dfrac{m_j}{m_i} u\;,&\quad
(\delta_{ij})_{LR}=w^{LR}_{ij} \dfrac{m_j}{m_{SUSY}} u\;.
\ea
\label{LFV:deltas}
\eeq
where for the mass insertion $(\delta_{ij})_{RL}$ we have also displayed the NLO contributions, in order to better compare our results with those of the effective Lagrangian approach. The explicit expression for the leading order coefficients $w^{CC'}_{ij}$ are listed in table \ref{table:coefficientsLFV}. Also the NLO coefficients $w^{'RL}_{ij} $ are dimensionless combinations of order one parameters. By substituting the mass insertions of eq. (\ref{LFV:deltas}) into the amplitudes $A_{L,R}^{ij}$ of eq. (\ref{LFV:ALAR}) and by using eq. (\ref{LFV:rij}), we get:
\beq
R^{SUSY}_{ij}= \frac{48\pi^3 \alpha}{G_F^2 M^4}\left[\vert w^{(0)}_{ij} u\vert^2+ 2 w^{(0)}_{ij} w^{(1)}_{ij} u^3+
\vert w^{(1)}_{ij} u^2\vert^2+\frac{m_j^2}{m_i^2} \vert w^{(2)}_{ij} u\vert^2\right]
\label{LFV:RSUSY}
\eeq
with $M= (4 \pi/g) m_{SUSY}$ and
\baq
w^{(0)}_{ij}&=&\dfrac{16 \pi^2}{g^2} a_{RL} w^{RL}_{ij}\;,\\[3mm]
w^{(1)}_{ij}&=&\dfrac{16 \pi^2}{g^2} \left(a_{LL} w^{LL}_{ij}+a_{RL}  w^{'RL}_{ij}\right)\;,\\[3mm]
w^{(2)}_{ij}&=&\dfrac{16 \pi^2}{g^2} \left(a_{RR} w^{RR}_{ij}+a_{LR} w^{LR}_{ij}\right)\;.
\label{LFV:wcoeff}
\eaq
The behaviour displayed in eq. (\ref{LFV:RSUSY}) differs from the one expected on the basis of the effective Lagrangian approach in the SUSY case, eq. (\ref{LFV:Eff:LFVsusy}).
This is due to the presence of the term $w^{(0)}_{ij} \propto w^{RL}_{ij}$. Assuming $w^{RL}_{ij}=0$ we recover what is expected from the effective Lagrangian approach
in the supersymmetric case, whereas when $w^{RL}_{ij}$ does not vanish, the leading order behaviour matches the prediction of the effective Lagrangian
approach in the generic, non-supersymmetric case, eq. (\ref{LFV:Eff:LFV}). As shown in table \ref{table:coefficientsLFV}, the coefficient $w^{RL}_{ij}$ is universal, namely it is independent from the flavour indices and it vanishes in two cases:
\begin{itemize}
\item[i)]
$c_F=c$, which reflects the alignment of the VEVs of the scalar
and auxiliary components of the flavon supermultiplet $\varphi_T$, see eqs. (\ref{LFV:vevs}) and (\ref{LFV:VEVsauxphiT}).
\item[ii)]
$\zeta=0$ which can be realised by special choices of the soft Supersymmetry breaking terms
in the flavon sector, i.e. the assumption of universal soft Supersymmetry breaking terms in the flavon potential.
\end{itemize}
In our model none of these possibilities is natural, see \cite{FHM_VEV}, and both require a tuning of the underlying parameters.
If $w^{RL}_{ij}=0$, the result expected from the effective Lagrangian approach in the supersymmetric case is obtained in a non-trivial way. Indeed, it is a consequence of a cancellation taking place
when going from the Lagrangian to the physical basis. In particular, for $w^{RL}_{ij}=0$, $R^{SUSY}_{ij}$ scales as $u^4$ and not as $u^2$
for $m_j=0$.

In the general case when $w^{RL}_{ij}$ is non-vanishing, the dominant contribution to $R_{ij}^{SUSY}$ regarding the expansion in $u$
is flavour independent and, at the leading order in the $u$ expansion,
we predict $R_{\mu e}=R_{\tau\mu}=R_{\tau e}$, at variance with the predictions
of most of the other models, where, for instance, $R_{\mu e}/R_{\tau \mu}$ can be much smaller than one \cite{Raidal,MLFV1,MLFVother,SUSYLFV+symmetries}.
If $w^{RL}_{ij}$ is non-vanishing, it is interesting to analyse the relative weight of the leading and subleading contributions to $R_{ij}$.
For this purpose we calculate the the numerical values of the functions $a_{CC'}$, in the limit $\mu=M_{1,2}=m_{SUSY}$:
\beq
a_{LL}\sim+(2.0\pm16.3)\;,\qquad\quad
a_{RL}=a_{LR}\sim+0.30\;,\qquad\quad
a_{RR}\sim-(0.5\div1.3)\;.
\eeq

As one can see, in this limit the dominant coefficient is  ${a}_{LL}$, which is larger than $a_{RL}=a_{LR}$ by a factor $7\div 54$,
and larger than $a_{RR}$ by a factor $-(4\div 13)$, depending on  $\tan\beta=2\div 15$. Assuming coefficients $w_{ij}^{CC'}$ of order one
in eqs. (\ref{LFV:wcoeff}), we see that the most important contributions in the amplitudes for the considered processes are $a_{RL} u$ and $a_{LL} u^2$. The ratio between the subleading and the leading one is $(a_{LL}/a_{RL}) u\approx (7\div 54) u$. When $u$ is close to its lower bound, which in our model requires a small value of $\tan\beta$, the leading contribution clearly dominates over the subleading one. However, for $u$ close to $0.05$, which allows to consider larger values of $\tan\beta\approx 15$, the non-leading contribution can be as large as the leading one and can even dominate over it. The transition between the two regimes occurs towards larger values of $u$.

The numerical dominance of the coefficient $a_{LL}$ has also another consequence: for vanishing $w^{RL}_{ij}$, $R_{ij}$ is dominated by the contributions of $a_{LL} w^{LL}_{ij}$, whose values are not universal, but expected to be of the same order of magnitude for all channels. Thus even when $w^{RL}_{ij}=0$, we predict $R_{\mu e}\approx R_{\tau\mu}\approx R_{\tau e}$.

\section{Numerical analysis}
\label{Sec:LFV:Numerical}
\setcounter{footnote}{3}

In this section we perform a numerical study of the normalised branching ratios $R_{ij}$ and of the deviation $\delta a_\mu$  of the anomalous magnetic moment of the muon from the Standard Model value. We use the full one-loop results for the branching ratios of the radiative decays
as well as for $\delta a_\mu$. These can be found in \cite{deltaamu_susy,HisanoFukuyama,Arganda,g2BR} and are displayed in appendix \ref{AppE:one-loop} for convenience.

\subsection{Framework}
\label{Sec:LFV:Framework}
\setcounter{footnote}{3}

As discussed in the preceding sections, in our model the flavour symmetry $A_4 \times Z_3 \times U(1)_{FN} \times U(1)_R$ constrains not only the mass matrices of leptons, but also those of sfermions. These are given at the high energy scale $\La_f \approx \La_L$, which we assume to be close to $10^{16}$ GeV, the supersymmetric grand unification scale. The flavour symmetry does not fix the soft supersymmetric mass scale $m_{SUSY}$. It also does not constrain the parameters involved in the gaugino as well as the Higgs(ino) sector. These are fixed by
our choice of a SUGRA framework in which $m_{SUSY}$ is the common soft mass scale for all scalar particles and $m_{1/2}$ the common mass scale of the gauginos. Thus, at the scale $\La_f \approx \La_L$ we have 
\beq
M_1 (\La_L) = M_2 (\La_L) = m_{1/2} \; .
\eeq
Effects of RG running lead at low energies (at the scale $m_Z$ of the $Z$ mass) to the following masses for gauginos
\beq
M_1(m_Z)\simeq\dfrac{\alpha_1(m_Z)}{\alpha_1(\Lambda_L)}M_1(\Lambda_L)\qquad\qquad
M_2(m_Z)\simeq\dfrac{\alpha_2(m_Z)}{\alpha_2(\Lambda_L)}M_2(\Lambda_L)\;,
\eeq
where $\alpha_i=g_i^2/4\pi$ ($i=1,2$) and according to gauge coupling unification at $\La_f\approx \La_L$, $\alpha_1(\Lambda_L)=\alpha_2(\Lambda_L)\simeq 1/25$. Concerning the effects of the RG running on the soft mass terms, as we have seen in section \ref{Sec:LFV:Phys} these are small or can be absorbed into our parametrisation of the soft mass terms. Thus, in the contributions $(\hat{m}_{eRL}^2)_{1,2}$ to the RL block we take $m_{SUSY}$ as input parameter. Nevertheless, we explicitly take into account the RG effect on the average mass scale of the LL block, $m_L^2$, and in the RR block, $m_R^2$,
\beq
\begin{array}{rcl}
m_L^2(m_Z)&\simeq& m_L^2(\Lambda_L)+0.5M_2^2(\Lambda_L)+0.04M_1^2(\Lambda_L) \simeq m_{SUSY}^2 +0.54 m_{1/2}^2 \; ,\\[3mm]
m_R^2(m_Z)&\simeq& m_R^2(\Lambda_L)+0.15M_1^2(\Lambda_L) \simeq m_{SUSY}^2 +0.15 m_{1/2}^2 \; .
\end{array}
\eeq
The parameter $\mu$ is fixed through the requirement of correct electroweak symmetry breaking: 
\beq
|\mu|^2\simeq-\dfrac{m_Z^2}{2}+m_{SUSY}^2\dfrac{1+0.5\tan^2\beta}{\tan^2\beta-1}+m_{1/2}^2\dfrac{0.5+3.5 \tan^2\beta}{\tan^2\beta-1}\;,
\label{LFV:defmuSUGRA}
\eeq
so that $\mu$ is determined by $m_{SUSY}$, $m_{1/2}$  and $\tan\beta$ up to its sign. We recall that in our model the low-energy parameter $\tan\beta$ is related to the size of the expansion parameter $u$, the mass of the $\tau$ lepton and the $\tau$ Yukawa coupling $y_\tau$ in as in the following
\beq
u=\dfrac{\tan\beta}{|y_\tau|}\dfrac{\sqrt2 m_\tau}{v}\approx0.01\dfrac{\tan\beta}{|y_\tau|}\;,
\label{LFV:tanb&u&yt}
\eeq
as we have already pointed out in eq. (\ref{AFTBM:tanb&u&yt}). For this reason, requiring $1/3 \lesssim |y_\tau| \lesssim 3$
constrains $\tan\beta$ to lie in the range $2 \lesssim \tan\beta \lesssim 15$. As already commented, the lower bound $\tan\beta =2$
is almost excluded experimentally, since such low values of $\tan\beta$ usually lead to a mass for the lightest Higgs below the LEP2 bound
of $114.4$ GeV \cite{mhbound}. \footnote{This bound assumes that the Higgs is SM-like. For the case of generic MSSM Higgs the bound is much lower, $91.0$ GeV \cite{MSSMmhbound}.}

In our numerical analysis the parameters are the following: the two independent mass scales $m_{SUSY}$ and $m_{1/2}$, the sign of the parameter $\mu$ and the parameters of the slepton mass matrices shown in section \ref{Sec:LFV:Phys} in the physical basis. We recall that the results of section \ref{Sec:LFV:Phys} have been obtained under the assumption that the parameters are real and we keep working under the same assumption here.  We also assume that the parameters on the diagonal of the slepton mass matrices $(\hat{m}_{(e,\nu)LL}^2)_K$ and $(\hat{m}_{eRR}^2)_K$, $n_0$ and $n^c_{1,2,3}$, are positive in order to favour positive definite square-masses, to avoid electric-charge breaking minima and further sources of electroweak symmetry breaking. The absolute value of the ${\cal O}(1)$ parameters is varied between $1/2$ and $2$. We will choose some representative values for $u$ in the allowed range $0.007 \lesssim u \lesssim 0.05$. The other expansion parameter $t$ is fixed to be $0.05$. In the analysis of the normalised branching ratios $R_{ij}$ we fix $\tan\beta$ and $u$ and then we derive the Yukawa couplings $y_e$, $y_\mu$ and $y_\tau$. When discussing the anomalous magnetic moment of the muon instead we vary $y_\tau$ between $1/3$ and $3$ and calculate $\tan\beta$ by using eq. (\ref{LFV:tanb&u&yt}). Having determined $\tan\beta$, the Yukawa couplings $y_e$ and $y_\mu$ can be computed.

The allowed region of the parameter space is determined by performing several tests. We check whether the mass of the lightest chargino is above 100 GeV \cite{PDG08}, whether the lightest neutralino is lighter than the lightest charged slepton, whether the lower bounds for the charged slepton masses are obeyed \cite{PDG08} and whether the masses of all sleptons are positive. The constraint on the mass of the lightest chargino implies a lower bound on $m_{1/2}$ which slightly depends on the sign of $\mu$. In our plots for $R_{ij}$ we also show the results for points of the parameter space that do not respect the chargino mass bound. For low values of $m_{SUSY}$, e.g. $m_{SUSY}=100$ GeV, the requirement that the lightest neutralino is lighter than the lightest charged slepton is equivalent to the requirement that the parameters in the diagonal entries of the slepton mass matrices $(\hat{m}_{(e,\nu)LL}^2)_K$ and $(\hat{m}_{eRR}^2)_K$ are larger than one.
For larger values of $m_{SUSY}$, e.g. $m_{SUSY}=1000$ GeV, this requirement does not affect our analysis anymore. We note that masses of charginos and neutralinos  are essentially independent from the ${\cal O}(1)$ parameters of the slepton mass matrices and thus their masses fulfill with very good accuracy (better for larger $m_{1/2}$)
\beq
M_{ \widetilde{\chi}^0_1} \approx 0.4 m_{1/2} \;,\qquad
M_{ \widetilde{\chi}^0_2} \approx M_{ \widetilde{\chi}^-_1} \approx 0.8 m_{1/2} \;,\qquad
M_{ \widetilde{\chi}^0_3} \approx M_{ \widetilde{\chi}^0_4} \approx  M_{ \widetilde{\chi}^-_2} \approx |\mu| \;.
\eeq
For the slepton masses we find certain ranges which depend on our choice of the ${\cal O}(1)$ parameters.

\subsection{Results for Radiative Leptonic Decays}
\label{Sec:LFV:ResultsDecays}
\setcounter{footnote}{3}

We first discuss the results for the branching ratio of the decay $\mu\to e\gamma$. This branching ratio is severely constrained
by the result of the MEGA experiment \cite{MEGA}
\beq
R_{\mu e} \approx BR(\mu\to e\gamma) < 1.2 \times 10^{-11}
\eeq
and will be even more constrained by the on-going MEG experiment \cite{MEG} which will probe the regime
\beq
R_{\mu e}\approx BR(\mu\to e\gamma) \gtrsim 10^{-13} \; .
\eeq
We explore the parameter space of the model by considering two different values of the expansion parameter $u$, $u=0.01$ and $u=0.05$, two different values of $\tan\beta$, $\tan\beta=2$ and $\tan\beta=15$, as well as two different values of the mass scale $m_{SUSY}$, $m_{SUSY}=100$ GeV and $m_{SUSY}=1000$ GeV. We show our results in scatter plots in figure \ref{LFV:ScatterBR} choosing $m_{1/2}$ to be $m_{1/2} \lesssim 1000$ GeV. All plots shown in figure \ref{LFV:ScatterBR} are generated for $\mu>0$.

\begin{figure}
 \centering
\subfigure[$\tan\beta=2$, $u=0.01$ and $m_{SUSY}=100$ GeV.]
   {\includegraphics[width=7.8cm]{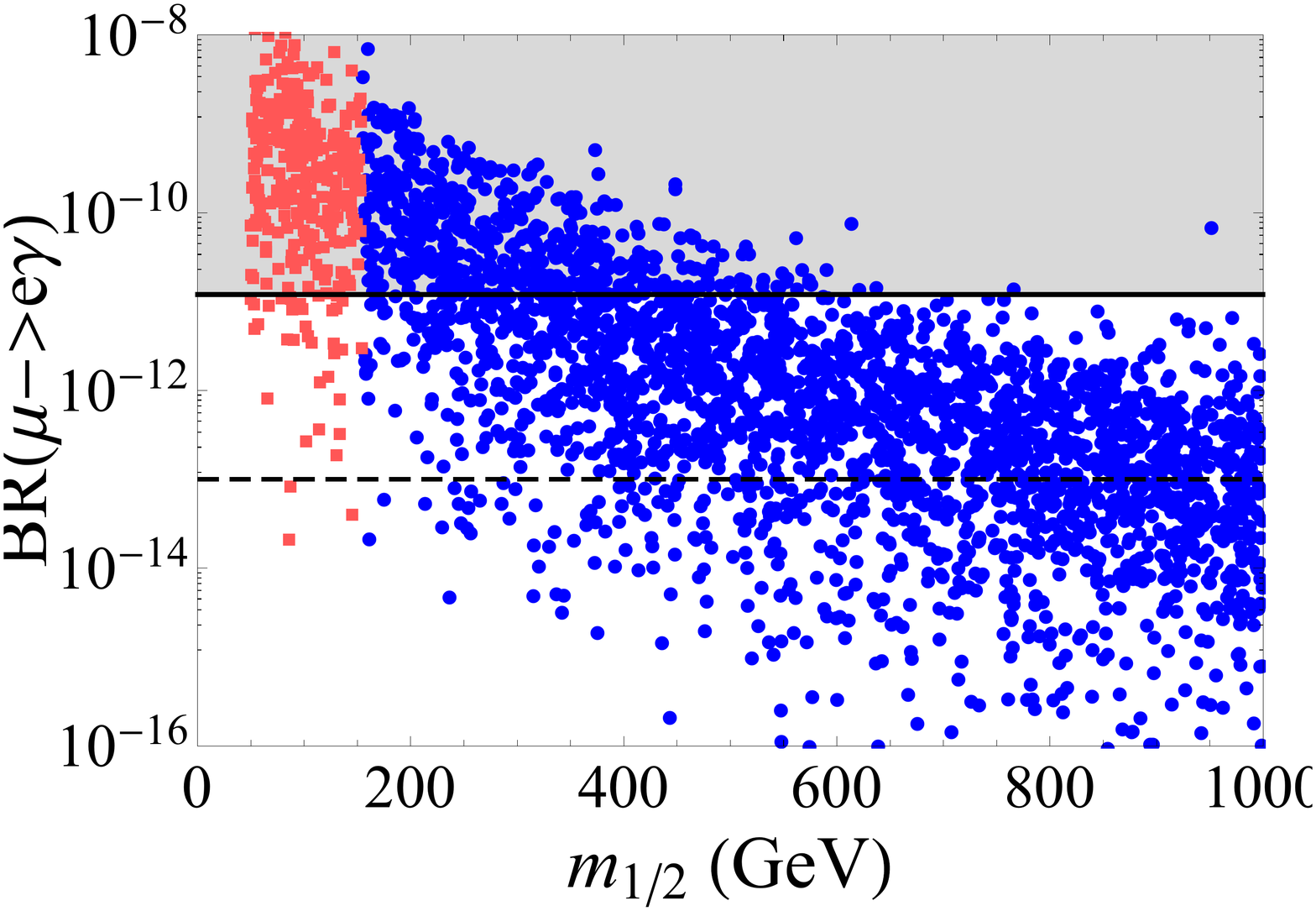}}
\subfigure[$\tan\beta=2$, $u=0.01$ and $m_{SUSY}=1000$ GeV.]
   {\includegraphics[width=7.8cm]{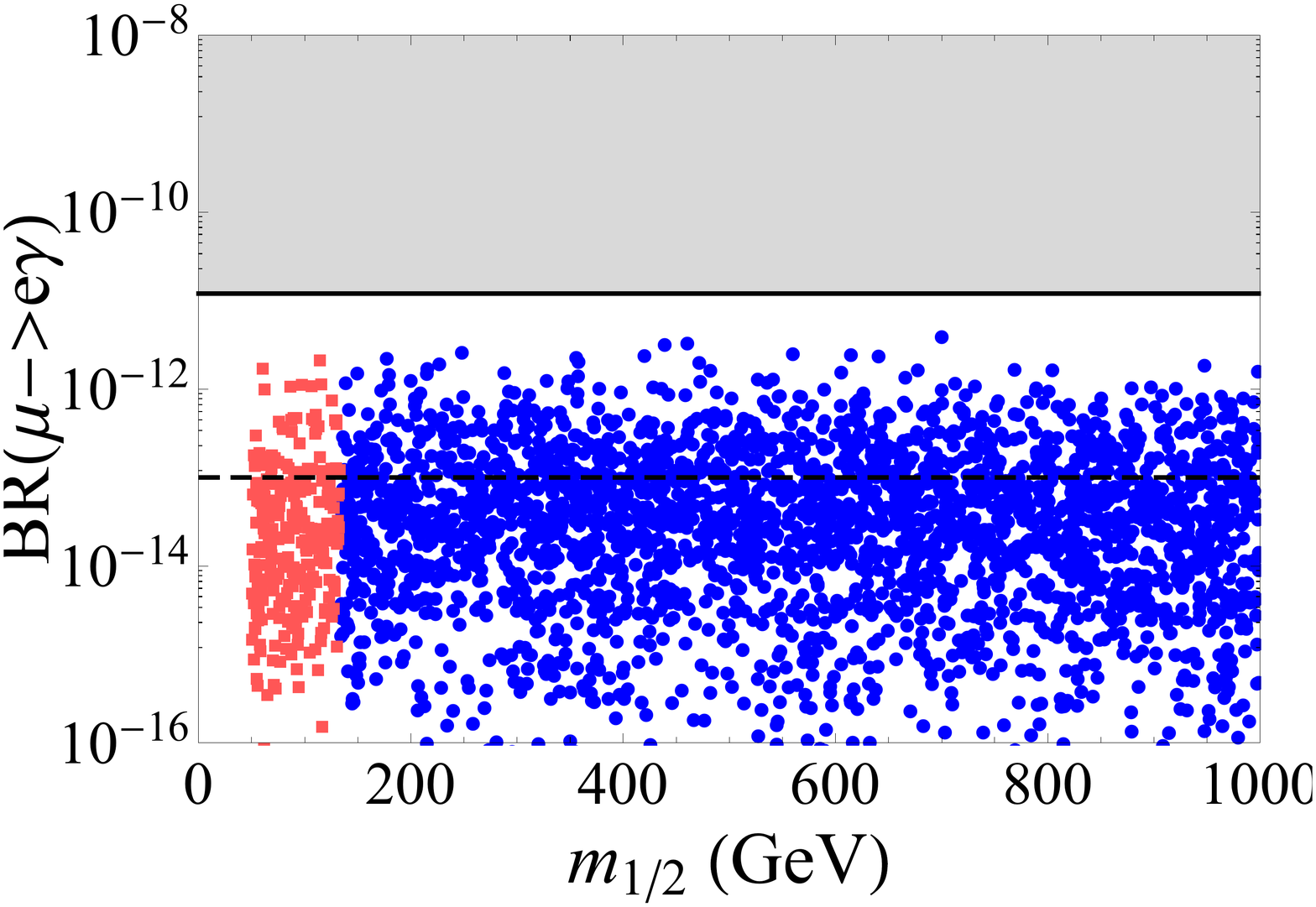}}
\subfigure[$\tan\beta=2$, $u=0.05$ and $m_{SUSY}=100$ GeV.]
   {\includegraphics[width=7.8cm]{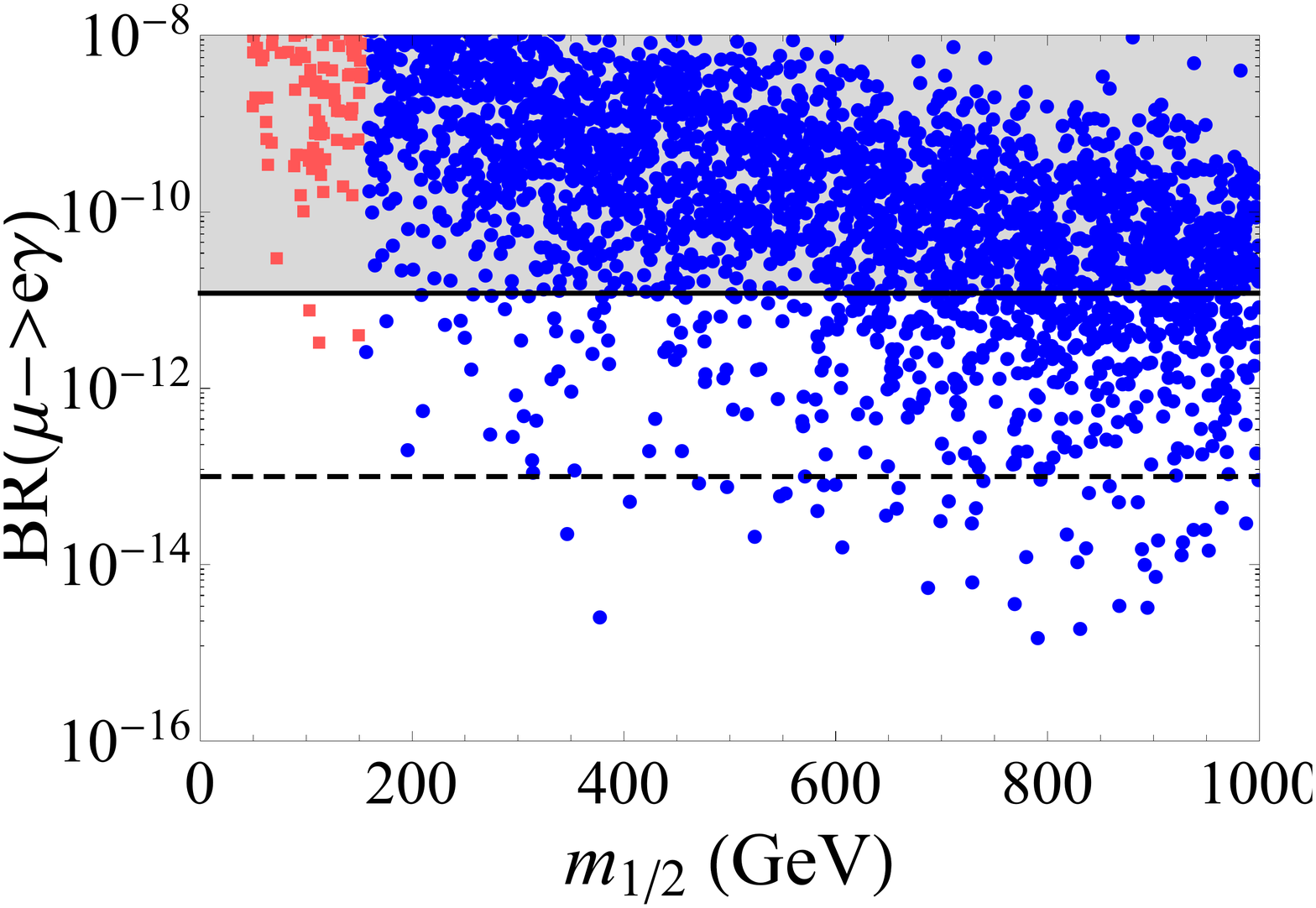}}
\subfigure[$\tan\beta=2$, $u=0.05$ and $m_{SUSY}=1000$ GeV.]
   {\includegraphics[width=7.8cm]{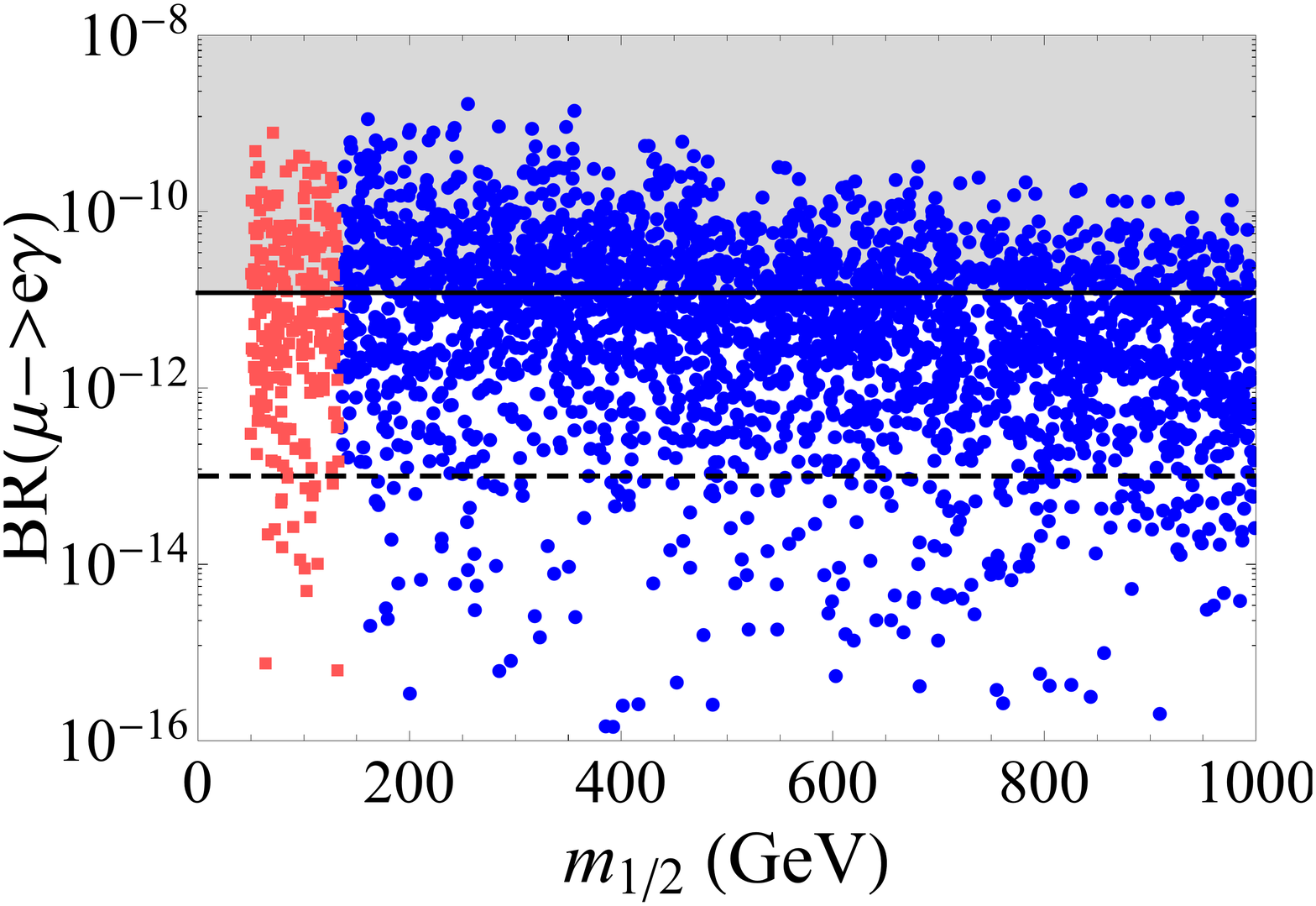}}
\subfigure[$\tan\beta=15$, $u=0.05$ and $m_{SUSY}=100$ GeV.]
   {\includegraphics[width=7.8cm]{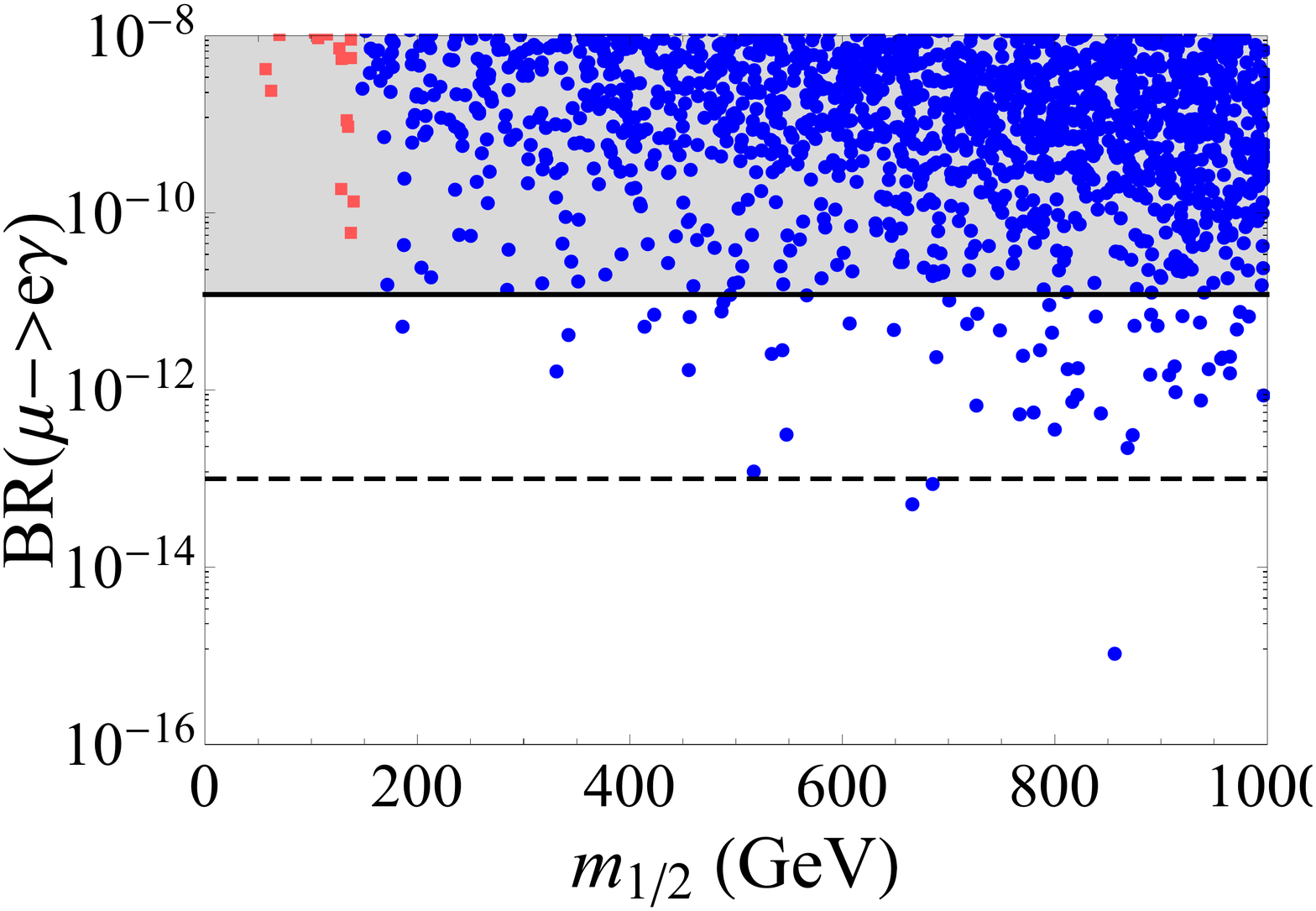}}
\subfigure[$\tan\beta=15$, $u=0.05$ and $m_{SUSY}=1000$ GeV.]
   {\includegraphics[width=7.8cm]{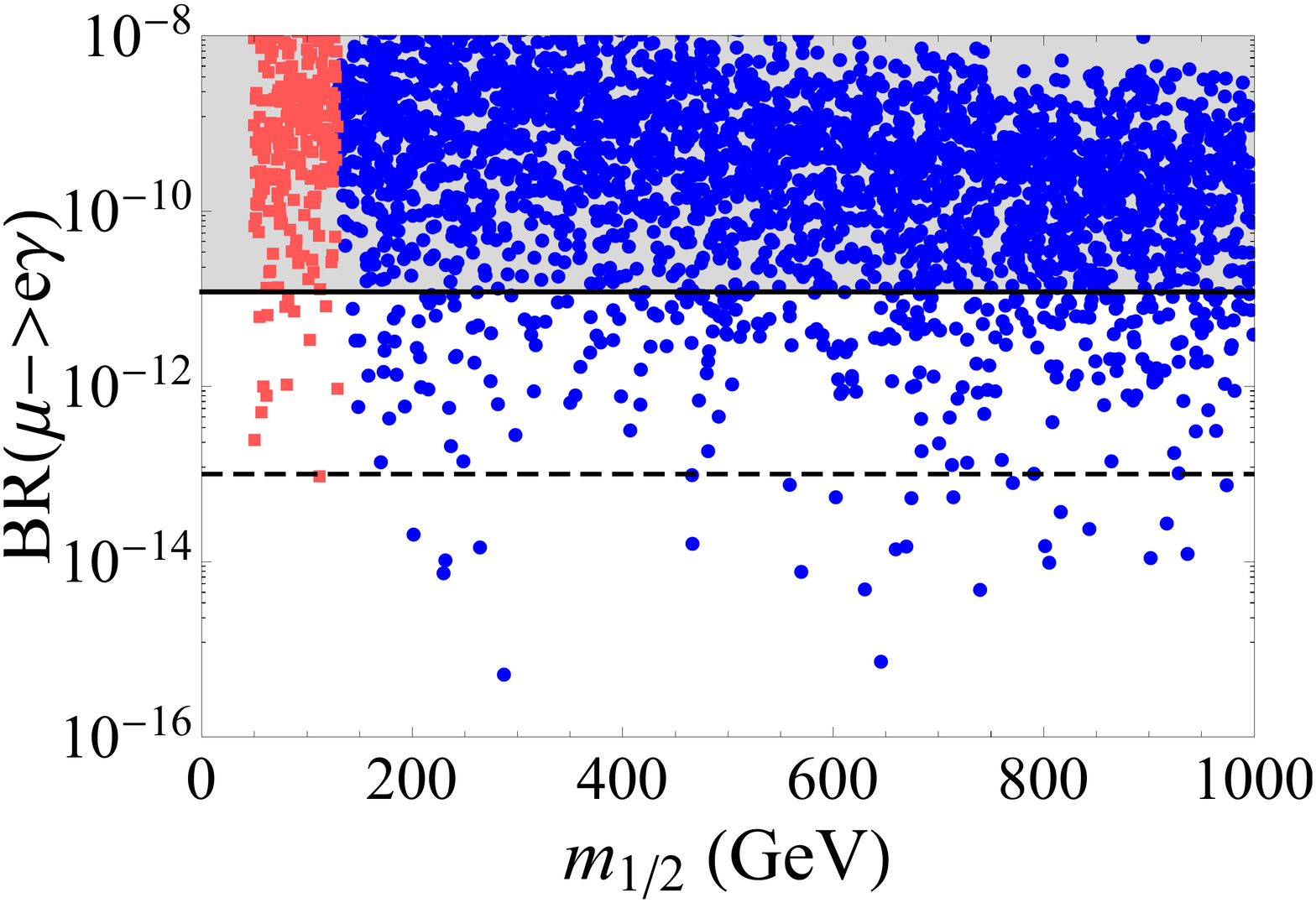}}
\vspace{-0.3cm}
\caption{\it Scatter plots of $BR(\mu\to e \gamma)$ as a function of $m_{1/2}$, for different values of $\tan\beta$, $u$ and $m_{SUSY}$.
The red (dark gray) points correspond to points in which the mass of the lightest chargino is below the limit coming from direct searches.
The horizontal lines show the current MEGA bound (continuous line) and the prospective MEG bound (dashed line).}
\label{LFV:ScatterBR}
\end{figure}

As one can see from figure \ref{LFV:ScatterBR}(a), for very low $\tan\beta=2$, small $u=0.01$, small $m_{SUSY}=100$ GeV the experimental upper limit from the MEGA experiment on $BR(\mu\to e\gamma)$ can be passed in almost all parameter space of our model for values of $m_{1/2}$ as small as $450$ GeV. For $m_{SUSY}=100$ GeV and $m_{1/2}=450$ GeV the sparticle masses are rather light: the lightest neutralino has a mass of $175$ GeV, the lightest chargino of $350$ GeV, the masses of the right-handed (charged) sleptons vary between $175$ and $285$ GeV and the masses of the left-handed sleptons are in the range $(250\div 500)$ GeV. Thus, especially the right-handed sleptons are expected to be detected at LHC. In a model also including quarks (and hence squarks) we find for the squarks that they can have masses $\gtrsim 700$ GeV and gluinos with masses of about $1000$ GeV, all accessible at LHC. To pass the prospective bound coming from the MEG experiment in a sizable portion of parameter space of our model $m_{1/2}$ has to be chosen larger, $m_{1/2} \gtrsim 600$ GeV. Then, however, the masses of the sleptons might only be detected at LHC in case of right-handed sleptons. As one can see, values of $m_{1/2} \lesssim 155$ GeV are excluded due to the constraint on the lightest chargino mass. Studying the same value of $\tan\beta$ and $u$, but taking $m_{SUSY}$ to be as large as $1000$ GeV, we can see from figure \ref{LFV:ScatterBR}(b) that now the bound set by the MEGA experiment on $BR(\mu\to e\gamma)$ is respected in the whole parameter space of our model for all values of $m_{1/2}$. Also the foreseen limit of the MEG experiment can only exclude a smaller part of the parameter space of the model for all values of $m_{1/2}$. In this setup, the prospects for detecting supersymmetric particles at LHC are the best for gauginos due to the possible low value of $m_{1/2}$. The slepton masses are expected to be roughly $m_{SUSY}$ and thus too large to allow for a detection at LHC.

Increasing the value of the expansion parameter $u$ from $u=0.01$ to $u=0.05$, as done in figure \ref{LFV:ScatterBR}(c) and \ref{LFV:ScatterBR}(d), increases also the branching ratio of the decay $\mu\to e\gamma$ by approximately two orders of magnitude, since the different contributions to the branching ratio scale at least with $u^2$, as analysed in section \ref{Sec:LFV:MI}. For this reason for low values of $m_{SUSY}=100$ GeV, $m_{1/2}$ has to take values $m_{1/2} \gtrsim 600$ GeV in order for the result of $BR(\mu\to e\gamma)$ to be compatible with the MEGA bound at least in some portion of the parameter space of our model. For the point $(m_{SUSY},m_{1/2})=(100 \, \rm GeV, 600 \, GeV)$ the sparticle spectrum is characterised as follows: the lightest neutralino has a mass of $240$ GeV, the lightest chargino of $470$ GeV, right-handed sleptons between $250$ and $350$ GeV and left-handed sleptons generally above $300$ GeV. For this reason
there still exists the possibility to detect right-handed sleptons at LHC. Concerning gluinos and squarks these are expected to have masses between $1000$ and $1500$ GeV so that they also can be detected at LHC. As one can see from figure \ref{LFV:ScatterBR}(c) the on-going
MEG experiment can probe nearly the whole parameter space of the model for $\tan\beta=2$, $u=0.05$ and $m_{SUSY}=100$ GeV for values
of $m_{1/2} \lesssim 1000$ GeV. Increasing the parameter $m_{SUSY}$ to $1000$ GeV shows that applying the existing bound on $BR(\mu\to e\gamma)$ of $1.2 \times 10^{-11}$ cannot exclude small values of $m_{1/2}$. The situation changes, if the expected bound from the MEG experiment is employed, because then values of $m_{1/2}$ smaller than $1000$ GeV become disfavoured.

Finally, we show in figure \ref{LFV:ScatterBR}(e) and \ref{LFV:ScatterBR}(f) the results obtained for $\tan\beta=15$. We remind that this value is the largest possible one of $\tan\beta$ in our model. Requiring that the $\tau$ Yukawa coupling does not become too large entails that $\tan\beta=15$ fixes the expansion parameter $u$ to take a value close to its upper limit, $u=0.05$. The value of $BR(\mu\to e \gamma)$ is thus enhanced through $\tan\beta$ as well as $u$. This is clearly shown in  figure \ref{LFV:ScatterBR}(e) and \ref{LFV:ScatterBR}(f), because for a low value of $m_{SUSY}=100$ GeV already the MEGA bound practically excludes almost the whole parameter space of our model
for all values of $m_{1/2} \lesssim 1000$ GeV.  Increasing the mass parameter $m_{SUSY}$ to $1000$ GeV slightly improves the situation, because now there exists a marginal probability to pass the MEGA bound. Again, however, the MEG experiment can probe all parameter space of our model for $m_{1/2} \lesssim 1000$ GeV. Thus, for $m_{SUSY} \lesssim 1000$ GeV and $m_{1/2} \lesssim 1000$ GeV  the parameter space of our model is already severely constrained for moderate values of $\tan\beta$ which entail large $u \approx 0.05$ by the bound coming from the MEGA collaboration, but surely will be conclusively probed by the MEG experiment. Choosing $\mu<0$ hardly affects the results presented here apart from slightly decreasing the lower bound on $m_{1/2}$ coming from the chargino mass bound. Thus, all statements made also apply for $\mu<0$.

In summary, the current bound on $BR(\mu\to e\gamma)$ prefers regions in the parameter space of our model with small $u$ or small $\tan\beta$, as long as the SUGRA mass parameters should be chosen smaller than $1000$ GeV. The foreseen MEG bound strongly favours regions in which $u$ is small for $m_{SUSY}$ and $m_{1/2}$ being not too large. The fact that smaller values of $u$ are preferred has consequences also for the expectations of the detection prospects for the reactor mixing angle $\theta_{13}$, because this angle scales with $u$: it might thus not be possible to detect $\theta_{13}$ with the reactor and neutrino beam experiments under preparation \cite{NeutrinoData,Theta13future}.

Concerning the radiative $\tau$ decays, $\tau\to \mu\gamma$ and $\tau\to e\gamma$, the result found in the MI approximation that
the branching ratios of these decays are of the same order of magnitude as $BR(\mu\to e\gamma)$ is essentially confirmed in a numerical
analysis. Due to the random parameters differences up to two orders of magnitude are expected and found, especially for the case of larger
$\tan\beta$: looking at table \ref{table:coefficientsLFV}, when the parameters $c_F$ and $c$ are of opposite sign the leading contributions are suppressed and the non-universal NLO terms become relevant. However, it is still highly improbable that the decays $\tau\to \mu\gamma$ and $\tau\to e\gamma$ could be detected at a SuperB factory, assuming a prospective limit of $BR(\tau\to \mu \gamma)$, $BR(\tau\to e \gamma) \gtrsim 10^{-9}$ \cite{SuperB}.

\subsection{Results for Anomalous Magnetic Moment of the Muon}
\label{Sec:LFV:ResultsAmu}
\setcounter{footnote}{3}

As is well known, the value found for the anomalous magnetic moment of the muon \cite{ExperimentalBoundsMDMmu}
\beq a_\mu^{EXP}=116592080(63)\times 10^{-11}
\eeq
shows a 3.4 $\sigma$ deviation
\beq \delta a_\mu=a_\mu^{EXP}-a_\mu^{SM}=+302(88)\times 10^{-11}
\label{LFV:Da}
\eeq
from the value expected in the Standard Model
\beq a_\mu^{SM}=116591778(61)\times 10^{-11} \; .
\eeq
Thus, it might be interesting to consider the case in which this deviation is attributed to the presence of supersymmetric particles with masses of a few hundred GeV. The one-loop contribution to the anomalous magnetic moment of the muon in supersymmetric extensions of the Standard Model has been studied by several authors \cite{deltaamu_susy}.

\begin{figure}[h!]
 \centering
\subfigure[$u=0.01$.]
   {\includegraphics[width=7.8cm]{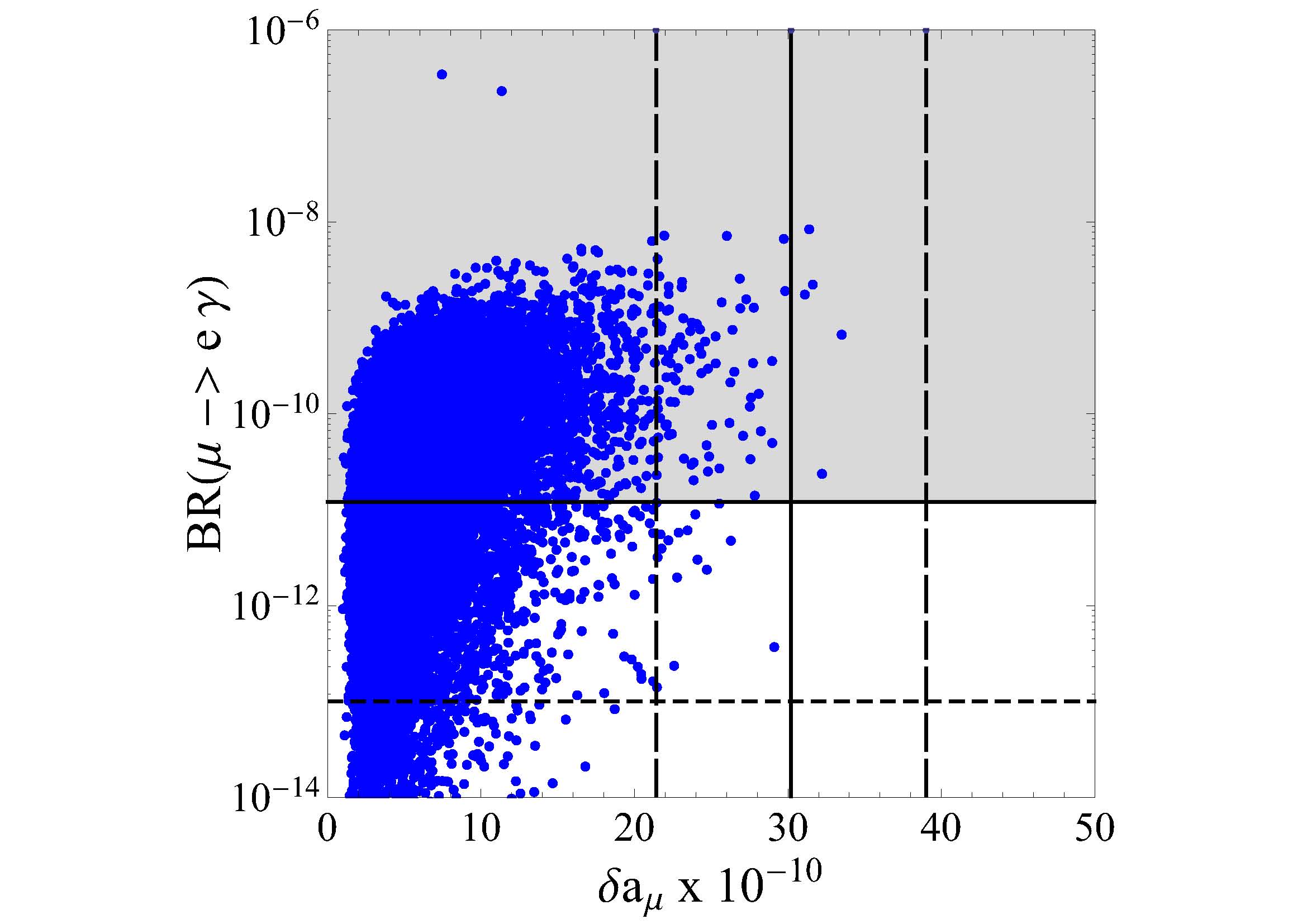}}
\subfigure[$u=0.05$.]
   {\includegraphics[width=7.8cm]{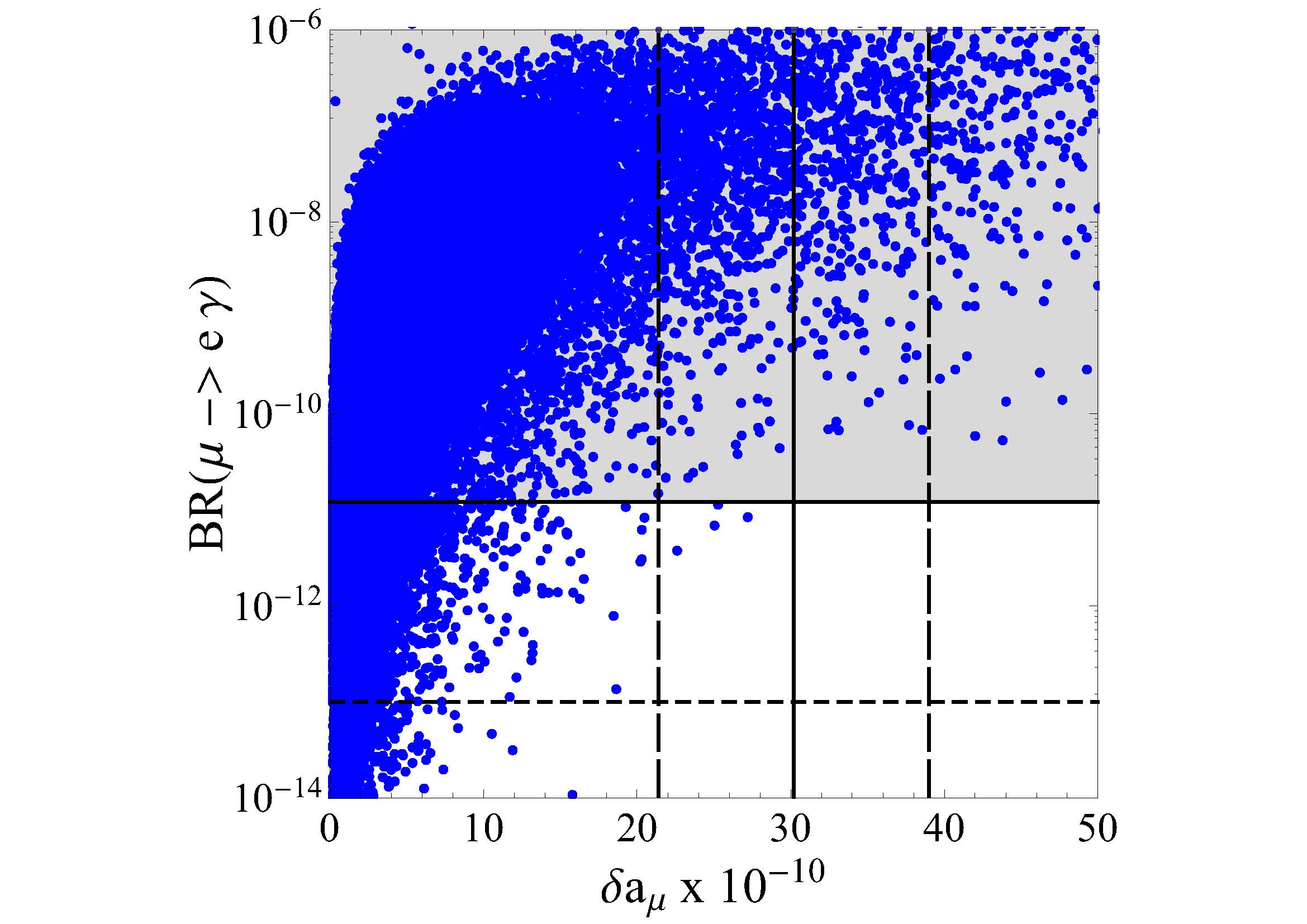}}
\vspace{-0.3cm}
\caption{\it Scatter plots in the plane $BR(\mu\to e\gamma)-\delta a_\mu$, for values of $u=0.01,\,0.05$. The value of $\tan\beta$ is fixed through the relation with the $\tau$ Yukawa coupling, which lies in the interval $[1/3,3]$. The values of $m_{SUSY}$ and of $m_{1/2}$ are chosen between $10$ and $300$ GeV for the left panel and between $10$ and $1000$ GeV in the right one. The horizontal lines correspond to the MEGA (continuous line) and the MEG bounds (dashed line); the vertical lines correspond to the measurements on $\delta a_\mu$: the continuous one is the best fit value and the dashed ones correspond to the $3\sigma$ boundaries.}
\label{LFV:ScatterBrG2}
\end{figure}

We study the compatibility between the requirement that $\delta a_\mu$ is explained by the exchange of relatively light supersymmetric particles
and the experimental upper limit on $BR(\mu\to e\gamma)$ coming from the MEGA experiment. We choose again two different values of $u$, $u=0.01$ and $u=0.05$. To better explore the parameter space we vary the $\tau$ Yukawa coupling between $1/3$ and $3$ and fix the value of $\tan\beta$ through the relation given in eq. (\ref{LFV:tanb&u&yt}). As a consequence in the plot for $u=0.01$, see figure \ref{LFV:ScatterBrG2}(a), $2 \lesssim \tan\beta \lesssim 3$ holds and for $u=0.05$ $\tan\beta$ takes values $2\lesssim \tan\beta \lesssim 15$. Similarly, $m_{SUSY}$ and $m_{1/2}$ are chosen to lie in intervals $[10 \, \rm GeV, 300 \, GeV]$ and $[10 \, \rm GeV, 1000 \, GeV]$ for $u=0.01$ and $u=0.05$, respectively. The different choice of intervals is due to the fact that values of a few hundred GeV for $m_{SUSY}$ and $m_{1/2}$ are disfavoured by the existing limit on $BR(\mu\to e\gamma)$ when $u=0.05$. As one can clearly see from figure \ref{LFV:ScatterBrG2}, in almost the whole parameter space of our model it is not natural to reproduce the observed deviation of the muon anomalous magnetic moment and at the same time to respect the existing bound on the branching ratio of $\mu\to e \gamma$.

This kind of incompatibility is well known in constrained supersymmetric theories, because the explanation of the $3.4 \sigma$ discrepancy necessitates small values of $m_{SUSY}$ and $m_{1/2}$ and larger values of $\tan\beta$, which in turn enhance the branching ratio of the radiative LFV decays. Thus, we have to conclude that either there exist further sources of contributions to the anomalous magnetic moment of the muon beyond those present in our model, or - as is also discussed in the literature - the theoretical value $a_\mu^{SM}$ found in the Standard Model is closer to $a_\mu^{EXP}$ so that the possible discrepancy becomes less than $100 \times 10^{-11}$, a value which could well be explained in our model.

\section{Conclusions of the Chapter}
\label{Sec:LFV:Conclusion}
\setcounter{footnote}{3}

In this chapter we studied a series of phenomena involving flavour transitions in order to find new observables, not related to neutrino oscillations, which can be useful to test the numerous flavour models present in literature. The introduction of new physics, supersymmetric or not, corresponding to an energy scale at about $M=1\div10$ TeV is an interesting possibility: a solution to the hierarchy problem, a successful gauge coupling unification, the existence of a Dark Matter candidate and an explanation of the observed discrepancy in the anomalous magnetic moment of the muon might be found if this intermediate energy scale is introduced.

We pursed the analysis considering a set of flavour models, based on the symmetry group $A_4\times Z_3\times U(1)_{FN}$, originally proposed in order to describe lepton masses and mixing angles, through the introduction the Weinberg operator  (for alternative studies with the type I See-Saw see \cite{LFVA4+SeeSaw}). The introduction of new physics in this class of models allows to study a set of new low-energy observables, as leptonic MDMs, EDMs and LFV transitions like $\mu\to e \gamma$, $\tau\to\mu\gamma$ and $\tau\to e \gamma$. We have constructed the effective low-energy Lagrangian that describes these observables. Such an effective Lagrangian is invariant under the flavour symmetry, and all flavour breaking effects are encoded in the dependence on a set of flavons $\varphi$ which develop VEVs. These also control lepton masses and mixing angles. The dominant operators are obtained by expanding the Lagrangian in powers of $\varphi/\Lambda_f$ and by keeping the first few terms in the expansion. The leading contributions have dimension six and are suppressed by two powers of a new scale $M$. Apart from this energy scale, all the relevant information needed to predict MDMs, EDMs and LFV transitions is contained in a dimensionless matrix $\mathcal{\hat M}$, whose elements can be computed up to unknown order-one coefficients from our Lagrangian.
The strongest bound comes from the EDM of the electron: $M>80$ TeV. A lower value for $M$ can be tolerated in the presence of a cancellation in the imaginary part of $\mathcal{\hat{M}}_{ee}$, perhaps related to CP-conservation in the considered sector of the theory.
This problem is also present in MFV\cite{MFV,MLFV1,MLFVother}, where one simply assumes that $\mathcal{\hat{M}}_{ee}$ is real.
Coming to LFV dipole transitions, we have found that in the general case the branching ratios for $\mu\to e \gamma$, $\tau\to \mu\gamma$ and $\tau \to\ e \gamma$ are all expected to be of the same order, at variance with MFV. Given the present limit on $BR(\mu\to e \gamma)$, this implies that $\tau\to \mu\gamma$ and $\tau \to\ e \gamma$ have rates much smaller than the present (and near future) sensitivity. The absolute values of these branching ratios depend on the flavon VEVs that in our class of models are determined by a parameter $u$ in the range $0.001<u<0.05$. In the general case, for $BR(\mu\to e \gamma)<1.2\times 10^{-11}~(10^{-13})$ we get $M>10~(30)$ TeV if $u=0.001$ and $M>70~(200)$ TeV for $u=0.05$. The anomalous MDM of the muon $a_\mu$ and its deviation from the SM expectation provide the indication for a lower scale $M$,
of the order of few TeV, which would also be of great interest for LHC. In order to reconcile this possibility with the results derived from the LFV dipole transitions, we have reconsidered the matrix $\mathcal{\hat M}$ in a supersymmetric context, where additional constraints have to be applied. The operators describing $\mu\to e \gamma$, $\tau\to \mu\gamma$ and $\tau \to\ e \gamma$ flip the lepton chirality. By assuming that in a supersymmetric theory the only sources of chirality flips are the fermion masses and the sfermion mass terms of left-right type, we find that a cancellation takes place in the elements of $\mathcal{\hat M}$ below the diagonal. As a result the limits on the scale $M$ become less severe.
For $BR(\mu\to e \gamma)<1.2\times 10^{-11}~(10^{-13})$ we get $M>0.7~(2)$ TeV if $u=0.001$ and $M>14~(48)$ TeV for $u=0.05$.
At variance with the non-supersymmetric case there is a range of values of the parameter $u$ for which the scale $M$ can be sufficiently small to allow for an explanation of the observed discrepancy in $a_\mu$, without conflicting with the present bound on $\mu\to e \gamma$. Since in our framework $\theta_{13}$ is comparable to $u$, the present limit on $BR(\mu\to e \gamma)$ together with the existing discrepancy
in $a_\mu$ point to a rather small value for $\theta_{13}$, of the order of few percents in radians, close to but probably just below the sensitivity expected in future experiments at reactors or with high intensity neutrino beams.

Subsequently, we have extended the original model by including Supersymmetry breaking terms consistent with all symmetry requirements. Our model is an effective theory, valid at energy scales below a cutoff $\Lambda_f \approx \Lambda_L$, where we have derived the spectrum of supersymmetric particles, in the slepton sector, under the assumption that the Supersymmetry breaking scale is larger than $\Lambda_f$. It provides an example of a model in which the slepton mass matrices at the scale $\Lambda_f$ are not universal. Left-handed sleptons are approximately universal, with a small departure from universality controlled by $u$, the flavour symmetry breaking parameter which runs in a small range around few per cent and has a size similar to the reactor mixing angle $\theta_{13}$. Right-handed sleptons have soft masses of the same order, but the relative difference among them is expected to be of order one. Off-diagonal elements in both sectors, as well as in the RL block, are small and have a characteristic pattern in terms of powers of $u$.
This structure is maintained by the effects coming from the RG running from $\Lambda_f \approx \Lambda_L$ down to the electroweak scale.

We got the slepton mass matrices to compute the normalised branching ratios $R_{ij}$ for the transitions $\mu\to e \gamma$, $\tau\to\mu\gamma$ and $\tau\to e \gamma$. At variance with other models based on flavour symmetries we found $R_{\mu e}\approx R_{\tau\mu}\approx R_{\tau e}$ and, given the present limit on $R_{\mu e}$, these rare $\tau$ decays are practically unobservable in our model. On a more theoretical side, the scaling $R_{ij}\propto u^2$, found in the MI approximation, violates an expectation based on an effective Lagrangian approach, which suggested $R_{ij}\propto u^4$ in the limit of massless final charged lepton. We have identified the source of such a violation in a single, flavour independent, contribution of the RL block of the slepton mass matrix. Such a contribution originates from the VEVs of the auxiliary components of the flavon supermultiplets. We have classified the conditions under which this universal contribution is absent.

In a numerical analysis of $R_{\mu e}$ we found that already the current bound from the MEGA experiment requires the parameter $u$ to be small or $\tan\beta$ to be small for supersymmetric mass parameters $m_{SUSY}$ and $m_{1/2}$ below $1000$ GeV, to guarantee detection of sparticles at LHC. Applying the prospective MEG bound tightens the parameter space of our model even more to small $u$ and $\tan\beta$ or requires mass parameters $m_{SUSY}$ and $m_{1/2}$ above $1000$ GeV. Furthermore, we showed that the deviation of the experimentally observed value of the magnetic moment of the muon from the Standard Model one cannot be naturally explained in our framework, for $BR(\mu\to e\gamma)$ below the current bound. The maximal value of $\delta a_{\mu}$ in our model is around $100 \times 10^{-11}$ for $BR(\mu\to e\gamma) \lesssim 10^{-11}$.

\clearpage{\pagestyle{empty}\cleardoublepage}

\newpage
\chapter{Leptogenesis}
\label{Sec:Leptogenesis}
\setcounter{equation}{0}
\setcounter{footnote}{3}

From a theoretical perspective the smallness of neutrino masses can be well understood within the See-Saw mechanism, in which Standard Model is extended by adding new heavy states. As we have already seen in the previous chapters, a flavour symmetry can act on the flavour structure of the Majorana and Dirac mass matrices, explaining the observed lepton mixing pattern. Introducing right-handed neutrinos with large Majorana masses,  the type I See-Saw naturally contains all the necessary ingredients for a dynamical generation of a cosmic lepton asymmetry through the decays of these heavy singlet states (leptogenesis): ($a$) Lepton number violation arising from the Majorana mass terms of the new fermionic fields; ($b$) CP-violating sources from complex Yukawa couplings; ($c$) departure from thermal equilibrium in the hot primeval plasma at the time the singlet neutrinos start decaying. This lepton asymmetry is then reprocessed into a baryon asymmetry through $B+L$ violating anomalous electroweak processes \cite{Kuzmin} thus yielding an explanation to the origin of the baryon asymmetry of the Universe \cite{Hinshaw} i.e.  baryogenesis through leptogenesis (for a recent review see \cite{DavidsonNardiNir}).

In the lepton sector a source of CP violation, e.g. neutrino oscillations, is already present, but it has been generically shown that baryon asymmetry is insensitive to the low-energy CP-violating phases \cite{Branco,Davidson}. However, it is possible to identify a class of models in which a connection between CP violation responsible for leptogenesis and CP violation observable at low energies can be established thanks to some flavour effects. This is linked to the additional constraints on the parameter space coming from the flavour sector, when a flavour symmetry is implemented in a model. As pointed out in \cite{JM_A4Lepto}, in the context of the Altarelli-Feruglio model with type I See-Saw the CP-violating asymmetry ($\epsilon_{\nu^c}$) vanishes in the limit of exact tribimaximal mixing, with leptogenesis becoming viable only when deviations from this pattern are taken into account.  The explicit structure of the corrections responsible for these deviations are model-dependent and therefore whether a connection between $\epsilon_{\nu^c}$ and low-energy parameters can be established will depend on the particular realisation.

In this chapter we extend upon the work in \cite{JM_A4Lepto}. In particular, we study the viability of leptogenesis in the context of
models based on an arbitrary flavour symmetry leading to mass-independent mixing textures. When there is only type I See-Saw, dealing with three right-handed neutrinos, independently of the nature of the underlying symmetry, we conclude that $\epsilon_{\nu^c}=0$ in the limit of exact symmetry. Under these conditions, only deviations coming from the flavour symmetry breaking yield $\epsilon_{\nu^c}\neq 0$. It is important to note that this result is not in general valid in the presence of other types of See-Saws.

On the same context, a different analysis but with very similar conclusions has been proposed in \cite{BBFN_Lepto}. In particular it is interesting to report one of the results therein: if the three right-handed neutrinos transform under the flavour symmetry as an irreducible representation then, in the limit of the exact symmetry, the neutrino Yukawa is proportional to a unitary matrix and as a result the CP asymmetry vanishes. It is then possible to find a relation between this statement and ours one: indeed, when the three right-handed neutrinos transform as an irreducible representation of the flavour symmetry it is possible that the lepton mixing matrix is a mass-independent texture, in the limit of the exact symmetry; as a result our conclusion can be seen an alternative formulation of the statement in \cite{BBFN_Lepto}.

\section{The Basic Framework}
\label{Sec:Lep:basis}
\setcounter{footnote}{3}

In this section we establish a more suitable notation, commonly used to deal with leptogenesis. The type I See-Saw Lagrangian is given by
\beq
\LL= (Y_e)_{ij}\overline{\ell}_i\,H\,e^c_j + (Y_\nu)_{i \alpha}\overline{\ell}_i \widetilde{H} \nu^c_\alpha  + \dfrac{1}{2}(M_R)_{\al \be}\nu^c_\al\nu^c_\be+\hc\;,
\label{Lep:eq:Lagrangian}
\eeq
where as usual $\ell$ are the lepton $SU(2)_L$ doublets, $e^c$ are the complex conjugate charged lepton $SU(2)_L$ singlets and $H$ is the
Higgs $SU(2)_L$ doublet. Latin indices $i,j\dots$ label the lepton flavour of the left-handed species, whereas Greek indices $\al,\be\dots$ refer to the right-handed species. $Y_e$, $Y_\nu$ and $M_R$ are $3\times3$ matrices in flavour space.

The effective light neutrino masses originate by the usual type I See-Saw and, after the electroweak symmetry breaking, the mass matrix is  given by
\beq
m_\nu=-m_D \, M_R^{-1}\, m_D^T\;,
\label{Lep:eq:lightNeu-MM}
\eeq
where $m_D=Y_\nu\,v/\sqrt2$. From now on we assume that in the physical basis, $m_\nu$ is exactly diagonalised by a mass-independent mixing matrix $U_0$ and therefore
\beq
\hat{m}_\nu=P\,U_0^T\, m_\nu\,U_0\,P\;,
\label{Lep:eq:diagonalisation}
\eeq
where $P$ accounts for the low-energy Majorana phases. In general, $m_D$ as well as $M_R$ ($M_R=M_R^T$) are complex matrices which can be diagonalised as follows
\beq
\hat{m}_D=U_L^\dag  \, m_D\, U_R \;,\qquad\qquad
\hat{M}_R=V_R^T \, M_R\, V_R \;,
\label{Lep:def}
\eeq
with $U_L,U_R,V_R$ $3\times 3$ unitary matrices, characterised in general by 3 rotation angles and 6 phases.

According to eq. (\ref{Lep:def}) the effective neutrino mass matrix in (\ref{Lep:eq:lightNeu-MM}) can be written as
\beq
m_\nu= - U_L\,\hat{m}_D \, (U_R^\dag\,V_R)\, \hat{M}_R^{-1}\, (V_R^T U_R^*)\,\hat{m}_D\,U_L^T\,.
\label{Lep:ss}
\eeq
The requirement of having exact $U_0$ diagonalisation can be written according to eqs. (\ref{Lep:eq:diagonalisation}) and (\ref{Lep:ss}), requiring that
\beq
\hat{m}_\nu= -  P\,(U_0^T U_L)\,\hat{m}_D \, (U_R^\dag\,V_R)\,\hat{M}_R^{-1}\, (V_R^T U_R^*)\,\hat{m}_D\,(U_L^T U_0)\,P
\label{Lep:ssdiag}
\eeq
is diagonal and real. It is useful to introduce the notation of the Dirac neutrino mass matrix in the basis in which the right-handed neutrino mass matrix $\hat{M}_R$ is real and diagonal:
\beq
m_D^R\equiv m_DV_R\,.
\label{Lep:eq:mDR_def}
\eeq

\subsection{General Remarks on Leptogenesis}
\label{Sec:Lep:gen-rem-lepto}
\setcounter{footnote}{3}

Our discussion will be entirely devoted to ``unflavoured'' leptogenesis scenarios: in the framework of flavour symmetry models the heavy singlet neutrinos typically have masses above $10^{13}$~GeV and for $T\gtrsim 10^{12}$ GeV lepton flavours are indistinguishable \cite{Nardi,Abada}. In the standard thermal leptogenesis scenario singlet neutrinos $\nu^c$ are produced by scattering processes after inflation. Subsequent out-of-equilibrium decays of these heavy states generate CP-violating asymmetries given by \cite{DavidsonNardiNir,Covi}
\beq
\epsilon_{\nu^c_\al} \;=\;\dfrac{\Gamma_\alpha-\ov{\Gamma}_\alpha}{\Gamma_\alpha+\ov{\Gamma}_\alpha}\;
=\; \dfrac{1}{4v^2 \pi (m_D^{R\,\dagger} \;m_D^R)_{\al\al}} \sum_{\be\neq \al}\im\left[\left((m_D^{R\,\dagger} \;m_D^R)_{\be\al}\right)^2\right] f(z_\be)\;,
\label{Lep:eq:cp-asymm}
\eeq
where $\Gamma_\alpha$ and $\ov{\Gamma}_\alpha$ are the total $\nu^c_\alpha$-decay widths into leptons and anti-leptons, respectively, $z_\be=M_\be^2/M_\al^2$ and the loop function can be expressed as
\begin{equation}
f(z_\be) = \sqrt{z_\be} \left[ \dfrac{2 - z_\be}{1 - z_\be} - (1 + z_\be)\; \log\left(\dfrac{1 + z_\be}{z_\be}\right) \right]\;.
\label{Lep:eq:loop-func}
\end{equation}
Depending on the singlet neutrino mass spectrum the loop function can be further simplified. In the hierarchical limit ($M_\al\ll M_\be$) and in the case of an almost degenerate heavy neutrino spectrum ($z_\be=1+\delta_\be$, $\delta_\be\ll 1$), this function becomes respectively
\beq
f(z_\be) \to -\dfrac{3}{2\sqrt{z_\be}}\;,\qquad\qquad f(1+\delta_\be)\simeq -\dfrac{1}{\delta_\be}\;.
\label{Lep:eq:loop-function}
\eeq
In any case, as can be seen from eq. (\ref{Lep:eq:cp-asymm}), whether the CP-violating asymmetry vanishes will be determined by the Yukawa coupling combination $m_D^{R\,\dagger} m_D^R$.

\mathversion{bold}
\section[CP asymmetry and Exact $U_0$ Mixing Without any FS]{CP asymmetry and Exact $U_0$ Mixing Without any Flavour Symmetry}
\label{Sec:Lep:unfamiliar}
\setcounter{footnote}{3}
\mathversion{normal}

While the $U_0$ mixing pattern can be well understood as a consequence of an underlying flavour symmetry, in principle it might be that it arises from a random set of parameters (though quite unlikely). For completeness, in this section we consider this possibility and study the consequences on the CP-violating asymmetry. We focus on the tribimaximal mixing pattern, but similar analysis can be done with the same conclusions dealing with other mass-independent textures.

When the neutrino mixing angles are fixed to satisfy the tribimaximal mixing pattern and in addition to the measured mass squared differences we have a set of eight constraints on the parameter space:
\beq
m_{\nu_{12}}=m_{\nu_{13}}\;,\qquad
m_{\nu_{22}}=m_{\nu_{33}}\;,\qquad
m_{\nu_{11}}=m_{\nu_{22}}+m_{\nu_{23}}-m_{\nu_{12}}\;,
\label{Lep:cond1}
\eeq
yielding six constraints (from the real and imaginary parts of the mass matrix entries); the atmospheric and solar mass scales provide the remaining two.

To determine the effect of such constraints on $\epsilon_{\nu^c}$ it is practical to use a parametrisation of $m_D$ which ensures that the tribimaximal mixing and the correct neutrino masses occur. In the basis in which the right-handed neutrino mass matrix is diagonal and real it is convenient to introduce the orthogonal complex matrix $R$ defined by the so-called Casas-Ibarra parametrisation \cite{CasasIbarra}, namely
\beq
R^*= (\hat{m}_\nu)^{-1/2} \,U_0^T\,m_D^R\, (\hat{M}_R)^{-1/2}\;.
\label{Lep:eq:casas-ibarra}
\eeq
All the low-energy observables are contained in the leptonic mixing matrix $U_0$ and in the diagonal and real light neutrino mass matrix $\hat{m}_\nu$. The matrix $R$ turns out to be very useful in expressing the CP-violating asymmetry parameter. Considering for simplicity the case of hierarchical right-handed neutrinos ($M_1 \ll M_2 \ll M_3$ - thus validating the approximation in \eq{Lep:eq:loop-function}), \eq{Lep:eq:cp-asymm} can be rewritten as
\begin{equation}
\epsilon_{\nu^c_\al} = -\dfrac{3 M_\al}{8 \pi v^2} \dfrac{\Im \left[\sum_j m_j^2 R_{j\al}^2\right]} {\sum_j m_j |R_{j\al}|^2}\;,
\label{Lep:eq:cp-asymm-CI}
\end{equation}
where $m_j\equiv(\hat m_\nu)_{jj}$. Once the right-handed neutrino mass spectrum and low-energy observables are fixed, random values of $m^R_D$ correspond to random values of $R$. It is shown by eq. (\ref{Lep:eq:cp-asymm-CI}) that leptogenesis is completely insensitive to low-energy lepton mixing and CP-violating phases \cite{Branco} \footnote{This statement is in general also true in flavoured leptogenesis \cite{Davidson}.} and therefore the viability of leptogenesis is not at all related with the tribimaximal mixing or in general with any accidental mixing pattern considered. The CP-violating asymmetry is determined by the values of the entries of $R$ which are arbitrary in the absence of any flavour symmetry, and consequently $\epsilon_{\nu^c}\neq 0$ in general and its absolute value depends upon the heavy fermionic singlet masses, the light neutrino masses and $R$.

\mathversion{bold}
\section{Implications of Flavour Symmetries on $\epsilon_{\nu^c_\al}$}
\label{Sec:Lep:exTBFS}
\setcounter{footnote}{3}
\mathversion{normal}

We consider now the case in which an underlying flavour symmetry enforces an exact mixing pattern. It will be evident throughout the proof that it holds for any mixing pattern where the mixing matrix consists purely of numbers, but we will assume tribimaximal mixing for definiteness.

Within the case considered the transformation properties of $\ell$ and $\nu^c$ under the flavour symmetry group $G_f$ determine the structure of $m_D$ and $M_R$ (which are no longer arbitrary). Indeed, these matrices can be regarded as form-diagonalisable matrices \cite{LV_Theorem}, i.e. the parameters which determine their eigenvalues are completely independent from the parameters that define their diagonalising
matrices. Accordingly, vanishing off-diagonal elements of $\hat{m}_\nu$ in eq. (\ref{Lep:ssdiag}) can arise only if
\begin{equation}
U_{TB}^T U_L= P_L \, O_{D_i} \quad \mbox{and}\quad U_R^\dag\,V_R=O^\dag_{D_i}\,P_R\,O_{R_{m}}\;,
\label{Lep:eq:rot-mat-relations}
\end{equation}
where $P_{L,R}=\text{diag}(e^{i\alpha^{R,L}_1},e^{i\alpha^{R,L}_2}, e^{i \alpha^{R,L}_3})$ whereas $O_{D_i}$ and $O_{R_m}$ are respectively unitary and orthogonal matrices that arbitrarily rotate the $i$ and $m$ degenerate eigenvalues of $m_D$ and $M_R$ such that if $m_D$ ($M_R$) has no degenerate eigenvalues $O_{D_i}=\unity$ ($O_{R_m}=\unity$). Note that the requirement of having canonical kinetic terms in addition to preserving the $m$-fold degeneracy of the right-handed neutrino mass matrix enforces $O_{R_m}$ to be real. It is easy to understand the conditions given in \eq{Lep:eq:rot-mat-relations} by the use of a \emph{reductio ad absurdum}. Let us consider for simplicity the case without any degeneracy in the eigenvalues of $\hat{m}_D$ and $\hat{M}_R$: $O_{D_i}=\unity$ and $O_{R_m}=\unity$. If the products $U_{TB}^TU_L$ and $U_R^\dag V_R$ are not diagonal, but simply unitary matrices with non-vanishing off-diagonal entries, then the right-hand side of eq. \eqref{Lep:ssdiag} is in general a matrix whose entries are linear combinations of the mass eigenvalues of $\hat{m}_D$ and of $\hat{M}_R$. In order to have $\hat{m}_\nu$ diagonal, the off-diagonal entries must vanish and this is possible only if the respective linear combinations cancel out. However, there are no a priori reasons to have such cancellations, since it corresponds to have well-defined relationships between the eigenvalues of $\hat{m}_D$ and of $\hat{M}_R$, which is, in other words, a fine-tuning. Avoiding this possibility, the only solution is to consider \eq{Lep:eq:rot-mat-relations}.

As shown in \cite{ABMMM_Lepto}, under the condition in \eq{Lep:eq:rot-mat-relations} the $m_D^R$ can be written as 
\beq
m_D^R= U_{TB}\,D\, \hat{v}\, O_{R_m}\;,
\label{Lep:vCI1}
\eeq
with $\hat{v}$ a diagonal real matrix and $D$ a diagonal unitary matrix which contains all the phases $\alpha^{R,L}_i$. It is straightforward to recover from eq. (\ref{Lep:vCI1}) the following $R^*$ matrix:
\beq
R^*=\hat{m}_\nu^{-1/2}\,  \hat{v}\, \hat{M}_R^{-1/2} \;.
\label{Lep:ourR}
\eeq
By comparing \eq{Lep:ourR} with the Casas-Ibarra parametrisation given in \eq{Lep:eq:casas-ibarra} we deduce that in the case of exact tribimaximal mixing the matrix $R$ is real and according to \eq{Lep:eq:cp-asymm-CI} the CP-violating asymmetry vanishes.

Note that so far we did not refer to any specific model realisation and we have assumed just exact tribimaximal diagonalisation of $m_\nu$ within the context of type I See-Saw. We not only confirm the result in \cite{JM_A4Lepto}, but also extend it to any possible flavour symmetry responsible for the exact tribimaximal scheme. It is also straightforward to check (by replacing $U_{TB}$ with a mass-independent mixing matrix $U_0$) that the matrix $R$ still turns out to be real for other exact mixing schemes as long as they are mass-independent. Note also that although we have only considered three right-handed neutrinos our result is absolutely generalisable to models with either two right-handed neutrinos or more than three such as \cite{moreNR}. On the other side, the proof does not hold however in the presence of additional degrees of freedom, e.g. in models involving type I and type II See-Saw.

An important consequence of our proof is that if the $U_0$ mixing pattern is due to any underlying flavour symmetry in a type I See-Saw
scenario, the viability of leptogenesis depends upon possible departures from the exact pattern. In the context of models based on
discrete flavour symmetries that predict $U_0$ mixing at the leading order this is achieved through NLO corrections. Since the size of the
deviations from $U_0$ mixing are not arbitrary, in principle one might expect the CP-violating asymmetry to be constrained by low-energy
observables such as $\theta_{13}$ and/or the CP-violating phases. However we have shown that in the general case the combination of NLO corrections that produce $\epsilon_{\nu^c}\neq0$ is not directly related with any low-energy observable. Consequently, while we conclude that general model-independent NLO corrections guarantee a non-vanishing CP-violating asymmetry, correlations among low-energy observables in the leptonic sector and $\epsilon_{\nu^c}$ cannot be established unless the nature of the corrections is well known, i.e. once the flavour model realisation has
been specified \cite{FelipeSerodio_Lepto,HMP_Lepto,BBFN_Lepto,JM_A4Lepto}.

\section{Conclusions of the Chapter}
\label{Sec:Lep:Conclusions}
\setcounter{footnote}{3}

In this chapter we considered under rather general conditions the possibility of links between low-energy observables and high-energy
parameters that are relevant for leptogenesis: in the most general case no such connections can be recovered. The situation can improve considering  flavour constraints. When simply assuming, in particular without introducing an underlying flavour symmetry, for the lepton mixing matrix a mass-independent texture, such as the well-known tribimaximal or bimaximal patterns, we conclude that it is in general possible to obtain leptogenesis, but it is not sufficient to provide a link between the different type of parameters.
  
On the contrary, it is possible to improve the situation considering the more natural case where mass-independent mixing patterns originate from an underlining flavour symmetry. We confirmed the results of \cite{JM_A4Lepto}, for which the CP-violating asymmetry vanishes in the limit of exact tribimaximal mixing in the case of unflavoured leptogenesis and only type I See-Saw. We generalised this conclusion into a model-independent proof that is valid for any flavour symmetries which impose mass-independent mixing textures. On the other hand, in order to have viable leptogenesis, the model has to require NLO corrections lifting the exact mixing, or alternatively independent contributions to the CP asymmetries such as those that naturally arise from an interplay between different See-Saws.

\clearpage{\pagestyle{empty}\cleardoublepage}

\clearpage{\pagestyle{empty}\cleardoublepage}
\newpage
\chapter{Summary and Final Remarks}
\label{Sec:Conclusions}
\setcounter{equation}{0}
\setcounter{footnote}{3}

Here we present a very brief concluding summary, while detailed remarks can be found at the end of each chapter.

The thesis collects the results of a series of projects which investigate on the use of flavour symmetries, in particular discrete and non-Abelian, to account for the flavour problem. The thesis itself can be seen as a unique subject developed in several directions: we first deal with model building and subsequently we analyse the impact of the underlying flavour symmetries on flavor violation and on leptogenesis.

In the first part, we presented a series of flavour models which produce phenomenologically successful patterns for fermion masses and mixings: while in the quark sector the mixing matrix shows a Wolfenstein-type structure in all the realisations in which quarks are considered, in the lepton sector we focussed on two distinct textures, the tribimaximal and the bimaximal schemes. They answer to two different requirements: in the tribimaximal models the reactor angle is almost vanishing while in the bimaximal realisations it can reach values close to its present upper bound. The future experiments on $\nu_e$ appearance will be able to discriminate between these two proposals.

In the second part of the thesis, we focus on flavour models based on the group $A_4$, analysing their predictions for some rare decays in the lepton sector, such as $\mu\to e\gamma$, $\tau\to e\gamma$ and $\tau\to\mu\gamma$, and the possibility to explain the discrepancy between the Standard Model prediction and the experimental measurement of the anomalous magnetic moment of the muon, through the presence of new physics at $1\div10$ TeV. We first adopt an effective operator approach a la MLFV in which these observables are described by six-dimensional operators invariant under the flavour symmetry; thereafter we identify the kind of new physics with Supersymmetry and we introduce a complete set of Supersymmetry breaking terms consistent with the flavour symmetry. In this second study, we found that the present and future experimental bounds on $\mu\to e\gamma$ represent strong constraints on the models, forbidding parts of the parameter space, but they do not exclude possible observations of supersymmetric particles at LHC. Furthermore, the normalised branching rations for $\mu\to e\gamma$, $\tau\to e\gamma$ and $\tau\to\mu\gamma$ are found to be of the same order of magnitude and, given the present limit on $\mu\to e\gamma$, we can conclude that the $\tau$ decays are practically unobservable. Regarding the anomalous magnetic moment of the muon, the deviation of the experimentally observed value from the Standard Model prediction cannot be naturally explained in our framework, for $BR(\mu\to e\gamma)$ below the current bound, indeed the maximal value of such a deviation is around $100 \times 10^{-11}$ for $BR(\mu\to e\gamma) \lesssim 10^{-11}$.

Finally in chapter \ref{Sec:Leptogenesis} we presented an argument for which $\epsilon$, the CP-violating parameter relevant for leptogenesis, is vanishing when the leptonic mixing matrix develops a mass-independent texture in the exact symmetry phase. Only by allowing symmetry breaking contributions, $\epsilon$ receives positive corrections and leptogenesis represents a viable explanation of the baryon asymmetry of the universe.

\clearpage{\pagestyle{empty}\cleardoublepage}

\newpage
\chapter*{Acknowledgments}
\addcontentsline{toc}{chapter}{Acknowledgments}
\thispagestyle{empty}

{\it I am deeply grateful to Ferruccio Feruglio for his teachings, clear explanations, encouragements and for having transmitted to me his professional and genuine enthusiasm in doing physics over these years. Together with Ferruccio, I warmly thank Guido Altarelli for his stimulating collaboration and formative hospitality at CERN.}\\

{\it Then I would like to remember Reinier de Adelhart Toorop, Diego Aristizabal Sierra, Federica Bazzocchi, Claudia Hagedorn, Yin Lin, Ivo de Medeiros Varzielas, Stefano Morisi and Alessio Paris for many fruitful collaborations.}\\

{\it I also thank all the members of the theory group for suggestions and advice: in particular, many thanks to Gianguido Dall'Agata, Massimo Passera, Stefano Rigolin and Fabio Zwirner.}\\

{\it Finally I warmly thank the other Ph.D. students and my office mates, in particular Alessandra Albano, Alessandra Cagnazzo, Francesca Catino, Angela Fava, Alessandra Gnecchi, Elena Moretti and Mia Tosi.}\\

\clearpage{\pagestyle{empty}\cleardoublepage}
\newpage
\renewcommand{\chaptermark}[1]{\markboth{{\sc Appendix\ \thechapter.\ #1}}{}}
\appendix

\chapter{Group Theory Details}
\label{AppendixA}
\lhead[\fancyplain{}{\bfseries\leftmark}]{\fancyplain{}{\bfseries\thepage}}
\rhead[\fancyplain{}{\bfseries\thepage}]{\fancyplain{}{\bfseries\rightmark}}
\renewcommand{\theequation}{A.\;\arabic{equation}}
\setcounter{equation}{0}
\setcounter{footnote}{3}
\setcounter{chapter}{1}
\setcounter{section}{0}

In this appendix we report the character tables and the Clebsch-Gordan coefficients of the $A_4$, $T'$ and $S_4$ discrete groups. Notice that two distinct description have been reported for the group $S_4$: this corresponds to two distinct but equivalent choices of the generators, which help in the model building.

In the character tables, $C_{i}$ are the classes of the group, $^{\circ} C_{i}$ is the order of the $i ^{\mathrm{th}}$ class, i.e. the number of distinct elements contained in this class, $^{\circ} h_{C_{i}}$ is the order of the elements $A$ in the class $C_{i}$, i.e. the smallest integer ($>0$) for which the equation $A ^{^{\circ} h_{C_{i}}}= \mathbb{1}$ holds. Furthermore the tables contain one representative $\rm G$ for each class $C_{i}$ given as product of the generators $S$ and $T$ of the group.

\mathversion{bold}
\section{The Group $A_4$}
\label{AppA:A4}
\setcounter{footnote}{3}
\mathversion{normal}

\begin{table}[ht]
\begin{center}
\begin{tabular}{l|cccc|}
&\multicolumn{4}{|c|}{classes} \\
\cline{2-5}
& $C_{1}$ & $C_{2}$ & $C_{3}$ & $C_{4}$ \\
\cline{1-5}
\rule[0.15in]{0cm}{0cm} $\rm G$  &$\rm \mathbb{1}$ & $S$ & $T^2$ & $T$ \\
\cline{1-5}
$^{\circ} C_{i}$ & 1 & 3 & 4 & 4 \\
\cline{1-5}
$^{\circ} h_{C_{i}}$ & 1 & 2 & 3 & 3 \\
\hline
${\bf1}$ & 1 & 1 & 1 & 1 \\[3pt]
${\bf1}'$ & 1 & 1 & $\om$ & $\om^2$ \\[3pt]
${\bf1}''$ & 1 & 1 & $\om^2$ & $\om$ \\[3pt]
${\bf3}$ & 3 & -1 & 0 & 0 \\
\hline
\end{tabular}
\end{center}
\caption{\it Character table of the group  $A_4$. $\omega$  is the third root of unity, i.e. $\omega= e^{\frac{2 \pi i}{3}} =  -\frac{1}{2} + i \frac{\sqrt{3}}{2}$.}
\label{AppA:table:A4chartab}
\end{table}

The group $A_4$ is generated by two elements $S$ and $T$ obeying the relations\cite{GroupRepresentations}:
\beq
S^2=(ST)^3=T^3=1\;.
\eeq
It has three independent one-dimensional representations, $\bf1$, $\bf1'$ and $\bf1''$ and one three-dimensional representation $\bf3$.
The one-dimensional representations are given by:
\beq
\begin{array}{lll}
{\bf1} & S=1 & T=1 \\[3mm]
{\bf1'} & S=1 & T=e^{i 4 \pi/3} \equiv \omega^2\\[3mm]
{\bf1''}& S=1 & T=e^{i 2\pi/3} \equiv\omega\\[3mm]
\end{array}
\eeq
The three-dimensional representation, in a basis where the generator $T$ is diagonal, is given by:
\beq
T=\left(
        \begin{array}{ccc}
        1 & 0 & 0 \\
        0 & \omega^2 & 0 \\
        0 & 0 & \omega \\
        \end{array}
    \right),\qquad\qquad
S=\dfrac{1}{3}
    \left(
        \begin{array}{ccc}
        -1 & 2 & 2 \\
        2 & -1 & 2 \\
        2 & 2 & -1 \\
        \end{array}
    \right)\;.
\label{ST}
\eeq

We now report the multiplication rules between the various representations. In the following we use $\alpha=(\alpha_1,\,\alpha_2,\,\alpha_3)$ to indicate the elements of the first representation of the product and $\beta=(\beta_1,\,\beta_2,\,\beta_3)$ to indicate those of the second representation. Moreover $a,b=0,\pm1$ and we denote ${\bf1}^0\equiv{\bf1}$, ${\bf1}^1\equiv{\bf1}^\prime$, ${\bf1}^{-1}\equiv{\bf1}^{\prime\prime}$ and similarly for the doublet representations. On the right-hand side the sum $a+b$ is modulo 3.

We start with all the multiplication rules which include the one-dimensional representations:
\beq
\begin{array}{l}
{\bf1}\times {\bf r}={\bf r}\times{\bf1}={\bf r}\qquad\text{with ${\bf r}$ any representation}\;,\\[3mm]
{\bf1}^a\times{\bf1}^b={\bf1}^b\times{\bf1}^a={\bf1}^{a+b}\sim\alpha\beta\;,\\[3mm]
{\bf1}^\prime\times{\bf3}={\bf3}\sim\left(\begin{array}{c}
                                        \alpha\beta_3 \\
                                        \alpha\beta_1 \\
                                        \alpha\beta_2\\
                                \end{array}\right)\;,\qquad
{\bf1}^{\prime\prime}\times{\bf3}={\bf3}\sim\left(\begin{array}{c}
                                        \alpha\beta_2 \\
                                        \alpha\beta_3 \\
                                        \alpha\beta_1\\
                                \end{array}\right)\;.
\end{array}
\eeq

The multiplication rule with the three-dimensional representation is
\beq
\begin{array}{l}
{\bf3}\times{\bf3}={\bf3}_S+{\bf3}_A+{\bf1}+{\bf1}'+{\bf1}''\quad\text{with}\quad\!\left\{
        \begin{array}{l}
        {\bf1}\;\,\sim\alpha_1\beta_1+\alpha_2\beta_3+\alpha_3\beta_2\;,\\[3mm]
        {\bf1}'\;\sim\alpha_3\beta_3+\alpha_1\beta_2+\alpha_2\beta_1\;,\\[3mm]
        {\bf1}''\,\sim\alpha_2\beta_2+\alpha_1\beta_3+\alpha_3\beta_1\;,\\[3mm]
        {\bf3}_S\sim\dfrac{1}{3}\left(\begin{array}{c}
                                     2\alpha_1\beta_1-\alpha_2\beta_3-\alpha_3\beta_2\\
                                     2\alpha_3\beta_3-\alpha_1\beta_2-\alpha_2\beta_1\\
                                     2\alpha_2\beta_2-\alpha_1\beta_3-\alpha_3\beta_1\\
                                     \end{array}
                               \right)\\[3mm]
        {\bf3}_A\sim\dfrac{1}{2}\left(\begin{array}{c}
                                 \alpha_2\beta_3-\alpha_3\beta_2\\
                                 \alpha_1\beta_2-\alpha_2\beta_1\\
                                 \alpha_3\beta_1-\alpha_1\beta_3\\
                                \end{array}\right)
\end{array}\right.
\ea
\eeq
Note that due to the choice of complex representation matrices for the real representation ${\bf3}$ the conjugate $\al^*$ of $\al \sim {\bf3}$ does not transform as ${\bf3}$, but rather $(\al_1^{\star},\, \al_3^*,\, \al_2^*)$ transforms as triplet under $A_4$. The reason for this is that $T^*= U_{23}^T\,T\,U_{23}$ and $S^*=U_{23}^T\,S\,U_{23}=S$ where $U_{23}$ is the matrix which exchanges the 2nd and 3rd row and column.

\mathversion{bold}
\section{The Group $T'$}
\label{AppA:Tp}
\setcounter{footnote}{3}
\mathversion{normal}

\begin{table}[ht]
\begin{center}
\begin{tabular}{l|ccccccc|}
&\multicolumn{7}{|c|}{classes}\\ 
\cline{2-8}
&$C_{1}$&$C_{2}$&$C_{3}$&$C_{4}$&$C_{5}$&$C_{6}$&$C_{7}$\\
\cline{1-8}
\rule[0.15in]{0cm}{0cm} $\rm G$  &$\rm \mathbb{1}$&$\rm \mathbb{R}$&$S$&$S T \mathbb{R}$&$T^{2}$&$T$&$(S T)^{2} \mathbb{R}$\\
\cline{1-8}
$^{\circ} C_{i}$ & 1 & 1 & 6 & 4 & 4 & 4 & 4 \\
\cline{1-8}
$^{\circ} h_{C_{i}}$ & 1 & 2 & 4 & 6 & 3 & 3 & 6 \\
\hline
$\bf1$ & 1 & 1 & 1 & 1 & 1 & 1 & 1 \\
$\bf1'$ & 1 & 1 & 1 & $\omega$ & $\omega^{2}$ & $\omega$ & $\omega^{2}$ \\
$\bf1''$ & 1 & 1 & 1 & $\omega ^{2}$ & $\omega$ & $\omega ^{2}$ & $\omega$ \\
$\bf2$ & 2 & $-2$ & 0 & 1 & $-1$ & $-1$ & 1 \\
$\bf2'$ & 2 & $-2$ & 0 & $\omega$ & $-\omega^{2}$ & $-\omega$ & $\omega ^{2}$ \\
$\bf2''$ & 2 & $-2$ & 0 & $\omega ^{2} $ & $-\omega$ & $-\omega^{2}$ & $\omega$ \\
$\bf3$ & 3 & 3 & $-1$ & 0 & 0 & 0 & 0 \\
\hline
\end{tabular}
\end{center}
\caption{\it Character table of the group  $T^{\prime}$. $\omega$  is the third root of unity, i.e. $\omega= e^{\frac{2 \pi i}{3}} =  -\frac{1}{2} + i \frac{\sqrt{3}}{2}$.}
\label{AppA:table:Tpchartab}
\end{table}

The matrices $S$ and $T$ representing the generator depend on the representations of the group:
\begin{equation}
\begin{array}{ccccc}
{\bf1} &\qquad& S=1 &\quad& T=1\\[3mm]
{\bf1}^\prime &\qquad& S=1 &\quad& T=\omega\\[3mm]
{\bf1}^{\prime\prime} &\qquad& S=1 &\quad& T=\omega^2\\[3mm]
{\bf2} &\qquad& S=A_1 &\quad& T=\omega A_2\\[3mm]
{\bf2}^\prime &\qquad& S=A_1 &\quad& T=\omega^2 A_2\\[3mm]
{\bf2}^{\prime\prime} &\qquad& S=A_1 &\quad& T=A_2\\[3mm]
{\bf3} &\qquad& S=\dfrac{1}{3}\left(\begin{array}{ccc}
                                -1 & 2\omega & 2\omega^2 \\
                                2\omega^2 & -1 & 2\omega \\
                                2\omega & 2\omega^2 & -1 \\
                                \end{array}\right)
        &\quad& T=\left(\begin{array}{ccc}
                            1 & 0 & 0 \\
                            0 & \omega & 0 \\
                            0 & 0 & \omega^2 \\
                        \end{array}\right)
\end{array}
\end{equation}
where we have used the matrices
\begin{equation}
A_1=-\dfrac{1}{\sqrt{3}}\left(\begin{array}{cc}
                                    i & \sqrt2e^{i\pi/12} \\
                                    -\sqrt2e^{-i\pi/12} & -i \\
                                \end{array}\right)\;,\qquad
A_2=\left(\begin{array}{cc}
              \omega & 0 \\
              0 & 1 \\
            \end{array}\right)\;.
\end{equation}

The multiplication rules involving one- and three-dimensional representations are equal to those ones of the $A_4$ group and we report here only the rules which deal with the two-dimensional representations:
\bea
{\bf1}^a\times{\bf2}^b={\bf2}^b\times{\bf1}^a={\bf2}^{a+b}&&\hspace{-5mm}\sim\left(\begin{array}{c}
                                                \alpha\beta_1 \\
                                                \alpha\beta_2 \\
                                        \end{array}\right)\\[3mm]
{\bf2}\times{\bf2}={\bf2}^\prime\times{\bf2}^{\prime\prime}={\bf2}^{\prime\prime}\times{\bf2}^\prime={\bf3}+{\bf1}&&
\text{with}\quad\left\{\begin{array}{ll}
                    {\bf3}\sim\left(\begin{array}{c}
                        \dfrac{1-i}{2}(\alpha_1\beta_2+\alpha_2\beta_1) \\
                        i\alpha_1\beta_1 \\
                        \alpha_2\beta_2 \\
                    \end{array}\right)\\[3mm]
                    {\bf1}\sim\alpha_1\beta_2-\alpha_2\beta_1\\
                    \end{array}
            \right.\\[3mm]
{\bf2}\times{\bf2}^\prime={\bf2}^{\prime\prime}\times{\bf2}^{\prime\prime}={\bf3}+{\bf1}^\prime&&
\text{with}\quad\left\{\begin{array}{ll}
                    {\bf3}\sim\left(\begin{array}{c}
                        \alpha_2\beta_2 \\
                        \dfrac{1-i}{2}(\alpha_1\beta_2+\alpha_2\beta_1) \\
                        i\alpha_1\beta_1 \\
                    \end{array}\right)\\[3mm]
                    {\bf1}^\prime\sim\alpha_1\beta_2-\alpha_2\beta_1
                    \end{array}
            \right.\\[3mm]
{\bf2}\times{\bf2}^{\prime\prime}={\bf2}^\prime\times{\bf2}^\prime={\bf3}+{\bf1}^{\prime\prime}&&
\text{with}\quad\left\{\begin{array}{ll}
                    {\bf3}\sim\left(\begin{array}{c}
                        i\alpha_1\beta_1 \\
                        \alpha_2\beta_2\\
                        \dfrac{1-i}{2}(\alpha_1\beta_2+\alpha_2\beta_1) \\
                    \end{array}\right)\\[3mm]
                    {\bf1}''\sim\alpha_1\beta_2-\alpha_2\beta_1
                    \end{array}
            \right.\\[3mm]
{\bf2}\times{\bf3}={\bf2}+{\bf2}^\prime+{\bf2}^{\prime\prime}&&
\text{with}\quad\left\{\begin{array}{ll}
                    {\bf2}\;\sim\;\left(\begin{array}{c}
                                (1+i)\alpha_2\beta_2+\alpha_1\beta_1 \\
                                (1-i)\alpha_1\beta_3-\alpha_2\beta_1 \\
                            \end{array}\right)\\[3mm]
                    {\bf2}'\sim\;\left(\begin{array}{c}
                                (1+i)\alpha_2\beta_3+\alpha_1\beta_2 \\
                                (1-i)\alpha_1\beta_1-\alpha_2\beta_2 \\
                            \end{array}\right)\\[3mm]
                    {\bf2}''\sim\left(\begin{array}{c}
                            (1+i)\alpha_2\beta_1+\alpha_1\beta_3 \\
                            (1-i)\alpha_1\beta_2-\alpha_2\beta_3 \\
                             \end{array}\right)
                    \end{array}
            \right.\\[3mm]
{\bf2}^\prime\times{\bf3}={\bf2}+{\bf2}^\prime+{\bf2}^{\prime\prime}&&
\text{with}\quad\left\{\begin{array}{ll}
                    {\bf2}\;\,\sim\left(\begin{array}{c}
                                (1+i)\alpha_2\beta_1+\alpha_1\beta_3 \\
                                (1-i)\alpha_1\beta_2-\alpha_2\beta_3 \\
                            \end{array}\right)\\[3mm]
                    {\bf2}^\prime\,\sim\left(\begin{array}{c}
                                (1+i)\alpha_2\beta_2+\alpha_1\beta_1 \\
                                (1-i)\alpha_1\beta_3-\alpha_2\beta_1 \\
                            \end{array}\right)\\[3mm]
                    {\bf2}^{\prime\prime}\sim\left(\begin{array}{c}
                                (1+i)\alpha_2\beta_3+\alpha_1\beta_2 \\
                                (1-i)\alpha_1\beta_1-\alpha_2\beta_2 \\
                            \end{array}\right)
                    \end{array}
            \right.\\[3mm]
{\bf2}^{\prime\prime}\times{\bf3}={\bf2}+{\bf2}^\prime+{\bf2}^{\prime\prime}&&
\text{with}\quad\left\{\begin{array}{ll}
                    {\bf2}\;\,\sim\left(\begin{array}{c}
                                    (1+i)\alpha_2\beta_3+\alpha_1\beta_2 \\
                                    (1-i)\alpha_1\beta_1-\alpha_2\beta_2 \\
                                \end{array}\right)\\[3mm]
                    {\bf2}^\prime\,\sim\left(\begin{array}{c}
                                    (1+i)\alpha_2\beta_1+\alpha_1\beta_3 \\
                                    (1-i)\alpha_1\beta_2-\alpha_2\beta_3 \\
                                \end{array}\right)\\[3mm]
                    {\bf2}^{\prime\prime}\sim\left(\begin{array}{c}
                                    (1+i)\alpha_2\beta_2+\alpha_1\beta_1 \\
                                    (1-i)\alpha_1\beta_3-\alpha_2\beta_1 \\
                                \end{array}\right)
                    \end{array}
            \right.
\eea

\mathversion{bold}
\section[The Group $S_4$ -- I Version]{The Group $\mathbf{S_4}$ -- I Version}
\label{AppA:S4}
\setcounter{footnote}{3}
\mathversion{normal}

\begin{table}[ht]
\begin{center}
\begin{tabular}{l|ccccc|}
&\multicolumn{5}{|c|}{classes}\\
\cline{2-6}
& $C_{1}$ & $C_{2}$ & $C_{3}$ & $C_{4}$ & $C_{5}$ \\
\cline{1-6}
\rule[0.15in]{0cm}{0cm} $\rm G$  &$\rm \mathbb{1}$& $S^2$ & $T$ & $ST^{2}$ & $S$ \\
\cline{1-6}
$^{\circ} C_{i}$ & 1 & 3 & 8 & 6 & 6 \\
\cline{1-6}
$^{\circ} h_{C_{i}}$ & 1 & 2 & 3 & 2 & 4 \\
\hline
$\bf1$ & 1 & 1 & 1 & 1 & 1 \\[3pt]
$\bf1'$ & 1 & 1 & 1 & $-1$ & $-1$ \\[3pt]
$\bf2$ & 2 & 2 & $-1$ & 0 & 0 \\[3pt]
$\bf3$ & 3 & $-1$ & 0 & 1 & $-1$ \\[3pt]
$\bf3'$ & 3 & $-1$ & 0 & $-1$ & 1 \\
\hline
\end{tabular}
\end{center}
\caption{\it Character table of the group $S_4$ -- I version.}
\label{AppA:table:S4chartabI}
\end{table}

The generators, $S$ and $T$, obey to the following rules
\beq
S^4= T^3=  (ST^2)^2=1
\eeq
and can be written in the different representations as
\begin{equation}
\begin{array}{ccccc}
{\bf1} &\qquad& S=1 &\quad& T=1\\[3mm]
{\bf1}^\prime &\qquad& S=-1 &\quad& T=1\\[3mm]
{\bf2} &\qquad& S=\left(
                    \begin{array}{cc}
                      0 & 1 \\
                      1 & 0 \\
                    \end{array}
                  \right)
        &\quad& T=\left(
             \begin{array}{cc}
               \omega & 0 \\
               0 & \omega^2 \\
             \end{array}
           \right)\\[3mm]
{\bf3} &\qquad& S=\dfrac{1}{3}\left(\begin{array}{ccc}
                                -1 & 2\omega & 2\omega^2 \\
                                2\omega & 2\omega^2 & -1 \\
                                2\omega^2 & -1 & 2\omega \\
                                \end{array}\right)
        &\quad& T=\left(\begin{array}{ccc}
                            1 & 0 & 0 \\
                            0 & \omega^2 & 0 \\
                            0 & 0 & \omega \\
                        \end{array}\right)\\[3mm]
{\bf3'} &\qquad& S=\dfrac{1}{3}\left(\begin{array}{ccc}
                                1 & -2\omega & -2\omega^2 \\
                                -2\omega & -2\omega^2 & 1 \\
                                -2\omega^2 & 1 & -2\omega \\
                                \end{array}\right)
        &\quad& T=\left(\begin{array}{ccc}
                            1 & 0 & 0 \\
                            0 & \omega^2 & 0 \\
                            0 & 0 & \omega \\
                        \end{array}\right)
\end{array}
\end{equation}

The $24$ elements define different subgroups of $S_4$: the elements of $\mathcal{C}_{2,4}$ define two different sets of $Z_2$ subgroups, corresponding to $S^2$ and $S T^2$ respectively, those of the class $\mathcal{C}_{4}$ a set of $Z_3$ Abelian discrete symmetries associated to $T$ and those belonging to $\mathcal{C}_{5}$ a set of $Z_4$ Abelian discrete symmetries corresponding to $S$. From the three relations that define the group $S_4$ we see that it contains also a non-Abelian subgroup, $S_3$. Indeed defining $S'= S^2$ and using $S^2  T S^2= T^2$ we get the relations that define $S_3$, namely
\begin{equation}
T^3=  S^{\prime2}= (S' T)^2=1\;.
\end{equation}

We now report the Clebsch-Gordan coefficients for our basis. We start with all the multiplication rules which include the one-dimensional representations:
\beq
\begin{array}{lcl}
{\bf1}\times{\bf r}&=&{\bf r}\times{\bf1}={\bf r}\quad\text{with ${\bf r}$ any representation}\;,\\[5mm]
{\bf1}'\times{\bf1}'&=&{\bf1}\sim\alpha\beta\;,\qquad\qquad\quad\;\;
{\bf1}'\times{\bf2}\;=\;{\bf2}\sim\left(\begin{array}{c}
                    \alpha\beta_1 \\
                    -\alpha\beta_2 \\
            \end{array}\right)\;,\\[3mm]
{\bf1}'\times{\bf3}&=&{\bf3}'\sim\left(\begin{array}{c}
                    \alpha\beta_1 \\
                    \alpha\beta_2 \\
                    \alpha\beta_3 \\
                \end{array}\right)\;,\qquad
{\bf1}'\times{\bf3}'\;=\;{\bf3}\sim\left(\begin{array}{c}
                            \alpha\beta_1 \\
                            \alpha\beta_2 \\
                            \alpha\beta_3 \\
                    \end{array}\right)\;.
\end{array}
\eeq
The multiplication rules with the two-dimensional representation are the following:
\beq
\begin{array}{ll}
{\bf2}\times{\bf2}={\bf1}+{\bf1}'+{\bf2}&\quad
\text{with}\quad\left\{\begin{array}{l}
                    {\bf1}\sim\alpha_1\beta_2+\alpha_2\beta_1\\[3mm]
                    {\bf1}'\sim\alpha_1\beta_2-\alpha_2\beta_1\\[3mm]
                    {\bf2}\sim\left(\begin{array}{c}
                        \alpha_2\beta_2 \\
                        \alpha_1\beta_1 \\
                    \end{array}\right)
                    \end{array}
            \right.\\[3mm]
{\bf2}\times{\bf3}={\bf3}+{\bf3}'&\quad
\text{with}\quad\left\{\begin{array}{l}
                    {\bf3}\sim\left(\begin{array}{c}
                        \alpha_1\beta_2+\alpha_2\beta_3 \\
                        \alpha_1\beta_3+\alpha_2\beta_1 \\
                        \alpha_1\beta_1+\alpha_2\beta_2 \\
                    \end{array}\right)\\[3mm]
                    {\bf3}'\sim\left(\begin{array}{c}
                        \alpha_1\beta_2-\alpha_2\beta_3\\
                        \alpha_1\beta_3-\alpha_2\beta_1 \\
                        \alpha_1\beta_1-\alpha_2\beta_2 \\
                    \end{array}\right)
                    \end{array}
            \right.\\[3mm]
{\bf2}\times{\bf3}'={\bf3}+{\bf3}'&\quad
\text{with}\quad\left\{\begin{array}{l}
                    {\bf3}\sim\left(\begin{array}{c}
                        \alpha_1\beta_2-\alpha_2\beta_3\\
                        \alpha_1\beta_3-\alpha_2\beta_1 \\
                        \alpha_1\beta_1-\alpha_2\beta_2 \\
                    \end{array}\right)\\[3mm]
                    {\bf3}'\sim\left(\begin{array}{c}
                        \alpha_1\beta_2+\alpha_2\beta_3 \\
                        \alpha_1\beta_3+\alpha_2\beta_1 \\
                        \alpha_1\beta_1+\alpha_2\beta_2 \\
                    \end{array}\right)
                    \end{array}
            \right.
\end{array}
\eeq
The multiplication rules with the three-dimensional representations are the following:
\beq
\begin{array}{c}
{\bf3}\times{\bf3}={\bf3}'\times{\bf3}'={\bf1}+{\bf2}+{\bf3}+{\bf3}'
\quad\;\text{with}\quad\left\{
\begin{array}{l}
{\bf1}\sim\alpha_1\beta_1+\alpha_2\beta_3+\alpha_3\beta_2 \\[3mm]
{\bf2}\sim\left(
     \begin{array}{c}
       \al_2\be_2+\al_1\be_3+\al_3\be_1 \\
       \al_3\be_3+\al_1\be_2+\al_2\be_1 \\
     \end{array}
   \right)\\[3mm]
{\bf3}\sim\left(\begin{array}{c}
         2\al_1\be_1-\alpha_2\beta_3-\alpha_3\beta_2 \\
         2\al_3\be_3-\alpha_1\beta_2-\alpha_2\beta_1 \\
         2\al_2\be_2-\alpha_1\beta_3-\alpha_3\beta_1 \\
        \end{array}\right)\\[3mm]
{\bf3}'\sim\left(\begin{array}{c}
         \alpha_2\beta_3-\alpha_3\beta_2 \\
         \alpha_1\beta_2-\alpha_2\beta_1 \\
         \alpha_3\beta_1-\alpha_1\beta_3 \\
    \end{array}\right)
\end{array}\right.
\end{array}
\eeq
\beq
\begin{array}{ll}
{\bf3}\times{\bf3}'={\bf1}'+{\bf2}+{\bf3}+{\bf3}'\quad\;\text{with}\quad\left\{
\begin{array}{l}
{\bf1}'\sim\alpha_1\beta_1+\alpha_2\beta_3+\alpha_3\beta_2\\[3mm]
{\bf2}\sim\left(
     \begin{array}{c}
       \al_2\be_2+\al_1\be_3+\al_3\be_1 \\
       -\al_3\be_3-\al_1\be_2-\al_2\be_1 \\
     \end{array}
   \right)\\[3mm]
{\bf3}\sim\left(\begin{array}{c}
         \alpha_2\beta_3-\alpha_3\beta_2 \\
         \alpha_1\beta_2-\alpha_2\beta_1 \\
         \alpha_3\beta_1-\alpha_1\beta_3 \\
    \end{array}\right)\\[3mm]
{\bf3}'\sim\left(\begin{array}{c}
         2\al_1\be_1-\alpha_2\beta_3-\alpha_3\beta_2 \\
         2\al_3\be_3-\alpha_1\beta_2-\alpha_2\beta_1 \\
         2\al_2\be_2-\alpha_1\beta_3-\alpha_3\beta_1 \\
    \end{array}\right)
\end{array}\right.
\end{array}
\eeq

\section[The Group $S_4$ -- II Version]{The Group $\mathbf{S_4}$ -- II Version}
\label{AppA:S4BM}
\setcounter{footnote}{3}

\begin{table}[ht]
\begin{center}
\begin{tabular}{l|ccccc|}
&\multicolumn{5}{|c|}{classes}\\
\cline{2-6}
& $C_{1}$ & $C_{2}$ & $C_{3}$ & $C_{4}$ & $C_{5}$ \\
\cline{1-6}
\rule[0.15in]{0cm}{0cm} $\rm G$  &$\rm \mathbb{1}$& $T^2$ & $ST$ & $S$ & $T$ \\
\cline{1-6}
$^{\circ} C_{i}$ & 1 & 3 & 8 & 6 & 6 \\
\cline{1-6}
$^{\circ} h_{C_{i}}$ & 1 & 2 & 3 & 2 & 4 \\
\hline
$\bf1$ & 1 & 1 & 1 & 1 & 1 \\[3pt]
$\bf1'$ & 1 & 1 & 1 & $-1$ & $-1$ \\[3pt]
$\bf2$ & 2 & 2 & $-1$ & 0 & 0 \\[3pt]
$\bf3$ & 3 & $-1$ & 0 & 1 & $-1$ \\[3pt]
$\bf3'$ & 3 & $-1$ & 0 & $-1$ & 1 \\
\hline
\end{tabular}
\end{center}
\caption{\it Character table of the group $S_4$ -- II version.}
\label{AppA:table:S4chartabII}
\end{table}

In order to realise the bimaximal mixing, we find that it is useful to define group generators different from those used to study the tribimaximal pattern: the two new operators $S$ and $T$ satisfy to
\beq
T^4=S^2=(ST)^3=(TS)^3=1\;.
\eeq
Explicit forms of $S$ and $T$ in each of the irreducible representations can be simply obtained:
\begin{equation}
\begin{array}{ccccc}
{\bf1} &\qquad& S=1 &\quad& T=1\\[3mm]
{\bf1}^\prime &\qquad& S=-1 &\quad& T=1\\[3mm]
{\bf2} &\qquad& S=\dfrac{1}{2}\left(
                    \begin{array}{cc}
                      -1 & \sqrt3 \\
                      \sqrt3 & -1 \\
                    \end{array}
                  \right)
        &\quad& T=\left(
             \begin{array}{cc}
               1 & 0 \\
               0 & -1 \\
             \end{array}
           \right)\\[3mm]
{\bf3} &\qquad& S=\left(\begin{array}{ccc}
                                0 & -1/\sqrt2 & -1/\sqrt2 \\
                                -1/\sqrt2 & 1/2 & -1/2 \\
                                -1/\sqrt2 & -1/2 & 1/2 \\
                                \end{array}\right)
        &\quad& T=\left(\begin{array}{ccc}
                            -1 & 0 & 0 \\
                            0 & -i & 0 \\
                            0 & 0 & i \\
                        \end{array}\right)\\[3mm]
{\bf3'} &\qquad& S=\left(\begin{array}{ccc}
                                0 & 1/\sqrt2 & 1/\sqrt2 \\
                                1/\sqrt2 & -1/2 & 1/2 \\
                                1/\sqrt2 & 1/2 & -1/2 \\
                                \end{array}\right)
        &\quad& T=\left(\begin{array}{ccc}
                            1 & 0 & 0 \\
                            0 & i & 0 \\
                            0 & 0 & -i \\
                        \end{array}\right)
\end{array}
\end{equation}

We recall here the multiplication table for $S_4$ and we list the Clebsch-Gordan coefficients in our basis. We start with all the multiplication rules which include the one-dimensional representations:
\beq
\begin{array}{l}
{\bf1}\times {\bf r}={\bf r}\times{\bf1}={\bf r}\quad\text{with ${\bf r}$ any representation}\;,\\[5mm]
{\bf1}'\times{\bf1}'={\bf1}\sim\alpha\beta\;,\qquad\qquad\quad\;
{\bf1}'\times{\bf2}={\bf2}\sim\left(\begin{array}{c}
                    \alpha\beta_2 \\
                    -\alpha\beta_1 \\
            \end{array}\right)\;,\\[3mm]
{\bf1}'\times{\bf3}={\bf3}'\sim\left(\begin{array}{c}
                    \alpha\beta_1 \\
                    \alpha\beta_2 \\
                    \alpha\beta_3\\
                    \end{array}\right)\;,\qquad
{\bf1}'\times{\bf3}'={\bf3}\sim\left(\begin{array}{c}
                            \alpha\beta_1 \\
                            \alpha\beta_2 \\
                            \alpha\beta_3\\
                    \end{array}\right)\;.
\end{array}
\eeq
The multiplication rules with the two-dimensional representation are the following ones:
\beq
\begin{array}{ll}
{\bf2}\times{\bf2}={\bf1}+{\bf1}'+{\bf2}&\quad
\rm{with}\quad\left\{\begin{array}{l}
                    {\bf1}\sim\alpha_1\beta_1+\alpha_2\beta_2\\[3mm]
                    {\bf1}'\sim\alpha_1\beta_2-\alpha_2\beta_1\\[3mm]
                    {\bf2}\sim\left(\begin{array}{c}
                        \alpha_2\beta_2-\alpha_1\beta_1 \\
                        \alpha_1\beta_2+\alpha_2\beta_1\\
                    \end{array}\right)
                    \end{array}
            \right.\\[3mm]
{\bf2}\times{\bf3}={\bf3}+{\bf3}^\prime&\quad
\rm{with}\quad\left\{\begin{array}{l}
                    {\bf3}\sim\left(\begin{array}{c}
                        \alpha_1\beta_1\\
                        \frac{\sqrt3}{2}\alpha_2\beta_3-\frac{1}{2}\alpha_1\beta_2 \\
                        \frac{\sqrt3}{2}\alpha_2\beta_2-\frac{1}{2}\alpha_1\beta_3 \\
                    \end{array}\right)\\[3mm]
                    {\bf3}'\sim\left(\begin{array}{c}
                        -\alpha_2\beta_1\\
                        \frac{\sqrt3}{2}\alpha_1\beta_3+\frac{1}{2}\alpha_2\beta_2 \\
                        \frac{\sqrt3}{2}\alpha_1\beta_2+\frac{1}{2}\alpha_2\beta_3 \\
                    \end{array}\right)\\
                    \end{array}
            \right.\\[3mm]
{\bf2}\times{\bf3}'={\bf3}+{\bf3}^\prime&\quad
\rm{with}\quad\left\{\begin{array}{l}
                    {\bf3}\sim\left(\begin{array}{c}
                       -\alpha_2\beta_1\\
                        \frac{\sqrt3}{2}\alpha_1\beta_3+\frac{1}{2}\alpha_2\beta_2 \\
                        \frac{\sqrt3}{2}\alpha_1\beta_2+\frac{1}{2}\alpha_2\beta_3 \\
                    \end{array}\right)\\[3mm]
                    {\bf3}'\sim\left(\begin{array}{c}
                        \alpha_1\beta_1\\
                        \frac{\sqrt3}{2}\alpha_2\beta_3-\frac{1}{2}\alpha_1\beta_2 \\
                        \frac{\sqrt3}{2}\alpha_2\beta_2-\frac{1}{2}\alpha_1\beta_3 \\
                    \end{array}\right)\\
                    \end{array}
            \right.
\end{array}
\eeq

The multiplication rules involving the three-dimensional representations are:
\beq
\begin{array}{l}
{\bf3}\times{\bf3}={\bf3}'\times{\bf3}'={\bf1}+{\bf2}+{\bf3}+{\bf3}'\quad\text{with}\quad\!\left\{
        \begin{array}{l}
        {\bf1}\sim\alpha_1\beta_1+\alpha_2\beta_3+\alpha_3\beta_2\\[3mm]
        {\bf2}\sim\left(
             \begin{array}{c}
               \alpha_1\beta_1-\frac{1}{2}(\alpha_2\beta_3+\alpha_3\beta_2)\\
               \frac{\sqrt3}{2}(\alpha_2\beta_2+\alpha_3\beta_3)\\
             \end{array}
           \right)\\[3mm]
        {\bf3}\sim\left(\begin{array}{c}
                 \alpha_3\beta_3-\alpha_2\beta_2\\
                 \alpha_1\beta_3+\alpha_3\beta_1\\
                 -\alpha_1\beta_2-\alpha_2\beta_1\\
                \end{array}\right)\\[3mm]
        {\bf3}'\sim\left(\begin{array}{c}
                 \alpha_3\beta_2-\alpha_2\beta_3\\
                 \alpha_2\beta_1-\alpha_1\beta_2\\
                 \alpha_1\beta_3-\alpha_3\beta_1\\
            \end{array}\right)
\end{array}\right.
\ea
\eeq
\beq
\ba{l}
{\bf3}\times{\bf3}'={\bf1}'+{\bf2}+{\bf3}+{\bf3}'\quad\;\text{with}\quad\left\{
        \begin{array}{l}
        {\bf1}'\sim\alpha_1\beta_1+\alpha_2\beta_3+\alpha_3\beta_2\\[3mm]
        {\bf2}\sim\left(
             \begin{array}{c}
             \frac{\sqrt3}{2}(\alpha_2\beta_2+\alpha_3\beta_3)\\
             -\alpha_1\beta_1+\frac{1}{2}(\alpha_2\beta_3+\alpha_3\beta_2)\\
             \end{array}
           \right)\\[3mm]
        {\bf3}\sim\left(\begin{array}{c}
                 \alpha_3\beta_2-\alpha_2\beta_3\\
                 \alpha_2\beta_1-\alpha_1\beta_2\\
                 \alpha_1\beta_3-\alpha_3\beta_1\\
            \end{array}\right)\\[3mm]
        {\bf3}'\sim\left(\begin{array}{c}
                 \alpha_3\beta_3-\alpha_2\beta_2\\
                 \alpha_1\beta_3+\alpha_3\beta_1\\
                 -\alpha_1\beta_2-\alpha_2\beta_1\\
            \end{array}\right)
        \end{array}\right.
\end{array}
\eeq

\clearpage{\pagestyle{empty}\cleardoublepage}
\chapter{Vacuum Alignment}
\label{AppendixB}
\lhead[\fancyplain{}{\bfseries\leftmark}]{\fancyplain{}{\bfseries\thepage}}
\rhead[\fancyplain{}{\bfseries\thepage}]{\fancyplain{}{\bfseries\rightmark}}
\renewcommand{\theequation}{B.\;\arabic{equation}}
\setcounter{equation}{0}
\setcounter{footnote}{3}
\setcounter{section}{0}

\mathversion{bold}
\section{The Group $T'$}
\label{AppB:Tp}
\setcounter{footnote}{3}
\mathversion{normal}

Here we discuss particular aspects regarding the vacuum alignment in the $T'$-based model. To this purpose we should complete the definition of the superpotential $w$ by specifying the last term in eq. (\ref{TpTBM:fullw}). This is the term responsible for the spontaneous symmetry breaking of $T'$ and it includes a new set of fields, the ``driving'' fields, which are gauge singlets and transform non-trivially only under the flavour symmetry as shown in table \ref{AppB:table:Tpflavoncharges}. All these fields do not develop VEV and their $F$-terms are the determining equations for the VEVs of the flavons.

\begin{table}[ht!]
\centering
\begin{tabular}{|c||c|c|c|c|c||c|c|c|c|c|}
\hline
&&&&&&&&&&\\[-9pt]
Field & $\varphi_T$ & $\varphi_S$ & $\eta$ & $\xi$,$\tilde{\xi}$ & $\xi''$ & $\varphi^0_T$ & $\varphi^0_S$ & $\eta^0$ & $\xi^0$ & $ \xi^{\prime0}$ \\
&&&&&&&&&&\\[-9pt]
\hline
&&&&&&&&&&\\[-9pt]
$T^{\prime}$ & $\bf3$ & $\bf3$ & $\bf2'$ & $\bf1$ & $\bf1''$ & $\bf3$ & $\bf3$ & $\bf2''$ & $\bf1$ & $\bf1'$ \\[3pt]
$Z_3$ & $1$ & $\omega$ & $1$ & $\omega$ & $1$ &  $1$ & $\omega$ & $1$ & $\omega$ & $1$ \\ [3pt]
$U(1)_R$ & $0$ & $0$ & $0$ & $0$ & $0$ & $2$ & $2$ & $2$ & $2$ & $2$ \\ [3pt]
\hline
\end{tabular}
\caption{\it The transformation rules of flavons and driving fields under the symmetries associated to the groups $T'$, $Z_3$ and $U(1)_{R}$.}
\label{AppB:table:Tpflavoncharges}
\end{table}

Driving fields have $R$-charge $2$ and this prevents them to directly couple to Standard Model fermions. Moreover all the terms in the driving superpotential $w_d$ can only be linear in the driving fields:
\beq
\begin{split}
w_d\,=&\phantom{+} M \,(\varphi^0_T\,\varphi_T)+g\,(\varphi^0_T\,\varphi_T\,\varphi_T)+g_{7}\,\xi''\,(\varphi^0_T\,\varphi_T)^{\prime}+ g_{8}\,(\varphi^0_T\,\eta\,\eta)+\\[3mm]
&+g_{1}\,(\varphi^0_S\,\varphi_S\,\varphi_S)+g_{2}\,\tilde{\xi}\,(\varphi^0_S\,\varphi_S)+\\[3mm]
&+g_{3}\,\xi^0\,(\varphi_S\,\varphi_S)+g_{4}\,\xi^0\,\xi^2 +g_{5}\,\xi^0\,\xi\,\tilde{\xi}+g_{6}\,\xi^0\,\tilde{\xi}^2+\\[3mm]
&+M_{\eta}\,(\eta^0\,\eta)+g_9\,(\varphi_T\,\eta^0\,\eta)+\\[3mm]
&+M_{\xi}\,\xi^{\prime\,0}\,\xi'' +g_{10}\,\xi^{\prime\,0}\,(\varphi_T\,\varphi_T)^{\prime\,\prime}+\ldots
\end{split}
\label{TpTBM:DrivingSuperP}
\eeq

At this level there is no fundamental distinction between the singlets $\xi$ and $\tilde{\xi}$. Thus we are free to define $\tilde{\xi}$ as the combination that couples to $(\varphi^0_S \varphi_S)$ in the superpotential $w_d$. We notice that at the leading order there are no terms involving the Higgs fields $H_{u,d}$. We assume that the electroweak symmetry is broken by some mechanism, such as radiative effects when Supersymmetry is broken. It is interesting that at the leading order the electroweak scale does not mix with the potentially large scales of the VEVs. The scalar potential is given by:
\beq
V=\sum_i\left\vert\frac{\partial w}{\partial \phi_i}\right\vert^2 +m_i^2 \vert \phi_i\vert^2+...
\eeq
where $\phi_i$ denote collectively all the scalar fields of the theory, $m_i^2$ are soft masses and dots stand for $D$-terms for the fields charged under the gauge group and possible additional soft breaking terms. Since $m_i$ are expected to be much smaller than the mass scales involved in $w_d$, it makes sense to minimise $V$ in the supersymmetric limit and to account for soft breaking effects subsequently. Calculating the $F$-terms for the driving fields leads to two sets of equations
\beq
\begin{split}
\dfrac{\partial w}{\partial {\varphi^0_S}_1}\,=&\,\,
g_2\tilde{\xi}{\varphi_S}_1+\dfrac{2g_1}{3}({\varphi_S}_1^2-{\varphi_S}_2{\varphi_S}_3)=0\\[3mm]
\dfrac{\partial w}{\partial {\varphi^0_S}_2}\,=&\,\,g_2\tilde{\xi} {\varphi_S}_3+
\dfrac{2g_1}{3}({\varphi_S}_2^2-{\varphi_S}_1{\varphi_S}_3)=0\\[3mm]
\dfrac{\partial w}{\partial {\varphi^0_S}_3}\,=&\,\,g_2\tilde{\xi} {\varphi_S}_2+
\dfrac{2g_1}{3}({\varphi_S}_3^2-{\varphi_S}_1{\varphi_S}_2)=0\\[3mm]
\dfrac{\partial w}{\partial \xi^0}\,=&\,\,g_4 \xi^2+g_5 \xi \tilde{\xi}+g_6\tilde{\xi}^2 +g_3({\varphi_S}_1^2+2{\varphi_S}_2{\varphi_S}_3)=0
\end{split}
\label{TpTBM:neutrinomin}
\eeq
and
\beq
\begin{split}
\dfrac{\partial\,w}{\partial\,{\varphi^0_T}_1}\,=&\,\,M\,{\varphi_T}_1+\dfrac{2\,g}{3}\,({\varphi^2_T}_1-{\varphi_T}_2\,{\varphi_T}_3)+ g_7\,\xi''\,{\varphi_T}_2+i\,g_8\,\eta_1^2=0\\[3mm]
\dfrac{\partial\,w}{\partial\,{\varphi^0_T}_2}\,=&\,\,M\,{\varphi_T}_3+\dfrac{2\,g}{3}\,({\varphi^2_T}_2-{\varphi_T}_1\,{\varphi_T}_3)+ g_7\,\xi''\,{\varphi_T}_1+(1-i)\,g_8\,\eta_1\,\eta_2=0\\[3mm]
\dfrac{\partial\,w}{\partial\,{\varphi^0_T}_3}\,=&\,\,M\,{\varphi_T}_2+\dfrac{2\,g}{3}\,({\varphi^2_T}_3-{\varphi_T}_1\,{\varphi_T}_2)+ g_7\,\xi''\,{\varphi_T}_3+g_8\,\eta_2^2=0\\[3mm]
\dfrac{\partial\,w}{\partial\,\eta^0_1}\,=&\,\,-M_\eta\,\eta_2+g_9\,((1-i)\,\eta_1\,{\varphi_T}_3-\eta_2\,{\varphi_T}_1)=0\\[3mm]
\dfrac{\partial\,w}{\partial\,\eta^0_2}\,=&\,\,M_\eta\,\eta_1-g_9\,((1+i)\,\eta_2\,{\varphi_T}_2+\eta_1\,{\varphi_T}_1)=0\\[3mm]
\dfrac{\partial\,w}{\partial\,\xi^{\prime\,0}}\,=&\,\,M_\xi\,\xi''+g_{10}\,({\varphi^2_T}_2+2\,{\varphi_T}_1\,{\varphi_T}_3)=0
\end{split}
\label{TpTBM:chrgdmn}
\eeq

Concerning the first set of equations, (\ref{TpTBM:neutrinomin}), there are flat directions in the supersymmetric limit. We can enforce $\langle\tilde{\xi}\rangle=0$ by adding to the scalar potential a soft Supersymmetry breaking mass term
for the scalar field $\tilde{\xi}$, with $m^2_{\tilde{\xi}}>0$. In this case, in a finite portion of the parameter space, we find the solution
\beq
\ba{rcl}
\langle\tilde{\xi}\rangle&=&0\\[3mm]
\langle\xi\rangle&=&v_\xi\\[3mm]
\langle\varphi_S\rangle&=&(v_S,\,v_S,\,v_S)\;,\qquad v_S^2=-\dfrac{g_4}{3 g_3} v_\xi^2\;,
\ea
\label{TpTBM:solS}
\eeq
with $\xi$ undetermined. By choosing $m^2_{\varphi_S}, m^2_\xi<0$, then $\xi$ slides to a large scale, which we assume to be eventually
stabilised by one-loop radiative corrections in the Supersymmetry broken phase. The VEVs in (\ref{TpTBM:solS}) break $T'$ down to the subgroup $G_S$. It is remarkable that other two equivalent VEV configurations are allowed:
\beq
\langle\varphi_S\rangle=v_S(1,\,\omega^2,\,\omega)\;,\qquad\qquad
\langle\varphi_S\rangle=v_S(1,\,\omega,\,\omega^2)\;.
\eeq
These configurations break $T'$ down to different directions of $Z_4$, but they are equivalent to eq. (\ref{TpTBM:solS}), indeed they are obtained by acting on eq. \eqref{TpTBM:solS} with the elements of $T'$. Any of these solutions produces the same neutrino mass matrix: for example $\langle\varphi_S\rangle=v_S(1,\,\omega,\,\omega^2)$ and $\langle\varphi_S\rangle=v_S(1,\,1,\,1)$ are equivalent, being related by the local fields transformation $\nu_e\rightarrow\nu_e$, $\nu_\mu\rightarrow\omega^2\nu_\mu$ and $\nu_\tau\rightarrow\omega\nu_\tau$.

Concerning the last six equations, (\ref{TpTBM:chrgdmn}), by excluding the trivial solutions where all VEVs vanish, in the supersymmetric limit we find three classes of solutions. One class preserves the subgroup $G_S$, as for the set of VEVs given in (\ref{TpTBM:solS}). It is characterised by $\langle\xi''\rangle\ne 0$ and $\langle\eta\rangle=(0,\,0)$.
A representative VEV configuration in this class is:
\beq
\langle \xi'' \rangle= - \dfrac{M}{g_{7}}\;,\qquad
\langle \eta \rangle=(0,0)\;,\qquad
\langle \varphi _T \rangle =(v_T,v_T,v_T)\;,\qquad
v_T^2=\dfrac{M \, M_{\xi}}{3 \, g_{7} \, g_{10}}\;.
\label{TpTBM:cl1}
\eeq
The second class preserves a subgroup $Z_6$ generated by the elements $T$ and $\mathbb{R}$. It is characterised by $\langle\xi''\rangle=0$ and $\langle\eta\rangle=(0,\,0)$:
\beq
\langle \xi'' \rangle =0\;,\qquad
\langle \eta\rangle =(0,0)
\;,\qquad\langle \varphi_T\rangle =(v_T ,0,0)\;,\qquad
v_T=- \dfrac{3M}{2g}\;.
\label{TpTBM:cl21}
\eeq
The third class preserves the subgroup $G_T$. It is characterised by $\langle\xi''\rangle=0$ and $\langle\eta\rangle\ne 0$:
\beq
\ba{c}
\langle \xi'' \rangle =0\;,\qquad
\langle \eta\rangle = \pm(v_1 ,0)
\;,\qquad\langle \varphi_T\rangle =(v_T,0,0)\\[3mm]
\text{with}\qquad
v_1=\dfrac{1}{g_9 \, \sqrt{3 \, g_8}} \, \sqrt{i \, (2 \, M_{\eta} ^{2} \, g + 3 \, M \, M_{\eta} \, g_9)}\;,\qquad
v_T=\dfrac{M_{\eta}}{g_{9}}\;.
\ea
\label{TpTBM:cl22}
\eeq
The three sets of minima in eqs. (\ref{TpTBM:cl1}), (\ref{TpTBM:cl21}) and (\ref{TpTBM:cl22}) are all degenerate in the supersymmetric limit and we will simply choose the one in eq. (\ref{TpTBM:cl22}). We have checked that, by adding soft masses $m_{\xi''}^2>0$, $m_\eta^2<0$, the desired vacuum is selected as the absolute minimum, thus reproducing the results in eqs. (\ref{TpTBM:love1}, \ref{TpTBM:love2}). In summary, we have shown that the VEVs in eqs. (\ref{TpTBM:love1}, \ref{TpTBM:love2}) represent a local minimum of the scalar potential of the theory in a finite portion of the parameter space, without any ad hoc relation among the parameters of the theory. As we will see below, these VEVs will be slightly perturbed by higher-order corrections induced by higher dimensional operators contributing to the driving potential $w_d$.
Such corrections will be important to achieve a realistic mass spectrum in the quark sector. Finally, concerning the numerical values of the VEVs, radiative corrections typically stabilise $v_\xi$ and $v_S$ well below the cutoff scale $\Lambda_f$. Similarly, mass parameters in the superpotential $w_d$ can be chosen in such a way that $v_1$ and $v_T$ are below $\Lambda_f$. It is not unreasonable to assume that all the VEVs are of the same order of magnitude:
\beq
VEV\approx \lambda^2 \Lambda_f\;.
\eeq

For the Froggatt-Nielsen (FN) field $\theta$ to acquire a VEV, we assume that the symmetry $U(1)_{FN}$ is gauged such that $\theta$ gets its VEV through a $D$-term. The corresponding potential is of the form:
\begin{equation}
V_{D, FN}=\dfrac{1}{2}(M_{FI}^2- g_{FN}\vert\theta\vert^2+...)^2
\label{TpTBM:Dterm}
\end{equation}
where $g_{FN}$ is the gauge coupling constant of $U(1)_{FN}$ and $M_{FI}^2$ denotes the contribution of the Fayet-Iliopoulos (FI) term.
Dots in eq. (\ref{TpTBM:Dterm}) represent e.g. terms involving the right-handed charged leptons $e^c$ and $\mu^c$ which are charged under $U(1)_{FN}$. These terms are however not relevant to calculate the VEV of the Froggatt-Nielsen field and we omit them in the present discussion.
$V_{D,FN}$ leads in the supersymmetric limit to:
\begin{equation}
|\langle\theta\rangle|^2= \dfrac{M_{FI}^2}{g_{FN}}
\end{equation}
which we parametrise as:
\begin{equation}
\dfrac{\langle\theta\rangle}{\Lambda_f}=t
\label{TpTBM:vevtheta}
\end{equation}
with $t$ being the second small symmetry breaking parameter in our model.

Before moving to discuss the higher-order terms in the superpotential, we comment on the $F$-terms of the flavons, since they determine the VEVs of the driving fields. In the unbroken Supersymmetry limit, we have
\beq
\ba{rcl}
\dfrac{\partial\,w_d}{\partial\,\varphi_{T1}}\, &=& \,M\,\varphi^0_{T1}\,+
\,\dfrac{2\,g}{3}\,(2\,\varphi^0_{T\,1}\,\varphi_{T1}\,-\,\varphi^0_{T\,2}\,\varphi_{T3}-
\varphi^0_{T\,3}\,\varphi_{T2})+\,g_7\,\varphi^0_{T2}\,\xi''\\[3mm]
&&\,-\,g_9\,(\eta_1^0\,\eta_2+\eta_2^0\,\eta_1)\,+\,2\,g_{10}\,\xi^{\prime\,0}\,\varphi_{T3}=0\\[3mm]

\dfrac{\partial\,w_d}{\partial\,\varphi_{T2}}\, &=& \,M\,\varphi^0_{T3}\,+
\,\dfrac{2\,g}{3}\,(2\,\varphi^0_{T\,2}\,\varphi_{T2}\,-\,\varphi^0_{T\,1}\,\varphi_{T3}-
\varphi^0_{T\,3}\,\varphi_{T1})+\,g_7\,\varphi^0_{T1}\,\xi''\\[3mm]
&&\,-(1+i)\,g_9\,\eta_2^0\,\eta_2\,+\,2\,g_{10}\,\xi^{\prime\,0}\,\varphi_{T2}=0\\[3mm]

\dfrac{\partial\,w_d}{\partial\,\varphi_{T3}}\, &=& \,M\,\varphi^0_{T2}\,+
\,\dfrac{2\,g}{3}\,(2\,\varphi^0_{T\,3}\,\varphi_{T3}\,-\,\varphi^0_{T\,1}\,\varphi_{T2}-
\varphi^0_{T\,2}\,\varphi_{T1})+\,g_7\,\varphi^0_{T3}\,\xi''\\[3mm]
&&\,+(1-i)\,g_9\,\eta_1^0\,\eta_1\,+\,2\,g_{10}\,\xi^{\prime\,0}\,\varphi_{T1}=0
\ea
\eeq

\beq
\hspace{-5mm}
\ba{rcl}
\dfrac{\partial\,w_d}{\partial\,\eta_1}\, &=& \,M_{\eta}\,\eta^0_2\,+\, g_8\,(2\,i\,\varphi^0_{T1}\,\eta_1\,+\, (1\,-\,i)\,\varphi^0_{T2}\,\eta_2)\,+\, g_9\,\left((1\,-\,i)\,\eta_1^0\,\varphi_{T3}\,-\,\eta_2^0\,\varphi_{T1}\right)=0\\[3mm]

\dfrac{\partial\,w_d}{\partial\,\eta_2}\, &=& -\,M_{\eta}\,\eta^0_1\,+\, g_8\,\left((1\,-\,i)\,\varphi^0_{T2}\,\eta_1\,+\, 2\,\varphi^0_{T3}\,\eta_2\right)\,-\, g_9\,\left(\eta_1^0\,\varphi_{T1}\,+(1\,+\,i)\,\eta_2^0\,\varphi_{T2}\right)=0
\ea
\eeq

\beq
\ba{rcl}
\dfrac{\partial\,w_d}{\partial\,\varphi_{S1}}\, &=&
\,\dfrac{2\,g_1}{3}\,(2\,\varphi^0_{S1}\,\varphi_{S1}\,-\,\varphi^0_{S2}\,\varphi_{S3}-
\varphi^0_{S3}\,\varphi_{S2})+\,g_2\,\varphi^0_{S1}\,\xit\,+ \,2\,g_3\,\xi^0\,\varphi_{S1}=0\\[3mm]

\dfrac{\partial\,w_d}{\partial\,\varphi_{S2}}\, &=&
\,\dfrac{2\,g_1}{3}\,(2\,\varphi^0_{S2}\,\varphi_{S2}\,-\,\varphi^0_{S1}\,\varphi_{S3}-
\varphi^0_{S3}\,\varphi_{S1})+\,g_2\,\varphi^0_{S3}\,\xit\,+ \,2\,g_3\,\xi^0\,\varphi_{S3}=0\\[3mm]

\dfrac{\partial\,w_d}{\partial\,\varphi_{S3}}\, &=&
\,\dfrac{2\,g_1}{3}\,(2\,\varphi^0_{S3}\,\varphi_{S3}\,-\,\varphi^0_{S1}\,\varphi_{S2}-
\varphi^0_{S2}\,\varphi_{S1})+\,g_2\,\varphi^0_{S2}\,\xit\,+ \,2\,g_3\,\xi^0\,\varphi_{S2}=0
\ea
\eeq

\beq
\ba{rcl}
\dfrac{\partial\,w_d}{\partial\,\xi} \,&=&\, 2\,g_4\,\xi^0\,\xi\,+\, g_5\,\xi^0\,\xit=0\\[3mm]

\dfrac{\partial\,w_d}{\partial\,\xit} \,&=&\, g_2\,\left(\varphi^0_{S1}\,\varphi_{S1}\,+\, \varphi^0_{S3}\,\varphi_{S2}\,+\, \varphi^0_{S2}\,\varphi_{S3}\right)\,+\, g_5\,\xi^0\,\xi\,+\, 2\,g_6\,\xi^0\,\xit=0\\[3mm]

\dfrac{\partial\,w_d}{\partial\,\xi''} \,&=&\, M _{\xi}\,\xi^{\prime0}\,+\, g_7\,\left(\varphi^0_{T2}\,\varphi_{T1}\,+\, \varphi^0_{T1}\,\varphi_{T2}\,+\, \varphi^0_{T3}\,\varphi_{T3}\right)=0\;.
\ea
\eeq
Notice that we did not consider the $F$-terms involving squarks and sleptons, since they do not develop VEV in order to preserve $S(3)_c$ and $U(1)_{em}$. It is easy to see from the equations above that each expression contain a driving field and as a result the minimum consists in the trivial solution in which all the VEV of the driving fields are vanishing. This result strictly holds in the exact supersymmetric limit. In section \ref{Sec:FlavourViolation} we comment on the fact that, when supersymmetric soft terms are considered, also the driving fields develop a VEV of the order of the soft breaking mass scale.

We now move to discuss the subleading contributions to the superpotential $w_d$, that is modified into
\beq
w_d+\delta w_d\;,
\eeq
where $w_d$ is the leading order contribution already introduced above, in which, for convenience, we have redefined
\beq
g_3\equiv3\,\tilde{g}_3^2\;,\qquad g_4\equiv-\tilde{g}_4^2\qquad\text{and}\qquad g_8\equiv i\,\tilde{g}_8^2\;.
\eeq
The remaining term, $\delta w_d$ is the most general quartic, $T'$-invariant polynomial linear in the driving fields and it is responsible for the corrections to the VEV alignment. We can parametrise the new VEVs of the flavons introducing shifts from the values in eqs. (\ref{TpTBM:solS}, \ref{TpTBM:cl22}):
\beq
\begin{array}{c}
\mean{\varphi_T}=(v_T+\delta v_{T1},\delta v_{T2},\delta v_{T3})\;,\qquad
\mean{\varphi_S}=(v_S+\delta v_{S1},v_S+\delta v_{S2},v_S+\delta v_{S3})\;\\[3mm]
\mean{\xi}=v_\xi\;,\qquad
\mean{\tilde{\xi}}=\delta v_{\tilde{\xi}} \;,\qquad
\mean{\eta}=(v_1+\delta v_1,\delta v_2)\;,\qquad\langle\xi''\rangle=\delta v_{\xi''}\;,
\end{array}
\eeq
where the corrections $\delta v_{Ti}$, $\delta v_{Si}$, $\delta v_{i}$, $\delta v_{\tilde{\xi}}$ and $\delta v_{\xi''}$ are independent of each other. Note that there also might be a correction to the VEV $v_\xi$, but we do not have to indicate this explicitly by the addition of a term $\delta v_\xi$, since $v_\xi$ is undetermined at tree-level anyway. The shifts can be determined studying in detail the term $\delta w_d$:
\beq
\delta w_d=\dd\frac{1}{\Lambda}\left(
\sum_{k=3}^{18} t_k I_k^T+
\sum_{k=1}^{15} s_k I_k^S+
\sum_{k=1}^{4} x_k I_k^X+
\sum_{k=1}^{4} n_k I_k^N+
\sum_{k=1}^{4} y_k I_k^Y
\right)
\eeq
where $t_k$, $s_k$, $x_k$, $n_k$ and $y_k$ are coefficients and $\{I_k^T,I_k^S,I_k^X,I_k^N,I_k^Y\}$ represent a basis of independent quartic invariants:
\beq
\begin{array}{ll}
I_3^T=(\varphi^0_T\varphi_T) (\varphi_T\varphi_T)&\quad
I_{11}^T=(\varphi^0_T\varphi_S) \xi^2\\
I_4^T=(\varphi^0_T\varphi_T)' (\varphi_T\varphi_T)''&\quad
I_{12}^T=(\varphi^0_T\varphi_S) \xi \tilde{\xi}\\
I_5^T=(\varphi^0_T\varphi_T)'' (\varphi_T\varphi_T)'&\quad
I_{13}^T=(\varphi^0_T\varphi_S) {\tilde{\xi}}^2\\
I_6^T=(\varphi^0_T\varphi_S) (\varphi_S\varphi_S)&\quad
I_{14}^T=(\varphi^0_T\varphi_T)''\xi''\xi''\\
I_7^T=(\varphi^0_T\varphi_S)' (\varphi_S\varphi_S)''&\quad
I_{15}^T=((\varphi_T\eta)_2(\varphi^0_T\eta)_2)\\
I_8^T=(\varphi^0_T\varphi_S)'' (\varphi_S\varphi_S)'&\quad
I_{16}^T=((\varphi_T\eta)_{2'}(\varphi^0_T\eta)_{2''})\\
I_9^T=\left(\varphi^0_T(\varphi_S\varphi_S)_S\right) \xi&\quad
I_{17}^T=\left((\eta\xi'')_2(\varphi^0_T\eta)_2\right)\\
I_{10}^T=\left(\varphi^0_T(\varphi_S\varphi_S)_S\right) \tilde{\xi}&\quad
I_{18}^T=(\left(\varphi_T\varphi_T)_S(\varphi^0_T\xi''\right)_3)
\end{array}
\eeq
\beq
\begin{array}{ll}
I_1^S=\left((\varphi^0_S\varphi_T)_S(\varphi_S\varphi_S)_S\right)&\quad
I_9^S=\left(\varphi^0_S(\varphi_T\varphi_S)_A\right) \tilde{\xi}\\
I_2^S=\left((\varphi^0_S\varphi_T)_A(\varphi_S\varphi_S)_S\right)&\quad
I_{10}^S=(\varphi^0_S\varphi_T) \xi^2\\
I_3^S=(\varphi^0_S\varphi_T) (\varphi_S\varphi_S)&\quad
I_{11}^S=(\varphi^0_S\varphi_T) \xi \tilde{\xi}\\
I_4^S=(\varphi^0_S\varphi_T)' (\varphi_S\varphi_S)''&\quad
I_{12}^S=(\varphi^0_S\varphi_T) {\tilde{\xi}}^2\\
I_5^S=(\varphi^0_S\varphi_T)'' (\varphi_S\varphi_S)'&\quad
I_{13}^S=(\varphi^0_S\varphi_S)' \xi \xi''\\
I_6^S=\left(\varphi^0_S(\varphi_T\varphi_S)_S\right) \xi&\quad
I_{14}^S=(\varphi^0_S\varphi_S)' \tilde{\xi} \xi''\\
I_7^S=\left(\varphi^0_S(\varphi_T\varphi_S)_S\right) \tilde{\xi}&\quad
I_{15}^S=\left((\varphi_S\xi'')_3(\varphi^0_S\varphi_S\right)_S)\\
I_8^S=\left(\varphi^0_S(\varphi_T\varphi_S)_A\right) \xi&\quad
\end{array}
\eeq
\beq
\begin{array}{ll}
I_1^X=\xi^0 \left(\varphi_T(\varphi_S\varphi_S)_S\right)&\quad
I_3^X=\xi^0 (\varphi_T\varphi_S) \tilde{\xi}\\
I_2^X=\xi^0 (\varphi_T\varphi_S) \xi&\quad
I_4^X=\xi^0 (\varphi_S\varphi_S)' \xi''
\end{array}
\eeq
\beq
\begin{array}{ll}
I_1^N=(\eta^0\eta)(\varphi_T\varphi_T)&\quad
I_3^N=((\eta\xi'')_2(\eta^0\varphi_T)_2)\\
I_2^N=((\eta\varphi_T)_2(\eta^0\varphi_T)_2)&\quad
I_4^N=\left((\eta^0\eta)_3(\eta\eta)_3\right)
\end{array}
\eeq
\beq
\begin{array}{ll}
I_1^Y=\xi^{\prime\,0}(\varphi_T\varphi_T)\xi''&\quad
I_3^Y=\xi^{\prime\,0}(\varphi_S\varphi_S)''\xi\\
I_2^Y=((\eta\varphi_T)_{2'}(\xi^{\prime\,0}\eta)_{2''})&\quad
I_4^Y=\xi^{\prime\,0}(\varphi_S\varphi_S)''\tilde{\xi}\;.
\end{array}
\eeq
We only take into account terms which are at most linear in $\delta v$ and no terms of the order $\cO(\delta v/\Lambda_f)$. If we plug in the VEVs $v_{T}$ and $v_1$, the equations for the shifts of the VEVs take the following form:
\bea
&&\hspace{-1cm}\begin{split}
&\dfrac{\tilde{g}_4\, v_{\xi}^3}{3\,\tilde{g}_3\,\Lambda_f}\,\left(t_{11}+ \dfrac{\tilde{g}^2_4}{3\,\tilde{g}_3^2}\,\left(t_6+t_7+t_8\right)\,\right)+\dfrac{t_3}{\Lambda_f}\,v_T^3+ (1-i)\,\dfrac{t_{16}}{\Lambda_f}\,v_1^2\,v_T+\\
&\qquad\qquad\qquad
-2\,v_T\,\left(\dfrac{2\,g\,v_T}{3}+M\right)\,\dfrac{\delta v_1}{v_1}+ \left(M+\dfrac{4\,g\,v_T}{3}\right)\,\delta v_{T1}=0
\end{split}\label{AppB:deltavT1}\\[3mm]
&&\hspace{-1cm}\dfrac{\tilde{g}_4\, v_{\xi}^3}{3\,\tilde{g}_3\,\Lambda_f }\,\left(t_{11}+ \dfrac{\tilde{g}^2_4}{3\,\tilde{g}_3^2}\,\left(t_6+t_7+t_8\right)\,\right)+\left(M-\dfrac{2\,g\,v_T}{3}\right)\,\delta v_{T2}=0
\label{AppB:deltavT2}\\[3mm]
&&\hspace{-1cm}\begin{split}
&\dfrac{\tilde{g}_4\, v_{\xi}^3}{3\,\tilde{g}_3\,\Lambda_f }\,\left(t_{11}+ \dfrac{\tilde{g}^2_4}{3\,\tilde{g}_3^2}\,\left(t_6+t_7+t_8\right)\,\right)+g_7\,v_T\,\delta v_{\xi''}+\\
&\qquad\qquad\qquad
+(1+i)\,v_T\,\left(\dfrac{2\,g\,v_T}{3}+M\right)\,\dfrac{\delta v_2}{v_1}+ \left(M-\dfrac{2\,g\,v_T}{3}\right)\,\delta v_{T3}=0
\end{split}\label{AppB:deltavT3}\\[3mm]
&&\hspace{-1cm}\left(\dfrac{9\,\tilde{g}_3\,s_{10}}{\tilde{g}_4}
+\dfrac{3\,\tilde{g}_4\,s_3}{\tilde{g}_3}+ 2\,s_6\right)\,\dfrac{v_T\, v_{\xi}}{\Lambda_f}+ 3\,g_2\,\delta v_{\tilde{\xi}}+ 2\,g_1\,\left(2\,\delta v_{S1}-\delta v_{S2}-\delta v_{S3}\right)=0
\label{AppB:deltavS1}\\[3mm]
&&\hspace{-1cm}\left(\dfrac{3\,\tilde{g}_4\,s_4}{\tilde{g}_3}-s_6- \dfrac{3}{2}\,s_8\right)\,\dfrac{v_T\, v_{\xi}}{\Lambda_f}+ 3\,g_2\delta v_{\tilde{\xi}}+ 2\,g_1\,\left(2\,\delta v_{S2}-\delta v_{S1}-\delta v_{S3}\right)=0
\label{AppB:deltavS2}\\[3mm]
&&\hspace{-1cm}\left(\dfrac{3\,\tilde{g}_4\,s_5}{\tilde{g}_3}-s_6+ \dfrac{3}{2}\,s_8\right)\,\dfrac{v_T\, v_{\xi}}{\Lambda_f}+ 3\,g_2\,\delta v_{\tilde{\xi}}+2\,g_1\,\left(2\,\delta v_{S3}-\delta v_{S1}-\delta v_{S2}\right)=0
\label{AppB:deltavS3}\\[3mm]
&&\hspace{-1cm}\dfrac{x_2\,v_T\, v_{\xi}}{3\,\tilde{g}_3\,\Lambda_f}+\dfrac{g_5}{\tilde{g}_4}\,\delta v_{\tilde{\xi}}+
2\,\tilde{g}_3\,\left(\delta v_{S1}+\delta v_{S2}+\delta v_{S3}\right)=0
\label{AppB:deltau}\\[3mm]
&&\hspace{-1cm}v_T\,\delta v_2-\dfrac{1}{2}(1-i)\,v_1\,\delta v_{T3}=0
\label{AppB:deltav2}\\[3mm]
&&\hspace{-1cm}-\dfrac{1}{2\,\Lambda_f }\,(1+i)\,n_4\,v_1^2+\dfrac{n_1}{\Lambda_f}\,v_T^2+g_9\,\delta v_{T1}=0
\label{AppB:deltav1}\\[3mm]
&&\hspace{-1cm}\dfrac{\tilde{g}_4^2\,y_3\,v_{\xi}^3}{3\,\tilde{g}_3^2\,\Lambda_f}+ M_{\xi}\,\delta v_{\xi''}+ 2\,g_{10}\,v_T\,\delta v_{T3}=0
\label{AppB:deltaupr2}
\eea
The typical order of all the shifts is $VEV^2/\Lambda_f\sim\lambda^4\Lambda_f$, considering that $VEV/\Lambda_f\sim\lambda^2$. As a result the relative size of a shift compared to a non-vanishing VEV is $\lambda^2$. Thereby, it is reasonable that all masses, $M$, $M_\xi$ and $M_\eta$, are of the order of the VEVs, since they are (at least partly) correlated to the VEVs, as one can read off eqs. (\ref{TpTBM:solS}, \ref{TpTBM:cl22}).

\newpage
\mathversion{bold}
\section[The Group $S_4$ -- I Version]{The Group $\mathbf{S_4}$ -- I Version}
\label{AppB:S4}
\setcounter{footnote}{3}
\mathversion{normal}

In the following we present the mechanism to get the particular VEV alignment used in the previous sections. In table \ref{AppendixB:table:flavon_transformation} we illustrate all the flavon and the driving fields of the model. In order to distinguish between the matter fields, the flavons and the driving fields we use the $U(1)_R$, under which the fields have quantum number 1, 0 and 2 respectively.

\begin{table}[h]
\centering
\begin{tabular}{|c||c|c|c|c||c|c|c||c|c|}
  \hline
  &&&&&&&&& \\[-0,3cm]
  & $\Upsilon$ & $\varphi$ & $\Upsilon^0$ & $\varphi^0$ & $\psi$ & $\eta$ & $\psi^0$ & $\xi'$ & $\xi'^0$ \\
  &&&&&&&&& \\[-0,3cm]
  \hline
  &&&&&&&&& \\[-0,3cm]
  $S_4$ & $\bf3$ & $\bf2$ & $\bf3'$ & $\bf2$ & $\bf3$ & $\bf2$ & $\bf3$ & $\bf1'$ & $\bf1'$ \\
  &&&&&&&&& \\[-0,3cm]
  $Z_5$ & $\om^3$ & $\om^3$ & $\om^4$ & $\om^4$ & $\om^2$ & $\om^2$ & $\om$ & 1 & 1 \\
  &&&&&&&&& \\[-0,3cm]
  $U(1)_R$ & $0$ & $0$ & $2$ & $2$ & $0$ & $0$ & $2$ & 0 & 2 \\
  \hline
\end{tabular}
\caption{\it Transformation properties of the flavons and the driving fields.}
\label{AppendixB:table:flavon_transformation}
\end{table}

The driving superpotential is given by
\bac{rcl}
w_d&=&g_1(\Upsilon^0\Upsilon\varphi)+g_2(\varphi^0\Upsilon\Upsilon)+g_3(\varphi^0\varphi\varphi)+\\[0.3cm]
&&+f_1(\psi^0\psi\psi)+f_2(\psi^0\psi\eta)+\\[0.3cm]
&&+M_{\xi'}\xi^{\prime\,0}\xi'+H_1\xi^{\prime\,0}(\eta\varphi)'\;.
\label{S4TBM:eq:wd:driving}
\eac
The equations for the minimum of the scalar potential are obtained deriving $w_d$ by the driving fields:
\beq
\begin{split}
\dfrac{\partial w_d}{\partial\Upsilon^0_1}\;=&\;g_1(\varphi_1\Upsilon_2-\varphi_2\Upsilon_3)=0\\[3mm]
\dfrac{\partial w_d}{\partial\Upsilon^0_2}\;=&\;g_1(\varphi_1\Upsilon_1-\varphi_2\Upsilon_2)=0\\[3mm]
\dfrac{\partial w_d}{\partial\Upsilon^0_3}\;=&\;g_1(\varphi_1\Upsilon_3-\varphi_2\Upsilon_1)=0\;,\\[3mm]
\end{split}
\label{S4TBM:eq:wd:Neutrinos1}
\eeq
\beq
\begin{split}
\dfrac{\partial w_d}{\partial\varphi^0_1}\;=&\;g_2(\Upsilon_3^2+2\Upsilon_1\Upsilon_2)+g_3\varphi_1^2=0\\[3mm]
\dfrac{\partial w_d}{\partial\varphi^0_2}\;=&\;g_2(\Upsilon_2^2+2\Upsilon_1\Upsilon_3)+g_3\varphi_2^2=0\;,\\[3mm]
\end{split}
\label{S4TBM:eq:wd:Neutrinos2}
\eeq
\beq
\begin{split}
\dfrac{\partial w_d}{\partial\psi^0_1}\;=&\;2f_1(\psi_1^2-\psi_2\psi_3)+f_2(\eta_1\psi_2+\eta_2\psi_3)=0\\[3mm]
\dfrac{\partial w_d}{\partial\psi^0_2}\;=&\;2f_1(\psi_2^2-\psi_1\psi_3)+f_2(\eta_1\psi_1+\eta_2\psi_2)=0\\[3mm]
\dfrac{\partial w_d}{\partial\psi^0_3}\;=&\;2f_1(\psi_3^2-\psi_1\psi_2)+f_2(\eta_1\psi_3+\eta_2\psi_1)=0\;,\\[3mm]
\end{split}
\label{S4TBM:eq:wd:ChargedLeptons}
\eeq
\beq
\label{S4TBM:eq:wd:Quarks}
\dfrac{\partial w_d}{\partial\xi^{\prime\,0}}\;=\;M_{\xi'}\xi'+H_1(\eta_1\varphi_2-\eta_2\varphi_1)=0\;.
\eeq

The equations can be divided into almost separated groups. The first five equations in (\ref{S4TBM:eq:wd:Neutrinos1}, \ref{S4TBM:eq:wd:Neutrinos2}) are satisfied by the alignment
\beq
\mean{\Upsilon}=(v_\Upsilon,\,v_\Upsilon,\,v_\Upsilon)\;,\qquad
\mean{\varphi}=(v_\varphi,\,v_\varphi)\;,
\label{S4TBM:vev:neutrinos}
\eeq
which is a stable solution of the scalar potential, with
\beq
v_\Upsilon^2=-\dfrac{g_3}{3g_2}v_\varphi^2\;,\qquad\qquad v_\varphi\;\text{undetermined}\,.
\eeq
The three equations in (\ref{S4TBM:eq:wd:ChargedLeptons}), almost separated from the others, are satisfied by two different patterns: the first is
\beq \mean{\psi}=(0,\,v_\psi,\,0)\;,\qquad
\mean{\eta}=(0,\,v_\eta)
\label{S4TBM:vev:leptons}
\eeq
with
\beq
v_\psi=-\dfrac{f_2}{2f_1}v_\eta\;,\qquad v_\eta\text{ undetermined}
\eeq
and the second is
\beq
\mean{\psi}=(v_\psi,\,v_\psi,\,v_\psi)\;,\qquad
\mean{\eta}=(v_\eta,\,-v_\eta)\;,
\label{S4TBM:vev:leptonsno}
\eeq
with $v_\eta$ and $v_\psi$ undetermined. Only the first solution provides the results presented in the previous sections and we need of some soft masses in order to discriminate it as the lowest minimum of the scalar potential. We manage in doing it, considering some $Z_5$-breaking soft terms involving $\psi$ and $\eta$, which in the most general form can be written as
\beq
m^2_\psi |\psi|^2+ m^2_\eta |\eta|^2 +\tilde{m}^2_\psi \psi \psi  + \tilde{m}^2_\eta \eta \eta\,.
\eeq
Assuming that $m^2_{\psi,\eta} <0$ the first two terms stabilise the potential for both the vacuum configurations. On the other hand the last two terms vanish for the first vacuum configuration and get a value different from zero in the second one.  With a suitable choice of the soft parameters, these contributions can be positive, distinguishing the two configurations of VEVs and assuring that one in eq. (\ref{S4TBM:vev:leptons}) as the setting with the corresponding lowest minimum.

Acting on the configurations of eq. (\ref{S4TBM:vev:neutrinos}) or eq. (\ref{S4TBM:vev:leptons}) with elements of the flavour symmetry group $S_4$, we can generate other minima of the scalar potential. These new minima are physically equivalent to those of the original sets, but it is not restrictive to analyse the model by choosing as local minimum exactly those ones in eqs. (\ref{S4TBM:vev:neutrinos}) and (\ref{S4TBM:vev:leptons}) (it is possible to show that the different scenarios are related by field redefinitions in a similar way as in the Altarelli-Feruglio model of section \ref{Sec:AFTBM} and in the $T'$ based-model of section \ref{Sec:TpTBM}).

The last equation (\ref{S4TBM:eq:wd:Quarks}) connects all the sectors and fixes the VEV of $\xi'$
\beq
\mean{\xi'}=v_{\xi'}=\dfrac{H_1}{M_{\xi'}}v_\eta v_\varphi\;.
\eeq

For the flavon field $\theta$, related to the Froggatt-Nielsen symmetry, the non vanishing VEV is determined by the $D$-term associated with the $U(1)_{FN}$ symmetry: the mechanism works exactly in the same way as in the $T'$ model and we referee to section \ref{AppB:Tp} for the details.

When we consider the higher dimensional operators some corrections are introduced in the VEV alignment. The part of the superpotential depending on the driving fields $\Upsilon^0$, $\varphi^0$, $\psi^0$ and $\xi'^0$ is modified into
\beq
w_d+\delta w_d\;,
\eeq
where $w_d$ is the leading order contribution studied above. The remaining part, $\delta w_d$, is the most general quartic, $S_4$-invariant polynomial linear in the driving fields:
\beq
\delta w_d=\dfrac{1}{\La_f}\left(\sum_{i=1}^5x_iI_i^{\Upsilon^0}+\sum_{i=1}^{6}w_iI_i^{\varphi^0}+\sum_{i=1}^7s_iI_i^{\psi^0}+
\sum_{i=1}^2v_iI_i^{\xi'^0}\right)
\eeq
where $x_i$, $w_i$, $s_i$ and $v_i$ are coefficients and $\left\{I_i^{\Upsilon^0},\;I_i^{\varphi^0},\;I_i^{\psi^0},\;I_i^{\xi'^0}\right\}$ represents a basis of independent quartic invariants,
\bac{ll}
I_1^{\Upsilon^0}=(\Upsilon^0(\Upsilon\varphi)_3)'\xi'\qquad\qquad&
I_4^{\Upsilon^0}=((\Upsilon^0\eta)_3(\psi\psi)_3)\\
I_2^{\Upsilon^0}=(\Upsilon^0(\Upsilon\Upsilon)_3)'\xi'\qquad\qquad&
I_5^{\Upsilon^0}=((\Upsilon^0\psi)_2(\eta\eta)_2)\\
I_3^{\Upsilon^0}=((\Upsilon^0\psi)_2(\psi\psi)_2)&\\
\\
I_1^{\varphi^0}=(\varphi^0(\Upsilon\Upsilon)_2)'\xi'\qquad\qquad&
I_4^{\varphi^0}=(\varphi^0\eta)(\psi\psi)\\
I_2^{\varphi^0}=(\varphi^0(\varphi\varphi)_2)'\xi'\qquad\qquad&
I_5^{\varphi^0}=(\varphi^0\eta)(\eta\eta)\\
I_3^{\varphi^0}=((\varphi^0\eta)_2(\psi\psi)_2)\qquad\qquad&\\
\\
I_1^{\psi^0}=((\psi^0\psi)_2\eta)'\xi'\qquad\qquad&
I_4^{\psi^0}=((\psi^0\varphi)_3(\Upsilon\Upsilon)_3)\\
I_2^{\psi^0}=((\psi^0\Upsilon)_2(\Upsilon\Upsilon)_2)\qquad\qquad&
I_5^{\psi^0}=((\psi^0\Upsilon)_2(\varphi\varphi)_2)\\
I_3^{\psi^0}=(\psi^0\Upsilon)(\Upsilon\Upsilon)\qquad\qquad&
I_6^{\psi^0}=(\psi^0\Upsilon)(\varphi\varphi)\\
\\
I_1^{\xi'^0}=\xi'^0\xi'\xi'\xi'\qquad\qquad&
I_2^{\xi'^0}=\xi'^0\xi'(\varphi\eta)\\
I_2^{\xi'^0}=\xi'^0\xi'(\Upsilon\psi)\qquad\qquad\;.&\\
\eac
The new minimum for $\Upsilon$, $\varphi$, $\psi$, $\eta$ and $\xi'$ is obtained by searching for the zeros of the $F$-terms, the first derivative of $w_d+\delta w_d$, associated to the driving fields $\Upsilon^0$, $\varphi^0$, $\psi^0$ and $\xi'^0$. We look for a solution that perturbs eq. (\ref{S4TBM:vev:neutrinos}, \ref{S4TBM:vev:leptons}) to first order in the $1/\La_f$ expansion: denoting the general flavon field with $\Phi$, we can write the new VEVs as
\beq
\mean{\Phi_i}=\mean{\Phi_i}^{(LO)}+\de\Phi_i\;.
\eeq
The minimum conditions become equations in the unknown $\de\Phi_i$, $v_\varphi$ and $v_\eta$. By keeping only the first order in the expansion, we see that the equations can be separated into different groups: the first five concern only the neutrino sector, the second three only the charged lepton one and the last one connects the two sectors. Finally all the perturbations are non vanishing, apart $\de\eta_1$ and $\de\eta_2$ and one of the perturbations in the neutrino sector, which remain undetermined. On the other hand the NLO terms fixes the relation between $v_\varphi$ and $v_\eta$. We can conclude that the VEV alignment in eq. (\ref{S4TBM:vev:neutrinos}, \ref{S4TBM:vev:leptons}) is stable under the NLO corrections and the deviations are of relative order $u$ with respect to the leading order results.

\mathversion{bold}
\section[The Group $S_4$ -- II Version]{The Group $\mathbf{S_4}$ -- II Version}
\label{AppB:S42}
\setcounter{footnote}{3}
\mathversion{normal}

In this section we show that the superpotential $w_d$ given by
\beq
\begin{split}
w_d\;=&\;M_\varphi\Lambda_f (\varphi_\nu^0\varphi_\nu)+g_1\left(\varphi_\nu^0(\varphi_\nu\varphi_\nu)_3\right)+g_2\left(\varphi_\nu^0\varphi_\nu\right)\xi_\nu+\\[3mm]
&+M_\xi^2\La^2\xi_\nu^0+M'_\xi\Lambda_f\xi_\nu^0\xi_\nu+g_3\xi_\nu^0\xi_\nu\xi_\nu+g_4\xi_\nu^0(\varphi_\nu\varphi_\nu)+\\[3mm]
&+f_1\left(\psi_\ell^0(\varphi_\ell\varphi_\ell)_2\right)+f_2\left(\psi_\ell^0(\chi_\ell\chi_\ell)_2\right) +f_3\left(\psi_\ell^0(\varphi_\ell\chi_\ell)_2\right)+\\[3mm]
&+f_4\left(\chi_\ell^0(\varphi_\ell\chi_\ell)_{3'}\right)+\dots
\end{split}
\label{AFM:wd}\;,
\eeq
has an isolated minimum that corresponds to the VEVs in eqs. (\ref{AFM:vev:charged:best}) and (\ref{AFM:vev:neutrinos}). At the leading order, the equations for the minimum of the potential can be divided into two decoupled parts: one for the neutrino sector and one for the charged lepton sector. Given this separation, for shorthand we can omit the indexes $\ell$ and $\nu$ for flavons and  driving fields. It is easy to see by explicit computation that the driving fields have vanishing VEVs in the limit of exact Supersymmetry. In the neutrino sector, the equations for the vanishing of the derivatives of $w_d$ with respect to each component of the driving fields are:
\beq
\ba{c}
M_\varphi\Lambda_f\varphi_1+g_1(\varphi_3^2-\varphi_2^2)+g_2\xi\varphi_1=0\\[3mm]
M_\varphi\Lambda_f\varphi_3-2g_1\varphi_1\varphi_2+g_2\xi\varphi_3=0\\[3mm]
M_\varphi\Lambda_f\varphi_2+2g_1\varphi_1\varphi_3+g_2\xi\varphi_2=0\\
\\
M_\xi^2\La_f^2+M'_\xi\Lambda_f\xi+g_3\xi^2+g_4(\varphi_1^2+2\varphi_2\varphi_3)=0\;.
\ea
\eeq
A solution to these equations is given in eqs. (\ref{AFM:vev:neutrinos}, \ref{AFM:CD}).
For the charged lepton sector the equations are:
\beq
\ba{c}
f_1(\varphi_1^2-\varphi_2\varphi_3)+f_2(\chi_1^2-\chi_2\chi_3)+\dfrac{\sqrt{3}}{2}f_3(\varphi_2\chi_2+\varphi_3\chi_3)=0\\[3mm]
\dfrac{\sqrt{3}}{2}f_1(\varphi_2^2+\varphi_3^2)+\dfrac{\sqrt{3}}{2}f_2\left(\chi_2^2+\chi_3^2\right)+f_3\left[-\varphi_1\chi_1+ \dfrac{1}{2}\left(\varphi_2\chi_3+\varphi_3\chi_2\right)\right]=0\\
\\
f_4(\varphi_3\chi_3-\varphi_2\chi_2)=0\\[3mm]
f_4(-\varphi_1\chi_2-\varphi_2\chi_1)=0\\[3mm]
f_4(\varphi_1\chi_3+\varphi_3\chi_1)=0\;.
\ea
\label{AFM:Feq:wd:charged}
\eeq
There are two independent solutions of this set of equations. One is given in eqs. (\ref{AFM:vev:charged:best}, \ref{AFM:AB}).
A second solution is given by:
\beq
\dfrac{\mean{\varphi_\ell}}{\La_f}=\left(
                     \begin{array}{c}
                       -\sqrt{2} \\
                       1 \\
                       1 \\
                     \end{array}
                   \right)A\;,\qquad
\qquad
\dfrac{\mean{\chi_\ell}}{\La_f}=\left(
                     \begin{array}{c}
                       \sqrt{2} \\
                       1 \\
                       1 \\
                     \end{array}
                   \right)B\;,
\label{AFM:Fvev:charged:wrong}
\eeq
where the factors $A$ and $B$ should obey to the relation
\beq
f_1A^2+f_2B^2+\sqrt{3}f_3AB=0\;.
\label{AFM:ABbis}
\eeq
At this level we assume that some additional input neglected so far in the analysis, such as for instance some choice of the
soft Supersymmetry breaking parameters, selects the first solution as the the lowest minimum of the scalar potential.

Notice the existence of a flat direction related to an arbitrary, common rescaling of $A$ and $B$: if we indicate with $m^2_{\varphi_\ell}$ and $m^2_{\chi_\ell}$ the soft masses of the two flavons $\varphi_\ell$ and $\chi_\ell$, we can assume $m^2_{\varphi_\ell},m^2_{\chi_\ell}<0$ and then $\mean{\varphi_\ell}$ and $\mean{\chi_\ell}$ slide to a large scale, which we assume to be possibly stabilised by one-loop radiative corrections, fixing in this way $A$ and $B$.

It is important to note that the stability of the alignment in eqs. (\ref{AFM:vev:charged:best}) and (\ref{AFM:vev:neutrinos}), under small perturbations can be proven. If one introduces small parameters in the VEVs of the fields as follows
\beq
\dfrac{\mean{\varphi_\ell}}{\La_f}=\left(
                     \begin{array}{c}
                       x_1 \\
                       1 \\
                       x_2 \\
                     \end{array}
                   \right)A\;,\qquad
\qquad
\dfrac{\mean{\chi_\ell}}{\La_f}=\left(
                     \begin{array}{c}
                       y_1 \\
                       y_2 \\
                       1 \\
                     \end{array}
                   \right)B\;,\qquad
\qquad
\dfrac{\mean{\varphi_\nu}}{\La_f}=\left(
                     \begin{array}{c}
                       z \\
                       1 \\
                       -1 \\
                     \end{array}
                   \right)C\;,
\eeq
it is only a matter of a simple algebra to show that, for small $(z,\,x_1,\,x_2,\,y_1,\,y_2)$, the only solution minimising the scalar potential in the supersymmetric limit is indeed that one with \mbox{$(z,\,x_1,\,x_2,\,y_1,\,y_2)=(0,\,0,\,0,\,0,\,0)$}.

Given the symmetry of the superpotential $w_d$, starting from the field configurations of eqs. (\ref{AFM:vev:neutrinos}, \ref{AFM:CD}),
(or (\ref{AFM:vev:charged:best}, \ref{AFM:AB})) and acting on them with elements of the flavour symmetry group $S_4\times Z_4$, we can generate other minima of the scalar potential. Some of them are
\beq
\mean{\varphi_\nu}\propto\left(
                   \begin{array}{c}
                     1 \\
                     0 \\
                     0 \\
                   \end{array}
                 \right)\;,\qquad\qquad
\mean{\varphi_\nu}\propto\left(
                   \begin{array}{c}
                     0 \\
                     1 \\
                     1 \\
                   \end{array}
                 \right)\;.
\eeq
The new minima are however physically equivalent to those of the original set and it is not restrictive to analyse the model by choosing as local minimum the specific field configuration discussed so far.

We can now consider the higher-order terms and their effects io the flavon VEV alignments. The superpotential term $w_d$, linear in the driving fields $\psi_\ell^0$, $\chi_\ell^0$, $\xi_\nu^0$ and $\varphi_\nu^0$, is modified into:
\beq
w_d+\Delta w_d\;,
\eeq
where $\Delta w_d$ is the NLO contribution, suppressed by one power of $1/\Lambda_f$ with respect to $w_d$.  The corrective term $\Delta w_d$ is given by the most general quartic, $S_4\times Z_4 \times U(1)_{FN}$-invariant polynomial linear in the driving fields, and can be obtained by inserting an additional flavon field in all the leading order terms. The $Z_4$-charges prevent any addition of the flavons $\varphi_\ell$ and $\chi_\ell$ at the NLO, while a factor of $\xi_\nu$ or $\varphi_\nu$ can be added to all the leading order terms. The full expression of $\Delta w_d$ is the following:
\beq
\Delta w_d=\frac{1}{\La_f}\left(\sum_{i=1}^3x_iI_i^{\xi_\nu^0}+\sum_{i=1}^{5}w_iI_i^{\varphi_\nu^0}+\sum_{i=1}^6s_iI_i^{\psi_\ell^0}+
\sum_{i=1}^5v_iI_i^{\chi_\ell^0}\right)
\eeq
where $x_i$, $w_i$, $s_i$ and $v_i$ are coefficients and $\left\{I_i^{\xi_\nu^0},\;I_i^{\varphi_\nu^0},\;I_i^{\psi_\ell^0},\;I_i^{\chi_\ell^0}\right\}$ represent a basis of independent quartic invariants:
\bac{ll}
I_1^{\xi_\nu^0}=\xi_\nu^0\xi_\nu\xi_\nu\xi_\nu\qquad\qquad&
I_3^{\xi_\nu^0}=\xi_\nu^0\xi_\nu(\varphi_\nu\varphi_\nu)\\
I_2^{\xi_\nu^0}=\xi_\nu^0(\varphi_\nu(\varphi_\nu\varphi_\nu)_3)\qquad\qquad&
\eac
\bac{ll}
I_1^{\varphi_\nu^0}=(\varphi_\nu^0\varphi_\nu)(\varphi_\nu\varphi_\nu)\qquad\qquad&
I_4^{\varphi_\nu^0}=\left(\varphi_\nu^0(\varphi_\nu\varphi_\nu)_3\right)\xi_\nu\\
I_2^{\varphi_\nu^0}=\left((\varphi_\nu^0\varphi_\nu)_2(\varphi_\nu\varphi_\nu)_2\right)\qquad\qquad& I_5^{\varphi_\nu^0}=(\varphi_\nu^0\varphi_\nu)\xi_\nu\xi_\nu\\
I_3^{\varphi_\nu^0}=\left((\varphi_\nu^0\varphi_\nu)_3(\varphi_\nu\varphi_\nu)_3\right)\qquad\qquad&
\eac
\bac{ll}
I_1^{\psi_\ell^0}=\left((\psi_\ell^0\varphi_\nu)_3(\varphi_\ell\chi_\ell)_3\right)\qquad\qquad&
I_4^{\psi_\ell^0}=\left(\psi_\ell^0(\varphi_\ell\varphi_\ell)_2\right)\xi_\nu\\
I_2^{\psi_\ell^0}=\left((\psi_\ell^0\varphi_\nu)_{3'}(\varphi_\ell\chi_\ell)_{3'}\right)\qquad\qquad&
I_5^{\psi_\ell^0}=\left(\psi_\ell^0(\chi_\ell\chi_\ell)_2\right)\xi_\nu\\
I_3^{\psi_\ell^0}=\left((\psi_\ell^0\varphi_\nu)_3(\chi_\ell\chi_\ell)_3\right)\qquad\qquad&
I_6^{\psi_\ell^0}=\left(\psi_\ell^0(\varphi_\ell\chi_\ell)_2\right)\xi_\nu\\
\eac
\bac{ll}
I_1^{\chi_\ell^0}=(\chi_\ell^0\varphi_\nu)'(\varphi_\ell\chi_\ell)'\qquad\qquad&
I_4^{\chi_\ell^0}=\left((\chi_\ell^0\varphi_\nu)_{3'}(\varphi_\ell\chi_\ell)_{3'}\right)\\
I_2^{\chi_\ell^0}=\left((\chi_\ell^0\varphi_\nu)_2(\varphi_\ell\chi_\ell)_2\right)\qquad\qquad&
I_5^{\chi_\ell^0}=\left(\chi_\ell^0(\varphi_\ell\chi_\ell)_{3'}\right)\xi_\nu\;.\\
I_3^{\chi_\ell^0}=\left((\chi_\ell^0\varphi_\nu)_3(\varphi_\ell\chi_\ell)_3\right)\qquad\qquad&
\eac

The new VEV configuration is obtained by imposing the vanishing of the first derivative of $w_d+\Delta w_d$ with respect to the driving fields $\xi_\nu^0$, $\varphi_\nu^0$, $\psi_\ell^0$ and $\chi_\ell^0$. We look for a solution that perturbs eqs. (\ref{AFM:vev:charged:best}) and (\ref{AFM:vev:neutrinos}) to first order in the $1/\La_f$ expansion: for all components of the flavons $\Phi=(\xi_\nu,~\varphi_\nu, ~\varphi_\ell, ~\chi_\ell)$, we denote the shifted VEVs by
\beq
\langle \Phi \rangle=\langle \Phi \rangle_{LO}+\delta \Phi
\eeq
where $\langle \Phi \rangle_{LO}$ are given by eqs. (\ref{AFM:vev:charged:best}) and (\ref{AFM:vev:neutrinos}).

After some straightforward algebra the results can be described as follows. In the neutrino sector the shifts $\delta \xi_\nu,~\delta \varphi_\nu$ turn out to be proportional to the leading order VEVs $\langle \Phi \rangle_{LO}$ and can be absorbed in a redefinition of the parameters $C$ and $D$. Instead, in the charged lepton sector, the shifts $\delta \varphi_\ell, ~\delta  \chi_\ell$ have a non trivial structure, so that the leading order texture is modified:
\beq
\mean{\varphi_\ell}=\left(
                     \begin{array}{c}
                       {\delta \varphi_\ell}_1\\
                       A' \La_f \\
                       0\\
                     \end{array}
                   \right)\qquad
\qquad\mean{\chi_\ell}=\left(
                     \begin{array}{c}
                       {\delta \chi_\ell}_1 \\
                       0 \\
                       B' \La_f \\
                     \end{array}
                   \right)
\label{AFM:vev:charged:nlo}
\eeq
where $A'$ and $B'$ satisfy a relation similar to that in eq. (\ref{AFM:AB}) and the shifts ${\delta \varphi_\ell}_1$ and ${\delta \chi_\ell}_1$ are proportional to $v'v\La_f$, that are, in other words, suppressed by a factor $v'$ with respect to the leading order entries $A\La_f$ and $B\La_f$, respectively.

\clearpage{\pagestyle{empty}\cleardoublepage}
\newpage
\chapter{Renormalisation Group Equations}
\label{AppendixC}
\lhead[\fancyplain{}{\bfseries\leftmark}]{\fancyplain{}{\bfseries\thepage}}
\rhead[\fancyplain{}{\bfseries\thepage}]{\fancyplain{}{\bfseries\rightmark}}
\renewcommand{\theequation}{C.\;\arabic{equation}}
\setcounter{equation}{0}
\setcounter{footnote}{3}
\setcounter{section}{0}

In order to calculate the evolution of the fermion mass matrix from the cutoff of the low-energy theory down to the electroweak energy scale, the renormalisation group equations for all the parameters have to be solved simultaneously. We use the notation defined in section \ref{Sec:Running}, where a superscript $(n)$ denotes a quantity between the $n$th and the $(n+1)$th mass threshold. When all the right-handed neutrinos are integrated out, the renormalisation group equations can be recovered by setting the neutrino Yukawa coupling $Y_\nu$ to zero, while in the full theory above the highest See-Saw scale, the superscript $(n)$ has to be omitted.

In the following, $t:=\ln(\mu/\mu_0)$ and $Y_{u(d)}$ is the Yukawa coupling for the up- (down-) quarks, in the GUT normalisation, such that $g_2=g$ and $g_1=\sqrt{5/3} g'$.

In the MSSM context the 1-loop renormalisation group equations for the renormalisation group equations for $\accentset{(n)}{Y_e}$, $\accentset{(n)}{Y}_\nu$, $\accentset{(n)}{M}$, $\accentset{(n)}{\kappa}$, $\accentset{(n)}{Y_d}$, and $\accentset{(n)}{Y_u}$ are given by
\beq
\label{LMP:EqRGEMSSM}
\hspace{-4mm}
\ba{ccl}
16\pi^2 \dfrac{\D}{\D t}\accentset{(n)}{Y_e} & = & Y_e\left\{ 3Y_e^\dagger Y_e +\accentset{(n)}{Y}_\nu ^\dagger  \accentset{(n)}{Y}_\nu + \Tr\Big[3Y_d^\dagger Y_d +Y_e^\dagger Y_e\Big] - \dfrac{9}{5}g_1^2 - 3g_2^2\right\}\;,\\[3mm]
16\pi^2 \dfrac{\D}{\D t}\accentset{(n)}{Y_\nu} &=& \accentset{(n)}{Y}_\nu \left\{ 3 \accentset{(n)}{Y}^\dagger_\nu \accentset{(n)}{Y}_\nu + Y_e^\dagger Y_e + \Tr\Big[3Y_u^\dagger Y_u + \accentset{(n)}{Y}^{\dagger}_\nu \accentset{(n)}{Y}_\nu\Big] - \dfrac{3}{5} g_1^2 - 3 g_2^2 \right\}\;,\\[3mm]
16\pi^2 \dfrac{\D}{\D t} \accentset{(n)}{M}_R &=& \vphantom{\dfrac{1}{2}} 2\Big(\accentset{(n)}{Y}_\nu \accentset{(n)}{Y}^\dagger_\nu\Big) \accentset{(n)}{M}_R + 2\accentset{(n)}{M}_R\Big(\accentset{(n)}{Y}_\nu \accentset{(n)}{Y}^\dagger_\nu\Big)^T\;,\\[3mm]
16\pi^2 \dfrac{\D}{\D t}\accentset{(n)}{\kappa} & = & \Big[\accentset{(n)}{Y}^\dagger_\nu \accentset{(n)}{Y}_\nu + Y_e^\dagger Y_e \Big]^T \accentset{(n)}{\kappa} + \accentset{(n)}{\kappa} \Big[\accentset{(n)}{Y}^\dagger_\nu\accentset{(n)}{Y}_\nu + Y_e^\dagger Y_e \Big] + 2 \Tr\Big[3 Y_u^\dagger Y_u + \accentset{(n)}{Y}^{\dagger}_\nu \accentset{(n)}{Y}_\nu \Big]\accentset{(n)}{\kappa}+\\[3mm]
&&-\dfrac{6}{5} g_1^2 \accentset{(n)}{\kappa}- 6 g_2^2 \accentset{(n)}{\kappa}\;,\\[3mm]
16\pi^2 \dfrac{\D}{\D t}\accentset{(n)}{Y_d}& = & Y_d\left\{ 3Y_d^\dagger Y_d + Y_u^\dagger Y_u + \Tr\Big[3 Y_d^\dagger Y_d + Y_e^\dagger Y_e\Big] - \dfrac{7}{15}g_1^2 - 3g_2^2 - \dfrac{16}{3}g_3^2 \right\}\;,\\[3mm]
16\pi^2 \dfrac{\D}{\D t} \accentset{(n)}{Y_u} & = & Y_u\left\{ Y_d^\dagger Y_d + 3 Y_u^\dagger Y_u + \Tr\Big[3Y_u^\dagger Y_u + \accentset{(n)}{Y}_\nu ^\dagger  \accentset{(n)}{Y}_\nu\Big] - \dfrac{13}{15}g_1^2- 3g_2^2 - \dfrac{16}{3}g_3^2\right\}\;.
\ea
\eeq


In the Standard Model extended by singlet neutrinos, the renormalisation group equations for the same quantities are given by
\beq
\label{LMP:EqRGESM}
\hspace{-4mm}
\ba{ccl}
16\pi^2 \dfrac{\D}{\D t}\accentset{(n)}{Y_e} & = & Y_e \left\{ \dfrac{3}{2} Y_e^\dagger Y_e -\dfrac{3}{2} \accentset{(n)}{Y}_\nu^\dagger \accentset{(n)}{Y}_\nu +\Tr\left[ 3\,Y_u^\dagger Y_u + 3\,Y_d^\dagger Y_d + \accentset{(n)}{Y}_\nu^\dagger \accentset{(n)}{Y}_\nu + Y_e^\dagger Y_e\right] -\dfrac{9}{4} g_1^2 - \dfrac{9}{4} g_2^2\right\}\;,\\[3mm]
16\pi^2 \dfrac{\D}{\D t}\accentset{(n)}{Y_\nu} & = & \accentset{(n)}{Y}_\nu \left\{ \dfrac{3}{2} \accentset{(n)}{Y}_\nu^\dagger \accentset{(n)}{Y}_\nu - \dfrac{3}{2} Y_e^\dagger Y_e +\Tr\left[ 3\,Y_u^\dagger Y_u + 3\,Y_d^\dagger Y_d + \accentset{(n)}{Y}_\nu^\dagger \accentset{(n)}{Y}_\nu + Y_e^\dagger Y_e\right] -\dfrac{9}{20} g_1^2 -\dfrac{9}{4} g_2^2 \right\}\;,\\[3mm]
16\pi^2 \dfrac{\D}{\D t}\accentset{(n)}{M} &=& \Big(\accentset{(n)}{Y}_\nu \accentset{(n)}{Y}^\dagger_\nu\Big) \accentset{(n)}{M} + \accentset{(n)}{M} \Big(\accentset{(n)}{Y}_\nu \accentset{(n)}{Y}^\dagger_\nu\Big)^T \;,\\[3mm]
16\pi^2\dfrac{\D}{\D t}\accentset{(n)}{\kappa} & = & \dfrac{1}{2} \Big[ \accentset{(n)}{Y}^\dagger_\nu \accentset{(n)}{Y}_\nu -3 Y_e^\dagger Y_e \Big]^T \accentset{(n)}{\kappa} +\dfrac{1}{2}\accentset{(n)}{\kappa} \Big[\accentset{(n)}{Y}^\dagger_\nu \accentset{(n)}{Y}_\nu -3 Y_e^\dagger Y_e \Big]+\\[3mm]
&&+2 \Tr\left[ 3\,Y_u^\dagger Y_u + 3\,Y_d^\dagger Y_d + \accentset{(n)}{Y}_\nu^\dagger \accentset{(n)}{Y}_\nu + Y_e^\dagger Y_e\right]
- 3 g_2^2\accentset{(n)}{\kappa} +\lambda_H\accentset{(n)}{\kappa}\;,\\[3mm]
16\pi^2 \dfrac{\D}{\D t}\accentset{(n)}{Y_d} & = & Y_d \left\{ \dfrac{3}{2} Y_d^\dagger Y_d - \dfrac{3}{2} Y_u^\dagger Y_u + \Tr\Big[3\,Y_u^\dagger Y_u + 3\,Y_d^\dagger Y_d + \accentset{(n)}{Y}_\nu^\dagger \accentset{(n)}{Y}_\nu + Y_e^\dagger Y_e \Big]+\right.\\[3mm]
&&\left.\qquad-\dfrac{1}{4} g_1^2 - \dfrac{9}{4} g_2^2 - 8g_3^2\right\}\;,\\[3mm]
16\pi^2 \dfrac{\D}{\D t} \accentset{(n)}{Y_u} & = & Y_u \left\{ \dfrac{3}{2} Y_u^\dagger Y_u - \dfrac{3}{2} Y_d^\dagger Y_d + \Tr\Big[3\,Y_u^\dagger Y_u + 3\,Y_d^\dagger Y_d + \accentset{(n)}{Y}_\nu^\dagger \accentset{(n)}{Y}_\nu + Y_e^\dagger Y_e \Big]+\right.\\[3mm]
&&\left.\qquad-\dfrac{17}{20} g_1^2 - \dfrac{9}{4} g_2^2 - 8g_3^2 \right\}\;,\\[3mm]
16\pi^2\dfrac{\D}{\D t}\accentset{(n)}{\lambda_H} & = & 6\lambda_H^2 -3\lambda_H \left(3g_2^2+\dfrac{3}{5} g_1^2\right) +3 g_2^4 +\dfrac{3}{2}\left(\dfrac{3}{5} g_1^2+g_2^2\right)^2 +\\[3mm]
&&+4\lambda_H \Tr\Big[3\,Y_u^\dagger Y_u + 3\,Y_d^\dagger Y_d + \accentset{(n)}{Y}_\nu^\dagger \accentset{(n)}{Y}_\nu + Y_e^\dagger Y_e \Big]+\\[3mm]
&&-8 \Tr\Big[ 3\,Y_u^\dagger Y_u\,Y_u^\dagger Y_u +  3\,Y_d^\dagger Y_d\,Y_d^\dagger Y_d + \accentset{(n)}{Y}_\nu^\dagger\accentset{(n)}{Y}_\nu \accentset{(n)}{Y}_\nu^\dagger\accentset{(n)}{Y}_\nu + Y_e^\dagger Y_e\,Y_e^\dagger Y_e \Big]\;.
\ea
\eeq
We use the convention that the Higgs self-interaction term in the Lagrangian is $-\lambda_H (H^\dagger H)^2/4$.

\clearpage{\pagestyle{empty}\cleardoublepage}
\newpage
\mathversion{bold}
\chapter{Mass Insertion and $1$-Loop Formulae}
\label{AppendixE}
\mathversion{normal}
\lhead[\fancyplain{}{\bfseries\leftmark}]{\fancyplain{}{\bfseries\thepage}}
\rhead[\fancyplain{}{\bfseries\thepage}]{\fancyplain{}{\bfseries\rightmark}}
\renewcommand{\theequation}{D.\;\arabic{equation}}
\setcounter{equation}{0}
\setcounter{footnote}{3}
\setcounter{section}{0}

\section{Mass Insertion Formulae}
\label{AppE:MI}
\setcounter{footnote}{3}

The ratios $R_{ij}$ can be expressed as:
\beq
R_{ij}= \frac{48\pi^3 \alpha}{G_F^2 m_{SUSY}^4}\left(\vert A_L^{ij} \vert^2+\vert A_R^{ij} \vert^2 \right)\;.
\label{AppE:LFV:rij}
\eeq
At the leading order, the amplitudes $A_L^{ij}$ and $A_R^{ij}$ are given by:
\bea
A_L^{ij}&=&a_{LL} (\delta_{ij})_{LL} + a_{RL} \frac{m_{SUSY}}{m_i} (\delta_{ij})_{RL}\nn\\
A_R^{ij}&=&a_{RR} (\delta_{ij})_{RR} + a_{LR} \frac{m_{SUSY}}{m_i} (\delta_{ij})_{LR}
\label{AppE:LFV:ALAR}
\eea
where $a_{CC'}$ $(C,C'=L,R)$ are dimensionless functions of the ratios $M_{1,2}/m_{SUSY}$, $\mu/m_{SUSY}$ and of $\tan\theta_W$. Their typical size is one tenth of $g^2/(16\pi^2)$, $g$ being the $SU(2)_L$ gauge coupling constant. In our conventions their explicit expression is given by:
\beq
\hspace{-1cm}
\ba{rcl}
a_{LL}&=&\dfrac{g^2}{16\pi^2}\left[ f_{1n}(a_2)+f_{1c}(a_2)+
\dfrac{M_2\mu\tan\beta}{M_2^2-\mu^2}\Big(f_{2n}(a_2,b)+f_{2c}(a_2,b)\Big)\right.\\
&&+\left.\tan^2\theta_W\left(f_{1n}(a_1)- \dfrac{M_1\mu\tan\beta}{M_1^2-\mu^2}f_{2n}(a_1,b)- M_1\left(\left(\dfrac{z_i}{y_i}+\zeta\right) m_{SUSY}-\mu\tan\beta\right)\dfrac{f_{3n}(a_1)}{m_{SUSY}^2}\right)\right]\\[3mm]
a_{RL}&=&\dfrac{g^2}{16\pi^2}\tan^2\theta_W\dfrac{M_1}{m_{SUSY}} 2 f_{2n}(a_1)\\[3mm]
a_{RR}&=&\dfrac{g^2}{16\pi^2}\tan^2\theta_W \left[4 f_{1n}(a_1)+ 2\dfrac{M_1\mu\tan\beta}{M_1^2-\mu^2}f_{2n}(a_1,b)- M_1\left(\left(\dfrac{z_i}{y_i}+\zeta\right) m_{SUSY}-\mu\tan\beta\right)\dfrac{f_{3n}(a_1)}{m_{SUSY}^2}\right]\\[3mm]
a_{LR}&=&\dfrac{g^2}{16\pi^2}\tan^2\theta_W\dfrac{M_1}{m_{SUSY}} 2 f_{2n}(a_1)\;,
\ea
\label{LFV:MIcoefficients}
\eeq
where $a_{1,2}=M^2_{1,2}/m_{SUSY}^2$, $b=\mu^2/m_{SUSY}^2$ and $f_{i(c,n)}(x,y)=f_{i(c,n)}(x)-f_{i(c,n)}(y)$. The functions $f_{in}(x)$ and $f_{ic}(x)$ are given by:
\beq
\ba{rcl}
f_{1n}(x)&=&(-17 x^3+9 x^2+9 x-1+6 x^2(x+3) \log x)/(24(1-x)^5)\\[3mm]
f_{2n}(x)&=&(-5 x^2+4 x+1+2x(x+2)\log x)/(4(1-x)^4)\\[3mm]
f_{3n}(x)&=&(1+9x -9x^2-x^3+6x(x+1) \log x)/(2(1-x)^5)\\[3mm]
f_{1c}(x)&=&(-x^3-9x^2+9x+1+6x(x+1) \log x)/(6(1-x)^5)\\[3mm]
f_{2c}(x)&=&(-x^2-4 x+5+2(2x+1)\log x)/(2(1-x)^4)\;.
\ea
\label{LFV:MIfunctions}
\eeq

\mathversion{bold}
\section{Notations and $1$-Loop Formulae for $R_{ij}$ and $\delta a_\mu^{SUSY}$}
\label{AppE:one-loop}
\setcounter{footnote}{3}
\mathversion{normal}

\newcommand{\ijg}{\ell_i\rightarrow \ell_j\gamma}
\newcommand{\ijnn}{\ell_i\rightarrow \ell_j\nu_i\overline{\nu}_j}
\newcommand{\mtl}{m_{\tilde{\ell}}}
\newcommand{\mtlx}{m_{\tilde{\ell}_X}}
\newcommand{\mtn}{m_{\tilde{\nu}}}
\newcommand{\mtnx}{m_{\tilde{\nu}_X}}
\newcommand{\mtca}{M_{\tilde{\chi}_A}}
\newcommand{\mtcn}{M_{\tilde{\chi}^0}}
\newcommand{\mtcna}{M_{\tilde{\chi}^0_A}}
\newcommand{\mtcca}{M_{\tilde{\chi}^-_A}}

In this part we fix the notations and we report the formulae for $R_{ij}$ and for $\delta a_\mu^{SUSY}$ which we have used in section \ref{Sec:FlavourViolation}. The main references are \cite{deltaamu_susy,Arganda,HisanoFukuyama,g2BR}. The mass matrix of the charginos is given by:

\beq
-\LL_m\supset\left(\overline{\widetilde{W}^-_R}\quad\overline{\widetilde{H}^-_{2R}}\right)M_c\left(
         \begin{array}{c}
           \widetilde{W}^-_L \\
           \widetilde{H}^-_{1L} \\
         \end{array}
       \right)+\hc
\eeq
with
\beq
M_c=
\left(
  \begin{array}{cc}
    M_2 & \sqrt{2}m_W\cos{\beta} \\
    \sqrt{2}m_W\sin{\beta} & \mu \\
  \end{array}
\right)\;.
\eeq
This matrix is diagonalised by $2\times 2$ rotation matrices $O_L$ and $O_R$ as:
\beq
O_R M_c O_L^T=\diag\left(M_{ \widetilde{\chi}^-_1}, M_{ \widetilde{\chi}^-_2}\right)\;,
\eeq
where the diagonalising matrices connect mass and interaction eigenstates in the following way:
\beq
\left(
 \begin{array}{c}
   \widetilde{\chi}^-_{1L} \\
   \widetilde{\chi}^-_{2L} \\
 \end{array}
\right)=O_L\left(
             \begin{array}{c}
               \widetilde{W}^-_L \\
               \widetilde{H}^-_{1L} \\
             \end{array}
            \right)\;,\qquad
\left(
 \begin{array}{c}
   \widetilde{\chi}^-_{1R} \\
   \widetilde{\chi}^-_{2R} \\
 \end{array}
\right)=O_R\left(
             \begin{array}{c}
               \widetilde{W}^-_R \\
               \widetilde{H}^-_{2R} \\
             \end{array}
            \right)
\eeq
and the mass eigenstates are written as $\widetilde{\chi}^-_A= \widetilde{\chi}^-_{AL}+ \widetilde{\chi}^-_{AR}$ ($A=1,2$) with masses $M_{ \widetilde{\chi}^-_A}$.

The neutralino mass matrix is given by:
\beq
-\LL_m\supset\dfrac{1}{2}\left(\begin{array}{cccc}\widetilde{B}_L&\widetilde{W}^0_L&
\widetilde{H}^0_{1L}&\widetilde{H}^0_{2L}\end{array}\right)M_N
\left(
 \begin{array}{c}
   \widetilde{B}_L \\
   \widetilde{W}^0_L \\
   \widetilde{H}^0_{1L} \\
   \widetilde{H}^0_{2L} \\
 \end{array}
\right)+\hc
\eeq
with
\beq
M_N=\left(
  \begin{array}{cccc}
    M_1 & 0 & -m_Z\sin{\theta_W}\cos{\beta} & m_Z\sin{\theta_W}\sin{\beta} \\
    0 & M_2 & m_Z\cos{\theta_W}\cos{\beta} & -m_Z\cos{\theta_W}\sin{\beta} \\
    -m_Z\sin{\theta_W}\cos{\beta} & m_Z\cos{\theta_W}\cos{\beta} & 0 & -\mu \\
    m_Z\sin{\theta_W}\sin{\beta} & -m_Z\cos{\theta_W}\sin{\beta} & -\mu & 0 \\
  \end{array}
\right)
\label{MN}
\eeq
We can diagonalise $M_N$ by a rotation matrix $O_N$:
\beq
O_N M_N O_N^T=\diag\left(M_{ \widetilde{\chi}^0_1}, M_{ \widetilde{\chi}^0_2}, M_{ \widetilde{\chi}^0_3}, M_{ \widetilde{\chi}^0_4}\right)\;,
\eeq
where $O_N$ connects mass and interaction eigenstates in the following way:
\beq
\left(
 \begin{array}{c}
   \widetilde{\chi}^0_{1L} \\
   \widetilde{\chi}^0_{2L} \\
   \widetilde{\chi}^0_{3L} \\
   \widetilde{\chi}^0_{4L} \\
 \end{array}
\right)=O_N\left(
 \begin{array}{c}
   \widetilde{B}_L \\
   \widetilde{W}^0_L \\
   \widetilde{H}^0_{1L} \\
   \widetilde{H}^0_{2L} \\
 \end{array}
\right)
\eeq
and the mass eigenstates are given by $\widetilde{\chi}^0_A= \widetilde{\chi}^0_{AL}+ \widetilde{\chi}^0_{AR}$ ($A=1,2,3,4$) with masses $M_{ \widetilde{\chi}^0_A}$.\\
The mass matrices for the charged sleptons and for sneutrinos are given by:
\beq
-\LL_m\supset\left(\ov{\tilde{\ell}}\quad\tilde{\ell}^c\right)
\hat{\cM}_e^2
\left(
  \begin{array}{c}
    \tilde{\ell} \\
    \ov{\tilde{\ell}^{c}} \\
  \end{array}
\right)+\ov{\tilde{\nu}}\,\hat{m}^2_{\nu LL}\tilde{\nu}
\eeq
with
\beq
\hat{\cM}_e^2=\left(
  \begin{array}{cc}
    \hat{m}^2_{eLL} & \hat{m}_{eLR}^2 \\
    \hat{m}_{eRL}^2 & \hat{m}^2_{e RR} \\
  \end{array}
\right)
\eeq
where $\hat{m}^2_{(e,\nu)LL}$ and $\hat{m}^2_{eRR}$ are in general hermitian matrices and $\hat{m}^2_{eLR}=\left(\hat{m}^2_{RL}\right)^\dag$.
We diagonalise the mass matrix $\hat{\cM}_e^2$ by a $6\times 6$ rotation matrix $U^{\tilde{\ell}}$ as:
\beq
U^{\tilde{\ell}}\,\hat{\cM}_e^2U^{\tilde{\ell}\,T}=\mtl^2
\eeq
where the mass eigenstates are:
\beq
\tilde{\ell}_X=U^{\tilde{\ell}}_{X,i}\tilde{\ell}_i+U^{\tilde{\ell}}_{X,i+3}\bar{\tilde{\ell}}^{c}_i
\eeq
with masses $m^2_{\tilde{\ell}_X}$ ($X=1,\ldots,6$).

\noindent Analogously, the sneutrino mass matrix is diagonalised by:
\beq
U^{\tilde{\nu}} \hat{m}_{\nu LL}^2U^{\tilde{\nu} T}=\mtn^2
\eeq
where the mass eigenstates are:
\beq
\tilde{\nu}_X=U^{\tilde{\nu}}_{X,i}\tilde{\nu}_i
\eeq
with masses $m^2_{\tilde{\nu}_X}$ ($X=1,2,3$).\\
\\
The normalised branching ratios, $R_{ij}$, for the LFV transitions $\ijg$ are:
\beq
R_{ij}=\dfrac{BR(\ijg)}{BR(\ijnn)}=\dfrac{48\pi^3\al}{G_F^2}\left(|A_2^L|^2+|A_2^R|^2\right)
\eeq
and the decay rates are given by:
\beq
\ba{ccl}
\Gamma(\ijnn)&=&\dfrac{G_F^2}{192\pi^3}m_i^5\;,\\[3mm]
\Gamma(\ijg)&=&\dfrac{e^2}{16\pi}m_i^5\left(|A_2^L|^2+|A_2^R|^2\right)\;.
\ea
\eeq
Each coefficient $A_2^{L,R}$ can be written as a sum of two terms:
\beq
A_2^{L,R}=A_2^{(n)L,R}+A_2^{(c)L,R}\;,
\eeq
where $A_2^{(n)L,R}$ and $A_2^{(c)L,R}$ stand for the contributions from the neutralino and from the chargino loops, respectively.
These coefficients are explicitly given by:
\beq
A_2^{(n)L}=\dfrac{1}{32\pi^2}\dfrac{1}{\mtlx^2}\bigg[N_{jAX}^L \bar{N}_{iAX}^{L} g_{1n}(x_{AX})+N_{jAX}^R \bar{N}_{iAX}^{R}\dfrac{m_j}{m_i} g_{1n}(x_{AX})
  +N_{jAX}^L \bar{N}_{iAX}^{R}\dfrac{\mtcna}{m_i} g_{2n}(x_{AX})\bigg]
\eeq
and $A_2^{(n)R}=A_2^{(n)L}\vert_{L\leftrightarrow R}$ with $x_{AX}=\mtcna^2 / \mtlx^2$, and
\beq
A_2^{(c)L}=-\dfrac{1}{32\pi^2}\dfrac{1}{\mtnx^2}\bigg[C_{jAX}^L \bar{C}_{iAX}^{L} g_{1c}(x_{AX})+C_{jAX}^R \bar{C}_{iAX}^{R}\dfrac{m_j}{m_i} g_{1c}(x_{AX})
 +C_{jAX}^L \bar{C}_{iAX}^{R}\dfrac{\mtcca}{m_i} g_{2c}(x_{AX})\bigg]
\eeq
and $A_2^{(c)R}=A_2^{(c)L}\vert_{L\leftrightarrow R}$ with $x_{AX}=\mtcca^2 / \mtnx^2$.\\
The terms $N_{iAX}$ and $C_{iAX}$ and the loop functions $g_{in}$ and $g_{ic}$ read as follows:
\beq
\hspace{-1.9mm}
\ba{rcl}
N_{iAX}^L &=& -\dfrac{g_2}{\sqrt{2}}\left\{\left[-(O_N)_{A,2}-(O_N)_{A,1}\tan{\theta_W}\right]
U^{\tilde{\ell}}_{X,i}+\dfrac{m_i}{m_W\cos{\beta}}(O_N)_{A,3}U^{\tilde{\ell}}_{X,i+3}\right\}\\[3mm]
N_{iAX}^R &=& -\dfrac{g_2}{\sqrt{2}}\left\{\dfrac{m_i}{m_W\cos{\beta}}(O_N)_{A,3}U^{\tilde{\ell}}_{X,i}+ 2(O_N)_{A,1}\tan{\theta_W}U^{\tilde{\ell}}_{X,i+3}\right\}\\[3mm]
C_{iAX}^L &=& -g_2(O_R)_{A,1}U^{\tilde{\nu}}_{X,i}\\[3mm]
C_{iAX}^R &=& g_2\dfrac{m_i}{\sqrt{2}m_W\cos{\beta}}(O_L)_{A,2}U^{\tilde{\nu}}_{X,i}
\ea
\label{AppE:functions:NC}
\eeq
and
\beq
\ba{rcl}
g_{1n}(x_{AX})&=&\dfrac{1}{6(1-x_{AX})^4}\left(1-6x_{AX}+3x_{AX}^2+2x_{AX}^3-6x_{AX}^2\ln{x_{AX}}\right)\\[3mm]
g_{2n}(x_{AX})&=&\dfrac{1}{(1-x_{AX})^3}\left(1-x_{AX}^2+2x_{AX}\ln{x_{AX}}\right)\\[3mm]
g_{1c}(x_{AX})&=&\dfrac{1}{6(1-x_{AX})^4}\left(2+3x_{AX}-6x_{AX}^2+x_{AX}^3+6x_{AX}\ln{x_{AX}}\right)\\[3mm]
g_{2c}(x_{AX})&=&\dfrac{1}{(1-x_{AX})^3}\left(-3+4x_{AX}-x_{AX}^2-2\ln{x_{AX}}\right)\;.
\ea
\label{AppE:functions}
\eeq
We note that the functions $f_{i n}$ and $f_{i c}$ , displayed in section \ref{Sec:LFV:MI_AnaliticResults}, are related to the loop functions $g_{in}$ and  $g_{ic}$, mostly through taking the first derivative.

The deviation of the muon anomalous magnetic moment, $\delta a_\mu^{SUSY}$, due to supersymmetric contributions can be written as:
\beq
\delta a_{\ell_i}^{SUSY}=\dfrac{g_{\ell_i}^{(n)}+g_{\ell_i}^{(c)}}{2}
\eeq
with
\beq
g_{\ell_i}^{(n,c)}=g_{\ell_i}^{(n,c)L}+g_{\ell_i}^{(n,c)R}\;.
\eeq
These terms are explicitly given by:
\beq
g_{\ell_i}^{(c)L}=\dfrac{1}{16\pi^2}\dfrac{m_i^2}{\mtnx^2}\bigg[C_{iAX}^L \bar{C}_{iAX}^{L} g_{1c}(x_{AX})+C_{iAX}^R \bar{C}_{iAX}^{R} g_{1c}(x_{AX})
 +C_{iAX}^L \bar{C}_{iAX}^{R}\dfrac{\mtcca}{m_i} g_{2c}(x_{AX})\bigg]
\eeq
and $g_{\ell_i}^{(c)R}=g_{\ell_i}^{(c)L}\vert_{L\leftrightarrow R}$ with $x_{AX}=\mtcca^2/\mtnx^2$, and
\beq
g_{\ell_i}^{(n)L}=-\dfrac{1}{16\pi^2}\dfrac{m_i^2}{\mtlx^2}\bigg[N_{iAX}^L \bar{N}_{iAX}^{L} g_{1n}(x_{AX})+N_{iAX}^R \bar{N}_{iAX}^{R} g_{1n}(x_{AX})
+N_{iAX}^L \bar{N}_{iAX}^{R}\dfrac{\mtcna}{m_i} g_{2n}(x_{AX})\bigg]
\eeq
and $g_{\ell_i}^{(n)R}=g_{\ell_i}^{(n)L}\vert_{L\leftrightarrow R}$ with $x_{AX}=\mtcna^2 / \mtlx^2$.
The terms $N_{iAX}$ and $C_{iAX}$ and the functions $g_{in}$ and $g_{ic}$
have been already introduced in eqs. (\ref{AppE:functions:NC}) and in eqs. (\ref{AppE:functions}).

\clearpage{\pagestyle{empty}\cleardoublepage}

\clearpage{\pagestyle{empty}\cleardoublepage}
\topmargin 0.5 cm

\newpage
\addtolength{\topmargin}{-2cm}
\pagestyle{plain}
\bibliographystyle{plainnat}

\end{document}